# How Did We Get Here? The Tangled History of the Second Law of Thermodynamics

Stephen Wolfram*

*An extensive survey is given of the historical origins of the Second Law of thermodynamics, illustrated by excerpts from many original sources, and with biographical information about key contributors. The major strands of conceptual development of the Second Law are traced, in particular showing how persistent beliefs about it emerged.*

*This is part 3 in a 3–part series about the Second Law:*
*Computational Foundations for the Second Law of Thermodynamics*
*A 50–Year Quest: My Personal Journey with the Second Law of Thermodynamics*
*How Did We Get Here? The Tangled History of the Second Law of Thermodynamics*

## The Basic Arc of the Story

As I've explained elsewhere, I think I now finally understand the Second Law of thermodynamics. But it's a new understanding, and to get to it I've had to overcome a certain amount of conventional wisdom about the Second Law that I at least have long taken for granted. And to check myself I've been keen to know just where this conventional wisdom came from, how it's been validated, and what might have made it go astray.

And from this I've been led into a rather detailed examination of the origins and history of thermodynamics. All in all, it's a fascinating story, that both explains what's been believed about thermodynamics, and provides some powerful examples of the complicated dynamics of the development and acceptance of ideas.

The basic concept of the Second Law was first formulated in the 1850s, and rather rapidly took on something close to its modern form. It began partly as an empirical law, and partly as something abstractly constructed on the basis of the idea of molecules, that nobody at the time knew for sure existed. But by the end of the 1800s, with the existence of molecules increasingly firmly established, the Second Law began to often be treated as an almost–mathematically–proven necessary law of physics. There were still mathematical loose ends,

A version of this document with immediately executable code is available at
writings.stephenwolfram.com/2023/01/how-did-we-get-here-the-tangled-history-of-the-second-law-of-thermodynamics/
Originally published January 31, 2023
Included in the book: S. Wolfram (2023), *The Second Law: Resolving the Mystery of the Second Law of Thermodynamics*.
*Email: s.wolfram@wolfram.com

as well as issues such as its application to living systems and to systems involving gravity. But the almost-universal conventional wisdom became that the Second Law must always hold, and if it didn't seem to in a particular case, then that must just be because there was something one didn't yet understand about that case.

There was also a sense that regardless of its foundations, the Second Law was successfully used in practice. And indeed particularly in chemistry and engineering it's often been in the background, justifying all the computations routinely done using entropy. But despite its ubiquitous appearance in textbooks, when it comes to foundational questions, there's always been a certain air of mystery around the Second Law. Though after 150 years there's typically an assumption that "somehow it must all have been worked out". I myself have been interested in the Second Law now for a little more than 50 years, and over that time I've had a growing awareness that actually, no, it hasn't all been worked out. Which is why, now, it's wonderful to see the computational paradigm—and ideas from our Physics Project—after all these years be able to provide solid foundations for understanding the Second Law, as well as seeing its limitations.

And from the vantage point of the understanding we now have, we can go back and realize that there were precursors of it even from long ago. In some ways it's all an inspiring tale—of how there were scientists with ideas ahead of their time, blocked only by the lack of a conceptual framework that would take another century to develop. But in other ways it's also a cautionary tale, of how the forces of "conventional wisdom" can blind people to unanswered questions and—over a surprisingly long time—inhibit the development of new ideas.

But, first and foremost, the story of the Second Law is the story of a great intellectual achievement of the mid-19th century. It's exciting now, of course, to be able to use the latest 21st-century ideas to take another step. But to appreciate how this fits in with what's already known we have to go back and study the history of what originally led to the Second Law, and how what emerged as conventional wisdom about it took shape.

## What Is Heat?

Once it became clear what heat is, it actually didn't take long for the Second Law to be formulated. But for centuries—and indeed until the mid-1800s—there was all sorts of confusion about the nature of heat.

That there's a distinction between hot and cold is a matter of basic human perception. And seeing fire one might imagine it as a disembodied form of heat. In ancient Greek times Heraclitus (~500 BC) talked about everything somehow being "made of fire", and also somehow being intrinsically "in motion". Democritus (~460–~370 BC) and the Epicureans had the important idea (that also arose independently in other cultures) that everything might be made of large numbers of a few types of tiny discrete atoms. They imagined these atoms moving around in the "void" of space. And when it came to heat, they seem to have correctly associated it with the motion of atoms—though they imagined it came from particular spherical "fire" atoms that could slide more quickly between other atoms, and they also thought that souls were the ultimate sources of motion and heat (at least in warm-blooded animals?), and were made of fire atoms.

And for two thousand years that's pretty much where things stood. And indeed in 1623 Galileo (1564–1642) (in his book *The Assayer*, about weighing competing world theories) was still saying:

> Those materials which produce heat in us and make us feel warmth, which are known by the general name of "fire," would then be a multitude of minute particles having certain shapes and moving with certain velocities. Meeting with our bodies, they penetrate by means of their extreme subtlety, and their touch as felt by us when they pass through our substance is the sensation we call "heat."

He goes on:

> Since the presence of fire-corpuscles alone does not suffice to excite heat, but their motion is needed also, it seems to me that one may very reasonably say that motion is the cause of heat… But I hold it to be silly to accept that proposition in the ordinary way, as if a stone or piece of iron or a stick must heat up when moved. The rubbing together and friction of two hard bodies, either by resolving their parts into very subtle flying particles or by opening an exit for the tiny fire-corpuscles within, ultimately sets these in motion; and when they meet our bodies and penetrate them, our conscious mind feels those pleasant or unpleasant sensations which we have named heat…

And although he can tell there's something different about it, he thinks of heat as effectively being associated with a substance or material:

> The tenuous material which produces heat is even more subtle than that which causes odor, for the latter cannot leak through a glass container, whereas the material of heat makes its way through any substance.

In 1620, Francis Bacon (1561–1626) (in his "update on Aristotle", *The New Organon*) says, a little more abstractly, if obscurely—and without any reference to atoms or substances:

> [It is not] that heat generates motion or that motion generates heat (though both are true in certain cases), but that heat itself, its essence and quiddity, is motion and nothing else.

But real progress in understanding the nature of heat had to wait for more understanding about the nature of gases, with air being the prime example. (It was actually only in the 1640s that any kind of general notion of gas began to emerge—with the word "gas" being invented by the "anti-Galen" physician Jan Baptista van Helmont (1580–1644), as a Dutch rendering of the Greek word "chaos", that meant essentially "void", or primordial formlessness.) Ever since antiquity there'd been Aristotle-style explanations like "nature abhors a vacuum" about what nature "wants to do". But by the mid-1600s the idea was emerging that there could be more explicit and mechanical explanations for phenomena in the natural world.

And in 1660 Robert Boyle (1627–1691)—now thoroughly committed to the experimental approach to science—published *New Experiments Physico-mechanicall, Touching the Spring of the Air and its Effects* in which he argued that air has an intrinsic pressure associated with it, which pushes it to fill spaces, and for which he effectively found Boyle's Law $PV =$ constant.

But what was air actually made of? Boyle had two basic hypotheses that he explained in rather flowery terms:

(23)

This Notion may perhaps be ſomewhat further explain'd, by conceiving the Air near the Earth to be ſuch a heap of little Bodies, lying one upon another, as may be reſembled to a Fleece of Wooll. For this (to omit other likeneſſes betwixt them) conſiſts of many ſlender and flexible Hairs; each of which, may indeed, like a little Spring, be eaſily bent or rouled up; but will alſo, like a Spring, be ſtill endeavouring to ſtretch it ſelf out again. For though both theſe Haires, and the Aerial Corpuſcles to which we liken them, do eaſily yield to externall preſſures; yet each of them (by vertue of its ſtructure) is endow'd with a Power or Principle of ſelf-Dilatation; by vertue whereof, though the hairs may by a Mans hand be bent and crouded cloſer together, and into a narrower room then ſuits beſt with the nature of the Body: Yet whil'ſt the compreſſion laſts, there is in the fleece they compoſe an endeavour outwards, whereby it continually thruſts againſt the hand that oppoſes its Expanſion. And upon the removall of the external preſſure, by opening the hand more or leſs, the compreſſed Wooll does, as it were, ſpontaneouſly expand or diſplay it ſelf towards

C 4 the

(24)

the recovery of its former more looſe and free condition, till the Fleece have either regain'd its former Dimenſions, or at leaſt, approach'd them as near as the compreſſing hand (perchance not quite open'd) will permit. This Power of ſelf-Dilatation, is ſomewhat more conſpicuous in a dry Spunge compreſſ'd, then in a Fleece of Wooll. But yet we rather choſe to imploy the latter, on this occaſion, becauſe it is not like a Spunge, an entire Body, but a number of ſlender and flexible Bodies, looſely complicated, as the Air it ſelf ſeems to be.

There is yet another way to explicate the Spring of the Air, namely, by ſuppoſing with that moſt ingenious Gentleman, Monſieur *Des Cartes*, That the Air is nothing but a Congeries or heap of ſmall and (for the moſt part) of flexible Particles; of ſeveral ſizes, and of all kinde of Figures which are raiſ'd by heat (eſpecially that of the Sun) into that fluid and ſubtle Etheriall Body that ſurrounds the Earth; and by the reſtleſſe agitation of that Celeſtial Matter wherein thoſe Particles ſwim, are ſo whirl'd round,

(25)

round, that each Corpuſcle endeavours to beat off all others from coming within the little Sphear requiſite to its motion about its own Center; and (in caſe any, by intruding into that Sphear ſhall oppoſe its free Rotation) to expell or drive it away: So that according to this Doctrine, it imports very little, whether the particles of the Air have the ſtructure requiſite to Springs, or be of any other form (how irregular ſoever) ſince their Elaſtical power is not made to depend upon their ſhape or ſtructure, but upon the vehement agitation, and (as it were) brandiſhing motion, which they receive from the fluid *Ether* that ſwiftly flows between them, and whirling about each of them (independently from the reſt) not onely keeps thoſe ſlender Aërial Bodies ſeparated and ſtretcht out (at leaſt, as far as the Neighbouring ones will permit) which otherwiſe, by reaſon of their flexibleneſs and weight, would flag or curl; but alſo makes them hit againſt, and knock away each other, and conſequently require more room, then that which if they were compreſſ'd, they would take up.

By

(26)

By theſe two differing ways, my Lord, may the Spring of the Air be explicated. But though the former of them be that, which by reaſon of its ſeeming ſomewhat more eaſie, I ſhall for the moſt part make uſe of in the following Diſcourſe: yet am I not willing to declare peremptorily for either of them, againſt the other. And indeed, though I have in another Treatiſe endeavoured to make it probable, that the returning of Elaſtical Bodies (if I may ſo call them) forcibly bent, to their former poſition, may be Mechanically explicated: Yet I muſt confeſs, that to determine whether the motion of Reſtitution in Bodies, proceed from this, That the parts of a Body of a peculiar Structure are put into motion by the bending of the ſpring, or from the endeavor of ſome ſubtle ambient Body, whoſe paſſage may be oppoſ'd or obſtructed, or elſe it's preſſure unequally reſiſted by reaſon of the new ſhape or magnitude, which the bending of a Spring may give the Pores of it: To determine this, I ſay, ſeems to me a matter of more difficulty, then at firſt ſight one would eaſily imagine it. Wherefore I ſhall decline medling with a ſubject, which is much more hard to be explica-ted,

His first hypothesis was that air might be like a “fleece of wool” made of “aerial corpuscles” (gases were later often called “aeriform fluids”) with a “power or principle of self-dilatation” that resulted from there being “hairs” or “little springs” between these corpuscles. But he had a second hypothesis too—based, he said, on the ideas of “that most ingenious gentleman, Monsieur Descartes”: that instead air consists of “flexible particles” that are “so whirled around” that “each corpuscle endeavors to beat off all others”. In this second hypothesis, Boyle’s “spring of the air” was effectively the result of particles bouncing off each other.

And, as it happens, in 1668 there was quite an effort to understand the “laws of impact” (that would for example be applicable to balls in games like croquet and billiards, that had existed since at least the 1300s, and were becoming popular), with John Wallis (1616–1703), Christopher Wren (1632–1723) and Christiaan Huygens (1629–1695) all contributing, and Huygens producing diagrams like:

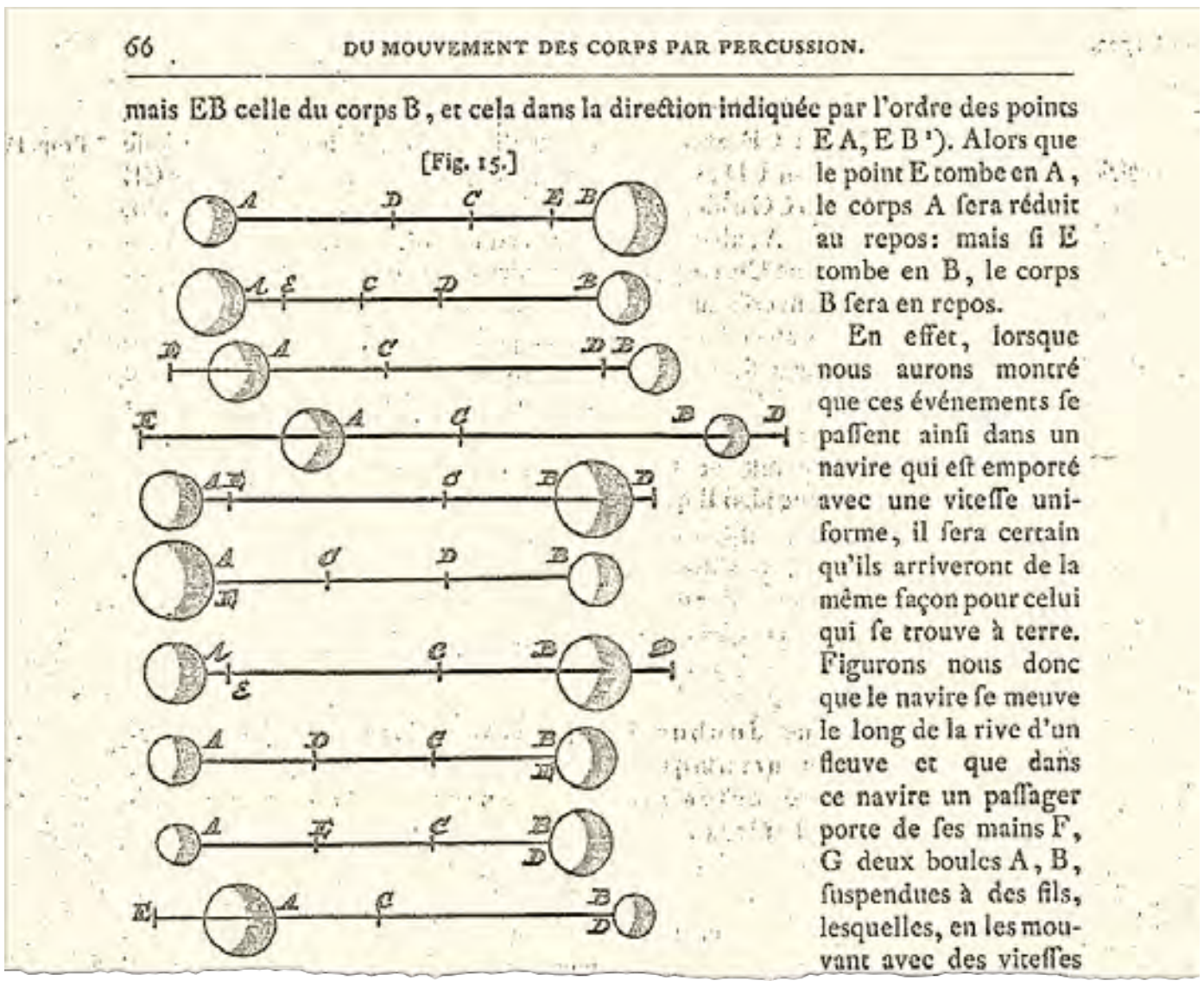
66 DU MOUVEMENT DES CORPS PAR PERCUSSION.

mais EB celle du corps B, et cela dans la direction indiquée par l’ordre des points EA, EB[1]). Alors que le point E tombe en A, le corps A sera réduit au repos: mais si E tombe en B, le corps B sera en repos.

[Fig. 15.]

En effet, lorsque nous aurons montré que ces événements se passent ainsi dans un navire qui est emporté avec une vitesse uniforme, il sera certain qu’ils arriveront de la même façon pour celui qui se trouve à terre. Figurons nous donc que le navire se meuve le long de la rive d’un fleuve et que dans ce navire un passager porte de ses mains F, G deux boules A, B, suspendues à des fils, lesquelles, en les mouvant avec des vitesses

But while some understanding developed of what amount to impacts between pairs of hard spheres, there wasn’t the mathematical methodology—or probably the idea—to apply this to large collections of spheres.

Meanwhile, in his 1687 *Principia Mathematica*, Isaac Newton (1642–1727), wanting to analyze the properties of self-gravitating spheres of fluid, discussed the idea that fluids could in effect be made up of arrays of particles held apart by repulsive forces, as in Boyle’s first hypothesis. Newton had of course had great success with his $1/r^2$ universal attractive force for gravity. But now he noted (writing originally in Latin) that with a $1/r$ repulsive force between particles in a fluid, he could essentially reproduce Boyle’s law:

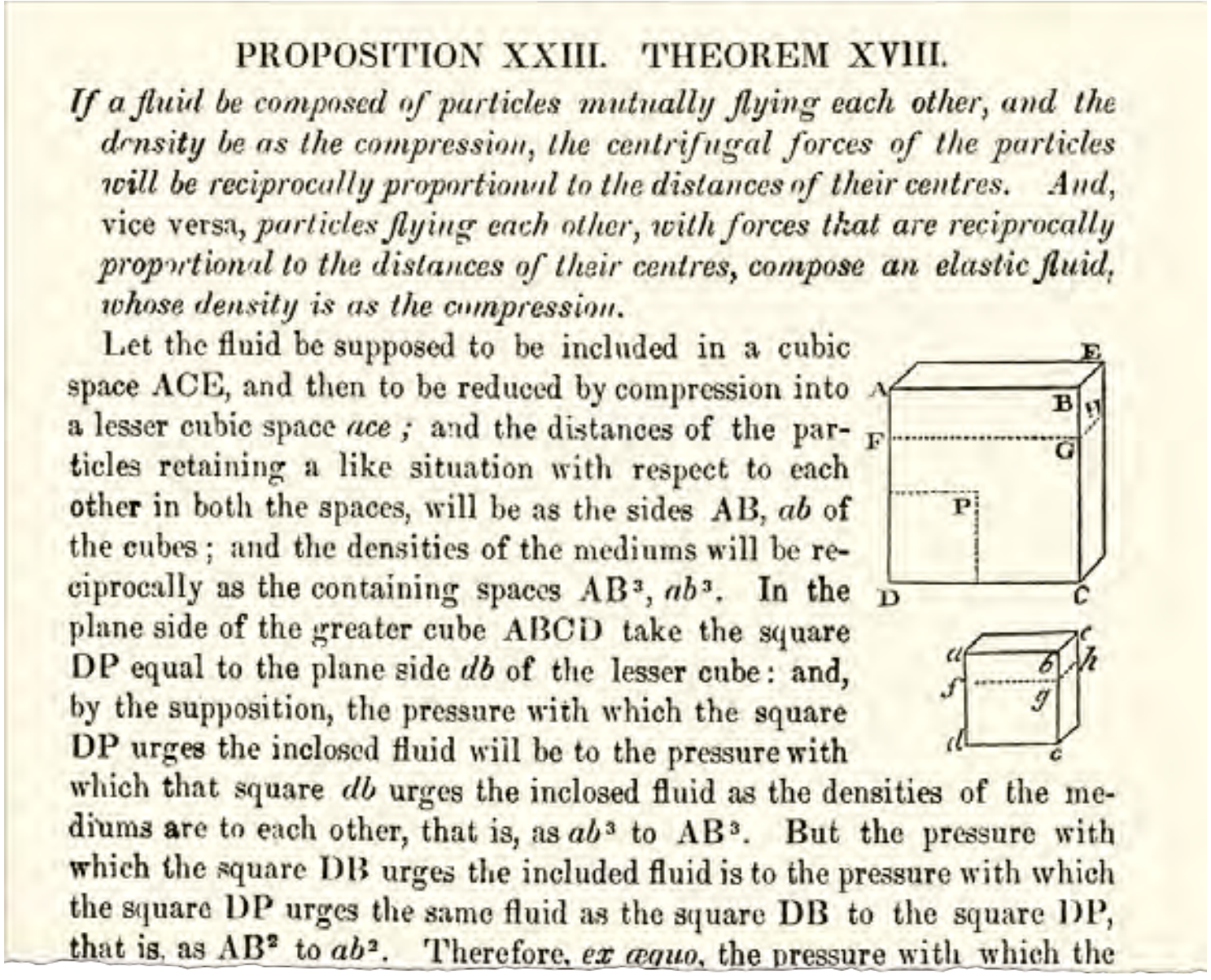

PROPOSITION XXIII. THEOREM XVIII.

*If a fluid be composed of particles mutually flying each other, and the density be as the compression, the centrifugal forces of the particles will be reciprocally proportional to the distances of their centres. And,* vice versa, *particles flying each other, with forces that are reciprocally proportional to the distances of their centres, compose an elastic fluid, whose density is as the compression.*

Let the fluid be supposed to be included in a cubic space ACE, and then to be reduced by compression into a lesser cubic space *ace*; and the distances of the particles retaining a like situation with respect to each other in both the spaces, will be as the sides AB, *ab* of the cubes; and the densities of the mediums will be reciprocally as the containing spaces $AB^3$, $ab^3$. In the plane side of the greater cube ABCD take the square DP equal to the plane side *db* of the lesser cube: and, by the supposition, the pressure with which the square DP urges the inclosed fluid will be to the pressure with which that square *db* urges the inclosed fluid as the densities of the mediums are to each other, that is, as $ab^3$ to $AB^3$. But the pressure with which the square DB urges the included fluid is to the pressure with which the square DP urges the same fluid as the square DB to the square DP, that is, as $AB^2$ to $ab^2$. Therefore, *ex æquo*, the pressure with which the

Newton discussed questions like whether one particle would "shield" others from the force, but then concluded:

> But whether elastic fluids do really consist of particles so repelling each other, is a physical question. We have here demonstrated mathematically the property of fluids consisting of particles of this kind, that hence philosophers may take occasion to discuss that question.

Well, in fact, particularly given Newton's authority, for well over a century people pretty much just assumed that this was how gases worked. There was one major exception, however, in 1738, when—as part of his eclectic mathematical career spanning probability theory, elasticity theory, biostatistics, economics and more—Daniel Bernoulli (1700–1782) published his book on hydrodynamics. Mostly he discusses incompressible fluids and their flow, but in one section he considers "elastic fluids"—and along with a whole variety of experimental results about atmospheric pressure in different places—draws the picture

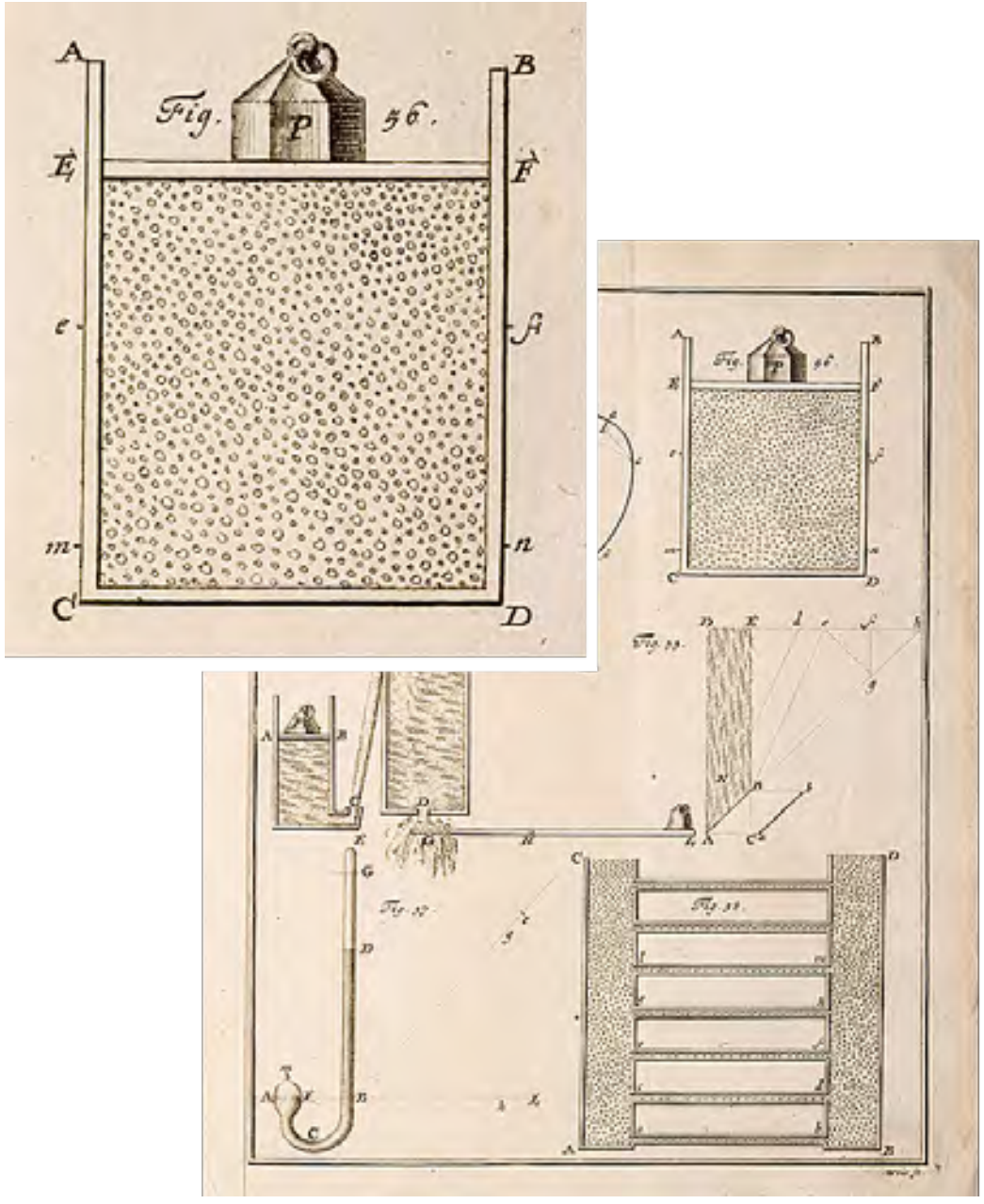

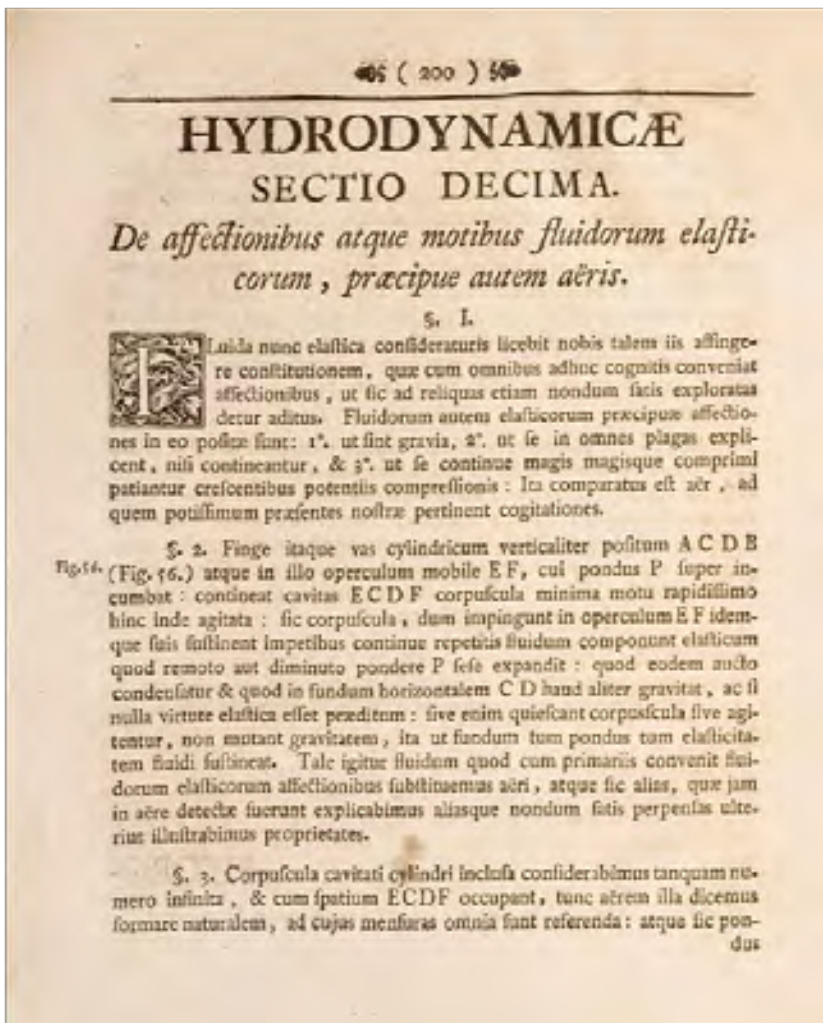
( 200 )

HYDRODYNAMICÆ

SECTIO DECIMA.

*De affectionibus atque motibus fluidorum elasticorum, præcipue autem aëris.*

§. I.

Fluida nunc elastica consideraturis licebit nobis talem iis affingere constitutionem, quæ cum omnibus adhuc cognitis conveniat affectionibus, ut sic ad reliquas etiam nondum satis exploratas detur aditus. Fluidorum autem elasticorum præcipuæ affectiones in eo positæ sunt: 1°. ut sint gravia, 2°. ut se in omnes plagas explicent, nisi contineantur, & 3°. ut se continue magis magisque comprimi patiantur crescentibus potentiis compressionis: Ita comparatus est aër, ad quem potissimum præsentes nostræ pertinent cogitationes.

§. 2. Finge itaque vas cylindricum verticaliter positum A C D B (Fig. 56.) atque in illo operculum mobile E F, cui pondus P super incumbat: contineat cavitas E C D F corpuscula minima motu rapidissimo hinc inde agitata: sic corpuscula, dum impingunt in operculum E F idemque suis sustinent impetibus continue repetitis fluidum componunt elasticum quod remoto aut diminuto pondere P sese expandit: quod eodem aucto condensatur & quod in fundum horizontalem C D haud aliter gravitat, ac si nulla virtute elastica esset præditum: sive enim quiescant corpuscula sive agitentur, non mutant gravitatem, ita ut fundum tum pondus tum elasticitatem fluidi sustineat. Tale igitur fluidum quod cum primariis convenit fluidorum elasticorum affectionibus substituemus aëri, atque sic alias, quæ jam in aëre detectæ fuerant explicabimus aliasque nondum satis perpensas ulterius illustrabimus proprietates.

§. 3. Corpuscula cavitati cylindri inclusa considerabimus tanquam numero infinita, & cum spatium ECDF occupant, tunc aërem illa dicemus formare naturalem, ad cujas mensuras omnia sunt referenda: atque sic pondus

and says

> Let the space ECDF contain very small particles in rapid motion; as they strike against the piston EF and hold it up by their impact, they constitute an elastic fluid which expands as the weight P is removed or reduced; but if P is increased it becomes denser and presses on the horizontal case CD just as if it were endowed with no elastic property.

Then—in a direct and clear anticipation of the kinetic theory of heat—he goes on:

> The pressure of the air is increased not only by reduction in volume but also by rise in temperature. As it is well known that heat is intensified as the internal motion of the particles increases, it follows that any increase in the pressure of air that has not changed its volume indicates more intense motion of its particles, which is in agreement with our hypothesis

But at the time, and in fact for more than a century thereafter, this wasn't followed up.

A large part of the reason seems to have been that people just assumed that heat ultimately had to have some kind of material existence; to think that it was merely a manifestation of microscopic motion was too abstract an idea. And then there was the observation of "radiant heat" (i.e. infrared radiation)—that seemed like it could only work by explicitly transferring some kind of "heat material" from one body to another.

But what was this "heat material"? It was thought of as a fluid—called caloric—that could suffuse matter, and for example flow from a hotter body to a colder. And in an echo of Democritus, it was often assumed that caloric consisted of particles that could slide between ordinary particles of matter. There was some thought that it might be related to the concept of phlogiston from the mid-1600s, that was effectively a chemical substance, for example participating in chemical reactions or being generated in combustion (through the "principle of fire"). But the more mainstream view was that there were caloric particles that would collect around ordinary particles of matter (often called “molecules”, after the use of that term by Descartes (1596–1650) in 1620), generating a repulsive force that would for example expand gases—and that in various circumstances these caloric particles would move around, corresponding to the transfer of heat.

To us today it might seem hacky and implausible (perhaps a little like dark matter, cosmological inflation, etc.), but the caloric theory lasted for more than two hundred years and managed to explain plenty of phenomena—and indeed was certainly going strong in 1825 when Laplace wrote his *A Treatise of Celestial Mechanics*, which included a successful computation of properties of gases like the speed of sound and the ratio of specific heats, on the basis of a somewhat elaborated and mathematicized version of caloric theory (that by then included the concept of “caloric rays” associated with radiant heat).

But even though it wasn’t understood what heat ultimately was, one could still measure its attributes. Already in antiquity there were devices that made use of heat to produce pressure or mechanical motion. And by the beginning of the 1600s—catalyzed by Galileo’s development of the thermoscope (in which heated liquid could be seen to expand up a tube)—the idea quickly caught on of making thermometers, and of quantitatively measuring temperature.

And given a measurement of temperature, one could correlate it with effects one saw. So, for example, in the late 1700s the French balloonist Jacques Charles (1746–1823) noted the linear increase of volume of a gas with temperature. Meanwhile, at the beginning of the 1800s Joseph Fourier (1768–1830) (science advisor to Napoleon) developed what became his 1822 *Analytical Theory of Heat*, and in it he begins by noting that:

> Heat, like gravity, penetrates every substance of the universe, its rays occupy all parts of space. The object of our work is to set forth the mathematical laws which this element obeys. The theory of heat will hereafter form one of the most important branches of general physics.

Later he describes what he calls the “Principle of the Communication of Heat”. He refers to “molecules”—though basically just to indicate a small amount of substance—and says

> When two molecules of the same solid are extremely near and at unequal temperatures, the most heated molecule communicates to that which is less heated a quantity of heat exactly expressed by the product of the duration of the instant, of the extremely small difference of the temperatures, and of certain function of the distance of the molecules.

then goes on to develop what’s now called the heat equation and all sorts of mathematics around it, all the while effectively adopting a caloric theory of heat. (And, yes, if you think of heat as a fluid it does lead you to describe its “motion” in terms of differential equations just like Fourier did. Though it’s then ironic that Bernoulli, even though he studied hydrodynamics, seemed to have a less “fluid-based” view of heat.)

# Heat Engines and the Beginnings of Thermodynamics

At the beginning of the 1800s the Industrial Revolution was in full swing—driven in no small part by the availability of increasingly efficient steam engines. There had been precursors of steam engines even in antiquity, but it was only in 1712 that the first practical steam engine was developed. And after James Watt (1736–1819) produced a much more efficient version in 1776, the adoption of steam engines began to take off.

Over the years that followed there were all sorts of engineering innovations that increased the efficiency of steam engines. But it wasn't clear how far it could go—and whether for example there was a limit to how much mechanical work could ever, even in principle, be derived from a given amount of heat. And it was the investigation of this question—in the hands of a young French engineer named Sadi Carnot (1796–1832)—that began the development of an abstract basic science of thermodynamics, and to the Second Law.

The story really begins with Sadi Carnot's father, Lazare Carnot (1753–1823), who was trained as an engineer but ascended to the highest levels of French politics, and was involved with both the French Revolution and Napoleon. Particularly in years when he was out of political favor, Lazare Carnot worked on mathematics and mathematical engineering. His first significant work—in 1778—was entitled *Memoir on the Theory of Machines*. The mathematical and geometrical science of mechanics was by then fairly well developed; Lazare Carnot's objective was to understand its consequences for actual engineering machines, and to somehow abstract general principles from the mechanical details of the operation of those machines. In 1803 (alongside works on the geometrical theory of fortifications) he published his *Fundamental Principles of [Mechanical] Equilibrium and Movement*, which argued for what was at one time called (in a strange foreshadowing of reversible thermodynamic processes) "Carnot's Principle": that useful work in a machine will be maximized if accelerations and shocks of moving parts are minimized—and that a machine with perpetual motion is impossible.

Sadi Carnot was born in 1796, and was largely educated by his father until he went to college in 1812. It's notable that during the years when Sadi Carnot was a kid, one of his father's activities was to give opinions on a whole range of inventions—including many steam engines and their generalizations. Lazare Carnot died in 1823. Sadi Carnot was by that point a well-educated but professionally undistinguished French military engineer. But in 1824, at the age of 28, he produced his one published work, *Reflections on the Motive Power of Fire, and on Machines to Develop That Power* (where by "fire" he meant what we would call heat):

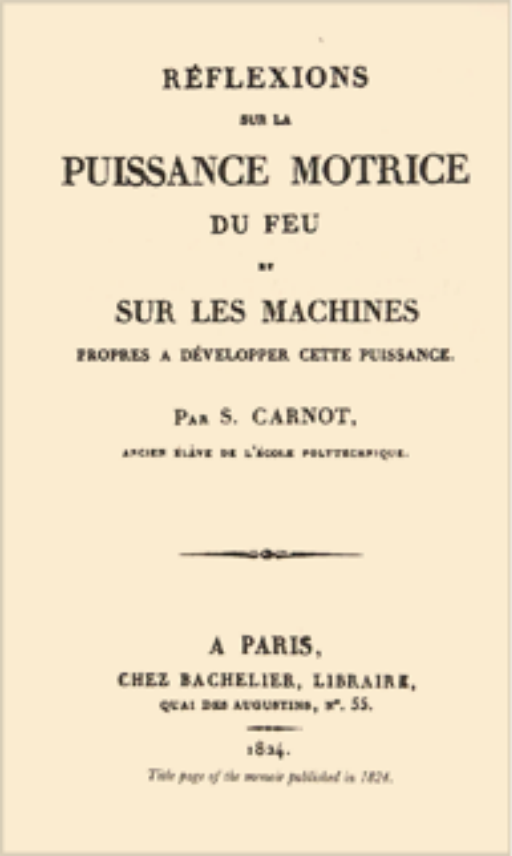
RÉFLEXIONS
SUR LA
PUISSANCE MOTRICE
DU FEU
ET
SUR LES MACHINES
PROPRES A DÉVELOPPER CETTE PUISSANCE.

PAR S. CARNOT,
ANCIEN ÉLÈVE DE L'ÉCOLE POLYTECHNIQUE.

A PARIS,
CHEZ BACHELIER, LIBRAIRE,
QUAI DES AUGUSTINS, Nº. 55.
1824.

*Title page of the memoir published in 1824.*

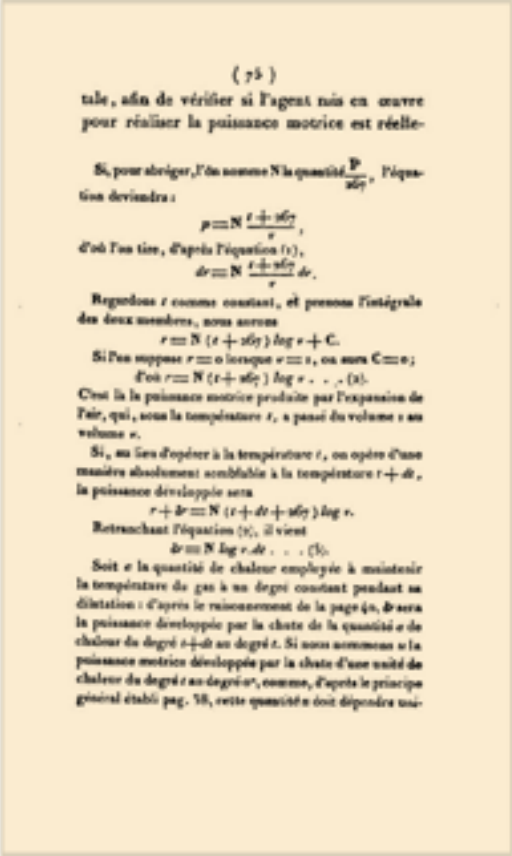

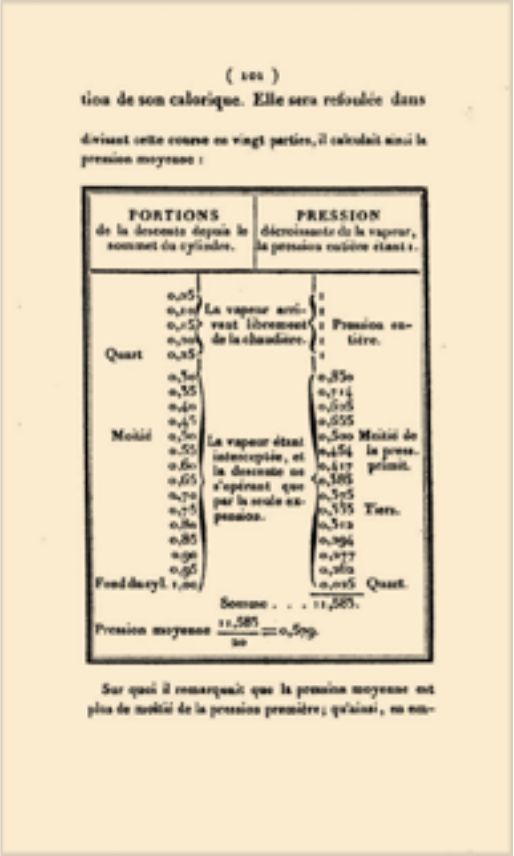

The style and approach of the younger Carnot's work is quite similar to his father's. But the subject matter turned out to be more fruitful. The book begins:

> Everyone knows that heat can produce motion. That it possesses vast motive-power none can doubt, in these days when the steam-engine is everywhere so well known… The study of these engines is of the greatest interest, their importance is enormous, their use is continually increasing, and they seem destined to produce a great revolution in the civilized world. Already the steam-engine works our mines, impels our ships, excavates our ports and our rivers, forges iron, fashions wood, grinds grain, spins and weaves our cloths, transports the heaviest burdens, etc.It appears that it must some day serve as a universal motor, and be substituted for animal power, water-falls, and air currents. …
>
> Notwithstanding the work of all kinds done by steam-engines, notwithstanding the satisfactory condition to which they have been brought to-day, their theory is very little understood, and the attempts to improve them are still directed almost by chance. …
>
> The question has often been raised whether the motive power of heat is unbounded, whether the possible improvements in steam–engines have an assignable limit, a limit which the nature of things will not allow to be passed by any means whatever; or whether, on the contrary, these improvements may be carried on indefinitely. We propose now to submit these questions to a deliberate examination.

Carnot operated very much within the framework of caloric theory, and indeed his ideas were crucially based on the concept that one could think about "heat itself" (which for him was caloric fluid), independent of the material substance (like steam) that was hot. But—like his father's efforts with mechanical machines—his goal was to develop an abstract "metamodel" of something like a steam engine, crucially assuming that the generation of unbounded heat or mechanical work (i.e. perpetual motion) in the closed cycle of the operation of the machine was impossible, and noting (again with a reflection of his father's work) that the system would necessarily maximize efficiency if it operated reversibly. And he then argued that:

> The production of motive power is then due in steam-engines not to an actual consumption of caloric, but to its transportation from a warm body to a cold body, that is, to its re–establishment of equilibrium…

In other words, what was important about a steam engine was that it was a "heat engine", that "moved heat around". His book is mostly words, with just a few formulas related to the behavior of ideal gases, and some tables of actual parameters for particular materials. But even though his underlying conceptual framework—of caloric theory—was not correct, the abstract arguments that he made (that involved essentially logical consequences of reversibility and of operating in a closed cycle) were robust enough that it didn't matter, and in particular he was able to successfully show that there was a theoretical maximum efficiency for a heat engine, that depended only on the temperatures of its hot and cold reservoirs of heat. But what's important for our purposes here is that in the setup Carnot constructed he basically ended up introducing the Second Law.

At the time it appeared, however, Carnot's book was basically ignored, and Carnot died in obscurity from cholera in 1832 (about 9 months after Évariste Galois (1811–1832)) at the age of 36. (The Sadi Carnot who would later become president of France was his nephew.) But in 1834, Émile Clapeyron (1799–1864)—a rather distinguished French engineering professor (and steam engine designer)—wrote a paper entitled "Memoir on the Motive Power of Heat". He starts off by saying about Carnot's book:

> The idea which serves as a basis of his researches seems to me to be both fertile and beyond question; his demonstrations are founded on the absurdity of the possibility of creating motive power or heat out of nothing. ...
>
> This new method of demonstration seems to me worthy of the attention of theoreticians; it seems to me to be free of all objection ...
>
> I believe that it is of some interest to take up this theory again; S. Carnot, avoiding the use of mathematical analysis, arrives by a chain of difficult and elusive arguments at results which can be deduced easily from a more general law which I shall attempt to prove...

Clapeyron's paper doesn't live up to the claims of originality or rigor expressed here, but it served as a more accessible (both in terms of where it was published and how it was written) exposition of Carnot's work, featuring, for example, for the first time a diagrammatic representation of a Carnot cycle

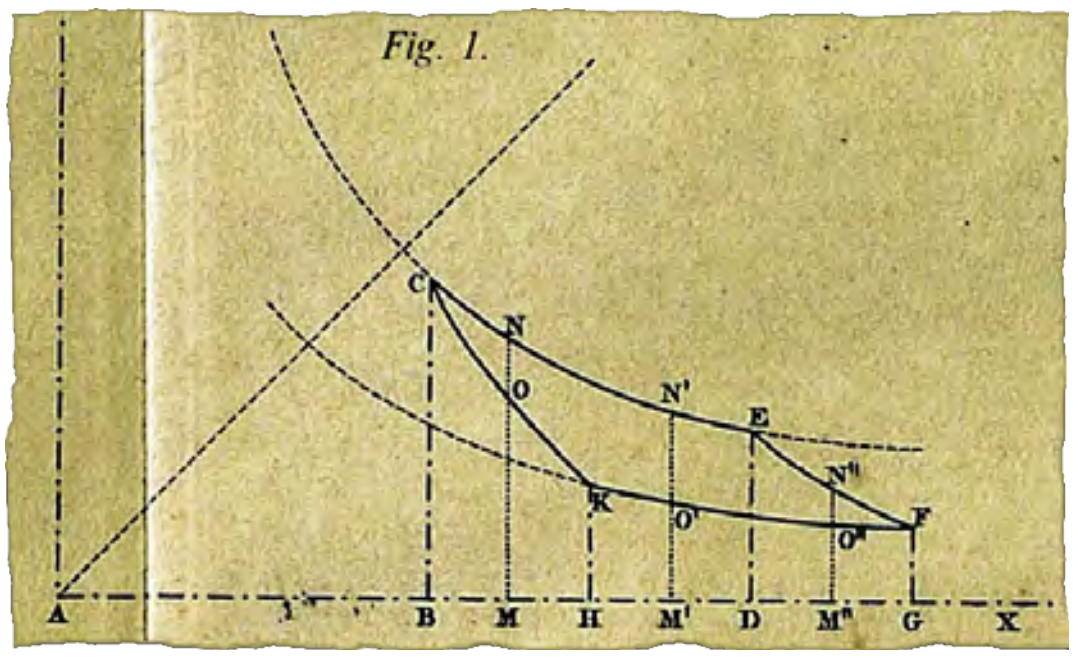

as well as notations like $Q$–for–heat that are still in use today:

Pendant le contact avec la source A, la température est restée constante; d'où il suit que les variations $dp$ et $dv$ de la pression et du volume sont liées par la relation

$$\frac{dT}{dp}\,dp + \frac{dT}{dv}\,dv = 0.$$

*XXIII*[e] *Cahier.* 23

Ces variations $dp$ et $dv$ en occasionent une dans la quantité absolue de chaleur Q, qui a pour expression

$$dQ = \frac{dQ}{dp}\,dp + \frac{dQ}{dv}\,dv = dv\left[\frac{dQ}{dv} - \frac{dQ}{dp}\,\frac{\left(\frac{dT}{dv}\right)}{\left(\frac{dT}{dp}\right)}\right];$$

telle est la quantité de chaleur consommée pour produire l'effet que

## The Second Law Is Formulated

One of the implications of Newton's Laws of Motion is that momentum is conserved. But what else might also be conserved? In the 1680s Gottfried Leibniz (1646–1716) suggested the quantity $m\,v^2$, which he called, rather grandly, *vis viva*—or, in English, "life force". And yes, in things like elastic collisions, this quantity did seem to be conserved. But in plenty of situations it wasn't. By 1807 the term "energy" had been introduced, but the question remained of whether it could in any sense globally be thought of as conserved.

It had seemed for a long time that heat was something a bit like mechanical energy, but the relation wasn't clear—and the caloric theory of heat implied that caloric (i.e. the fluid corresponding to heat) was conserved, and so certainly wasn't something that for example could be interconverted with mechanical energy. But in 1798 Benjamin Thompson (Count Rumford) (1753–1814) measured the heat produced by the mechanical process of boring a cannon, and began to make the argument that, in contradiction to the caloric theory, there was actually some kind of correspondence between mechanical energy and amount of heat.

It wasn't a very accurate experiment, and it took until the 1840s—with new experiments by the English brewer and "amateur" scientist James Joule (1818–1889) and the German physician Robert Mayer (1814–1878)—before the idea of some kind of equivalence between heat and mechanical work began to look more plausible. And in 1847 this was something William Thomson (1824–1907) (later Lord Kelvin)—a prolific young physicist recently graduated from the Mathematical Tripos in Cambridge and now installed as a professor of "natural philosophy" (i.e. physics) in Glasgow—began to be curious about.

But first we have to go back a bit in the story. In 1845 Kelvin (as we'll call him) had spent some time in Paris (primarily at a lab that was measuring properties of steam for the French government), and there he'd learned about Carnot's work from Clapeyron's paper (at first he couldn't get a copy of Carnot's actual book). Meanwhile, one of the issues of the time was a proliferation of different temperature scales based on using different kinds of thermometers based on different substances. And in 1848 Kelvin realized that Carnot's concept of a "pure heat engine"—assumed at the time to be based on caloric—could be used to define an "absolute" scale of temperature in which, for example, at absolute zero all caloric would have been removed from all substances:

> On an Absolute Thermometric Scale founded on Carnot's Theory of the Motive Power of Heat*, and calculated from Regnault's observations†. By Prof. W. Thomson, Fellow of St. Peter's College.
>
> The determination of temperature has long been recognized as a problem of the greatest importance in physical science. It has accordingly been made a subject of most careful attention, and, especially in late years, of very elaborate and refined experimental researches‡; and we are thus at present in possession of as complete a practical solution of the problem as can be desired, even for the most accurate investigations. The theory of thermometry is however as yet far from being in so satisfactory a state. The principle to be followed in constructing a thermometric scale might at first sight seem to be obvious, as it might appear that a perfect thermometer would indicate equal additions of heat, as corresponding to equal elevations of temperature, estimated by the numbered divisions of its scale. It is however now recognized (from the variations in the specific heats of bodies) as an experimentally demonstrated fact that
>
> * Published in 1824 in a work entitled *Réflexions sur la Puissance Motrice du Feu*, by M. S. Carnot. Having never met with the original work, it is only through a paper by M. Clapeyron, on the same subject, published in the *Journal de l'Ecole Polytechnique*, vol. xiv. 1834, and translated in the first volume of Taylor's Scientific Memoirs, that the author has become acquainted with Carnot's theory.—W. T.

Having found Carnot's ideas useful, Kelvin in 1849 wrote a 33-page summary of them (small world that it was then, the immediately preceding paper in the journal is "On the Theory of Rolling Curves", written by the then-17-year-old James Clerk Maxwell (1831–1879), while the one that follows is "Theoretical Considerations on the Effect of Pressure in Lowering the Freezing Point of Water" by James Thomson (1786–1849), engineering-oriented older brother of William):

> XXXVI.—*An Account of* CARNOT'S *Theory of the Motive Power of Heat;* with Numerical Results deduced from* REGNAULT'S *Experiments on Steam.*† By WILLIAM THOMSON, Professor of Natural Philosophy in the University of Glasgow.
>
> (Read January 2, 1849.)
>
> 1. The presence of heat may be recognised in every natural object; and there is scarcely an operation in nature which is not more or less affected by its all-pervading influence. An evolution and subsequent absorption of heat generally give rise to a variety of effects; among which may be enumerated, chemical combinations or decompositions; the fusion of solid substances; the vaporisation

He characterizes Carnot's work as being based not so much on physics and experiment, but on the "strictest principles of philosophy":

> medium of certain simple operations, may be clearly appreciated. Thus it is that CARNOT, in accordance with the strictest principles of philosophy, enters upon the investigation of the theory of the motive power of heat.

He doesn't immediately mention "caloric" (though it does slip in later), referring instead to a vaguer concept of "thermal agency":

> (2.) How may the amount of this thermal agency necessary for performing a given quantity of work be estimated?
>
> 3. In the following paper I shall commence by giving a short abstract of the reasoning by which CARNOT is led to an answer to the first of these questions; I

In keeping with the idea that this is more philosophy than experimental science, he refers to "Carnot's fundamental principle"—that after a complete cycle an engine can be treated as back in the "same state"—while adding the footnote that "this is tacitly assumed as an axiom":

> the sides of the boiler, and that heat is continually abstracted by the water employed for keeping the condenser cool. According to CARNOT's fundamental principle, the quantity of heat thus discharged, during a complete revolution (or double stroke) of the engine must be precisely equal to that which enters the water of the boiler;* provided the total mass of water and steam be invariable, and be restored to its primitive physical condition (which will be the case rigorously, if the condenser be kept cool by the external application of cold water, instead of by in-

> * So generally is CARNOT's principle tacitly admitted as an axiom, that its application in this case has never, so far as I am aware, been questioned by practical engineers.

In actuality, to say that an engine comes back to the same state is a nontrivial statement of the existence of some kind of unique equilibrium in the system, related to the Second Law. But in 1848 Kelvin brushes this off by saying that the "axiom" has "never, so far as I am aware, been questioned by practical engineers".

His next page is notable for the first-ever use of the term "thermo-dynamic" (then hyphenated) to discuss systems where what matters is "the dynamics of heat":

> II. On the measurement of Thermal Agency, considered with reference to its equivalent of mechanical affect.
>
> 12. A *perfect* thermo-dynamic engine of any kind, is a machine by means of which the greatest possible amount of mechanical effect can be obtained from a given thermal agency; and, therefore, if in any manner we can construct or imagine a perfect engine which may be applied for the transference of a given quantity of heat from a body at any given temperature, to another body, at a lower

That same page has a curious footnote presaging what will come, and making the statement that "no energy can be destroyed", and considering it "perplexing" that this seems incompatible with Carnot's work and its caloric theory framework:

* When "thermal agency" is thus spent in conducting heat through a solid, what becomes of the mechanical effect which it might produce? Nothing can be lost in the operations of nature—no energy can be destroyed. What effect then is produced in place of the mechanical effect which is lost? A perfect theory of heat imperatively demands an answer to this question; yet no answer can be given in the present state of science. A few years ago, a similar confession must have been made with reference to the mechanical effect lost in a fluid set in motion in the interior of a rigid closed vessel, and allowed to come to rest by its own internal friction; but in this case, the foundation of a solution of the difficulty has been actually found, in Mr JOULE's discovery of the generation of heat, by the internal friction of a fluid in motion. Encouraged by this example, we may hope that the very perplexing question in the theory of heat, by which we are at present arrested, will, before long, be cleared up.

It might appear, that the difficulty would be entirely avoided, by abandoning CARNOT's fundamental axiom; a view which is strongly urged by Mr JOULE (at the conclusion of his paper "On the Changes of Temperature produced by the Rarefaction and Condensation of Air." *Phil. Mag.*, May 1845, vol. xxvi.) If we do so, however, we meet with innumerable other difficulties—insuperable without farther experimental investigation, and an entire reconstruction of the theory of heat, from its foundation. It is in reality to experiment that we must look—either for a verification of CARNOT's axiom, and an explanation of the difficulty we have been considering; or for an entirely new basis of the Theory of Heat.

After going through Carnot's basic arguments, the paper ends with an appendix in which Kelvin basically says that even though the theory seems to just be based on a formal axiom, it should be experimentally tested:

*Appendix.*

(Read April 30, 1849.)

41. In p. 30, some conclusions drawn by CARNOT from his general reasoning were noticed; according to which it appears, that if the value of $\mu$ for any temperature is known, certain information may be derived with reference to the saturated vapour of any liquid whatever, and, with reference to any gaseous mass, without the necessity of experimenting upon the specific medium considered. Nothing in the whole range of Natural Philosophy is more remarkable than the establishment of general laws by such a process of reasoning. We have seen, however, that doubt may exist with reference to the truth of the axiom on which the entire theory is founded, and it therefore becomes more than a matter of mere curiosity to put the inferences deduced from it to the test of experience.

He proceeds to give some tests, which he claims agree with Carnot's results—and finally ends with a very practical (but probably not correct) table of theoretical efficiencies for steam engines of his day:

TABLE A. *Various Engines in which the temperature of the Boiler is* 140°, *and that of the Condenser* 30°.

*Theoretical Duty for each Unit of Heat transmitted,* 440 *foot-pounds.*

| CASES. | Work produced for each pound of coal consumed. | Work produced for each pound of water evaporated. | Work produced for each unit of heat transmitted. | Per centage of theoretical duty. |
|---|---|---|---|---|
| | Foot-Pounds. | Foot-Pounds. | Foot-Pounds. | |
| (1.) Fowey Consols Experiment, reported in 1845, | 1,330,734 | 156,556 | 253 | 57·5 |
| (2.) Taylor's Engine at the United Mines, working in 1840, | 1,042,553 | 122,653 | 198·4 | 45·1 |
| (3.) French Engines, according to contract, | * * * * | 98,427 | 159 | 36·1 |
| (4.) English Engines, according to contract, | 565,700 | 80,814 | 130·8 | 29·7 |
| (5.) Average actual performance of Cornish Engines, | 585,106 | 68,836 | 111·3 | 25·3 |
| (6.) Common Engines, consuming 12 lbs. of best coal per hour per horse-power, | 165,000 | 23,571 | 38·1 | 8·6 |
| (7.) Improved Engines with Expansion Cylinders, consuming an equivalent to 4 lbs. of best coal per horse-power per hour, | 495,000 | 70,710 | 114·4 | 26 |

But now what of Joule's and Mayer's experiments, and their apparent disagreement with the caloric theory of heat? By 1849 a new idea had emerged: that perhaps heat was itself a form of energy, and that, when heat was accounted for, the total energy of a system would always be conserved. And what this suggested was that heat was somehow a dynamical phenomenon, associated with microscopic motion—which in turn suggested that gases might indeed consist just of molecules in motion.

And so it was that in 1850 Kelvin (then still "William Thomson") wrote a long exposition "On the Dynamical Theory of Heat", attempting to reconcile Carnot's ideas with the new concept that heat was dynamical in origin:

> XV.—*On the Dynamical Theory of Heat, with numerical results deduced from* Mr JOULE'S *equivalent of a Thermal Unit, and* M. REGNAULT'S *Observations on Steam.* By WILLIAM THOMSON, M.A., Fellow of St Peter's College, Cambridge, and Professor of Natural Philosophy in the University of Glasgow.
>
> (Read 17th March 1851.)
>
> INTRODUCTORY NOTICE.
>
> 1. SIR HUMPHREY DAVY, by his experiment of melting two pieces of ice by rubbing them together, established the following proposition:—"The phenomena of repulsion are not dependent on a peculiar elastic fluid for their existence, or caloric does not exist." And he concludes that heat consists of a motion excited among the particles of bodies. "To distinguish this motion from others, and to signify the cause of our sensation of heat," and of the expansion or expansive pressure produced in matter by heat, "the name *repulsive* motion has been adopted."*
>
> 2. The Dynamical Theory of Heat, thus established by SIR HUMPHREY DAVY, is extended to radiant heat by the discovery of phenomena, especially those of the polarization of radiant heat, which render it excessively probable that heat propagated through vacant space, or through diathermane substances, consists of waves of transverse vibrations in an all-pervading medium.
>
> 3. The recent discoveries made by MAYER and JOULE,† of the generation of heat through the friction of fluids in motion, and by the magneto-electric excitation of galvanic currents, would, either of them be sufficient to demonstrate the immateriality of heat; and would so afford, if required, a perfect confirmation of SIR HUMPHREY DAVY'S views.
>
> 4. Considering it as thus established, that heat is not a substance, but a dynamical form of mechanical effect, we perceive that there must be an equivalence between mechanical work and heat, as between cause and effect. The first

He begins by quoting—presumably for some kind of "British-based authority"—an "anti-caloric" experiment apparently done by Humphry Davy (1778–1829) as a teenager, involving melting pieces of ice by rubbing them together, and included anonymously in a 1799 list of pieces of knowledge "principally from the west of England":

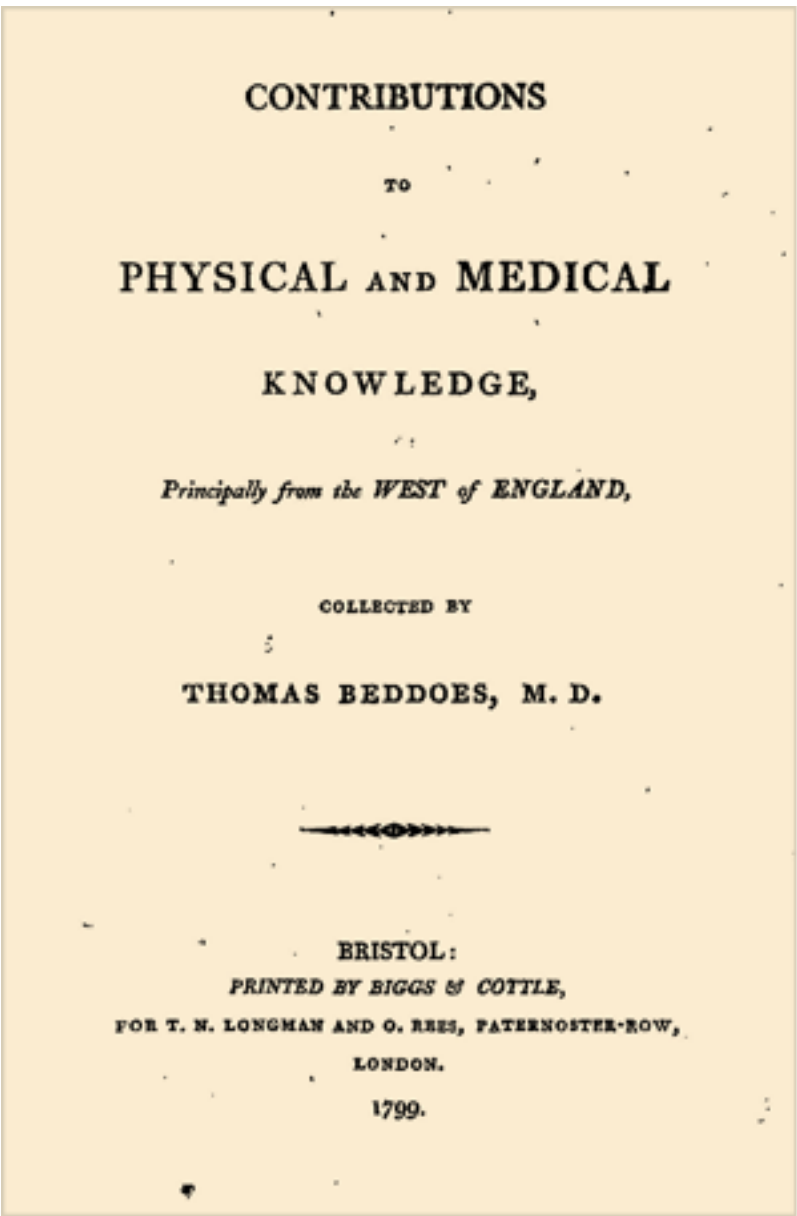

CONTRIBUTIONS

TO

PHYSICAL AND MEDICAL

KNOWLEDGE,

*Principally from the WEST of ENGLAND,*

COLLECTED BY

THOMAS BEDDOES, M. D.

BRISTOL:
*PRINTED BY BIGGS & COTTLE,*
FOR T. N. LONGMAN AND O. REES, PATERNOSTER-ROW,
LONDON.
1799.

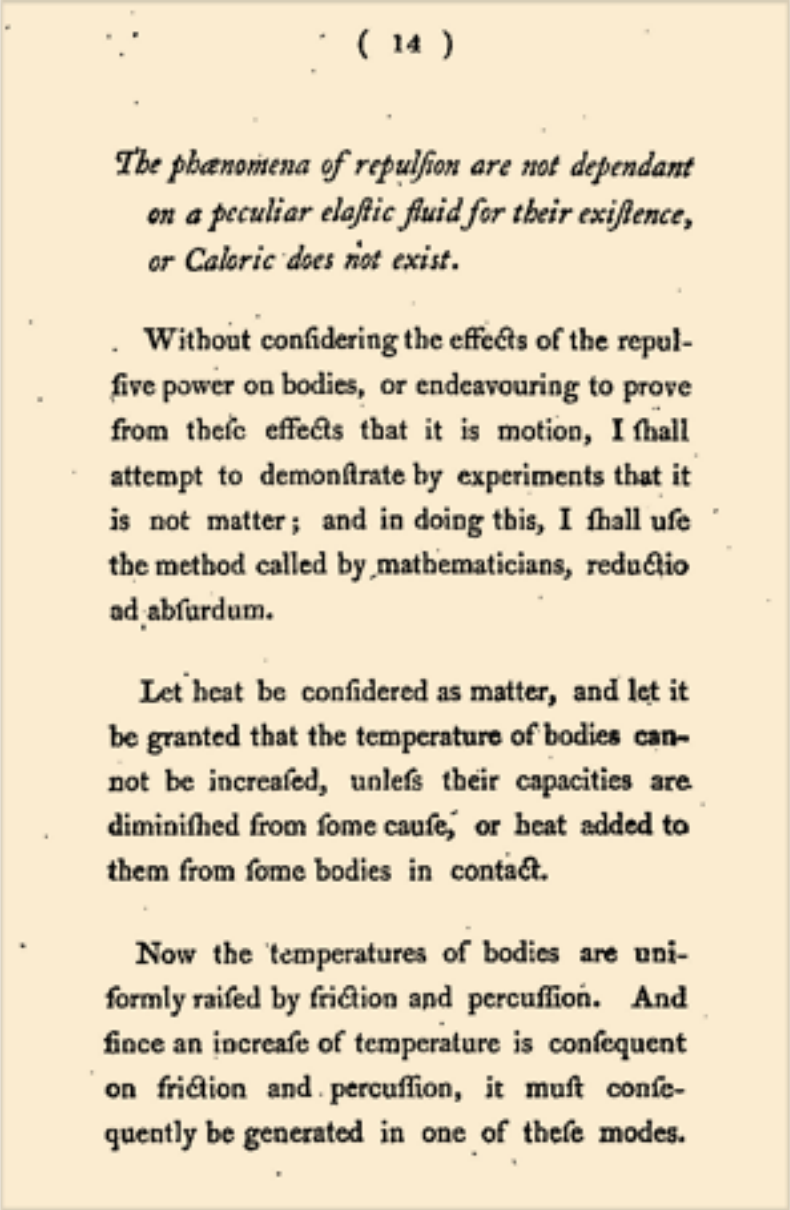

( 14 )

*The phænomena of repulſion are not dependant on a peculiar elaſtic fluid for their exiſtence, or Caloric does not exist.*

Without conſidering the effects of the repulſive power on bodies, or endeavouring to prove from theſe effects that it is motion, I ſhall attempt to demonſtrate by experiments that it is not matter; and in doing this, I ſhall uſe the method called by mathematicians, reductio ad abſurdum.

Let heat be conſidered as matter, and let it be granted that the temperature of bodies cannot be increaſed, unleſs their capacities are diminiſhed from ſome cauſe, or heat added to them from ſome bodies in contact.

Now the temperatures of bodies are uniformly raiſed by friction and percuſſion. And ſince an increaſe of temperature is conſequent on friction and percuſſion, it muſt conſequently be generated in one of theſe modes.

But soon Kelvin is getting to the main point:

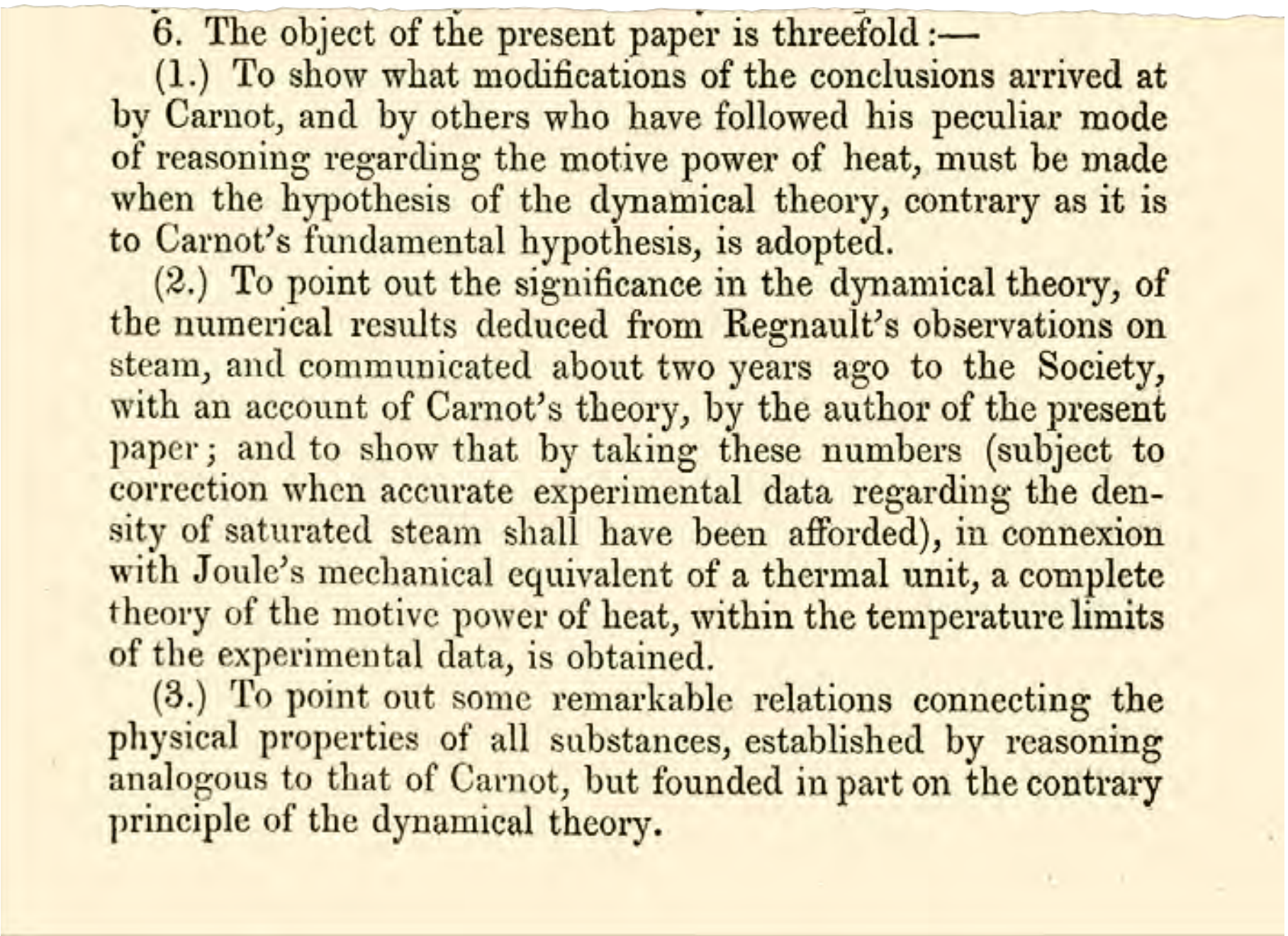

6. The object of the present paper is threefold:—

(1.) To show what modifications of the conclusions arrived at by Carnot, and by others who have followed his peculiar mode of reasoning regarding the motive power of heat, must be made when the hypothesis of the dynamical theory, contrary as it is to Carnot's fundamental hypothesis, is adopted.

(2.) To point out the significance in the dynamical theory, of the numerical results deduced from Regnault's observations on steam, and communicated about two years ago to the Society, with an account of Carnot's theory, by the author of the present paper; and to show that by taking these numbers (subject to correction when accurate experimental data regarding the density of saturated steam shall have been afforded), in connexion with Joule's mechanical equivalent of a thermal unit, a complete theory of the motive power of heat, within the temperature limits of the experimental data, is obtained.

(3.) To point out some remarkable relations connecting the physical properties of all substances, established by reasoning analogous to that of Carnot, but founded in part on the contrary principle of the dynamical theory.

And then we have it: a statement of the Second Law (albeit with some hedging to which we'll come back later):

> 12. The demonstration of the second proposition is founded on the following axiom :—
>
> *It is impossible, by means of inanimate material agency, to derive mechanical effect from any portion of matter by cooling it below the temperature of the coldest of the surrounding objects*.*

And there's immediately a footnote that basically asserts the "absurdity" of a Second-Law-violating perpetual motion machine:

> * If this axiom be denied for all temperatures, it would have to be admitted that a self-acting machine might be set to work and produce mechanical effect by cooling the sea or earth, with no limit but the total loss of heat from the earth and sea, or, in reality, from the whole material world.

But by the next page we find out that Kelvin admits he's in some sense been "scooped"—by a certain Rudolf Clausius (1822–1888), who we'll be discussing soon. But what's remarkable is that Clausius's "axiom" turns out to be exactly equivalent to Kelvin's statement:

> upon an axiom (§ 12) which I think will be generally admitted. It is with no wish to claim priority that I make these statements, as the merit of first establishing the proposition upon correct principles is entirely due to Clausius, who published his demonstration of it in the month of May last year, in the second part of his paper on the motive power of heat‡. I may be allowed to add, that I have given the demonstration exactly as it occurred to me before I knew that Clausius had either enunciated or demonstrated the proposition. The following is the axiom on which Clausius' demonstration is founded :—
>
> *It is impossible for a self-acting machine, unaided by any external agency, to convey heat from one body to another at a higher temperature.*
>
> It is easily shown, that, although this and the axiom I have used are different in form, either is a consequence of the other. The reasoning in each demonstration is strictly analogous to that which Carnot originally gave.

And what this suggests is that the underlying concept—the Second Law—is something quite robust. And indeed, as Kelvin implies, it's the main thing that ultimately underlies Carnot's results. And so even though Carnot is operating on the now-outmoded idea of caloric theory, his main results are still correct, because in the end all they really depend on is a certain amount of "logical structure", together with the Second Law (and a version of the First Law, but that's a slightly trickier story).

Kelvin recognized, though, that Carnot had chosen to look at the particular ("equilibrium thermodynamics") case of processes that occur reversibly, effectively at an infinitesimal rate. And at the end of the first installment of his exposition, he explains that things will be more complicated if finite rates are considered—and that in particular the results one gets in such cases will depend on things like having a correct model for the nature of heat.

Kelvin's exposition on the "dynamical nature of heat" runs to four installments, and the next two dive into detailed derivations and attempted comparison with experiment:

Hence equation (2) of § 20 becomes

$$\frac{\frac{dL}{dt}+c-h}{\gamma-\lambda}=\frac{1}{J}\frac{dp}{dt} \quad \ldots\ldots\ldots \quad (33).$$

Combining this with the conclusion (32) derived from the second fundamental proposition, we obtain

$$\frac{dL}{dt}+c-h=\frac{L\mu}{J} \quad \ldots\ldots\ldots \quad (34).$$

The former of these equations agrees precisely with one which was first given by Clausius, and the preceding investigation is substantially the same as the investigation by which he arrived at it. The second differs from another given by Clausius only in not implying any hypothesis as to the form of Carnot's function $\mu$.

57. If we suppose $\mu$ and L to be known for any temperature, equation (32) enables us to determine the value of $\frac{dp}{dt}$ for that temperature; and thence deducing a value of $dt$, we have

$$dt=\frac{\gamma-\lambda}{\mu L}dp \quad \ldots\ldots\ldots \quad (35);$$

which shows the effect of pressure in altering the "boiling-point" if the mixed medium be a liquid and its vapour, or the melting-point if it be a solid in contact with the same substance in the liquid state. This agrees with the conclusion arrived at by my elder brother in his Theoretical Investigation of the Effect of Pressure in Lowering the Freezing-Point of Water*. His result, obtained by taking as the value for $\mu$ that derived from Table I. of my former paper for the temperature 0°, is that the freezing-point is lowered by ·0075° Cent. by an additional atmosphere of pressure. Clausius, with the other data the same, obtains ·00733° as the lowering of temperature by the same additional pressure, which differs from my brother's result only from having been calculated from a formula which implies the hypothetical expression $J\frac{E}{1+Et}$ for $\mu$. It was by applying equation (33) to determine $\frac{dL}{dt}$ for the same case that Clausius arrived at the curious result regarding the latent heat of water under pressure mentioned above (§ 45).

58. Lastly, it may be remarked that every quantity which ap-

* Transactions, vol. xvi. part 5. His paper was republished, with some slight modifications, in the Cambridge and Dublin Mathematical Journal, new series, vol. v.—Nov. 1850.

pears in equation (33), except $h$, is known with tolerable accuracy for saturated steam through a wide range of temperature; and we may therefore use this equation to find $h$, which has never yet been made an object of experimental research. Thus we have

$$-h=\frac{\gamma-\lambda}{J}\frac{dp}{dt}-\left(\frac{dL}{dt}+c\right).$$

For the value of $\gamma$ the best data regarding the density of saturated steam that can be had must be taken. If for different temperatures we use the same values for the density of saturated steam (calculated according to the gaseous laws, and Regnault's observed pressure from $\frac{1}{1693\cdot 3}$, taken as the density at 100°), the values obtained for the first term of the second member of the preceding equation are the same as if we take the form

$$-h=\frac{L\mu}{J}-\left(\frac{dL}{dt}+c\right)$$

derived from (34), and use the values of $\mu$ shown in Table I. of my former paper. The values of $-h$ in the second column in the following table have been so calculated, with, besides, the following data afforded by Regnault from his observations on the total heat of steam, and the specific heat of water

$$\frac{dL}{dt}+c=\cdot 305.$$

$$L=606\cdot 5+\cdot 305t-(\cdot 00002t^2+\cdot 0000003t^3).$$

The values of $-h$ shown in the third column are those derived by Clausius from an equation which is the same as what (34) would become if $J\frac{E}{1+Et}$ were substituted for $\mu$.

| $t$. | $-h$ according to Table I. of "Account of Carnot's Theory." | $-h$ according to Clausius. |
|---|---|---|
| 0 | 1·863 | 1·916 |
| 50 | 1·479 | 1·465 |
| 100 | 1·174 | 1·133 |
| 150 | 0·951 | 0·879 |
| 200 | 0·780 | 0·676 |

59. From these results it appears, that through the whole range of temperatures at which observations have been made, the value of $h$ is negative; and, therefore, if a quantity of saturated vapour be compressed in a vessel containing no liquid water, heat must be continuously abstracted from it in order

But before Kelvin gets to publish part four of his exposition he publishes two other pieces. In the first, he's talking about sources of energy for human use (now that he believes energy is conserved):

2. On the Mechanical action of Radiant Heat or Light: On the Power of Animated Creatures over Matter: On the Sources available to Man for the production of Mechanical Effect. By Professor William Thomson.

*On the Mechanical Action of Radiant Heat or Light.*

It is assumed in this communication that the undulatory theory

He emphasizes that the Sun is—directly or indirectly—the main source of energy on Earth (later he'll argue that coal will run out, etc.):

> the following general conclusions :—
>
> 1. *Heat radiated from the sun* (sunlight being included in this term) *is the principal source of mechanical effect available to man.** From it is derived the whole mechanical effect obtained by means of animals working, water-wheels worked by rivers, steam-engines, and galvanic engines, and part at least of the mechanical effect obtained by means of windmills and the sails of ships not driven by the trade-winds.
>
> 2. The motions of the earth, moon, and sun, and their mutual attractions, constitute an important source of available mechanical effect. From them all, but chiefly, no doubt, from the earth's motion of rotation, is derived the mechanical effect of water-wheels driven by the tides. The mechanical effect so largely used in the

But he wonders how animals actually manage to produce mechanical work, noting that "the animal body does not act as a thermo–dynamic engine; and [it is] very probable that the chemical forces produce the external mechanical effects through electrical means":

> A curious inference is pointed out, that an animal would be sensibly less warm in going up-hill than in going down-hill, were the breathing not greater in the former case than in the latter.
>
> The application of Carnot's principle, and of Joule's discoveries regarding the heat of electrolysis and the calorific effects of magneto-electricity, is pointed out; according to which it appears nearly certain that, when an animal works against resisting force, there is not a *conversion of heat into external mechanical effect*, but the full thermal equivalent of the chemical forces is *never produced*; in other words that the animal body does not act as a *thermo-dynamic engine*; and very probable that the chemical forces produce the external mechanical effects through electrical means.

And then, by April 1852, he's back to thinking directly about the Second Law, and he's cut through the technicalities, and is stating the Second Law in everyday (if slightly ponderous) terms:

XLVII. *On a Universal Tendency in Nature to the Dissipation of Mechanical Energy.* *By* Prof. W. THOMSON*.

THE object of the present communication is to call attention to the remarkable consequences which follow from Carnot's proposition, that there is an absolute waste of mechanical energy available to man when heat is allowed to pass from one body to another at a lower temperature, by any means not fulfilling his criterion of a "perfect thermo-dynamic engine," established, on a new foundation, in the dynamical theory of heat. As it is most certain that Creative Power alone can either call into existence or annihilate mechanical energy, the "waste" referred to cannot be annihilation, but must be some transformation of energy†. To explain the nature of this transformation, it is convenient, in the first place, to divide *stores* of mechanical energy into two classes—*statical* and *dynamical*. A quantity of weights at a height, ready to descend and do work when wanted, an electrified body, a quantity of fuel, contain stores of mechanical energy of the statical kind. Masses of matter in motion, a volume of space through which undulations of light or radiant heat are passing, a body having thermal motions among its particles (that is, not infinitely cold), contain stores of mechanical energy of the dynamical kind.

The following propositions are laid down regarding the *dissipation* of mechanical energy from a given store, and the *restoration* of it to its primitive condition. They are necessary consequences of the axiom, "*It is impossible, by means of inanimate material agency, to derive mechanical effect from any portion of matter by cooling it below the temperature of the coldest of the surrounding objects.*" (Dynam. Th. of Heat, § 12.)

I. When heat is created by a reversible process (so that the mechanical energy thus spent may be *restored* to its primitive condition), there is also a transference from a cold body to a hot body of a quantity of heat bearing to the quantity created a definite proportion depending on the temperatures of the two bodies.

II. When heat is created by any unreversible process (such as friction), there is a *dissipation* of mechanical energy, and a full *restoration* of it to its primitive condition is impossible.

III. When heat is diffused by *conduction*, there is a *dissipation* of mechanical energy, and perfect *restoration* is impossible.

IV. When radiant heat or light is absorbed, otherwise than in

It's interesting to see his apparently rather deeply held Presbyterian beliefs manifest themselves here in his mention that "Creative Power" is what must set the total energy of the universe. He ends his piece with:

> The following general conclusions are drawn from the propositions stated above, and known facts with reference to the mechanics of animal and vegetable bodies :—
>
> 1. There is at present in the material world a universal tendency to the dissipation of mechanical energy.
>
> 2. Any *restoration* of mechanical energy, without more than an equivalent of dissipation, is impossible in inanimate material processes, and is probably never effected by means of organized matter, either endowed with vegetable life or subjected to the will of an animated creature.
>
> 3. Within a finite period of time past the earth must have been, and within a finite period of time to come the earth must again be, unfit for the habitation of man as at present constituted, unless operations have been, or are to be performed, which are impossible under the laws to which the known operations going on at present in the material world are subject.

In (2) the hedging is interesting. He makes the definitive assertion that what amounts to a violation of the Second Law "is impossible in inanimate material processes". And he's pretty sure the same is true for "vegetable life" (recognizing that in his previous paper he discussed the harvesting of sunlight by plants). But what about "animal life", like us humans? Here he says that "by our will" we can't violate the Second Law—so we can't, for example, build a machine to do it. But he leaves it open whether we as humans might have some innate ("God-given"?) ability to overcome the Second Law.

And then there's his (3). It's worth realizing that his whole paper is less than 3 pages long, and right before his conclusions we're seeing triple integrals:

> If the system of thermometry adopted* be such that $\mu = \frac{J}{t+\alpha}$, that is, if we agree to call $\frac{J}{\mu} - \alpha$ the *temperature* of a body, for which $\mu$ is the *value of Carnot's function* ($\alpha$ and J being constants), the preceding expression becomes
>
> $$T = \frac{\iiint c\, dx\, dy\, dz}{\iiint \frac{c}{t+\alpha}\, dx\, dy\, dz} - \alpha.$$
>
> The following general conclusions are drawn from the propositions stated above, and known facts with reference to the mechanics of animal and vegetable bodies :—

So what is (3) about? It's presumably something like a Second-Law-implies-heat-death-of-the-universe statement (but what's this stuff about the past?)—but with an added twist that there's something (God?) beyond the "known operations going on at present in the material world" that might be able to swoop in to save the world for us humans.

It doesn't take people long to pick up on the "cosmic significance" of all this. But in the fall of 1852, Kelvin's colleague, the Glasgow engineering professor William Rankine (1820–1872) (who was deeply involved with the First Law of thermodynamics), is writing about a way the universe might save itself:

> LVI. *On the Reconcentration of the Mechanical Energy of the Universe.* *By* William John Macquorn Rankine, *C.E., F.R.S.E. &c.**
>
> The following remarks have been suggested by a paper by Professor William Thomson of Glasgow, on the tendency which exists in nature to the dissipation or indefinite diffusion of mechanical energy originally collected in stores of power.
>
> * Communicated by the Author; having been read to the British Association for the Advancement of Science, Section A, at Belfast, on the 2nd of September 1852.

After touting the increasingly solid evidence for energy conservation and the First Law

> The experimental evidence is every day accumulating, of a law which has long been conjectured to exist,—that all the different kinds of physical energy in the universe are mutually convertible, —that the total amount of physical energy, whether in the form of visible motion and mechanical power, or of heat, light, magnetism, electricity, or chemical agency, or in other forms not yet understood, is unchangeably the transformations of its different portions from one of those forms of power into another, and their transference from one portion of matter to another, constituting the phænomena which are the objects of experimental physics.

he goes on to talk about dissipation of energy and what we now call the Second Law

> the state of heat. On the other hand, all visible motion is of necessity ultimately converted entirely into heat by the agency of friction. There is thus, in the present state of the known world, a tendency towards the conversion of all physical energy into the sole form of heat.
>
> Heat, moreover, tends to diffuse itself uniformly by conduction and radiation, until all matter shall have acquired the same temperature.
>
> There is, consequently, Professor Thomson concludes, so far as we understand the present condition of the universe, a tendency towards a state in which all physical energy will be in the state of heat, and that heat so diffused that all matter will be at the same temperature; so that there will be an end of all physical phænomena.

and the fact that it implies an "end of all physical phenomena", i.e. heat death of the universe. He continues:

> Vast as this speculation may seem, it appears to be soundly
>
> based on experimental data, and to represent truly the present condition of the universe, so far as we know it.

But now he offers a "ray of hope". He believes that there must exist a "medium capable of transmitting light and heat", i.e. an aether, "[between] the heavenly bodies". And if this aether can't itself acquire heat, he concludes that all energy must be converted into a radiant form:

> My object now is to point out how it is conceivable that, at some indefinitely distant period, an opposite condition of the world may take place, in which the energy which is now being diffused may be reconcentrated into foci, and stores of chemical power again produced from the inert compounds which are now being continually formed.
>
> There must exist between the atmospheres of the heavenly bodies a material medium capable of transmitting light and heat; and it may be regarded as almost certain, that this interstellar medium is perfectly transparent and diathermanous; that is to say, that it is incapable of converting heat, or light (which is a species of heat), from the radiant into the fixed or conductible form.
>
> If this be the case, the interstellar medium must be incapable of acquiring any temperature whatsoever; and all heat which arrives in the conductible form at the limits of the atmosphere of a star or planet, will there be totally converted, partly into ordinary motion, by the expansion of the atmosphere, and partly into the radiant form. The ordinary motion will again be converted into heat, so that *radiant heat* is the ultimate form to which all physical energy tends; and in this form it is, in the present condition of the world, diffusing itself from the heavenly bodies through the interstellar medium.

Now he supposes that the universe is effectively a giant drop of aether, with nothing outside, so that all this radiant energy will get totally internally reflected from its surface, allowing the universe to "[reconcentrate] its physical energies, and [renew] its activity and life"—and save it from heat death:

Let it now be supposed, that, in all directions round the visible world, the interstellar medium has bounds beyond which there is empty space.

If this conjecture be true, then on reaching those bounds the radiant heat of the world will be totally reflected, and will ultimately be reconcentrated into foci. At each of these foci the intensity of heat may be expected to be such, that should a star (being at that period an extinct mass of inert compounds) in the course of its motions arrive at that part of space, it will be vaporized and resolved into its elements; a store of chemical power being thus reproduced at the expense of a corresponding amount of radiant heat.

Thus it appears, that although, from what we can see of the known world, its condition seems to tend continually towards the equable diffusion, in the form of radiant heat, of all physical energy, the extinction of the stars, and the cessation of all phænomena, yet the world, as now created, may possibly be provided within itself with the means of reconcentrating its physical energies, and renewing its activity and life.

For aught we know, these opposite processes may go on together; and some of the luminous objects which we see in distant regions of space may be, not stars, but foci in the interstellar æther.

He ends with the speculation that perhaps "some of the luminous objects which we see in distant regions of space may be, not stars, but foci in the interstellar aether".

But independent of cosmic speculations, Kelvin himself continues to study the "dynamical theory of gases". It's often a bit unclear what's being assumed. There's the First Law (energy conservation). And the Second Law. But there's also reversibility. Equilibrium. And the ideal gas law ($P\,V = R\,T$). But it soon becomes clear that that's not always correct for real gases—as the Joule–Thomson effect demonstrates:

**LXXVI. *On the Thermal Effects experienced by Air in rushing through small Apertures.* *By* J. P. Joule *and* W. Thomson*.**

THE hypothesis that the heat evolved from air compressed and kept at a constant temperature is mechanically equivalent to the work spent in effecting the compression, assumed by Mayer as the foundation for an estimate of the numerical relation between quantities of heat and mechanical work, and adopted by Holtzmann, Clausius, and other writers, was made the subject of an experimental research by Mr. Joule†, and verified as at least approximately true for air at ordinary atmospheric temperatures. A theoretical investigation, founded on a conclusion of Carnot's‡, which requires no modification§ in the dynamical theory of heat, also leads to a verification of Mayer's hypo-

Kelvin soon returned to more cosmic speculations, suggesting that perhaps gravitation—

rather than direct "Creative Power"—might "in reality [be] the ultimate created antecedent of all motion...":

> Published speculations* were referred to, by which it is shown to be possible that the motions of the earth and of the heavenly bodies, and the heat of the sun, may all be due to gravitation; or, *that the potential energy of gravitation may be in reality the ultimate created antecedent of all motion, heat, and light at present existing in the universe.*
>
> * Prof. W. Thomson, "On the Mechanical Energies of the Solar System" (*Trans. Roy. Soc. Edinburgh*, April, 1854 [Art. LXVI. above]), and "On the Mechanical Antecedents of Motion, Heat, and Light" (*British Association Report*, Liverpool, 1854 [Art. LXIX. above]).

Not long after these papers Kelvin got involved with the practical "electrical" problem of laying a transatlantic telegraph cable, and in 1858 was on the ship that first succeeded in doing this. (His commercial efforts soon allowed him to buy a 126-ton yacht.) But he continued to write physics papers, which ranged over many different areas, occasionally touching thermodynamics, though most often in the service of answering a "general science" question—like how old the Sun is (he estimated 32,000 years from thermodynamic arguments, though of course without knowledge of nuclear reactions).

Kelvin's ideas about the inevitable dissipation of "useful energy" spread quickly—by 1854, for example, finding their way into an eloquent public lecture by Hermann von Helmholtz (1821–1894). Helmholtz had trained as a doctor, becoming in 1843 a surgeon to a German military regiment. But he was also doing experiments and developing theories about "animal heat" and how muscles manage to "do mechanical work", for example publishing an 1845 paper entitled "On Metabolism during Muscular Activity". And in 1847 he was one of the inventors of the law of conservation of energy—and the First Law of thermodynamics—as well as perhaps its clearest expositor at the time (the word "force" in the title is what we now call "energy"):

## Inhalt.

## Einleitung.

Vorliegende Abhandlung musste ihrem Hauptinhalte nach hauptsächlich für Physiker bestimmt werden, ich habe es daher vorgezogen, die Grundlagen derselben unabhängig von einer philosophischen Begründung rein in der Form einer physikalischen Voraussetzung hinzustellen, deren Folgerungen zu entwickeln, und dieselben in den verschiedenen Zweigen der Physik mit den erfahrungsmässigen Gesetzen der Naturerscheinungen zu vergleichen. Die Herleitung der

By 1854 Helmholtz was a physiology professor, beginning a distinguished career in physics, psychophysics and physiology—and talking about the Second Law and its implications. He began his lecture by saying that “A new conquest of very general interest has been recently made by natural philosophy”—and what he’s referring to here is the Second Law:

## INTERACTION OF NATURAL FORCES.

A NEW conquest of very general interest has been recently made by natural philosophy. In the following pages, I will endeavour to give a notion of the nature of this conquest. It has reference to a new and universal natural law, which rules the action of natural forces in their mutual relations towards each other, and is as influential on our theoretic views of natural processes as it is important in their technical applications.

Among the practical arts which owe their progress to the development of the natural sciences, from the conclusion of the middle ages downwards, practical mechanics, aided by the mathematical science which bears the same name, was one of the most prominent. The character of the art was, at the time referred to, naturally very different from its present one. Surprised and stimulated by its own success, it thought no problem beyond its power, and immediately attacked some of the most difficult and complicated. Thus it was attempted to build automaton figures which should perform the functions of men and animals. The wonder of the last century was

Having discussed the inability of "automata" (he uses that word) to reproduce living systems, he starts talking about perpetual motion machines:

From these efforts to imitate living creatures, another idea, also by a misunderstanding, seems to have developed itself, which, as it were, formed the new philosopher's stone of the seventeenth and eighteenth centuries. It was now the endeavour to construct a perpetual motion. Under this term was un-

First he disposes of the idea that perpetual motion can be achieved by generating energy from nothing (i.e. violating the First Law), charmingly including the anecdote:

magneto-electric machine, decomposed the water, and thus continually prepared its own fuel. This would certainly have been the most splendid of all discoveries; a perpetual motion which, besides the force which kept it going, generated light like the sun, and warmed all around it. The matter was by no means badly cogitated. Each practical step in the affair was known to be possible; but those which at that time were acquainted with the physical investigations which bear upon this subject could have affirmed, on the first hearing the report, that the matter was to be numbered among the numerous stories of the fable-rich America; and indeed, a fable it remained.

And then he's on to talking about the Second Law

> capacity for heat, and the expansion by heat of all bodies. It is not yet considered as actually proved, but some remarkable deductions having been drawn from it, and afterwards proved to be facts by experiment, it has attained thereby a great degree of probability. Besides the mathematical form in which the law was first expressed by Carnot, we can give it the following more general expression:—"Only when heat passes from a warmer to a colder body, and even then only partially, can it be converted into mechanical work."

and discussing how it implies the heat death of the universe:

> But the heat of the warmer bodies strives perpetually to pass to bodies less warm by radition and conduction, and thus to establish an equilibrium of temperature. At each motion of a terrestrial body, a portion of mechanical force passes by friction or collision into heat, of which only a part can be converted back again into mechanical force. This is also generally the case in every electrical and chemical process. From this, it follows that the first portion of the store of force, the unchangeable heat, is augmented by every natural process, while the second portion, mechanical, electrical, and chemical force, must be diminished; so that if the universe be delivered over to the undisturbed action of its physical processes, all force will finally pass into the form of heat, and all heat come into a state of equilibrium. Then all possibility of a further change would be at an end, and the complete cessation of all natural processes must set in. The life of men, animals, and plants, could not of course continue if the sun had lost its high temperature, and with it his light,—if all the components of the earth's surface had closed those combinations which their affinities demand. In short, the universe from that time forward would be condemned to a state of eternal rest.

He notes, correctly, that the Second Law hasn't been "proved". But he's impressed at how Kelvin was able to go from a "mathematical formula" to a global fact about the fate of the universe:

> These consequences of the law of Carnot are, of course, only valid, provided that the law, when sufficiently tested, proves to be universally correct. In the mean time there is little prospect of the law being proved incorrect. At all events we must admire the sagacity of Thomson, who, in the letters of a long known little mathematical formula, which only speaks of the heat, volume, and pressure of bodies, was able to discern consequences which threatened the universe, though certainly after an infinite period of time, with eternal death.

He ends the whole lecture quite poetically:

> Thus the thread which was spun in darkness by those who sought a perpetual motion has conducted us to a universal law of nature, which radiates light into the distant nights of the beginning and of the end of the history of the universe. To our own race it permits a long but not an endless existence; it threatens it with a day of judgment, the dawn of which is still happily obscured. As each of us singly must endure the thought of his death, the race must endure the same. But above the forms of life gone by, the human race has higher moral problems before it, the bearer of which it is, and in the completion of which it fulfils its destiny.

We've talked quite a bit about Kelvin and how his ideas spread. But let's turn now to Rudolf Clausius, who in 1850 at least to some extent "scooped" Kelvin on the Second Law. At that time Clausius was a freshly minted German physics PhD. His thesis had been on an ingenious but ultimately incorrect theory of why the sky is blue. But he'd also worked on elasticity theory, and there he'd been led to start thinking about molecules and their configurations in materials. By 1850 caloric theory had become fairly elaborate, complete with concepts like "latent heat" (bound to molecules) and "free heat" (able to be transferred). Clausius's experience in elasticity theory made him skeptical, and knowing Mayer's and Joule's results he decided to break with the caloric theory—writing his career–launching paper (translated from German in 1851, with Carnot's *puissance motrice* ["motive power"] being rendered as "moving force"):

> I. *On the Moving Force of Heat, and the Laws regarding the Nature of Heat itself which are deducible therefrom. By* R. Clausius*.
>
> THE steam-engine having furnished us with a means of converting heat into a motive power, and our thoughts being thereby led to regard a certain quantity of work as an equivalent for the amount of heat expended in its production, the idea of establishing theoretically some fixed relation between a quantity of heat and the quantity of work which it can possibly produce, from which relation conclusions regarding the nature of heat itself might be deduced, naturally presents itself. Already, indeed, have many instructive experiments been made with this view; I believe, however, that they have not exhausted the subject, but that, on the contrary, it merits the continued attention of physicists; partly because weighty objections lie in the way of the conclusions already drawn, and partly because other conclusions, which might render efficient aid towards establishing and completing the theory of heat, remain either entirely unnoticed, or have not as yet found sufficiently distinct expression.
>
> The most important investigation in connexion with this subject is that of S. Carnot†. Later still, the ideas of this author

The first installment of the English version of the paper gives a clear description of the ideal gas laws and the Carnot cycle, having started from a statement of the "caloric–busting" First Law:

1. *Deductions from the principle of the equivalence of heat and work.*

We shall forbear entering at present on the nature of the motion which may be supposed to exist within a body, and shall assume generally that a motion of the particles does exist, and that heat is the measure of their *vis viva*. Or yet more general, we shall merely lay down one maxim which is founded on the above assumption :—

*In all cases where work is produced by heat, a quantity of heat proportional to the work done is expended; and inversely, by the expenditure of a like quantity of work, the same amount of heat may be produced.*

Before passing on to the mathematical treatment of this maxim, a few of its more immediate consequences may be noticed, which

The general discussion continues in the second installment, but now there's a critical side comment that describes the "general deportment of heat, which every–where exhibits the tendency to annul differences of temperature, and therefore to pass from a warmer body to a colder one":

thus on the whole a transmission from B to A would take place.
Hence by repeating both these alternating processes, without expenditure of force or other alteration whatever, any quantity of heat might be transmitted from a *cold* body to a *warm* one; and this contradicts the general deportment of heat, which every-where exhibits the tendency to annul differences of temperature, and therefore to pass from a *warmer* body to a *colder* one.

From this it would appear that we are *theoretically* justified in

Clausius "has" the Second Law, as Carnot basically did before him. But when Kelvin quotes Clausius he does so much more forcefully:

which Clausius' demonstration is founded :—

*It is impossible for a self-acting machine, unaided by any external agency, to convey heat from one body to another at a higher temperature.*

It is easily shown that although this and the axiom I have But there it is: by 1852 the Second Law is out in the open, in at least two different forms. The path to reach it has been circuitous and quite technical. But in the end, stripped of its technical origins, the law seems somehow unsurprising and even obvious. For it's a matter of common experience that heat flows from hotter bodies to colder ones, and that motion is dissipated by friction into heat. But the point is that it wasn't until basically 1850 that the overall scientific framework existed to make it useful—or even really possible—to enunciate such observations as a formal scientific law.

Of course the fact that a law "seems true" based on common experience doesn't mean it'll always be true, and that there won't be some special circumstance or elaborate construction that will evade it. But somehow the very fact that the Second Law had in a sense been "technically hard won"—yet in the end seemed so "obvious"—appears to have given it a sense of inevitability and certainty. And it didn't hurt that somehow it seemed to have emerged from Carnot's work, which had a certain air of "logical necessity". (Of course, in reality, the Second Law entered Carnot's logical structure as an "axiom".) But all this helped set the stage for some of the curious confusions about the Second Law that would develop over the century that followed.

## The Concept of Entropy

In the first half of the 1850s the Second Law had in a sense been presented in two ways. First, as an almost "footnote-style" assumption needed to support the "pure thermodynamics" that had grown out of Carnot's work. And second, as an explicitly-stated-for-the-first-time—if "obvious"—"everyday" feature of nature, that was now realized as having potentially cosmic significance. But an important feature of the decade that followed was a certain progressive at-least-phenomenological "mathematicization" of the Second Law—pursued most notably by Rudolf Clausius.

In 1854 Clausius was already beginning this process. Perhaps confusingly, he refers to the Second Law as the "second fundamental theorem [*Hauptsatz*]" in the "mechanical theory of heat"—suggesting it's something that is proved, even though it's really introduced just as an empirical law of nature, or perhaps a theoretical axiom:

ON A MODIFIED FORM OF THE SECOND FUNDAMENTAL THEOREM IN THE MECHANICAL THEORY OF HEAT*.

IN my memoir "On the Moving Force of Heat, &c."†, I have shown that the theorem of the equivalence of heat and work, and Carnot's theorem, are not mutually exclusive, but that, by a small modification of the latter, which does not affect its principal part, they can be brought into accordance. With the exception of this indispensable change, I allowed the theorem of Carnot to retain its original form, my chief object then being, by the application of the two theorems to special cases, to arrive at conclusions which, according as they involved known or unknown properties of bodies, might suitably serve as proofs of the truth of the theorems, or as examples of their fecundity.

This form, however, although it may suffice for the deduction of the equations which depend upon the theorem, is incomplete, because we cannot recognize therein, with sufficient clearness, the real nature of the theorem, and its connexion with the first fundamental theorem. The modified form in the following pages will, I think, better fulfil this demand, and in its applications will be found very convenient.

He starts off by discussing the "first fundamental theorem", i.e. the First Law. And he emphasizes that this implies that there's a quantity *U* (which we now call "internal energy") that is a pure "function of state"—so that its value depends only on the state of a system, and not the path by which that state was reached. And as an "application" of this, he then points out that the overall change in *U* in a cyclic process (like the one executed by Carnot's heat engine) must be zero.

And now he's ready to tackle the Second Law. He gives a statement that at first seems somewhat convoluted:

*Theorem of the equivalence of transformations.*

Carnot's theorem, when brought into agreement with the first fundamental theorem, expresses a relation between two kinds of transformations, the transformation of heat into work, and the passage of heat from a warmer to a colder body, which may be regarded as the transformation of heat at a higher, into heat at a lower temperature. The theorem, as hitherto used, may be enunciated in some such manner as the following :—*In all cases where a quantity of heat is converted into work, and where the body effecting this transformation ultimately returns to its original condition, another quantity of heat must necessarily be transferred from a warmer to a colder body; and the magnitude of the last quantity of heat, in relation to the first, depends only upon the temperatures of the bodies between which heat passes, and not upon the nature of the body effecting the transformation.*

But soon he's deriving this from a more "everyday" statement of the Second Law (which, notably, is clearly not a "theorem" in any normal sense):

> This principle, upon which the whole of the following development rests, is as follows:—*Heat can never pass from a colder to a warmer body without some other change, connected therewith, occurring at the same time**. Everything we know concerning the interchange of heat between two bodies of different temperatures confirms this; for heat everywhere manifests a tendency to equalize existing differences of temperature, and therefore to pass in a contrary direction, *i. e.* from warmer to colder bodies. Without further explanation, therefore, the truth of the principle will be granted.

After giving a Carnot-style argument he's then got a new statement (that he calls "the theorem of the equivalence of transformations") of the Second Law:

> According to this, the second fundamental theorem in the mechanical theory of heat, which in this form might appropriately be called the *theorem of the equivalence of transformations*, may be thus enunciated:
>
> *If two transformations which, without necessitating any other permanent change, can mutually replace one another, be called equivalent, then the generation of the quantity of heat* Q *of the temperature* t *from work, has the equivalence-value*
>
> $$\frac{Q}{T},$$
>
> *and the passage of the quantity of heat* Q *from the temperature* $t_1$ *to the temperature* $t_2$, *has the equivalence-value*
>
> $$Q\left(\frac{1}{T_2}-\frac{1}{T_1}\right),$$
>
> *wherein* T *is a function of the temperature, independent of the nature of the process by which the transformation is effected.*

And there it is: basically what we now call entropy (even with the same notation of *Q* for heat and *T* for temperature)—together with the statement that this quantity is a function of state, so that its differences are "independent of the nature of the process by which the transformation is effected".

Pretty soon there's a familiar expression for entropy change:

> bodies; then the foregoing equation will assume the form
>
> $$N=\int\frac{dQ}{T}, \quad . \quad . \quad . \quad . \quad . \quad . \quad (11)$$
>
> wherein the integral extends over all the quantities of heat received by the several bodies.
>
> If the process is ***reversible,*** then, however complicated it may be, we can prove, as in the simple process before considered, ***that the transformations which occur must exactly cancel each other, so that their algebraical sum is zero.***

And by the next page he's giving what he describes as "the analytical expression" of the Second Law, for the particular case of reversible cyclic processes:

> Consequently the equation
>
> $$\int\frac{dQ}{T}=0 \quad . \quad . \quad . \quad . \quad . \quad . \quad . \quad (II)$$
>
> is the analytical expression, for all ***reversible cyclical processes,*** of the second fundamental theorem in the mechanical theory of heat.

A bit later he backs of out the assumption of reversibility, concluding that:

> ***The algebraical sum of all transformations occurring in a cyclical process can only be positive.***

(And, yes, with modern mathematical rigor, that should be "non-negative" rather than "positive".)

He goes on to say that if something has changed after going around a cycle, he'll call that an "uncompensated transformation"—or what we would now refer to as an irreversible change. He lists a few possible (now very familiar) examples:

> The different kinds of operations giving rise to uncompensated transformations are, as far as external appearances are concerned, rather numerous, even though they may not differ very essentially. One of the most frequently occurring examples is that of the transmission of heat by mere conduction, when two bodies of different temperatures are brought into immediate contact; other cases are the production of heat by friction, and by an electric current when overcoming the resistance due to imperfect conductibility, together with all cases where a force, in doing mechanical work, has not to overcome an equal resistance, and

Earlier in his paper he's careful to say that *T* is "a function of temperature"; he doesn't say it's actually the quantity we measure as temperature. But now he wants to determine what it is:

**In conclusion, we must direct our attention to the function T, which hitherto has been left quite undetermined; we shall not be able to determine it entirely without hypothesis, but by means of a very probable hypothesis it will be possible so to do.**

He doesn't talk about the ultimately critical assumption (effectively the Zeroth Law of thermodynamics) that the system is "in equilibrium", with a uniform temperature. But he uses an ideal gas as a kind of "standard material", and determines that, yes, in that case *T* can be simply the absolute temperature.

So there it is: in 1854 Clausius has effectively defined entropy and described its relation to the Second Law, though everything is being done in a very "heat–engine" style. And pretty soon he's writing about "Theory of the Steam–Engine" and filling actual approximate steam tables into his theoretical formulas:

**XXXIII.** ***On the Application of the Mechanical Theory of Heat to the Steam-engine.* By R. Clausius*.**

13. As indicated by equation (III), the above expression for $\frac{dQ}{dv}$ must be differentiated according to T, and the expression for $\frac{dQ}{dT}$ according to $v$. The magnitude M is constant, the magnitudes $u$, $\sigma$, $r$, $c$, and $h$ are all functions of T alone, and only the magnitude $m$ is a function of T and $v$, so that

$$\left.\begin{aligned}\frac{d}{dT}\left(\frac{dQ}{dv}\right)&=\frac{1}{u}\cdot\frac{dr}{dT}-\frac{r}{u^2}\cdot\frac{du}{dT}\\ \frac{d}{dv}\left(\frac{dQ}{dT}\right)&=\left(h-c-\frac{r}{u}\cdot\frac{du}{dT}\right)\frac{dm}{dv};\end{aligned}\right\} \quad . \quad (10)$$

or, substituting for $\frac{dm}{dv}$ its value $\frac{1}{u}$,

$$\frac{d}{dv}\left(\frac{dQ}{dT}\right)=\frac{h-c}{u}-\frac{r}{u^2}\cdot\frac{du}{dT}. \quad . \quad . \quad . \quad . \quad (11)$$

By substituting the expressions given in (10), (11), and (8) in (III) and (IV), we obtain the required equations, which represent the two principal theorems of the mechanical theory of heat as applied to vapours at their maximum density. These are

$$\frac{dr}{dT}+c-h=A\,.\,u\frac{dp}{dT}, \quad . \quad . \quad . \quad . \quad . \quad . \quad (V)$$

$$r=A\,.\,Tu\frac{dp}{dT}; \quad . \quad . \quad . \quad . \quad . \quad . \quad . \quad . \quad (VI)$$

and from a combination of both we have

$$\frac{dr}{dT}+c-h=\frac{r}{T}. \quad . \quad . \quad . \quad . \quad . \quad . \quad . \quad . \quad (12)$$

14. By means of these equations we will now treat a case, which in the following will so frequently occur, as to render it desirable at once to establish the results which have reference thereto.

Eliminating R from this equation by means of the former, we have the equation

$$(W)=\frac{W-(1-e)V\,.\,f}{1+\delta}, \quad . \quad . \quad . \quad . \quad (58)$$

by means of which, V being known, the useful work (W) can be deduced from the whole work W as soon as the quantities $f$ and $\delta$ are given.

I will not here enter into Pambour's method of finding the latter quantities, as this determination still rests upon a too insecure basis, and as friction is altogether foreign to the subject of this memoir.

Table containing the values, for steam, of $p$, its differential coefficient $\frac{dp}{dt}=g$, and the product $T\,.\,g$ expressed in millimetres of mercury.

| $t$ in Centigrade degrees. | $p$. | Δ. | $g$. | Δ. | T.$g$. | Δ. |
|---|---|---|---|---|---|---|
| 40° | 54·906 | 3·003 | 2·935 | 0·139 | 919 | 46 |
| 41 | 57·909 | 3·145 | 3·074 | 0·144 | 965 | 49 |
| 42 | 61·054 | 3·291 | 3·218 | 0·149 | 1014 | 50 |
| 43 | 64·345 | 3·444 | 3·367 | 0·155 | 1064 | 52 |
| 44 | 67·789 | 3·601 | 3·522 | 0·161 | 1116 | 55 |
| 45 | 71·390 | 3·766 | 3·683 | 0·167 | 1171 | 57 |
| 46 | 75·156 | 3·935 | 3·850 | 0·173 | 1228 | 59 |
| 47 | 79·091 | 4·112 | 4·023 | 0·180 | 1287 | 62 |
| 48 | 83·203 | 4·294 | 4·203 | 0·185 | 1349 | 64 |
| 49 | 87·497 | 4·483 | 4·388 | 0·193 | 1413 | 67 |
| 50 | 91·980 | 4·679 | 4·581 | 0·199 | 1480 | 69 |
| 51 | 96·659 | 4·882 | 4·780 | 0·207 | 1549 | 72 |
| 52 | 101·541 | 5·092 | 4·987 | 0·213 | 1621 | 74 |
| 53 | 106·633 | 5·309 | 5·200 | 0·221 | 1695 | 78 |
| 54 | 111·942 | 5·533 | 5·421 | 0·228 | 1773 | 80 |
| 55 | 117·475 | 5·766 | 5·649 | 0·237 | 1853 | 83 |
| 56 | 123·241 | 6·006 | 5·886 | 0·244 | 1936 | 87 |
| 57 | 129·247 | 6·254 | 6·130 | 0·252 | 2023 | 89 |
| 58 | 135·501 | 6·510 | 6·382 | 0·260 | 2112 | 93 |
| 59 | 142·011 | 6·775 | 6·642 | 0·269 | 2205 | 96 |
| 60 | 148·786 | 7·048 | 6·911 | 0·278 | 2301 | 100 |
| 61 | 155·834 | 7·330 | 7·189 | 0·286 | 2401 | 103 |
| 62 | 163·164 | 7·621 | 7·475 | 0·296 | 2504 | 107 |
| 63 | 170·785 | 7·922 | 7·771 | 0·305 | 2611 | 111 |
| 64 | 178·707 | 8·231 | 8·076 | 0·314 | 2722 | 114 |
| 65 | 186·938 | 8·550 | 8·390 | 0·325 | 2836 | 118 |
| 66 | 195·488 | 8·880 | 8·715 | 0·334 | 2954 | 123 |
| 67 | 204·368 | 9·218 | 9·049 | 0·344 | 3077 | 126 |
| 68 | 213·586 | 9·568 | 9·393 | 0·355 | 3203 | 131 |
| 69 | 223·154 | 9·928 | 9·748 | 0·365 | 3334 | 135 |
| 70 | 233·082 | 10·298 | 10·113 | 0·376 | 3469 | 139 |
| 71 | 243·380 | 10·680 | 10·489 | 0·387 | 3608 | 144 |
| 72 | 254·060 | 11·072 | 10·876 | 0·398 | 3752 | 149 |
| 73 | 265·132 | 11·476 | 11·274 | 0·410 | 3901 | 153 |
| 74 | 276·608 | 11·892 | 11·684 | 0·422 | 4054 | 159 |
| 75 | 288·500 | 12·320 | 12·106 | 0·433 | 4213 | 163 |
| 76 | 300·820 | 12·759 | 12·539 | 0·445 | 4376 | 168 |
| 77 | 313·579 | 13·210 | 12·984 | 0·458 | 4544 | 174 |
| 78 | 326·789 | 13·675 | 13·442 | 0·471 | 4718 | 179 |
| 79 | 340·464 | 14·152 | 13·913 | 0·484 | 4897 | 185 |
| 80 | 354·616 | 14·642 | 14·397 | 0·497 | 5082 | 190 |
| 81 | 369·258 | 15·146 | 14·894 | 0·511 | 5272 | 197 |
| 82 | 384·404 | 15·664 | 15·405 | 0·524 | 5469 | 202 |
| 83 | 400·068 | 16·194 | 15·929 | 0·538 | 5671 | 208 |
| 84 | 416·262 | 16·740 | 16·467 | 0·552 | 5879 | 214 |
| 85 | 433·002 | 17·299 | 17·019 | 0·577 | 6093 | 220 |
| 86 | 450·301 | 17·874 | 17·586 | 0·582 | 6313 | 227 |
| 87 | 468·175 | 18·463 | 18·168 | 0·597 | 6540 | 234 |
| 88 | 486·638 | | 18·765 | | 6774 | |

After a few years "off" (working, as we'll discuss later, on the kinetic theory of gases) Clausius is back in 1862 talking about the Second Law again, in terms of his "theorem of the equivalence of transformations":

> **XIII.** *On the Application of the Theorem of the Equivalence of Transformations to the Internal Work of a mass of Matter.* *By* Professor R. CLAUSIUS*.
>
> IN a memoir published in the year 1854†, wherein I sought to simplify to some extent the form of the developments I had previously published, I deduced, from my fundamental proposition *that heat cannot of itself pass from a colder into a warmer body*, a principle which is closely allied to, but does not entirely coincide with, the one first deduced by S. Carnot from considerations of a different class, based upon the older views of the nature of heat. It has reference to the circumstances under which work can be transformed into heat, and, conversely, heat converted into work; and I have called it the *Principle of the Equivalence of Transformations*. I did not, however, there communicate the entire proposition in the general form in which I had deduced it, but confined myself on that occasion to the publication of a part which can be treated separately from the rest, and is capable of more strict proof.

He's slightly tightened up his 1854 discussion, but, more importantly, he's now stating a result not just for reversible cyclic processes, but for general ones:

> The proposition respecting the equivalent values of the transformations may accordingly be stated thus:—*The algebraic sum of all the transformations occurring in a circular process can only be positive, or, as an extreme case, equal to nothing.*
>
> The mathematical expression for this proposition is as follows. Let $d$Q be an element of the heat given up by the body to any reservoir of heat during its modifications (heat which it may absorb from a reservoir being here reckoned as negative), and T the absolute temperature of the body at the moment of giving up this heat, then the equation
>
> $$\int \frac{dQ}{T} = 0 \quad . \quad . \quad . \quad . \quad . \quad . \quad . \quad (I.)$$
>
> must be true for every reversible circular process, and the relation
>
> $$\int \frac{dQ}{T} \geqq 0 \quad . \quad . \quad . \quad . \quad . \quad . \quad . \quad (Ia.)$$
>
> must hold good for every circular process which is in any way possible.

But what does this result really mean? Clausius claims that this "theorem admits of strict mathematical proof if we start from the fundamental proposition above quoted"—though it's not particularly clear just what that proposition is. But then he says he wants to find a "physical cause":

> § 2. Although the necessity of this proposition admits of strict mathematical proof if we start from the fundamental principle above quoted, it thereby nevertheless retains an abstract form, in which it is difficultly embraced by the mind, and we feel compelled to seek for the precise physical cause, of which this proposition is a consequence. Moreover, since there is no

A little earlier in the paper he said:

> I have delayed till now the publication of the remainder of my theorem, because it leads to a consequence which is considerably at variance with the ideas hitherto generally entertained of the heat contained in bodies, and I therefore thought it desirable to make still further trial of it. But as I have become more and more convinced in the course of years that we must not attach too great weight to such ideas, which in part are founded more upon usage than upon a scientific basis, I feel that I ought to hesitate no longer, but to submit to the scientific public the theorem of the equivalence of transformations in its complete form, with the principles which attach themselves to it. I venture to

So what does he think the "physical cause" is? He says that even from his first investigations he'd assumed a general law:

> *In all cases in which the heat contained in a body does mechanical work by overcoming a resistance, the magnitude of the resistance which it is capable of overcoming is proportional to the absolute temperature.*

What are these "resistances"? He's basically saying they are the forces between molecules in a material (which from his work on the kinetic theory of gases he now imagines exist):

> In order to understand the significance of this law, we require to consider more closely the processes by which heat can perform mechanical work. These processes always admit of being reduced to the alteration in some way or another of the arrangement of the constituent molecules of a body. For instance, bodies are expanded by heat, their molecules being thus separated from each other: in this case the mutual attractions of the molecules on the one hand, and on the other external opposing forces, in so far as any such are in operation, have to be overcome. Again,

He introduces what he calls the "disgregation" to represent the microscopic effect of adding heat:

> distances from one another. In order to be able to represent this mathematically, we will express the degree in which the molecules of a body are dispersed, by introducing a new magnitude, which we will call the *disgregation* of the body, and by help
>
> of which we can define the effect of heat as simply *tending to increase the disgregation.* The way in which a definite measure of

For ideal gases things are straightforward, including the proportionality of "resistance" to absolute temperature. But in other cases, it's not so clear what's going on. A decade later he identifies "disgregation" with average kinetic energy per molecule—which is indeed proportional to absolute temperature. But in 1862 it's all still quite muddy, with somewhat curious statements like:

> temperature, even water. To this it might perhaps be objected that, in other cases, the effect of increased temperature is to favour the union of two substances—that, for instance, hydrogen and oxygen do not combine at low temperatures, but do so easily at higher temperatures. I believe, however, that the heat exerts here only a secondary influence, contributing to bring the atoms into such relative positions that their inherent forces, by virtue of which they strive to unite, are able to come into operation. Heat itself can never, in my opinion, tend to produce combination, but only, and in every case, decomposition.

And then the main part of the paper ends with what seems to be an anticipation of the Third Law of thermodynamics:

> If we desired to cool a body down to the absolute zero of temperature, the corresponding alteration of disgregation, as shown by the foregoing formula, in which we should then have T=0, would be infinitely great. Herein lies a chief argument for supposing it to be impossible to produce such a degree of cold, by any alteration of the condition of a body, as to arrive at the absolute zero.

There's an appendix entitled "On Terminology" which admits that between Clausius's own work, and other people's, it's become rather difficult to follow what's going on. He agrees that the term "energy" that Kelvin is using makes sense. He suggests "energy of the body" for what he calls $U$ and we now call "internal energy". He suggests "heat of the body" or "thermal content of the body" for $Q$. But then he talks about the fact that these are measured in thermal units (say the amount of heat needed to increase the temperature of water by 1°), while mechanical work is measured in units related to kilograms and meters. He proposes therefore to introduce the concept of "ergon" for "work measured in thermal units":

> For this purpose, therefore, I will venture another proposition. Let heat and work continue to be measured each according to its most convenient unit, that is to say, heat according to the thermal unit, and work according to the mechanical one. But besides the work measured according to the mechanical unit, let another magnitude be introduced denoting *the work measured according to the thermal unit*, that is to say, *the numerical value of the work when the unit of work is that which is equivalent to the thermal unit.* For the work thus expressed a particular name is requisite. I propose to adopt for it the Greek word (ἔργον) *ergon**.

And pretty soon he's talking about the "interior ergon" and "exterior ergon", as well as concepts like "ergonized heat". (In later work he also tries to introduce the concept of "ergal" to go along with his development of what he called—in a name that did stick—the "virial theorem".)

But in 1865 he has his biggest success in introducing a term. He's writing a paper, he says, basically to clarify the Second Law, (or, as he calls it, "the second fundamental theorem"— rather confidently asserting that he will "prove this theorem"):

> ON SEVERAL CONVENIENT FORMS OF THE FUNDAMENTAL EQUATIONS OF THE MECHANICAL THEORY OF HEAT*.
>
> IN my former Memoirs on the Mechanical Theory of Heat, my chief object was to secure a firm basis for the theory, and I especially endeavoured to bring the second fundamental theorem, which is much more difficult to understand than the first, to its simplest and at the same time most general form, and to prove the necessary truth thereof. I have pursued special

Part of the issue he's trying to address is how the calculus is done:

> The more the mechanical theory of heat is acknowledged to be correct in its principles, the more frequently endeavours are made in physical and mechanical circles to apply it to different kinds of phenomena, and as the corresponding differential equations must be somewhat differently treated from the ordinarily occurring differential equations of similar forms, difficulties of calculation are frequently encountered which retard progress and occasion errors. Under these circumstances I believe I

The partial derivative symbol ∂ had been introduced in the late 1700s. He doesn't use it, but he does introduce the now-standard-in-thermodynamics subscript notation for variables that are kept constant:

> which is supposed to be constant during differentiation. Accordingly, we will write the two differential coefficients which denote the specific heat at constant volume, and the specific heat at constant pressure, in the following manner:—
>
> $$\left(\frac{dQ}{dT}\right)_v, \quad \text{and} \quad \left(\frac{dQ}{dT}\right)_p.$$

A little later, as part of the "notational cleanup", we see the variable *S*:

> present existing condition of the body, and not upon the way by which it reached the latter. Denoting this magnitude by S, we can write
>
> $$dS = \frac{dQ}{T}; \quad . \quad . \quad . \quad . \quad . \quad . \quad . \quad . \quad . \quad (59)$$
>
> or, if we conceive this equation to be integrated for any re-

And then—there it is—Clausius introduces the term "entropy", "Greekifying" his concept of "transformation":

> S is determined.
>
> We might call S the ***transformational content*** of the body, just as we termed the magnitude U its ***thermal and ergonal content***. But as I hold it to be better to borrow terms for important magnitudes from the ancient languages, so that they may be adopted unchanged in all modern languages, I propose to call the magnitude S the ***entropy*** of the body, from the Greek word τροπή, ***transformation***. I have intentionally formed the word ***entropy*** so as to be as similar as possible to the word ***energy***; for the two magnitudes to be denoted by these words are so nearly allied in their physical meanings, that a certain similarity in designation appears to be desirable.
>
> Before proceeding further, let us collect together, for the sake

His paper ends with his famous crisp statements of the First and Second Laws of thermodynamics—manifesting the parallelism he's been claiming between energy and entropy:

> ception of energy, we may express in the following manner the fundamental laws of the universe which correspond to the two fundamental theorems of the mechanical theory of heat.
>
> 1. ***The energy of the universe is constant.***
> 2. ***The entropy of the universe tends to a maximum.***

## The Kinetic Theory of Gases

We began above by discussing the history of the question of "What is heat?" Was it like a fluid—the caloric theory? Or was it something more dynamical, and in a sense more abstract? But then we saw how Carnot—followed by Kelvin and Clausius—managed in effect to sidestep the question, and come up with all sorts of "thermodynamic conclusions", by talking just about "what heat does" without ever really having to seriously address the question of "what heat is". But to be able to discuss the foundations of the Second Law—and what it says about heat—we have to know more about what heat actually is. And the crucial development that began to clarify the nature of heat was the kinetic theory of gases.

Central to the kinetic theory of gases is the idea that gases are made up of discrete molecules. And it's important to remember that it wasn't until the beginning of the 1900s that anyone knew for sure that molecules existed. Yes, something like them had been discussed ever since antiquity, and in the 1800s there was increasing "circumstantial evidence" for them. But nobody had directly "seen a molecule", or been able, for example, until about 1870, to even guess what the size of molecules might be. Still, by the mid-1800s it had become common for physicists to talk and reason in terms of ordinary matter at least effectively being made of up molecules.

But if a gas was made of molecules bouncing off each other like billiard balls according to the laws of mechanics, what would its overall properties be? Daniel Bernoulli had in 1738 already worked out the basic answer that pressure would vary inversely with volume, or in his notation, $\pi = P/s$ (and he even also gave formulas for molecules of nonzero size—in a precursor of van der Waals):

> Rationi D ad *d* aliam ſubſtituere poſſumus magis intelligibilem: nempe ſi putemus operculum EF pondere infinito depreſſum deſcendere usque in ſitum *mn*, in quo particulæ omnes ſe tangunt, atque lineam *m*C vocemus *m*, erit D ad *d* ut 1 ad $\sqrt[3]{m}$, quâ ratione ſubſtituta, erunt tandem vires aëris naturalis ECDF & compreſſi *e*CD*f* ut $s^{\frac{2}{3}} \times (\sqrt[3]{s} - \sqrt[3]{m})$ ad $1 - \sqrt[3]{m}$, ſeu ut $s - \sqrt[3]{mss}$ ad $1 - \sqrt[3]{m}$. Eſt igitur $\pi = \frac{1 - \sqrt[3]{m}}{s - \sqrt[3]{mss}} \times P$.
>
> §. 5. Ex omnibus phænomenis judicare poſſumus aërem naturalem admodum condenſari poſſe, & fere in ſpatiolum infinite parvum comprimi; facta igitur $m = o$, fit $\pi = \frac{P}{s}$, ita ut pondera comprimentia ſint fere in ratione inverſa ſpatiorum, quæ aër diverſimode compreſſus occupat;

Results like Bernouilli's would be rediscovered several times, for example in 1820 by John Herapath (1790–1868), a math teacher in England, who developed a fairly elaborate theory that purported to describe gravity as well as heat (but for example implied a $P\,V = a\,T^2$ gas law):

> *A Mathematical Inquiry into the Causes, Laws, and principal Phænomena of Heat, Gases, Gravitation, &c.* By John Herapath, Esq. (In a Letter to D. Gilbert, Esq. MP. VPRS. &c.)

Then there was the case of John Waterston (1811–1883), a naval instructor for the East India company, who in 1843 published a book called *Thoughts on the Mental Functions*, which included results on what he called the "*vis viva* theory of heat"—that he developed in more detail in a paper he wrote in 1846. But when he submitted the paper to the Royal Society it was rejected as "nonsense", and its manuscript was "lost" until 1891 when it was finally published (with an "explanation" of the "delay"):

> I. *On the Physics of Media that are Composed of Free and Perfectly Elastic Molecules in a State of Motion.*
>
> *By* J. J. WATERSTON.
>
> *Communicated by Captain* BEAUFORT, *R.N., F.R.S., &c.*
>
> Received December 11, 1845,—Read March 5, 1846.
>
> [PLATES 1, 2.]
>
> *Introduction by Lord* RAYLEIGH, *Sec.R.S.*
>
> THE publication of this paper after nearly half a century demands a word of explanation; and the opportunity may be taken to point out in what respects the

The paper had included a perfectly sensible mathematical analysis that included a derivation of the kinetic theory relation between pressure and mean-square molecular velocity:

This enables us conveniently to represent the relation between the density and the square root of the mean square molecular velocity of a medium while it is being dilated or compressed.

Fig. 3.

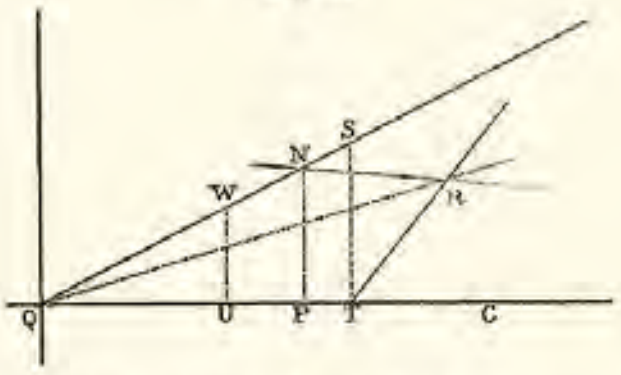

Take Q as the origin of co-ordinates, and let QP represent $v$ and PN the $\sqrt{\frac{1}{\Delta}}$. Join NQ. If the medium is compressed so that $\sqrt{\frac{1}{\Delta}}$ becomes TS, then shall $v$ become QT, and if it dilates so that $\sqrt{\frac{1}{\Delta}}$ becomes UW, then shall $v$ become QU.

throwing light upon the attitude then assumed by men of science in regard to this question, and in order to point a moral. The history of this paper suggests that highly speculative investigations, especially by an unknown author, are best brought before the world through some other channel than a scientific society, which naturally hesitates to admit into its printed records matter of uncertain value. Perhaps one

But with all these pieces of work unknown, it fell to a German high-school chemistry teacher (and sometime professor and philosophical/theological writer) named August Krönig (1822–1879) to publish in 1856 yet another "rediscovery", that he entitled "Principles of a Theory of Gases". He said it was going to analyze the "mechanical theory of heat", and once again he wanted to compute the pressure associated with colliding molecules. But to simplify the math, he assumed that molecules went only along the coordinate directions, at a fixed speed—almost anticipating a cellular automaton fluid:

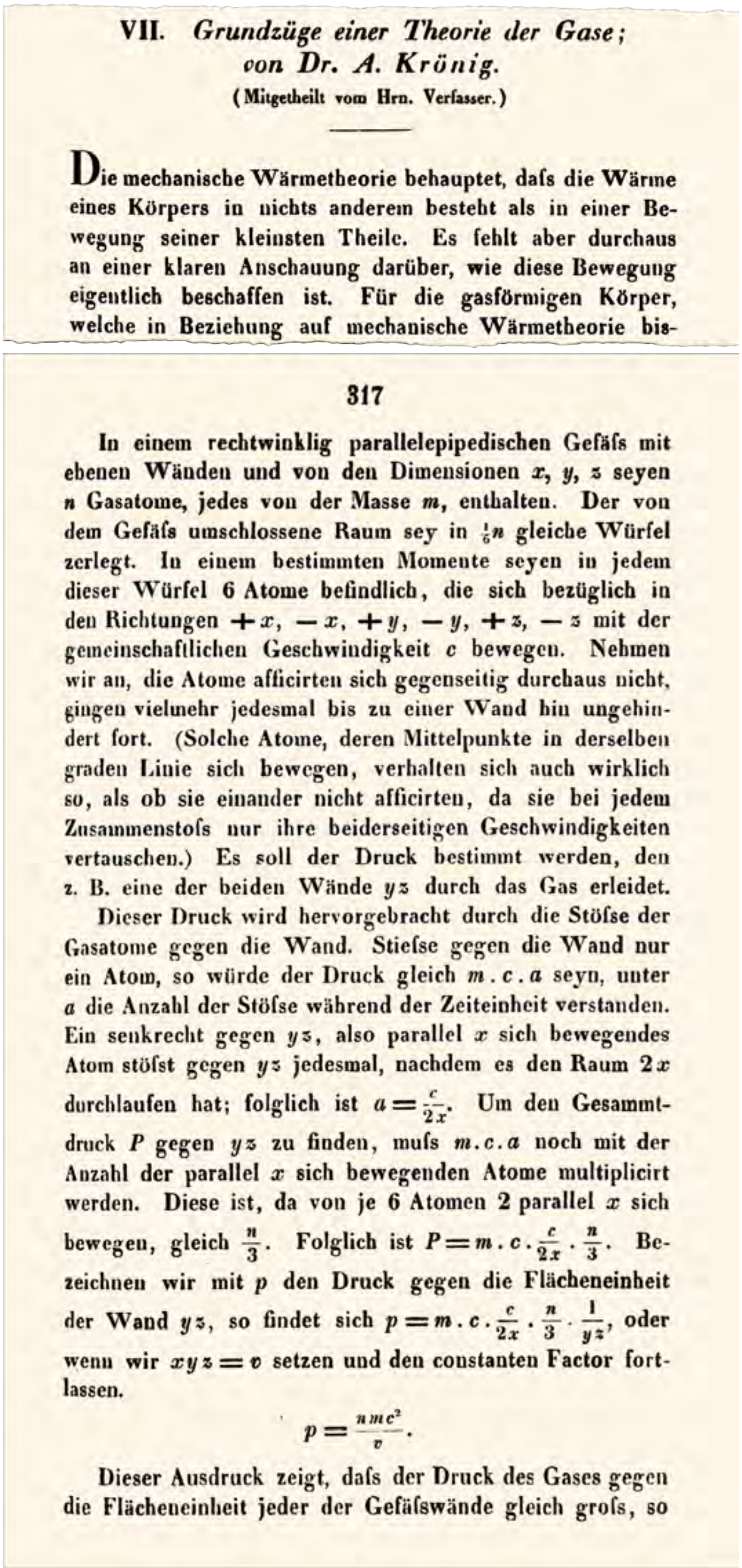

**VII.** ***Grundzüge einer Theorie der Gase; von Dr. A. Krönig.***

**(Mitgetheilt vom Hrn. Verfasser.)**

Die mechanische Wärmetheorie behauptet, dafs die Wärme eines Körpers in nichts anderem besteht als in einer Bewegung seiner kleinsten Theile. Es fehlt aber durchaus an einer klaren Anschauung darüber, wie diese Bewegung eigentlich beschaffen ist. Für die gasförmigen Körper, welche in Beziehung auf mechanische Wärmetheorie bis-

317

In einem rechtwinklig parallelepipedischen Gefäfs mit ebenen Wänden und von den Dimensionen $x$, $y$, $z$ seyen $n$ Gasatome, jedes von der Masse $m$, enthalten. Der von dem Gefäfs umschlossene Raum sey in $\frac{1}{6}n$ gleiche Würfel zerlegt. In einem bestimmten Momente seyen in jedem dieser Würfel 6 Atome befindlich, die sich bezüglich in den Richtungen $+x$, $-x$, $+y$, $-y$, $+z$, $-z$ mit der gemeinschaftlichen Geschwindigkeit $c$ bewegen. Nehmen wir an, die Atome afficirten sich gegenseitig durchaus nicht, gingen vielmehr jedesmal bis zu einer Wand hin ungehindert fort. (Solche Atome, deren Mittelpunkte in derselben graden Linie sich bewegen, verhalten sich auch wirklich so, als ob sie einander nicht afficirten, da sie bei jedem Zusammenstofs nur ihre beiderseitigen Geschwindigkeiten vertauschen.) Es soll der Druck bestimmt werden, den z. B. eine der beiden Wände $yz$ durch das Gas erleidet.

Dieser Druck wird hervorgebracht durch die Stöfse der Gasatome gegen die Wand. Stiefse gegen die Wand nur ein Atom, so würde der Druck gleich $m.c.a$ seyn, unter $a$ die Anzahl der Stöfse während der Zeiteinheit verstanden. Ein senkrecht gegen $yz$, also parallel $x$ sich bewegendes Atom stöfst gegen $yz$ jedesmal, nachdem es den Raum $2x$ durchlaufen hat; folglich ist $a=\frac{c}{2x}$. Um den Gesammtdruck $P$ gegen $yz$ zu finden, mufs $m.c.a$ noch mit der Anzahl der parallel $x$ sich bewegenden Atome multiplicirt werden. Diese ist, da von je 6 Atomen 2 parallel $x$ sich bewegen, gleich $\frac{n}{3}$. Folglich ist $P=m.c.\frac{c}{2x}.\frac{n}{3}$. Bezeichnen wir mit $p$ den Druck gegen die Flächeneinheit der Wand $yz$, so findet sich $p=m.c.\frac{c}{2x}.\frac{n}{3}.\frac{1}{yz}$, oder wenn wir $xyz=v$ setzen und den constanten Factor fortlassen.

$$p=\frac{nmc^2}{v}.$$

Dieser Ausdruck zeigt, dafs der Druck des Gases gegen die Flächeneinheit jeder der Gefäfswände gleich grofs, so

What ultimately launched the subsequent development of the kinetic theory of gases, however, was the 1857 publication by Rudolf Clausius (by then an increasingly established

German physics professor) of a paper entitled rather poetically "On the Nature of the Motion Which We Call Heat" ("Über die Art der Bewegung die wir Wärme nennen"):

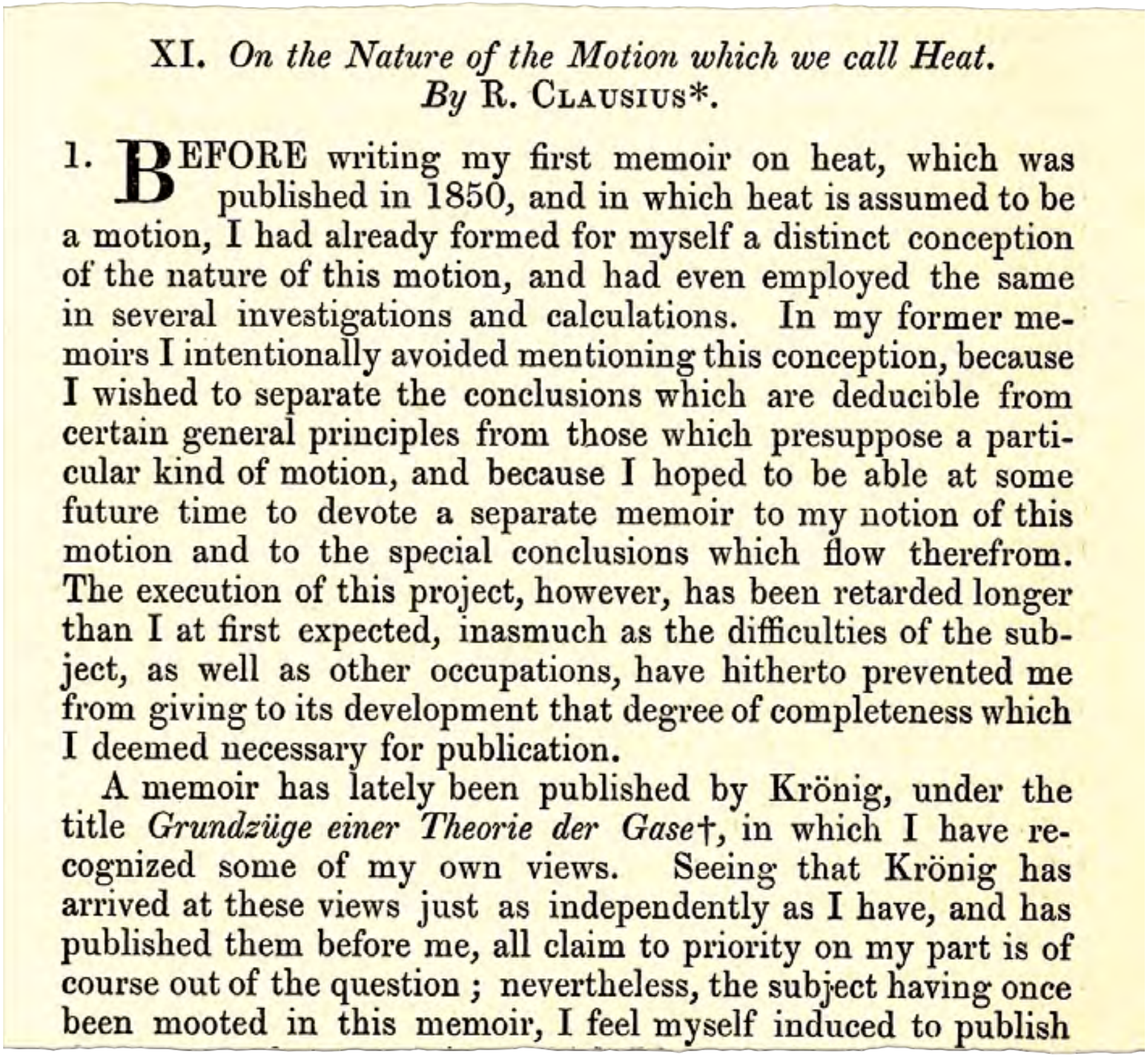

XI. *On the Nature of the Motion which we call Heat.*
*By* R. CLAUSIUS*.

1. BEFORE writing my first memoir on heat, which was published in 1850, and in which heat is assumed to be a motion, I had already formed for myself a distinct conception of the nature of this motion, and had even employed the same in several investigations and calculations. In my former memoirs I intentionally avoided mentioning this conception, because I wished to separate the conclusions which are deducible from certain general principles from those which presuppose a particular kind of motion, and because I hoped to be able at some future time to devote a separate memoir to my notion of this motion and to the special conclusions which flow therefrom. The execution of this project, however, has been retarded longer than I at first expected, inasmuch as the difficulties of the subject, as well as other occupations, have hitherto prevented me from giving to its development that degree of completeness which I deemed necessary for publication.

A memoir has lately been published by Krönig, under the title *Grundzüge einer Theorie der Gase*†, in which I have recognized some of my own views. Seeing that Krönig has arrived at these views just as independently as I have, and has published them before me, all claim to priority on my part is of course out of the question; nevertheless, the subject having once been mooted in this memoir, I feel myself induced to publish

It's a clean and clear paper, with none of the mathematical muddiness around Clausius's work on the Second Law (which, by the way, isn't even mentioned in this paper even though Clausius had recently worked on it). Clausius figures out lots of the "obvious" implications of his molecular theory, outlining for example what happens in different phases of matter:

In the *solid* state, the motion is such that the molecules move about certain positions of equilibrium without ever forsaking the same, unless acted upon by foreign forces. In solid bodies, therefore, the motion may be characterized as a vibrating one,

In the *liquid* state the molecules have no longer any definite position of equilibrium. They can turn completely around their centres of gravity; and the latter, too, may be moved completely out of its place. The separating action of the motion is not, however, sufficiently strong, in comparison to the mutual attraction between the molecules, to be able to separate the latter entirely. Although a molecule no longer adheres to definite neigh-

Lastly, in the *gaseous* state the motion of the molecules entirely transports them beyond the spheres of their mutual attraction, causing them to recede in right lines according to the ordinary laws of motion. If two such molecules come into collision during their motion, they will in general fly asunder again with the same vehemence with which they moved towards each other;

It takes him only a couple of pages of very light mathematics to derive the standard kinetic theory formula for the pressure of an ideal gas:

*which we call Heat.* 121

it would be reflected, and then commence a similar series of journeys to and fro, and so forth.

Lastly, there is no doubt that actually the greatest possible variety exists amongst the velocities of the several molecules. In our considerations, however, we may ascribe a certain mean velocity to all molecules. It will be evident from the following formulæ, that, in order to maintain an equal pressure, this mean velocity must be so chosen that with it the total *vis viva* of all the molecules may be the same as that corresponding to their actual velocities.

16. According to these assumptions, it is evident, that, during the unit of time, each molecule will strike the side under consideration just as often as during that time it can, by following its peculiar direction, travel from the side in question to the other and back again. Let $h$ be the distance between the large parallel sides, and $\vartheta$ the acute angle between the normal and the direction of motion; then $\frac{h}{\cos\vartheta}$ is the length of the path from one side to the other, and

$$\frac{u.\cos\vartheta}{2h} \quad . \quad . \quad . \quad . \quad . \quad . \quad . \quad . \quad (1)$$

the number of impulses given to the side, $u$ being the velocity of the molecule.

With respect to the directions of the several molecules, we must assume that on the average each direction is equally represented. From this it follows, that the number of molecules moving in directions which form with the normal angles included between $\vartheta$ and $\vartheta+d\vartheta$, has to the whole number of molecules the same ratio that the surface of the spherical zone, whose limiting circles correspond to the angles $\vartheta$ and $d\vartheta$, has to the surface of the hemisphere, that is, the ratio

$$2\pi \sin\vartheta d\vartheta : 2\pi.$$

Hence if $n$ represents the whole number of molecules, the number which corresponds to the angular interval between $\vartheta$ and $\vartheta+d\vartheta$ will be

$$n \sin\vartheta d\vartheta,$$

and the number of shocks imparted by them will be

$$\frac{nu}{2h}\cos\vartheta \sin\vartheta d\vartheta. \quad . \quad . \quad . \quad . \quad . \quad . \quad (2)$$

17. In order to determine the intensity of a shock, the whole velocity must be resolved into two components, one parallel and the other perpendicular to the side. Of these components, the first will not be affected by the shock, and will not enter into

122 Prof. Clausius *on the Nature of the Motion*

consideration in determining its intensity; the second, however, whose magnitude is represented by $u\cos\vartheta$, will be changed by the shock into an equal velocity in the opposite direction. The action of the side upon the molecule, therefore, consists in depriving it in one direction of the velocity $u\cos\vartheta$, calculated according to the normal, and of imparting to it an equal velocity in an opposite direction; in other words, of imparting to it a velocity of $2u.\cos\vartheta$ in the latter direction. Hence the quantity of motion imparted to the molecule will be

$$2mu.\cos\vartheta, \quad . \quad . \quad . \quad . \quad . \quad . \quad . \quad . \quad (3)$$

where $m$ is the mass of the molecule.

Applying this to all molecules which correspond to the interval between $\vartheta$ and $\vartheta+d\vartheta$, we obtain during the unit of time,

$$\frac{nu}{2n}\cos\vartheta \sin\vartheta d\vartheta$$

times the same action, hence the quantity of motion imparted to these molecules during the unit of time is

$$\frac{nmu^2}{h}\cos^2\vartheta . \sin\vartheta . d\vartheta. \quad . \quad . \quad . \quad . \quad . \quad . \quad (4)$$

Integrating this expression between the limits $\vartheta=0$ and $\vartheta=\frac{\pi}{2}$, we find the motion imparted by the side to all the molecules which strike against it during the unit of time to be

$$\frac{nmu^2}{3h}. \quad . \quad . \quad . \quad . \quad . \quad . \quad . \quad . \quad . \quad (5)$$

Let us now conceive the side to be capable of moving freely; then in order that it may not recede before the shocks of the molecules, it must be acted upon on the other side by a counter force, which latter may in fact be regarded as continuous, in consequence of the great number of shocks and the feebleness of each. The intensity of this force must be such as to enable it, during the unit of time, to generate the quantity of motion represented by the above expression. Since all forces, however, are measured by the quantity of motion they can produce in the unit of time, the above expression at once represents this force as well as the pressure exerted by the gas, the latter being equilibrated by the former.

If $a$ be the superficial area of the side, and $p$ the pressure on the unit of surface, then

$$p=\frac{mnu^2}{3ah}.$$

The product $ah$ here involved gives the volume of the vessel or

gas; hence representing the same by $v$, we have

$$p=\frac{mnu^2}{3v}. \quad . \quad . \quad . \quad . \quad . \quad . \quad . \quad . \quad (6)$$

He's implicitly assuming a certain randomness to the motions of the molecules, but he barely mentions it (and this particular formula is robust enough that average values are actually all that matter):

Let us consider one only of the two large sides; during the unit of time it is struck a certain number of times by molecules moving in all possible directions compatible with an approach towards the surface. We must first determine the number of such shocks, and how many correspond on the average to each direction.

Lastly, there is no doubt that actually the greatest possible variety exists amongst the velocities of the several molecules. In our considerations, however, we may ascribe a certain mean velocity to all molecules. It will be evident from the following for-

But having derived the formula for pressure, he goes on to use the ideal gas law to derive the relation between average molecular kinetic energy (which he still calls "*vis viva*") and absolute temperature:

*motion of the molecules**. But, according to Mariotte's and Gay-Lussac's laws,

$$pv = \text{T} \cdot \text{const.},$$

where T is the absolute temperature; hence

$$\frac{nmu^2}{2} = \text{T} \cdot \text{const.};$$

and, as before stated, the *vis viva* of the translatory motion is proportional to the absolute temperature.

From this he can do things like work out the actual average velocities of molecules in different gases—which he does without any mention of the question of just how real or not molecules might be. By knowing experimental results about specific heats of gases he also manages to determine that not all the energy ("heat") in a gas is associated with "translatory motion": he realizes that for molecules involving several atoms there can be energy associated with other (as we would now say) internal degrees of freedom:

Thus is corroborated what was before stated, that the *vis viva* of the translatory motion does not alone represent the whole quantity of heat in the gas, and that the difference is greater the greater the number of atoms of which the several molecules of the combination consist. We must conclude, therefore, that besides the translatory motion of the molecules as such, the constituents of these molecules perform other motions, whose *vis viva* also forms a part of the contained quantity of heat.

Clausius's paper was widely read. And it didn't take long before the Dutch meteorologist

(and effectively founder of the World Meteorological Organization) Christophorus Buys Ballot (1817–1890) asked why—if molecules were moving as quickly as Clausius suggested—gases didn't mix much more quickly than they're observed to do:

> X. *On the Mean Length of the Paths described by the separate Molecules of Gaseous Bodies on the occurrence of Molecular Motion: together with some other Remarks upon the Mechanical Theory of Heat.* *By* R. Clausius*.
>
> (1.) THE February Number of Poggendorff's *Annalen* contains a paper by Buijs-Ballot "On the Nature of the Motion which we call Heat and Electricity." Amongst the objections which the author there makes against the views advanced by Joule, Krönig, and myself concerning the molecular motion of gaseous bodies, the following deserves especial consideration. Attention is drawn to the circumstance that, if the molecules moved in straight lines, volumes of gases in contact would necessarily speedily mix with one another,—a result which does not actually take place. To prove that such mixture does

Within a few months, Clausius published the answer: the molecules didn't just keep moving in straight lines; they were constantly being deflected, to follow what we would now call a random walk. He invented the concept of a mean free path to describe how far on average a molecule goes before it hits another molecule:

> (4.) If, now, in a given space, we imagine a great number of molecules moving irregularly about amongst one another, and if we select one of them to watch, such a one would ever and anon impinge upon one of the other molecules, and bound off from it. We have now, therefore, to solve the question as to how great is the mean length of the path between two such impacts; or more exactly expressed, *how far on an average can the molecule move, before its centre of gravity comes into the sphere of action of another molecule.*

As a capable theoretical physicist, Clausius quickly brings in the concept of probability

> If, now, a point moves through this space in a straight line, let us suppose the space to be divided into parallel layers perpendicular to the motion of the point, and let us determine *how great is the probability that the point will pass freely through a layer of the thickness* x *without encountering the sphere of action of a molecule.*

and is soon computing the average number of molecules which will survive undeflected for a certain distance:

> According to equation (5), the number of points which either reach or pass the distance $x$ from the commencement of the motion is represented by
>
> $$Ne^{-\frac{\pi\rho^2}{\lambda^3}x};$$

Then he works out the mean free path $\lambda$ (and it's often still called $\lambda$):

> put:—*The mean length of path of a molecule is in the same pro*

> *portion to the radius of the sphere of action as the entire space occupied by the gas, to that portion of the space which is actually filled up by the spheres of action of the molecules.*

And he concludes that actually there's no conflict between rapid microscopic motion and large-scale "diffusive" motion:

> one is present, it is easy to convince oneself that the theory which explains the expansive force of gases does not lead to the conclusion that two quantities of gas bounding one another must mix with one another quickly and violently, but that only a comparatively small number of atoms can arrive quickly at a great distance, while the chief quantities only gradually mix at the surface of their contact.

Of course, he could have actually drawn a sample random walk, but drawing diagrams wasn't his style. And in fact it seems as if the first published drawing of a random walk was something added by John Venn (1834–1923) in the 1888 edition of his *Logic of Chance*—and, interestingly, in alignment with my computational irreducibility concept from a century later he used the digits of $\pi$ to generate his "randomness":

THE

LOGIC OF CHANCE

AN ESSAY

ON THE FOUNDATIONS AND PROVINCE OF THE THEORY OF PROBABILITY,

WITH ESPECIAL REFERENCE TO ITS LOGICAL BEARINGS AND ITS APPLICATION TO

MORAL AND SOCIAL SCIENCE, AND TO STATISTICS,

BY

JOHN VENN, Sc.D., F.R.S.,

FELLOW AND LECTURER IN THE MORAL SCIENCES, GONVILLE AND CAIUS COLLEGE, CAMBRIDGE.

LATE EXAMINER IN LOGIC AND MORAL PHILOSOPHY IN THE UNIVERSITY OF LONDON.

"So careful of the type she seems
So careless of the single life."

*THIRD EDITION, RE-WRITTEN AND ENLARGED.*

London:
MACMILLAN AND CO.
AND NEW YORK
1888

*[All Rights reserved.]*

118 *Randomness and its scientific treatment.* [CHAP. V.

character of the rows of figures displayed by the incommensurable or irrational ratios in question.

As it may interest the reader to see an actual specimen of such a path I append one representing the arrangement of the eight digits from 0 to 7 in the value of $\pi$. The data are taken from Mr Shanks' astonishing performance in the calculation of this constant to 707 places of figures (*Proc. of R. S.*, XXI. p. 319). Of these, after omitting 8 and 9, there remain 568; the diagram represents the course traced out by following the direction of these as the clue to our path. Many of the steps have of course been taken in opposite directions twice or oftener. The result seems to me to furnish a very fair graphical indication of randomness. I have compared it with corresponding paths furnished by rows of figures taken from logarithmic tables, and in other ways, and find the results to be much the same.

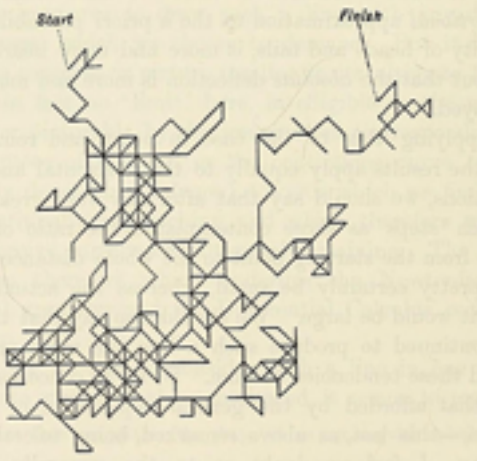

In 1859, Clausius's paper came to the attention of the then-28-year-old James Clerk Maxwell, who had grown up in Scotland, done the mathematical tripos in Cambridge, and was now back in Scotland as professor of "natural philosophy" at Aberdeen. Maxwell had already worked on things like elasticity theory, color vision, the mechanics of tops, the dynamics of the rings of Saturn and electromagnetism—having published his first paper (on geometry) at age 15. And, by the way, Maxwell was quite a "diagrammist"—and his early papers include all sorts of pictures that he drew:

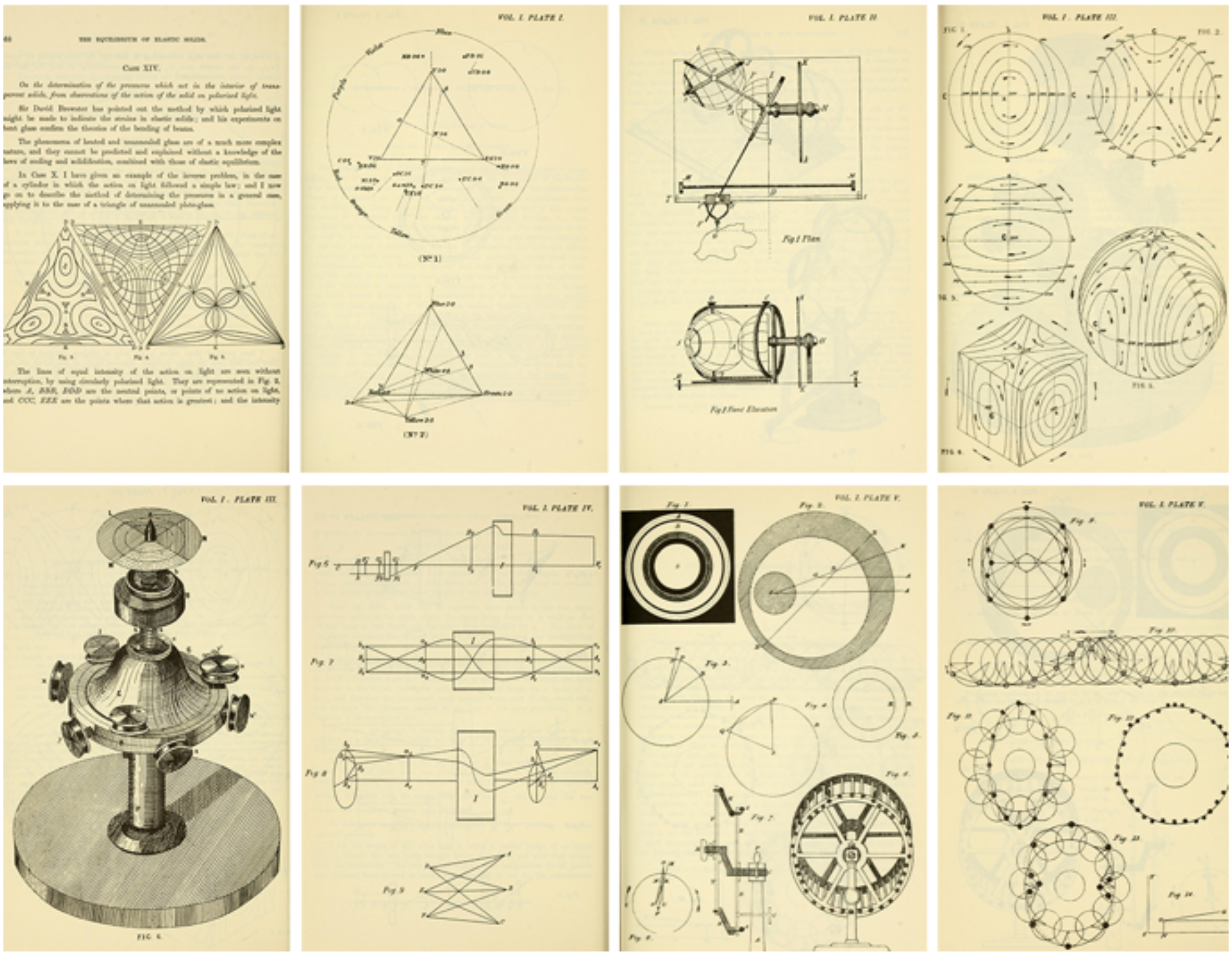

But in 1859 Maxwell applied his talents to what he called the "dynamical theory of gases":

V. *Illustrations of the Dynamical Theory of Gases.*—Part I. *On the Motions and Collisions of Perfectly Elastic Spheres. By* J. C. Maxwell, *M.A., Professor of Natural Philosophy in Marischal College and University of Aberdeen**.

So many of the properties of matter, especially when in the gaseous form, can be deduced from the hypothesis that their minute parts are in rapid motion, the velocity increasing with the temperature, that the precise nature of this motion becomes a subject of rational curiosity. Daniel Bernouilli, Herapath, Joule, Krönig, Clausius, &c. have shown that the relations between pressure, temperature, and density in a perfect gas can be explained by supposing the particles to move with uniform velocity in straight lines, striking against the sides of the containing vessel and thus producing pressure. It is not necessary to suppose each particle to travel to any great distance in the same straight line; for the effect in producing pressure will be the same if the particles strike against each other; so that the straight line described may be very short. M. Clausius has determined the mean length of path in terms of the average distance

* Communicated by the Author, having been read at the Meeting of the British Association at Aberdeen, September 21, 1859.

He models molecules as hard spheres, and sets about computing the "statistical" results of their collisions:

so that the probability is independent of $\phi$, that is, all directions of rebound are equally likely.

Prop. III. Given the direction and magnitude of the velocities of two spheres before impact, and the line of centres at impact; to find the velocities after impact.

Let O A, O B represent the velocities before impact, so that if there had been no action between the bodies they would have been at A and B at the end of a second. Join A B, and let G be their

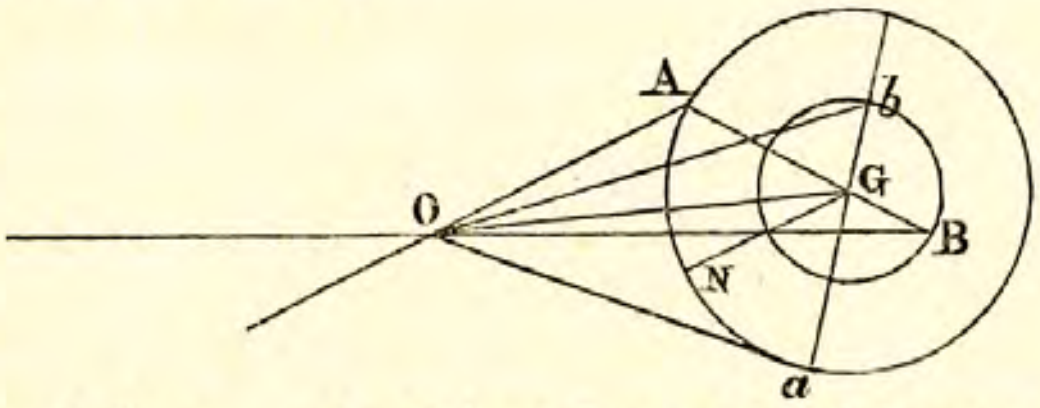

And pretty soon he's trying to compute distribution of their velocities:

Prop. IV. To find the average number of particles whose velocities lie between given limits, after a great number of collisions among a great number of equal particles.

Let N be the whole number of particles. Let $x$, $y$, $z$ be the components of the velocity of each particle in three rectangular directions, and let the number of particles for which $x$ lies between $x$ and $x+dx$ be $Nf(x)dx$, where $f(x)$ is a function of $x$ to be determined.

The number of particles for which $y$ lies between $y$ and $y+dy$ will be $Nf(y)dy$; and the number for which $z$ lies between $z$ and $z+dz$ will be $Nf(z)dz$, where $f$ always stands for the same function.

Now the existence of the velocity $x$ does not in any way affect that of the velocities $y$ or $z$, since these are all at right angles to each other and independent, so that the number of particles whose velocity lies between $x$ and $x+dx$, and also between $y$ and $y+dy$, and also between $z$ and $z+dz$, is

$$Nf(x)\,f(y)\,f(z)\,dx\,dy\,dz.$$

If we suppose the N particles to start from the origin at the same instant, then this will be the number in the element of volume $(dx\,dy\,dz)$ after unit of time, and the number referred to unit of volume will be

$$Nf(x)\,f(y)\,f(z).$$

But the directions of the coordinates are perfectly arbitrary, and therefore this number must depend on the distance from the origin alone, that is

$$f(x)\,f(y)\,f(z)=\phi(x^2+y^2+z^2).$$

Solving this functional equation, we find

$$f(x)=Ce^{Ax^2}, \quad \phi(r^2)=C^3e^{Ar^2}.$$

If we make A positive, the number of particles will increase with the velocity, and we should find the whole number of particles infinite. We therefore make A negative and equal to $-\frac{1}{\alpha^2}$, so that the number between $x$ and $x+dx$ is

$$NCe^{-\frac{x^2}{\alpha^2}}\,dx.$$

It's a somewhat unconvincing (or, as Maxwell himself later put it, "precarious") derivation (how does it work in 1D, for example?), but somehow it manages to produce what's now known as the Maxwell distribution:

> 2nd. The number whose actual velocity lies between $v$ and $v+dv$ is
>
> $$N\frac{4}{\alpha^3\sqrt{\pi}}v^2e^{-\frac{v^2}{\alpha^2}}dv. \quad . \quad . \quad . \quad . \quad . \quad . \quad (2)$$

Maxwell observes that the distribution is the same as for “errors ... in the ‘method of least squares’”:

> It appears from this proposition that the velocities are distributed among the particles according to the same law as the errors are distributed among the observations in the theory of the “method of least squares.” The velocities range from 0 to ∞, but the number of those having great velocities is comparatively small. In addition to these velocities, which are in all

Maxwell didn’t get back to the dynamical theory of gases until 1866, but in the meantime he was making a “dynamical theory” of something else: what he called the electromagnetic field:

> (3) The theory I propose may therefore be called a theory of the *Electromagnetic Field*, because it has to do with the space in the neighbourhood of the electric or magnetic bodies, and it may be called a *Dynamical* Theory, because it assumes that in that space there is matter in motion, by which the observed electromagnetic phenomena are produced.

Even though he’d worked extensively with the inverse square law of gravity he didn’t like the idea of “action at a distance”, and for example he wanted magnetic field lines to have some underlying “material” manifestation

> XXV. *On Physical Lines of Force.* *By* J. C. MAXWELL, *Professor of Natural Philosophy in King’s College, London**.
>
> PART I.—*The Theory of Molecular Vortices applied to Magnetic Phenomena.*

imagining that they might be associated with arrays of “molecular vortices”:

*Fig: 2.*

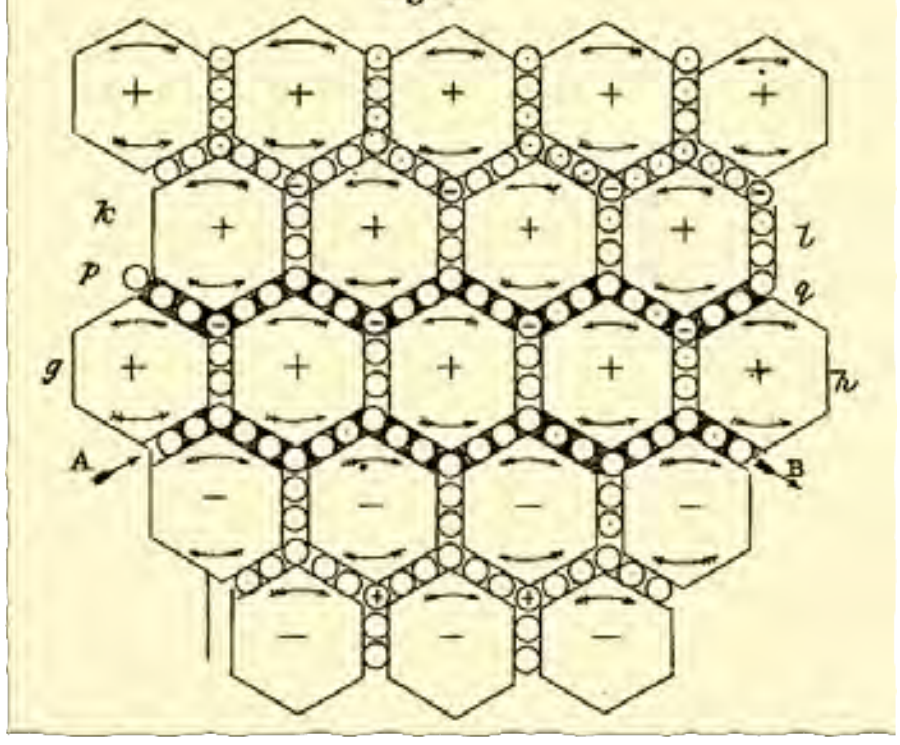

We now know, of course, that there isn't this kind of "underlying mechanics" for the electromagnetic field. But—with shades of the story of Carnot—even though the underlying framework isn't right, Maxwell successfully derives correct equations for the electromagnetic field—that are now known as Maxwell's equations:

Between these twenty quantities we have found twenty equations, viz.

| | | |
|---|---|---|
| Three equations of Magnetic Force | . . . . . . . . . . | (B) |
| „ Electric Currents | . . . . . . . . . . | (C) |
| „ Electromotive Force | . . . . . . . . . | (D) |
| „ Electric Elasticity | . . . . . . . . . | (E) |
| „ Electric Resistance | . . . . . . . . . | (F) |
| „ Total Currents | . . . . . . . . . . | (A) |
| One equation of Free Electricity | . . . . . . . . . . | (G) |
| „ Continuity | . . . . . . . . . . . . | (H) |

These equations are therefore sufficient to determine all the quantities which occur in them, provided we know the conditions of the problem. In many questions, however, only a few of the equations are required.

His statement of how the electromagnetic field "works" is highly reminiscent of the dynamical theory of gases:

(74) In speaking of the Energy of the field, however, I wish to be understood literally. All energy is the same as mechanical energy, whether it exists in the form of motion or in that of elasticity, or in any other form. The energy in electromagnetic phenomena is mechanical energy. The only question is, Where does it reside? On the old theories it resides in the electrified bodies, conducting circuits, and magnets, in the form of an unknown quality called potential energy, or the power of producing certain effects at a distance. On our theory it resides in the electromagnetic field, in the space surrounding the electrified and magnetic bodies, as well as in those bodies themselves, and is in two different forms, which may be described without hypothesis as magnetic polarization and electric polarization, or, according to a very probable hypothesis, as the motion and the strain of one and the same medium.

But he quickly and correctly adds:

(75) The conclusions arrived at in the present paper are independent of this hypothesis, being deduced from experimental facts of three kinds:—

And a few sections later he derives the idea of general electromagnetic waves

This wave consists entirely of magnetic disturbances, the direction of magnetization being in the plane of the wave. No magnetic disturbance whose direction of magnetization is not in the plane of the wave can be propagated as a plane wave at all.

Hence magnetic disturbances propagated through the electromagnetic field agree with light in this, that the disturbance at any point is transverse to the direction of propagation, and such waves may have all the properties of polarized light.

noting that there's no evidence that the medium through which he assumes they're propagating has elasticity:

> (100) The equations of the electromagnetic field, deduced from purely experimental evidence, show that transversal vibrations only can be propagated. If we were to go beyond our experimental knowledge and to assign a definite density to a substance which
>
> we should call the electric fluid, and select either vitreous or resinous electricity as the representative of that fluid, then we might have normal vibrations propagated with a velocity depending on this density. We have, however, no evidence as to the density of electricity, as we do not even know whether to consider vitreous electricity as a substance or as the absence of a substance.

By the way, when it comes to gravity he can't figure out how to make his idea of a "mechanical medium" work:

> The assumption, therefore, that gravitation arises from the action of the surrounding medium in the way pointed out, leads to the conclusion that every part of this medium possesses, when undisturbed, an enormous intrinsic energy, and that the presence of dense bodies influences the medium so as to diminish this energy wherever there is a resultant attraction.
>
> As I am unable to understand in what way a medium can possess such properties, I cannot go any further in this direction in searching for the cause of gravitation.

But in any case, after using it as an inspiration for thinking about electromagnetism, Maxwell in 1866 returns to the actual dynamical theory of gases, still feeling that he needs to justify looking at a molecular theory:

> IV. *On the Dynamical Theory of Gases.* *By* J. Clerk Maxwell, *F.R.S. L. & E*
>
> Received May 16,—Read May 31, 1866.
>
> Theories of the constitution of bodies suppose them either to be continuous and homogeneous, or to be composed of a finite number of distinct particles or molecules.
>
> In certain applications of mathematics to physical questions, it is convenient to suppose bodies homogeneous in order to make the quantity of matter in each differential element a function of the coordinates, but I am not aware that any theory of this kind has been proposed to account for the different properties of bodies. Indeed the properties of a body supposed to be a uniform *plenum* may be affirmed dogmatically, but cannot be explained mathematically.
>
> Molecular theories suppose that all bodies, even when they appear to our senses homogeneous, consist of a multitude of particles, or small parts the mechanical relations of which constitute the properties of the bodies. Those theories which suppose that the molecules are at rest relative to the body may be called statical theories, and those which suppose the molecules to be in motion, even while the body is apparently at rest, may be called dynamical theories.

And now he gives a recognizable (and correct, so far as it goes) derivation of the Maxwell distribution:

When the number of pairs of molecules which change their velocities from OA, OB to OA′ OB′ is equal to the number which change from OA′, OB′ to OA, OB, then the final distribution of velocity will be obtained, which will not be altered by subsequent exchanges. This will be the case when

$$f_1(a)f_2(b)=f_1(a')f_2(b') \qquad (22)$$

Now the only relation between $a$, $b$ and $a'$, $b'$ is

$$M_1a^2+M_2b^2=M_1a'^2+M_2b'^2, \qquad (23)$$

whence we obtain

$$f_1(a)=C_1e^{-\frac{a^2}{\alpha^2}}, \quad f_2(b)=C_2e^{-\frac{b^2}{\beta^2}}, \qquad (24)$$

where

$$M_1\alpha^2=M_2\beta^2. \qquad (25)$$

By integrating $\iiint C_1e^{-\frac{\xi^2+\eta^2+\zeta^2}{\alpha^2}}\, d\xi\, d\eta\, d\zeta$, and equating the result to $N_1$, we obtain the value of $C_1$. If, therefore, the distribution of velocities among $N_1$ molecules is such that the number of molecules whose component velocities are between $\xi$ and $\xi+d\xi$, $\eta$ and $\eta+d\eta$, and $\zeta$ and $\zeta+d\zeta$ is

$$dN_1=\frac{N_1}{\alpha^3\pi^{\frac{3}{2}}}e^{-\frac{\xi^2+\eta^2+\zeta^2}{\alpha^2}}\, d\xi\, d\eta\, d\zeta, \qquad (26)$$

then this distribution of velocities will not be altered by the exchange of velocities among the molecules by their mutual action.

He goes on to try to understand experimental results on gases, about things like diffusion, viscosity and conductivity. For some reason, Maxwell doesn't want to think of molecules, as he did before, as hard spheres. And instead he imagines that they have "action at a distance" forces, which basically work like hard squares if it's $r^{-5}$ force law:

The paths described by molecules about a centre of force S, repelling inversely as the fifth power of the distance, are given in the figure.

The molecules are supposed to be originally moving with equal velocities in parallel paths, and the way in which their deflections depend on the distance of the path from S is shown by the different curves in the figure.

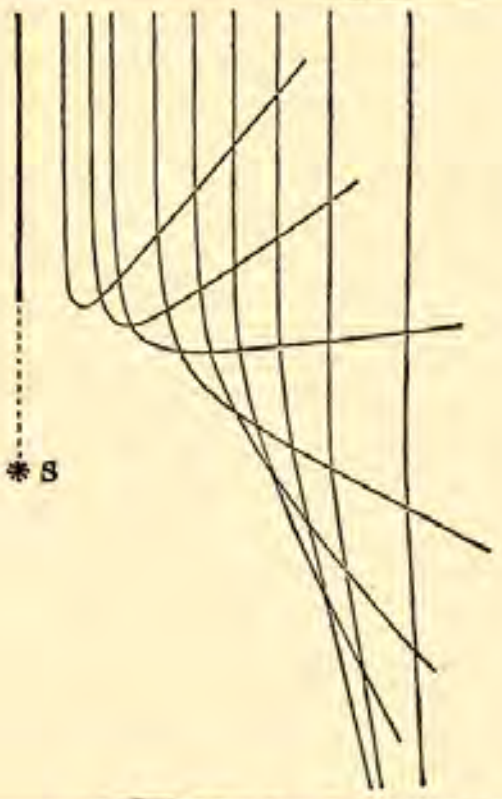

In the years that followed, Maxwell visited the dynamical theory of gases several more times. In 1871, a few years before he died at age 48, he wrote a textbook entitled *Theory of Heat*, which begins, in erudite fashion, discussing what "thermodynamics" should even be called:

conduction of electricity or the radiation of light. The science of heat has been called (by Dr. Whewell and others) Thermotics, and the theory of heat as a form of energy is called Thermodynamics. In the same way the theory of the equilibrium of heat might be called Thermostatics, and that of the motion of heat Thermokinematics.

Most of the book is concerned with the macroscopic "theory of heat"—though, as we'll discuss later, in the very last chapter Maxwell does talk about the "molecular theory", if in somewhat tentative terms.

## "Deriving" the Second Law from Molecular Dynamics

The Second Law was in effect originally introduced as a formalization of everyday observations about heat. But the development of kinetic theory seemed to open up the possibility that the Second Law could actually be proved from the underlying mechanics of molecules. And this was something that Ludwig Boltzmann (1844–1906) embarked on towards the end of his physics PhD at the University of Vienna. In 1865 he'd published his first paper ("On the Movement of Electricity on Curved Surfaces"), and in 1866 he published his second paper, "On the Mechanical Meaning of the Second Law of Thermodynamics":

**Über die mechanische Bedeutung des zweiten Hauptsatzes der Wärmetheorie.**

Von **Ludwig Boltzmann.**

(Vorgelegt in der Sitzung am 8. Februar 1866.)

Bereits längst ist die Identität des ersten Hauptsatzes der mechanischen Wärmetheorie mit dem Princip der lebendigen Kräfte

The introduction promises "a purely analytical, perfectly general proof of the Second Law". And what he seemed to imagine was that the equations of mechanics would somehow inevitably lead to motion that would reproduce the Second Law. And in a sense what computational irreducibility, rule 30, etc. now show is that in the end that's indeed basically how things work. But the methods and conceptual framework that Boltzmann had at his disposal were very far away from being able to see that. And instead what Boltzmann did was to use standard mathematical methods from mechanics to compute average properties of cyclic mechanical motions—and then made the somewhat unconvincing claim that combinations of these averages could be related (e.g. via temperature as average kinetic energy) to "Clausius's entropy":

wie aus dem Folgenden hervorgehen wird. Alsdann gelten ganz dieselben Schlüsse, die früher auf geschlossene Bahnen angewendet wurden, und man erhält wieder:

$$\iint \frac{\partial Q}{T}\, dk = 2\Sigma \log \frac{\int_{t_1}^{t_2} \frac{mc^2}{2}\, dt}{\int_{\tau_1}^{\tau_2} \frac{mc^2}{2}\, dt},$$

wenn $\tau_1$ und $\tau_2$ die Grenzen derjenigen Bahn sind, die dem zur untern

It's not clear how much this paper was read, but in 1871 Boltzmann (now a professor of mathematical physics in Graz) published another paper entitled simply "On the Priority of Finding the Relationship between the Second Law of Thermodynamics and the Principle of Least Action" that claimed (with some justification) that Clausius's then-newly-announced virial theorem was already contained in Boltzmann's 1866 paper.

But back in 1868—instead of trying to get all the way to Clausius's entropy—Boltzmann instead uses mechanics to get a generalization of Maxwell's law for the distribution of molecular velocities. His paper "Studies on the Equilibrium of [Kinetic Energy] between [Point Masses] in Motion" opens by saying that while analytical mechanics has in effect successfully studied the evolution of mechanical systems "from a given state to another", it's had little to say about what happens when such systems "have been left moving on their own for a long time". He intends to remedy that, and spends 47 pages—complete with elaborate diagrams and formulas about collisions between hard spheres—in deriving an exponential distribution of energies if one assumes "equilibrium" (or, more specifically, balance between forward and backward processes):

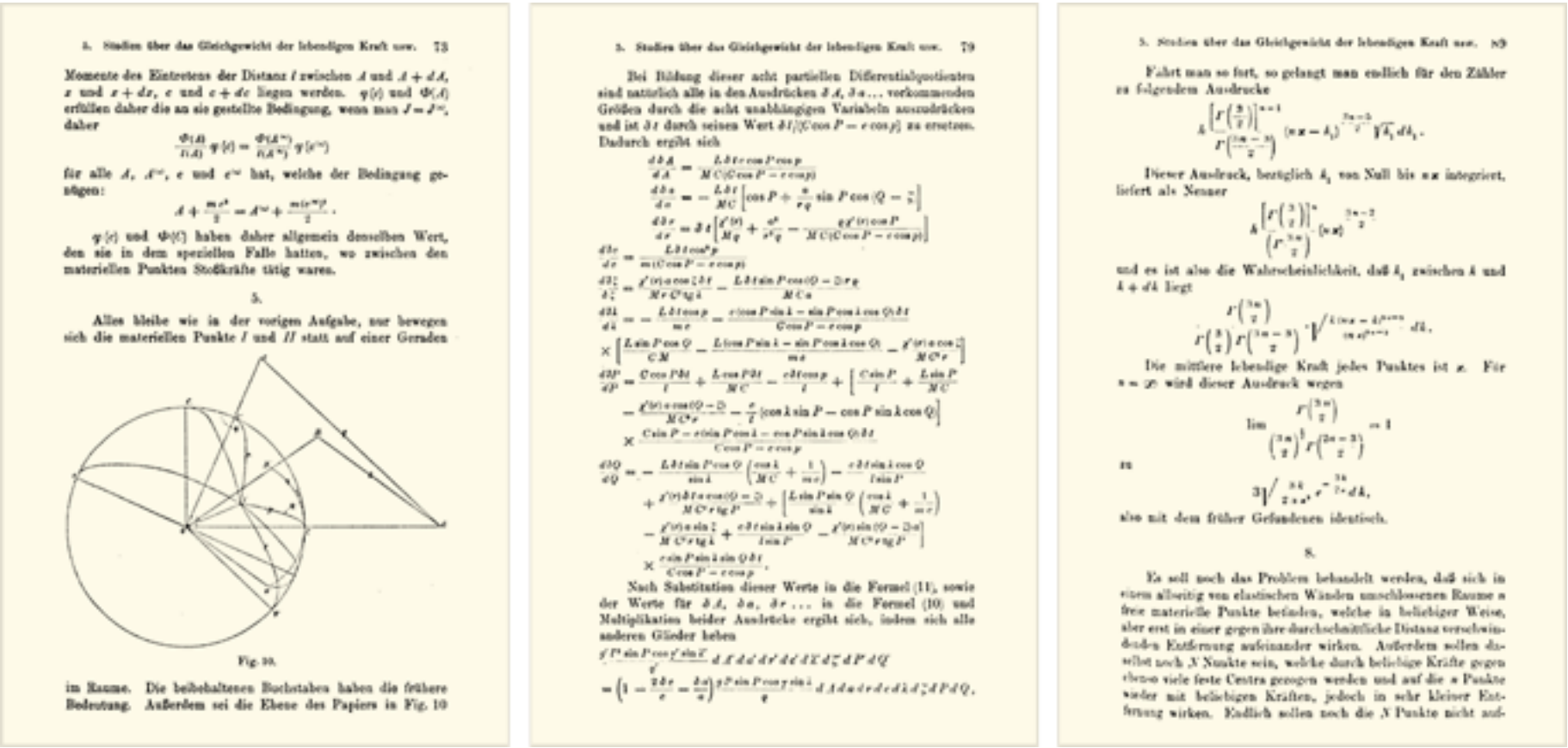

It's notable that one of the mathematical approaches Boltzmann uses is to discretize (i.e. effectively quantize) things, then look at the "combinatorial" limit. (Based on his later statements, he didn't want to trust "purely continuous" mathematics—at least in the context of discrete molecular processes—and wanted to explicitly "watch the limits happening".) But in the end it's not clear that Boltzmann's 1868 arguments do more than the few-line

functional–equation approach that Maxwell had already used. (Maxwell would later com plain about Boltzmann's "overly long" arguments.)

Boltzmann's 1868 paper had derived what the distribution of molecular energies should be "in equilibrium". (In 1871 he was talking about "equipartition" not just of kinetic energy, but also of energies associated with "internal motion" of polyatomic molecules.) But what about the approach to equilibrium? How would an initial distribution of molecular energies evolve over time? And would it always end up at the exponential ("Maxwell–Boltzmann") distribution? These are questions deeply related to a microscopic understanding of the Second Law. And they're what Boltzmann addressed in 1872 in his 22nd published paper "Further Studies on the Thermal Equilibrium of Gas Molecules":

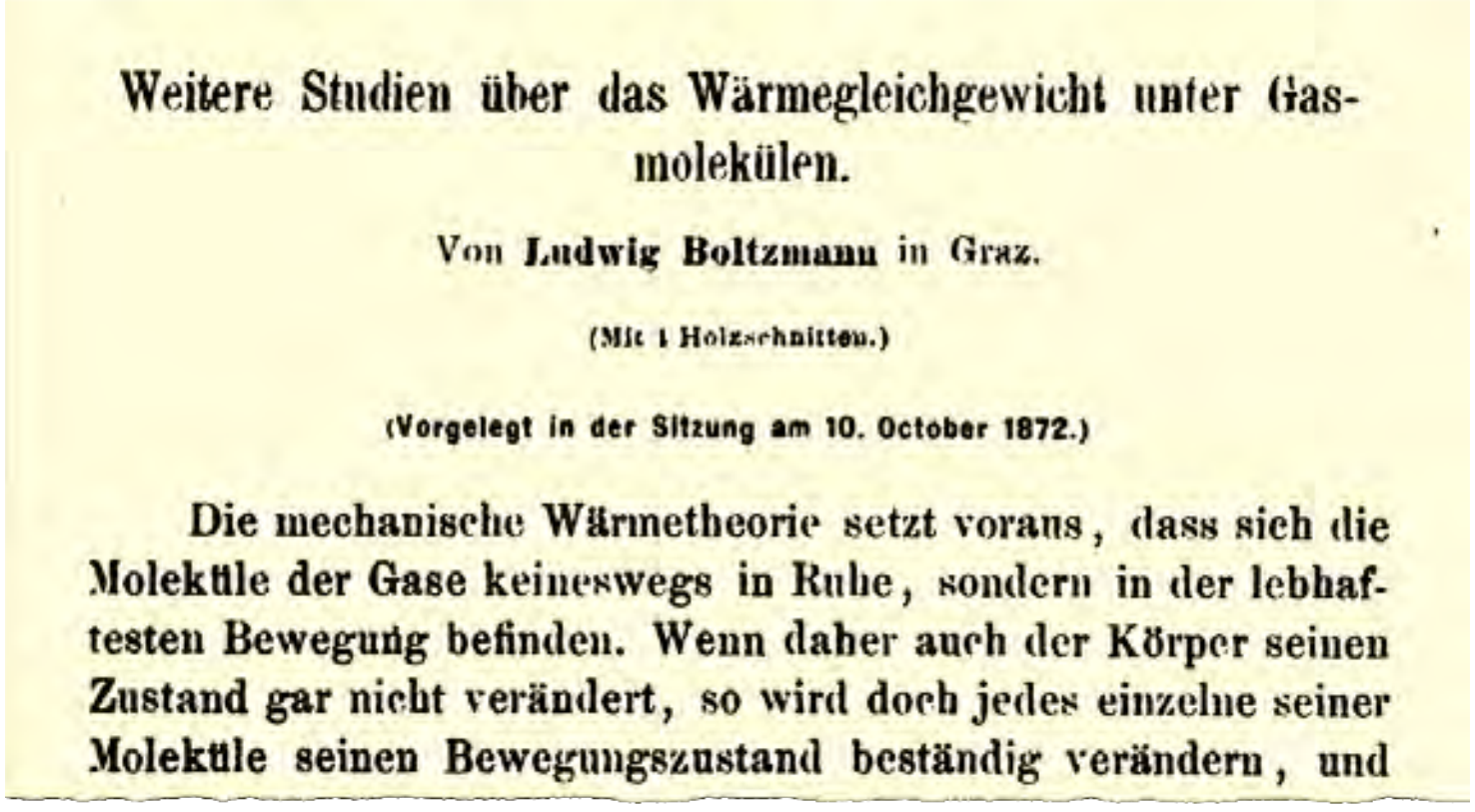
Weitere Studien über das Wärmegleichgewicht unter Gasmolekülen.

Von Ludwig Boltzmann in Graz.

(Mit 1 Holzschnitten.)

(Vorgelegt in der Sitzung am 10. October 1872.)

Die mechanische Wärmetheorie setzt voraus, dass sich die Moleküle der Gase keineswegs in Ruhe, sondern in der lebhaftesten Bewegung befinden. Wenn daher auch der Körper seinen Zustand gar nicht verändert, so wird doch jedes einzelne seiner Moleküle seinen Bewegungszustand beständig verändern, und

Boltzmann explains that:

> Maxwell already found the value $Av^2 e^{-Bv^2}$ [for the distribution of velocities] … so that the probability of different velocities is given by a formula similar to that for the probability of different errors of observation in the theory of the method of least squares. The first proof which Maxwell gave for this formula was recognized to be incorrect even by himself. He later gave a very elegant proof that, if the above distribution has once been established, it will not be changed by collisions. He also tries to prove that it is the only velocity distribution that has this property. But the latter proof appears to me to contain a false inference. It has still not yet been proved that, whatever the initial state of the gas may be, it must always approach the limit found by Maxwell. It is possible that there may be other possible limits. This proof is easily obtained, however, by the method which I am about to explain…

(He gives a long footnote explaining why Maxwell might be wrong, talking about how a sequence of collisions might lead to a "cycle of velocity states"—which Maxwell hasn't proved will be traversed with equal probability in each direction. Ironically, this is actually already an analog of where things are going to go wrong with Boltzmann's own argument.)

The main idea of Boltzmann's paper is not to assume equilibrium, but instead to write down an equation (now called the Boltzmann Transport Equation) that explicitly describes how the velocity (or energy) distribution of molecules will change as a result of collisions. He begins by defining infinitesimal changes in time:

**kül e, welche zur Zeit $t+\tau$ diese lebendige Kraft haben, also $f(x, t+\tau)dx$. Wir erhalten somit:**

$$f(x, t+\tau)dx = f(x, t)dx - \int dn + \int dv. \qquad 5)$$

He then goes through a rather elaborate analysis of velocities before and after collisions, and how to integrate over them, and eventually winds up with a partial differential equation for the time variation of the energy distribution (yes, he confusingly uses *x* to denote energy)—and argues that Maxwell's exponential distribution is a stationary solution to this equation:

$$\frac{\partial f(x,t)}{\partial t} = \int_0^\infty \int_0^{x+x'} \left[ \frac{f(\xi,t)}{\sqrt{\xi}} \frac{f(x+x'-\xi,t)}{\sqrt{x+x'-\xi}} - \frac{f(x,t)}{\sqrt{x}} \frac{f(x',t)}{\sqrt{x'}} \right] \times \qquad 16)$$

$$\times \sqrt{xx'}\, \psi(x, x', \xi) dx' d\xi.$$

**Dies ist die Fundamentalgleichung für die Veränderung der Function $f(x, t)$. Ich bemerke nochmal, dass die Wurzeln alle positiv zu nehmen sind, sowie auch $\psi$ und die $f$ wesentlich positive Grössen sind. Setzen wir für einen Augenblick**

$$f(x,t) = C\sqrt{x}e^{-hx}, \qquad 16\,a)$$

A few paragraphs further on, something important happens: Boltzmann introduces a function that here he calls *E*, though later he'll call it *H*:

$$E = \int_0^\infty f(x,t) \left\{ \log \left[ \frac{f(x,t)}{\sqrt{x}} \right] - 1 \right\} dx \qquad 17)$$

Ten pages of computation follow

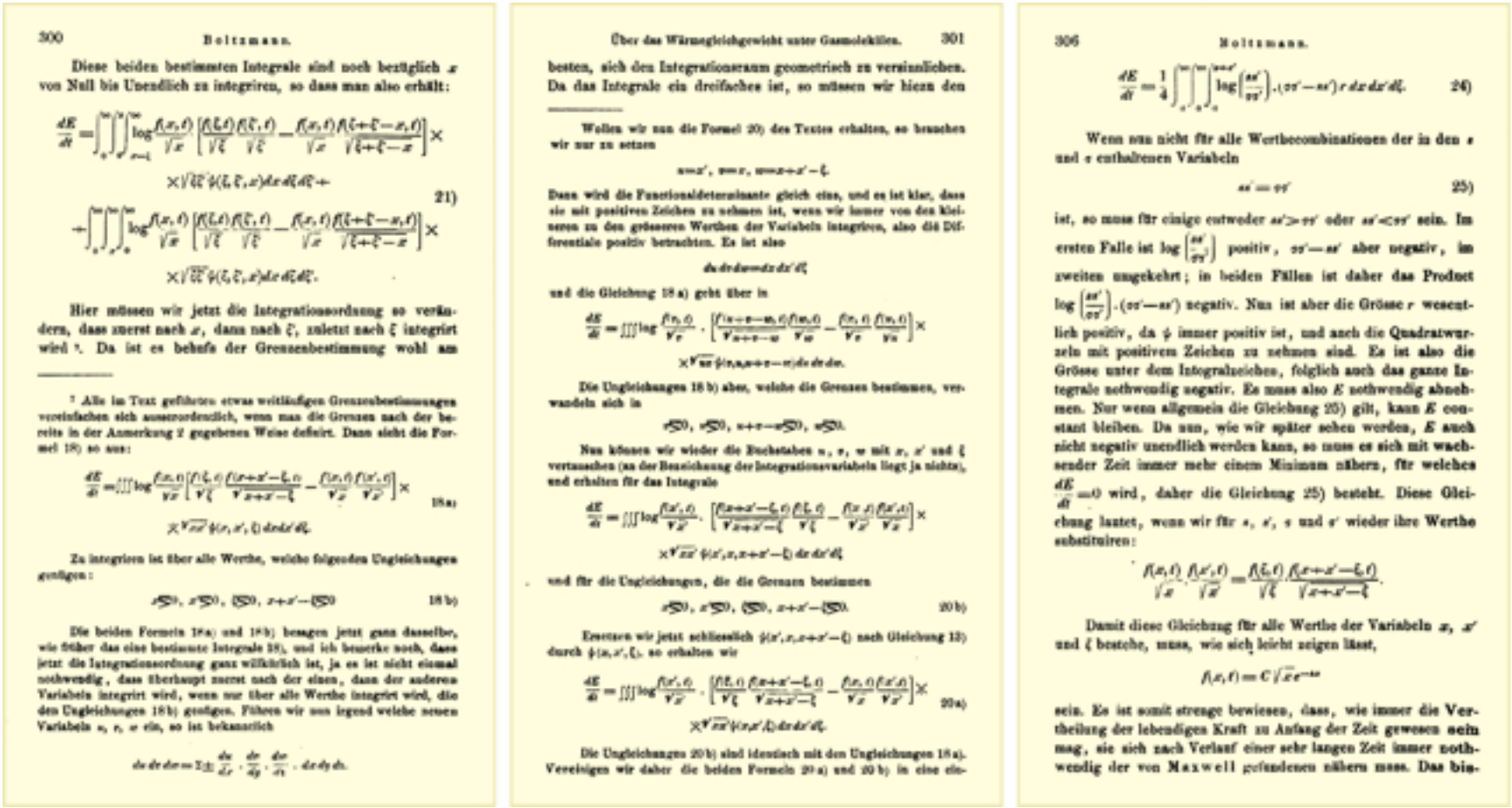

and finally Boltzmann gets his main result: if the velocity distribution evolves according to his equation, *H* can never increase with time, becoming zero for the Maxwell distribution. In other words, he is saying that he's proved that a gas will always ("monotonically") approach equilibrium—which seems awfully like some kind of microscopic proof of the Second Law.

But then Boltzmann makes a bolder claim:

> It has thus been rigorously proved that, whatever the initial distribution of kinetic energy may be, in the course of a very long time it must always necessarily approach the one found by Maxwell. The procedure used so far is of course nothing more than a mathematical artifice employed in order to give a rigorous proof of a theorem whose exact proof has not previously been found. It gains meaning by its applicability to the theory of polyatomic gas molecules. There one can again prove that a certain quantity *E* can only decrease as a consequence of molecular motion, or in a limiting case can remain constant. One can also prove that for the atomic motion of a system of arbitrarily many material points there always exists a certain quantity which, in consequence of any atomic motion, cannot increase, and this quantity agrees up to a constant factor with the value found for the well-known integral $\int dQ/T$ in my [1871] paper on the "Analytical proof of the 2nd law, etc.". We have therefore prepared the way for an analytical proof of the Second Law in a completely different way from those previously investigated. Up to now the object has been to show that $\int dQ/T = 0$ for reversible cyclic processes, but it has not been proved analytically that this quantity is always negative for irreversible processes, which are the only ones that occur in nature. The reversible cyclic process is only an ideal, which one can more or less closely approach but never completely attain. Here, however, we have succeeded in showing that $\int dQ/T$ is in general negative, and is equal to zero only for the limiting case, which is of course the reversible cyclic process (since if one can go through the process in either direction, $\int dQ/T$ cannot be negative).

In other words, he's saying that the quantity *H* that he's defined microscopically in terms of velocity distributions can be identified (up to a sign) with the entropy that Clausius defined as $dQ/T$. He says that he'll show this in the context of analyzing the mechanics of polyatomic molecules.

But first he's going to take a break and show that his derivation doesn't need to assume continuity. In a pre-quantum-mechanics pre-cellular-automaton-fluid kind of way he replaces all the integrals by limits of sums of discrete quantities (i.e. he's quantizing kinetic energy, etc.):

$$\int_0^\infty f(x,t)\,dx = \lim \varepsilon[f(\varepsilon,t)+f(2\varepsilon,t)+f(3\varepsilon,t)+\ldots f(p\varepsilon,t)]$$

$$\text{für } \lim \varepsilon = 0, \quad \lim p\varepsilon = \infty.$$

He says that this discrete approach makes everything clearer, and quotes Lagrange's derivation of vibrations of a string as an example of where this has happened before. But then he argues that everything works out fine with the discrete approach, and that *H* still decreases, with the Maxwell distribution as the only possible end point. As an aside, he makes a jab at Maxwell's derivation, pointing out that with Maxwell's functional equation:

> ... there are infinitely many other solutions, which are not useful however since $f(x)$ comes out negative or imaginary for some values of *x*. Hence, it follows very clearly that Maxwell's attempt to prove *a priori* that his solution is the only one must fail, since it is not the only one but rather it is the only one that gives purely positive probabilities, and therefore the only useful one.

But finally—after another aside about computing thermal conductivities of gases—Boltzmann digs into polyatomic molecules, and his claim about *H* being related to entropy. There's another 26 pages of calculations, and then we get to a section entitled "Solution of Equation (81) and Calculation of Entropy". More pages of calculation about polyatomic molecules ensue. But finally we're computing *H*, and, yes, it agrees with the Clausius result—but anticlimactically he's only dealing with the case of equilibrium for monatomic molecules, where we already knew we got the Maxwell distribution:

$$f^* = \frac{1}{V\left(\frac{4\pi T}{3m}\right)^{\frac{3}{2}}} e^{-\frac{3m}{4T}(u^2+v^2+w^2)},$$

daher

$$E^* = N \iint .\ . f^* \log f^*\, dx\, dy\, dz\, du\, dv\, dw =$$

$$= -N \log\left[V\left(\frac{4\pi T}{3m}\right)^{\frac{3}{2}}\right] - \frac{3}{2} N,$$

And now he decides he's not talking about polyatomic molecules anymore, and instead:

> In order to find the relation of the quantity [*H*] to the second law of thermodynamics in the form $\int dQ/T < 0$, we shall interpret the system of mass points not, as previously, as a gas molecule, but rather as an entire body.

But then, in the last couple of pages of his paper, Boltzmann pulls out another idea. He's discussed the concept that polyatomic molecules (or, now, whole systems) can be in many different configurations, or "phases". But now he says: "We shall replace [our] single system by a large number of equivalent systems distributed over many different phases, but which do not interact with each other". In other words, he's introducing the idea of an ensemble of states of a system. And now he says that instead of looking at the distribution just for a single velocity, we should do it for all velocities, i.e. for the whole "phase" of the system.

> [These distributions] may be discontinuous, so that they have large values when the variables are very close to certain values determined by one or more equations, and otherwise vanishingly small. We may choose these equations to be those that characterize visible external motion of the body and the kinetic energy contained in it. In this connection it should be noted that the kinetic energy of visible motion corresponds to such a large deviation from the final equilibrium distribution of kinetic energy that it leads to an infinity in *H*, so that from the point of view of the Second Law of thermodynamics it acts like heat supplied from an infinite temperature.

There are a bunch of ideas swirling around here. Phase-space density (*cf.* Liouville's equation). Coarse-grained variables. Microscopic representation of mechanical work. Etc. But the paper is ending. There's a discussion about *H* for systems that interact, and how there's an equilibrium value achieved. And finally there's a formula for entropy

$$E^* = \iint f^* \log f^*\, ds\, d\sigma = \log A - h \frac{\int \chi e^{-h\chi}\, d\sigma}{\int e^{-h\chi}\, d\sigma} - \frac{3r}{2}$$

that Boltzmann said "agrees ... with the expression I found in my previous [1871] paper".
So what exactly did Boltzmann really do in his 1872 paper? He introduced the Boltzmann Transport Equation which allows one to compute at least certain non-equilibrium proper ties of gases. But is his $f \log f$ quantity really what we can call "entropy" in the sense Clausius meant? And is it true that he's proved that entropy (even in his sense) increases? A century and a half later there's still a remarkable level of confusion around both these issues.

But in any case, back in 1872 Boltzmann's "minimum theorem" (now called his "*H* theorem") created quite a stir. But after some time there was an objection raised, which we'll discuss below. And partly in response to this, Boltzmann (after spending time working on microscopic models of electrical properties of materials—as well as doing some actual experiments) wrote another major paper on entropy and the Second Law in 1877:

Über die Beziehung zwischen dem zweiten Hauptsatze der mechanischen Wärmetheorie und der Wahrscheinlichkeitsrechnung, respective den Sätzen über das Wärmegleichgewicht.

Von dem c. M. **Ludwig Boltzmann** in Graz.

Eine Beziehung des zweiten Hauptsatzes zur Wahrscheinlichkeitsrechnung zeigte sich zuerst, als ich nachwies, dass ein analytischer Beweis desselben auf keiner anderen Grundlage

The translated title of the paper is "On the Relation between the Second Law of Thermodynamics and Probability Theory with Respect to the Laws of Thermal Equilibrium". And at the very beginning of the paper Boltzmann makes a statement that was pivotal for future discussions of the Second Law: he says it's now clear to him that an "analytical proof" of the Second Law is "only possible on the basis of probability calculations". Now that we know about computational irreducibility and its implications one could say that this was the point where Boltzmann and those who followed him went off track in understanding the true foundations of the Second Law. But Boltzmann's idea of introducing probability theory was effectively what launched statistical mechanics, with all its rich and varied consequences.

Boltzmann makes his basic claim early in the paper mit folgenden Worten: „Es ist klar, dass jede einzelne gleichförmige Zustandsvertheilung, welche bei einem bestimmten

Anfangszustande nach Verlauf einer bestimmten Zeit entsteht, ebenso unwahrscheinlich ist, wie eine einzelne noch so ungleichförmige Zustandsvertheilung, gerade so wie im Lottospiele jede einzelne Quinterne ebenso unwahrscheinlich ist, wie die Quinterne with the statement (quoting from a comment in a paper he'd written earlier the same year) that "it is clear" (always a dangerous thing to say!) that in thermal equilibrium all possible states of the system—say, spatially uniform and nonuniform alike—are equally probable

> … comparable to the situation in the game of Lotto where every single quintet is as improbable as the quintet 12345. The higher probability that the state distribution becomes uniform with time arises only because there are far more uniform than nonuniform state distributions…

He goes on:

> [Thus] it is possible to calculate the thermal equilibrium state by finding the probability of the different possible states of the system. The initial state will in most cases be highly improbable but from it the system will always rapidly approach a more probable state until it finally reaches the most probable state, i.e., that of thermal equilibrium. If we apply this to the Second Law we will be able to identify the quantity which is usually called entropy with the probability of the particular state...

He's talked about thermal equilibrium, even in the title, but now he says:

> ... our main purpose here is not to limit ourselves to thermal equilibrium, but to explore the relationship of the probabilistic formulation to the [Second Law].

He says his goal is to calculate probability distribution for different states, and he'll start with

> as simple a case as possible, namely a gas of rigid absolutely elastic spherical molecules trapped in a container with absolutely elastic walls. (Which interact with central forces only within a certain small distance, but not otherwise; the latter assumption, which includes the former as a special case, does not change the calculations in the least).

In other words, yet again he's going to look at hard sphere gases. But, he says:

> Even in this case, the application of probability theory is not easy. The number of molecules is not infinite, in a mathematical sense, yet the number of velocities each molecule is capable of is effectively infinite. Given this last condition, the calculations are very difficult; to facilitate understanding, I will, as in earlier work, consider a limiting case.

And this is where he "goes discrete" again—allowing ("cellular-automaton-style") only discrete possible velocities for each molecule:

I. Die Zahl der lebendigen Kräfte ist eine discrete.

Wir wollen zunächst annehmen, jedes Molekül sei nur im Stande, eine bestimmte endliche Anzahl von Geschwindigkeiten anzunehmen, z. B. die Geschwindigkeiten

$$0, \frac{1}{q}, \frac{2}{q}, \frac{3}{q}, \dots \frac{p}{q},$$

He says that upon colliding, two molecules can exchange these discrete velocities, but nothing more. As he explains, though:

> Even if, at first sight, this seems a very abstract way of treating the problem, it rapidly leads to the desired objective, and when you consider that in nature all infinities are but limiting cases, one assumes each molecule can behave in this fashion only in the limiting case where each molecule can assume more and more values of the velocity.

But now—much like in an earlier paper—he makes things even simpler, saying he's going to ignore velocities for now, and just say that the possible energies of molecules are "in an arithmetic progression":

Molekül anzunehmen im Stande ist, eine arithmetische Progression bildet, z. B. folgende:

$$0, \varepsilon, 2\varepsilon, 3\varepsilon, \dots p\varepsilon.$$

He plans to look at collisions, but first he just wants to consider the combinatorial problem of distributing these energies among *n* molecules in all possible ways, subject to the constraint of having a certain fixed total energy.  He sets up a specific example, with 7 molecules, total energy 7, and maximum energy per molecule 7—then gives an explicit table of all possible states (up to, as he puts it, "immaterial permutations of molecular labels"):

aufgeführte Reihe von Zustandsvertheilungen. Die Zahlen der ersten Colonne numeriren die verschiedenen Zustandsvertheilungen.

| | | 𝔓 |
|---|---|---|
| 1. | 0000007 | 7 |
| 2. | 0000016 | 42 |
| 3. | 0000025 | 42 |
| 4. | 0000034 | 42 |
| 5. | 0000115 | 105 |
| 6. | 0000124 | 210 |
| 7. | 0000133 | 105 |
| 8. | 0000223 | 105 |
| 9. | 0001114 | 140 |
| 10. | 0001123 | 420 |
| 11. | 0001222 | 140 |
| 12. | 0011113 | 105 |
| 13. | 0011122 | 210 |
| 14. | 0111112 | 42 |
| 15. | 1111111[1] | 1 |

Tables like this had been common for nearly two centuries in combinatorial mathematics books—like Jacob Bernoulli's (1655–1705) *Ars Conjectandi (*from 1713)

*Combinationum, ſeu Numerorum Figuratorum.*

*Exponentes Combinationum.*

*Numeri Rerum Combinandarum.*

| | I. | II. | III. | IV. | V. | VI. | VII. | VIII. | IX. | X. | XI. | XII. |
|---|---|---|---|---|---|---|---|---|---|---|---|---|
| 1. | 1 | 0 | 0 | 0 | 0 | 0 | 0 | 0 | 0 | 0 | 0 | 0 |
| 2. | 1 | 1 | 0 | 0 | 0 | 0 | 0 | 0 | 0 | 0 | 0 | 0 |
| 3. | 1 | 2 | 1 | 0 | 0 | 0 | 0 | 0 | 0 | 0 | 0 | 0 |
| 4. | 1 | 3 | 3 | 1 | 0 | 0 | 0 | 0 | 0 | 0 | 0 | 0 |
| 5. | 1 | 4 | 6 | 4 | 1 | 0 | 0 | 0 | 0 | 0 | 0 | 0 |
| 6. | 1 | 5 | 10 | 10 | 5 | 1 | 0 | 0 | 0 | 0 | 0 | 0 |
| 7. | 1 | 6 | 15 | 20 | 15 | 6 | 1 | 0 | 0 | 0 | 0 | 0 |
| 8. | 1 | 7 | 21 | 35 | 35 | 21 | 7 | 1 | 0 | 0 | 0 | 0 |
| 9. | 1 | 8 | 28 | 56 | 70 | 56 | 28 | 8 | 1 | 0 | 0 | 0 |
| 10. | 1 | 9 | 36 | 84 | 126 | 126 | 84 | 36 | 9 | 1 | 0 | 0 |
| 11. | 1 | 10 | 45 | 120 | 210 | 252 | 210 | 120 | 45 | 10 | 1 | 0 |
| 12. | 1 | 11 | 55 | 165 | 330 | 462 | 462 | 330 | 165 | 55 | 11 | 1 |

but this might have been the first place such a table had appeared in a paper about fundamental physics.

And now Boltzmann goes into an analysis of the distribution of states—of the kind that's now long been standard in textbooks of statistical physics, but will then have been quite unfamiliar to the pure–calculus–based physicists of the time:

Elemente unter einander gleich. Die Anzahl dieser Permutationen ist also bekanntlich

$$\mathfrak{P} = \frac{n!}{(w_0)!\,(w_1)!\dots} \qquad 3)$$

6b) darauf hinausläuft, dass wir an Stelle des Minimums des Ausdrucks 3) das Minimum von

$$\frac{\sqrt{2\pi}\left(\frac{n}{e}\right)^{n}}{\sqrt{2\pi}\left(\frac{w_0}{e}\right)^{w_0}\sqrt{2\pi}\left(\frac{w_1}{e}\right)^{w_1}\dots}$$

He derives the average energy per molecule, as well as the fluctuations:

käls, daher $\frac{\lambda}{n} = \frac{\mu}{\varepsilon}$, also jedenfalls ausserordentlich gross, wesshalb man hat

$$(\lambda+n-1)^{\lambda+n-1/2} = \lambda^{\lambda+n-1/2}\left[\left(1+\frac{n-1}{\lambda}\right)^{\lambda}\right]^{1+\frac{2n-1}{2\lambda}} = \lambda^{\lambda+n-1/2}.e^{n-1}.$$

Es ist also

$$J = \frac{1}{\sqrt{2\pi}}\frac{\lambda^{n-1}e^{n-1}}{(n-1)^{n-\frac{1}{2}}},$$

daher ist, abgesehen von verschwindenden Grössen

$$lJ = nl\frac{\lambda}{n} + n - l\lambda + \frac{1}{2}ln - 1 - \frac{1}{2}l(2\pi).$$

He says that "of course" the real interest is in the limit of an infinite number of molecules, but he still wants to show that for "moderate values" the formulas remain quite accurate. And then (even without Wolfram Language!) he's off finding (using Newton's method it seems) approximate roots of the necessary polynomials:

Dann verwandelt sich die Gleichung 12) in folgende:

$$6x^9 - 7x^8 + 2x - 1 = 0 \qquad 16)$$

woraus folgt:

$$x = \frac{1}{2} + \frac{7}{2}x^8 - 3x^9 \qquad 17)$$

Da $x$ nahe $= \frac{1}{2}$ ist, so können wir in den beiden letzten, ohnehin sehr kleinen Gliedern der rechten Seite $x = \frac{1}{2}$ setzen, und erhalten:

$$x = \frac{1}{2} + \frac{1}{2^9}(7-3) = \frac{1}{2} + \frac{1}{2^7} = 0{\cdot}5078125.$$

Just to show how it all works, he considers a slightly larger case as well:

0000011122345,

deren Ziffersumme in der That = 19 ist.

Die Anzahl der Permutationen, deren diese Complexion fähig ist, ist

$$\frac{13!}{5!\,3!\,2!} = \frac{13!}{4!\,3!\,2!} \cdot \frac{1}{5}.$$

Eine Complexion, deren Ziffernsumme ebenfalls = 19 ist, und von welcher man vermuthen könnte, dass sie sehr vieler Permutationen fähig sein wird, wäre folgende:

0000111222334.

Die Anzahl ihrer Permutationen ist

$$\frac{13!}{4!\,3!\,3!\,2!} = \frac{13!}{4!\,3!\,2!} \cdot \frac{1}{6},$$

also bereits kleiner als die Anzahl der Permutationen der ersten von uns aus der Annäherungsformel gefundenen Complexion. Ebenso überzeugt man sich, dass die Zahl der Permutationen der beiden Complexionen

0000111122335 und
0000111122344

kleiner ist. Dieselbe ist nämlich für beide Complexionen

$$\frac{13!}{4!\,4!\,2!\,2!} = \frac{13!}{4!\,3!\,2!} \cdot \frac{1}{8}.$$

Now he's computing the probability that a given molecule has a particular energy

$$w_s = \frac{n^2}{n+\lambda} \cdot \left(\frac{\lambda}{n+\lambda}\right)^s.$$

and determining that in the limit it's an exponential

$$w_s = \frac{n\varepsilon}{\mu} e^{-\frac{s\varepsilon}{\mu}}.$$

that is, as he says, "consistent with that known from gases in thermal equilibrium".
He claims that in order to really get a "mechanical theory of heat" it's necessary to take a continuum limit. And here he concludes that thermal equilibrium is achieved by maximizing the quantity Ω (where the "*l*" stands for log, so this is basically $f \log f$):

welche ein Maximum werden soll, den Werth erhält:

$$\Omega = -\int_{-\infty}^{+\infty}\int_{-\infty}^{+\infty}\int_{-\infty}^{+\infty} f(u,v,w)\, lf(u,v,w)\,du\,dv\,dw \qquad 34)$$

He explains that Ω is basically the log of the number of possible permutations, and that it's "of special importance", and he'll call it the "permutability measure". He immediately notes that "the total permutability measure of two bodies is equal to the sum of the permutability measures of each body". (Note that Boltzmann's Ω isn't the modern total-number-of-states Ω; confusingly, that's essentially the exponential of Boltzmann's Ω.)

He goes through some discussion of how to handle extra degrees of freedom in polyatomic molecules, but then he's on to the main event: arguing that Ω is (essentially) the entropy. It doesn't take long:

kules, so ist für den Zustand des Wärmegleichgewichtes

$$f(x,y,z,u,v,w) = \frac{N}{V\left(\frac{4\pi T}{3m}\right)^{\frac{3}{2}}}\cdot e^{-\frac{3m}{4T}(u^2+v^2+w^2)}.$$

Substituirt man diesen Werth in Gleichung 61), so erhält man

$$\Omega = \frac{3N}{2} + Nl\left[V\left(\frac{4\pi T}{3m}\right)^{\frac{3}{2}}\right] - NlN \qquad 62)$$

Versteht man nun unter $dQ$ das dem Gase zugeführte Wärmedifferentiale, so ist

$$dQ = NdT + pdV \qquad 63)$$

und

$$pV = \frac{2N}{3}.T \qquad 64)$$

$p$ ist der Druck, bezogen auf die Flächeneinheit. Die Entropie des Gases ist dann:

$$\int\frac{dQ}{T} = \frac{2}{3}N.l(V.T^{\frac{3}{2}}) + C.$$

Da hier $N$ als eine rein Constante anzusehen ist, so ist bei passender Bestimmung dieser Constante

$$\int\frac{dQ}{T} = \frac{2}{3}\Omega \qquad 65)$$

Basically he just says that in equilibrium the probability *f*(...) for a molecule to have a particular velocity is given by the Maxwell distribution, then he substitutes this into the formula for Ω, and shows that indeed, up to a constant, Ω is exactly the "Clausius entropy" ∫ *dQ*/*T*.

So, yes, in equilibrium Ω seems to be giving the entropy. But then Boltzmann makes a bit of a jump. He says that in processes that aren't reversible both "Clausius entropy" and Ω will increase, and can still be identified—and enunciates the general principle, printed in his paper in special doubled–spaced form:

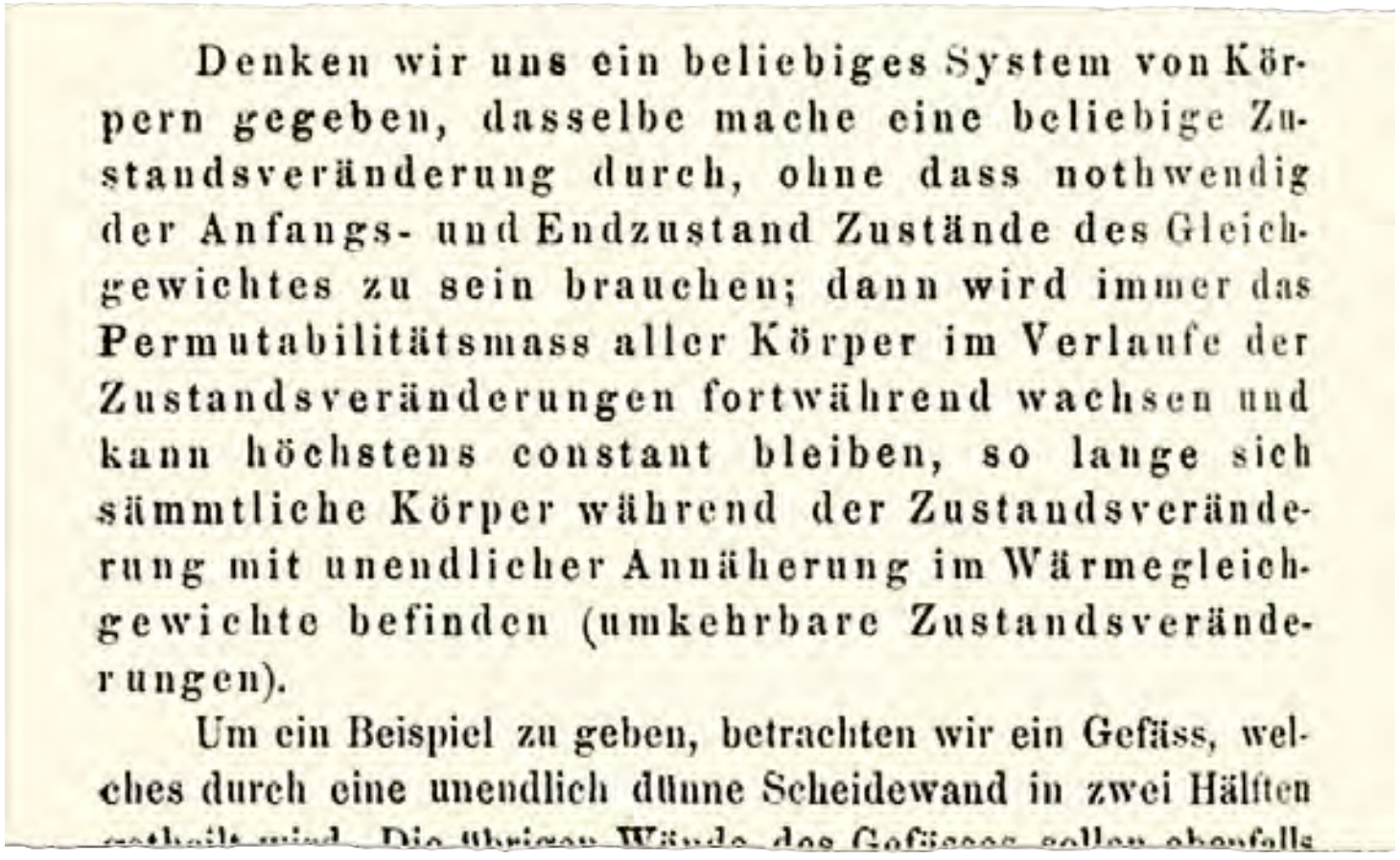

Denken wir uns ein beliebiges System von Körpern gegeben, dasselbe mache eine beliebige Zustandsveränderung durch, ohne dass nothwendig der Anfangs- und Endzustand Zustände des Gleichgewichtes zu sein brauchen; dann wird immer das Permutabilitätsmass aller Körper im Verlaufe der Zustandsveränderungen fortwährend wachsen und kann höchstens constant bleiben, so lange sich sämmtliche Körper während der Zustandsveränderung mit unendlicher Annäherung im Wärmegleichgewichte befinden (umkehrbare Zustandsveränderungen).

Um ein Beispiel zu geben, betrachten wir ein Gefäss, welches durch eine unendlich dünne Scheidewand in zwei Hälften getheilt wird. Die übrigen Wände des Gefässes sollen ebenfalls

> ... [In] any system of bodies that undergoes state changes ... even if the initial and final states are not in thermal equilibrium ... the total permutability measure for the bodies will continually increase during the state changes, and can remain constant only so long as all the bodies during the state changes remain infinitely close to thermal equilibrium (reversible state changes).

In other words, he's asserting that Ω behaves the same way entropy is said to behave according to the Second Law. He gives various thought experiments about gases in boxes with dividers, gases under gravity, etc. And finally concludes that, yes, the relationship of entropy to Ω "applies to the general case".

There's one final paragraph in the paper, though:

> Up to this point, these propositions may be demonstrated exactly using the theory of gases. If one tries, however, to generalize to liquid drops and solid bodies, one must dispense with an exact treatment from the outset, since far too little is known about the nature of the latter states of matter, and the mathematical theory is barely developed. But I have already mentioned reasons in previous papers, in virtue of which it is likely that for these two aggregate states, the thermal equilibrium is achieved when Ω becomes a maximum, and that when thermal equilibrium exists, the entropy is given by the same expression. It can therefore be described as likely that the validity of the principle which I have developed is not just limited to gases, but that the same constitutes a general natural law applicable to solid bodies and liquid droplets, although the exact mathematical treatment of these cases still seems to encounter extraordinary difficulties.

Interestingly, Boltzmann is only saying that it's "likely" that in thermal equilibrium his permutability measure agrees with Clausius's entropy, and he's implying that actually that's really the only place where Clausius's entropy is properly defined. But certainly his definition is more general (after all, it doesn't refer to things like temperature that are only properly defined in equilibrium), and so—even though Boltzmann didn't explicitly say it—one can imagine basically just using it as the definition of entropy for arbitrary cases. Needless to say, the story is actually more complicated, as we'll see soon.

But this definition of entropy—crispened up by Max Planck (1858–1947) and with different notation—is what ended up years later "written in stone" at Boltzmann's grave:

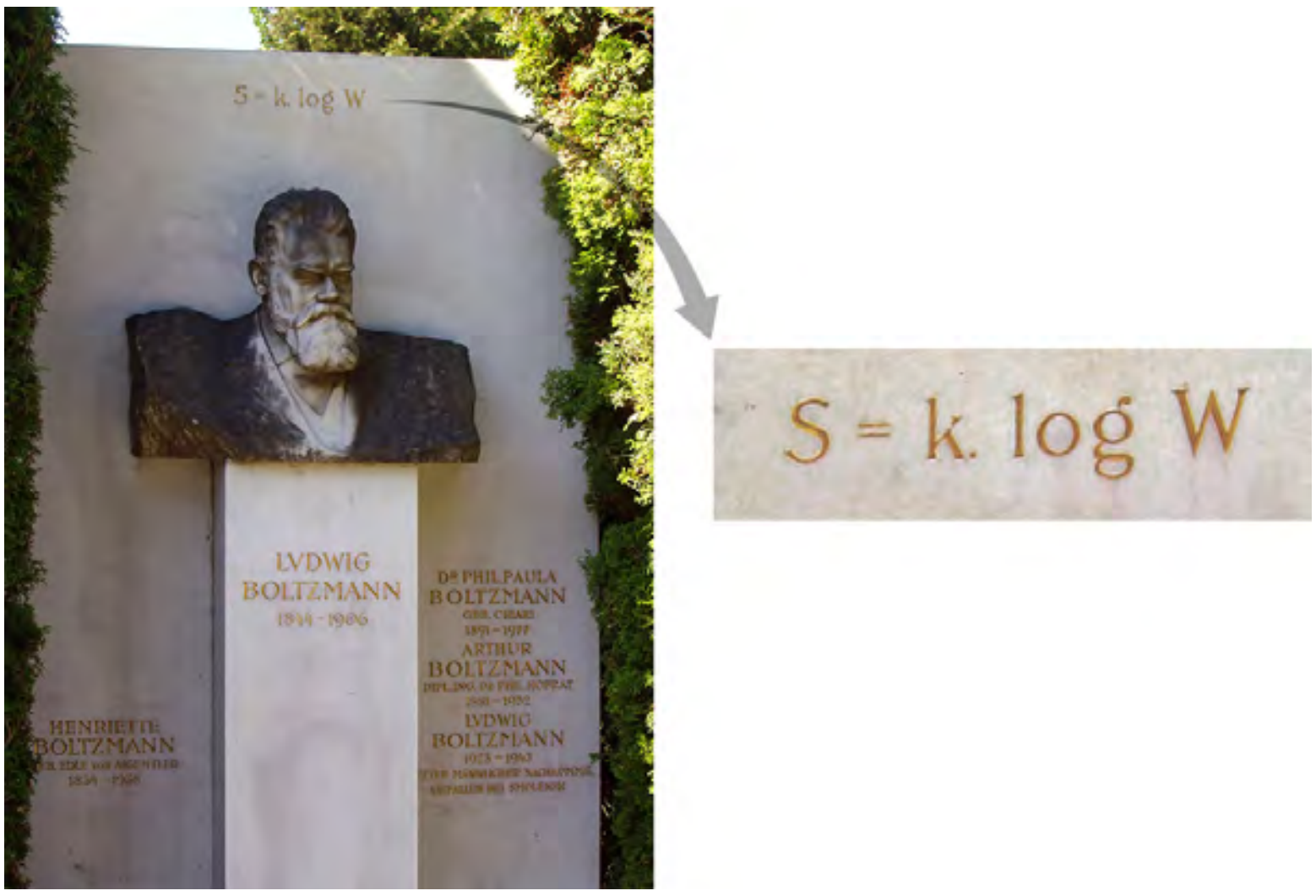

## The Concept of Ergodicity

In his 1877 paper Boltzmann had made the claim that in equilibrium all possible microscopic states of a system would be equally probable. But why should this be true? One reason could be that in its pure "mechanical evolution" the system would just successively visit all these states. And this was an idea that Boltzmann seems to have had—with increasing clarity—from the time of his very first paper in 1866 that purported to "prove the Second Law" from mechanics.

In modern times—with our understanding of discrete systems and computational rules—it's not difficult to describe the idea of "visiting all states". But in Boltzmann's time it was considerably more complicated. Did one expect to hit all the infinite possible infinitesimally separated configurations of a system? Or somehow just get close? The fact is that Boltzmann had certainly dipped his toe into thinking about things in terms of discrete quantities. But he didn't make the jump to imagining discrete rules, even though he certainly did know about discrete iterative processes, like Newton's method for finding roots.

Boltzmann knew about cases—like circular motion—where everything was purely periodic. But maybe when motion wasn't periodic, it'd inevitably "visit all states". Already in 1868 Boltzmann was writing a paper entitled "Solution to a Mechanical Problem" where he studies a single point mass moving in an $\alpha/r - \beta/r^2$ potential and bouncing elastically off a line—and manages to show that it visits every position with equal probability. In this paper he's just got traditional formulas, but by 1871, in "Some General Theorems about Thermal Equilibrium"—computing motion in the same potential as before—he's got a picture:

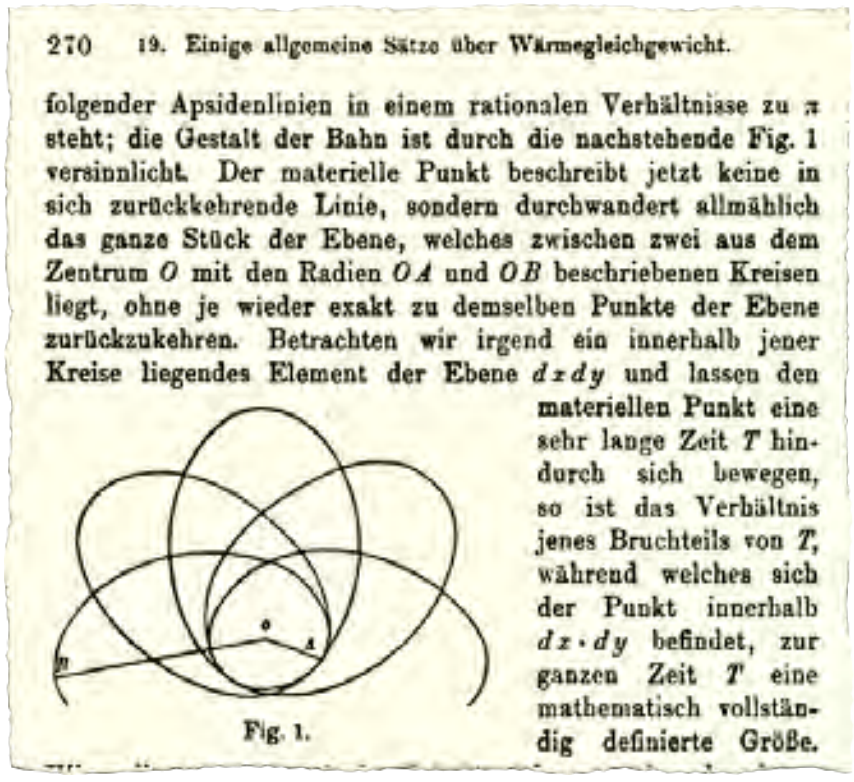

270 19. Einige allgemeine Sätze über Wärmegleichgewicht.

folgender Apsidenlinien in einem rationalen Verhältnisse zu $\pi$ steht; die Gestalt der Bahn ist durch die nachstehende Fig. 1 versinnlicht. Der materielle Punkt beschreibt jetzt keine in sich zurückkehrende Linie, sondern durchwandert allmählich das ganze Stück der Ebene, welches zwischen zwei aus dem Zentrum $O$ mit den Radien $OA$ und $OB$ beschriebenen Kreisen liegt, ohne je wieder exakt zu demselben Punkte der Ebene zurückzukehren. Betrachten wir irgend ein innerhalb jener Kreise liegendes Element der Ebene $dx\,dy$ und lassen den materiellen Punkt eine sehr lange Zeit $T$ hindurch sich bewegen, so ist das Verhältnis jenes Bruchteils von $T$, während welches sich der Punkt innerhalb $dx \cdot dy$ befindet, zur ganzen Zeit $T$ eine mathematisch vollständig definierte Größe.

Fig. 1.

Boltzmann probably knew about Lissajous figures—cataloged in 1857

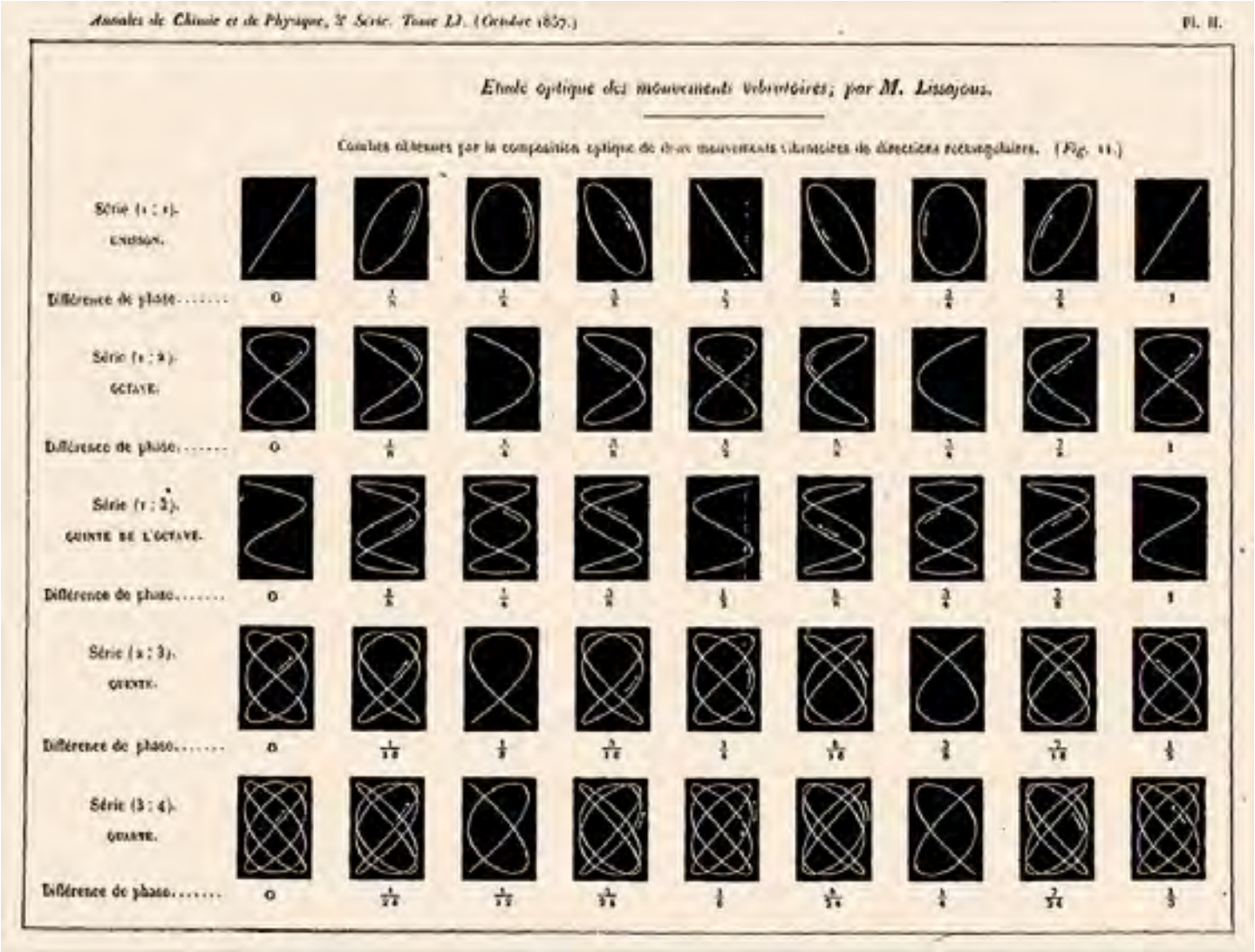

and the fact that in this case a rational ratio of $x$ and $y$ periods gives a periodic overall curve while an irrational one always gives a curve that visits every position might have led him to suspect that all systems would either be periodic, or would visit every possible configuration (or at least, as he identified in his paper, every configuration that had the same values of "constants of the motion", like energy).

In early 1877 Boltzmann returned to the same question, including as one section in his "Remarks on Some Problems in the Mechanical Theory of Heat" more analysis of the same potential as before, but now showing a diversity of more complicated pictures that almost seem to justify his rule-30-before-its-time idea that there could be "pure mechanics" that would lead to "Second Law" behavior:

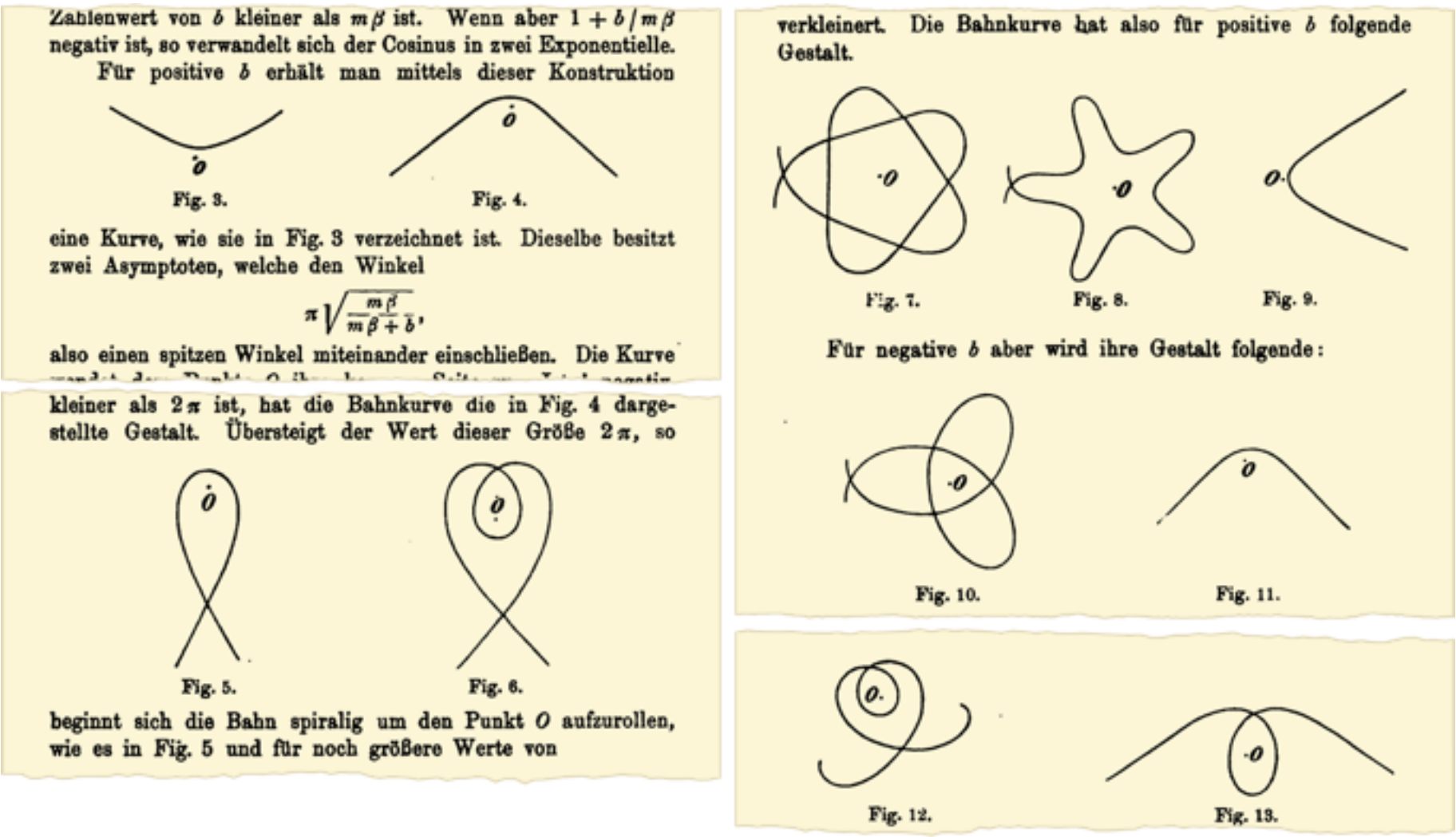

Zahlenwert von $b$ kleiner als $m\beta$ ist. Wenn aber $1 + b / m\beta$ negativ ist, so verwandelt sich der Cosinus in zwei Exponentielle.

Für positive $b$ erhält man mittels dieser Konstruktion

Fig. 3. Fig. 4.

eine Kurve, wie sie in Fig. 3 verzeichnet ist. Dieselbe besitzt zwei Asymptoten, welche den Winkel

$$\pi \sqrt{\frac{m\beta}{m\beta + b}},$$

also einen spitzen Winkel miteinander einschließen. Die Kurve kleiner als $2\pi$ ist, hat die Bahnkurve die in Fig. 4 dargestellte Gestalt. Übersteigt der Wert dieser Größe $2\pi$, so beginnt sich die Bahn spiralig um den Punkt $O$ aufzurollen, wie es in Fig. 5 und für noch größere Werte von verkleinert. Die Bahnkurve hat also für positive $b$ folgende Gestalt.

Fig. 5. Fig. 6.

Fig. 7. Fig. 8. Fig. 9.

Für negative $b$ aber wird ihre Gestalt folgende:

Fig. 10. Fig. 11.

Fig. 12. Fig. 13.

In modern times, of course, it's easy to solve those equations of motion, and typical results obtained for an array of values of parameters are:

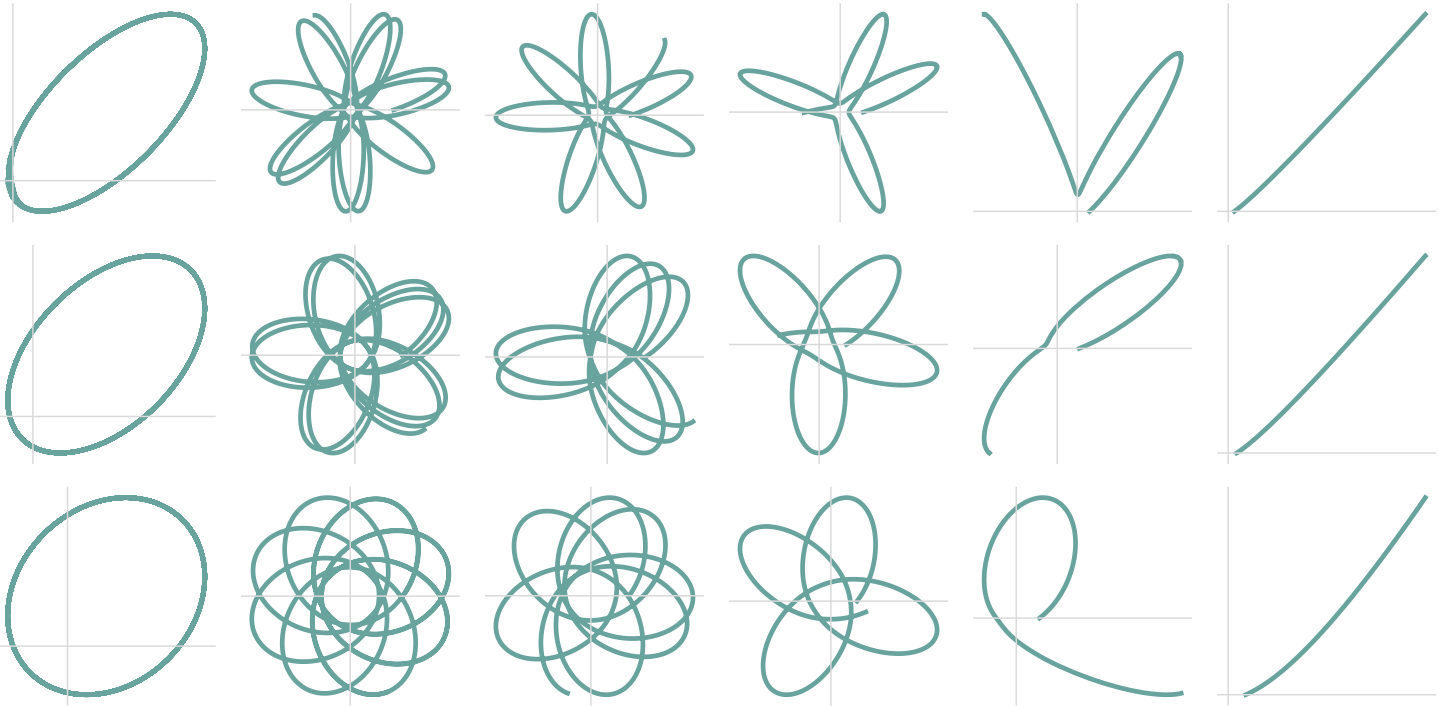

Boltzmann returned to these questions in 1884, responding to Helmholtz's analysis of what he was calling "monocyclic systems". Boltzmann used the same potential again, but now with a name for the "visit-all-states" property: isodic. Meanwhile, Boltzmann had introduced the name "ergoden" for the collection of all possible configurations of a system with a given energy (what would now be called the microcanonical ensemble). But somehow, quite a few years later, Boltzmann's student Paul Ehrenfest (1880–1933) (along with Tatiana Ehrenfest-Afanassjewa (1876–1964)) would introduce the term "ergodic" for Boltzmann's isodic. And "ergodic" is the term that caught on. And in the twentieth century there was all sorts of development of "ergodic theory", as we'll discuss a bit later.

But back in the 1800s people continued to discuss the possibility that what would become called ergodicity was somehow generic, and would explain why all states would somehow be equally probable, why the Maxwell distribution of velocities would be obtained, and ultimately why the Second Law was true. Maxwell worked out some examples. So did Kelvin. But it remained unclear how it would all work out, as Kelvin (now with many letters after his name) discussed in a talk he gave in 1900 celebrating the new century:

I. *Nineteenth Century Clouds over the Dynamical Theory of Heat and Light* *. *By* The Right. Hon. Lord KELVIN, *G.C.V.O., D.C.L., LL.D., F.R.S., M.R.I.* †.

§ 1. THE beauty and clearness of the dynamical theory, which asserts heat and light to be modes of motion, is at present obscured by two clouds. I. The first

The dynamical theory of light didn't work out. And about the dynamical theory of heat, he quotes Maxwell (following Boltzmann) in one of his very last papers, published in 1878, as saying, in reference to what amounts to a proof of the Second Law from underlying dynamics:

"The only assumption which is necessary for the direct "proof is that the system, if left to itself in its actual state of

"motion, will, sooner or later, pass [infinitely nearly *] "through every phase which is consistent with the equation "of energy" (p. 714) and, again (p. 716).

I have never seen validity in the demonstration ‡ on which Maxwell founds this statement, and it has always seemed to me exceedingly improbable that it can be true. If true, it would be very wonderful, and most interesting in pure mathematical dynamics. Having been published by Boltzmann and Maxwell it would be worthy of most serious attention, even without consideration of its bearing on thermo-dynamics. But, when we consider its bearing

Kelvin talks about exploring test cases:

mathematics. Ten years ago *, I suggested a number of test-cases, some of which have been courteously considered by Boltzmann; but no demonstration either of the truth or untruth of the doctrine as applied to any one of them has hitherto been given. A year later, I suggested what seemed to me a decisive test-case disproving the doctrine; but my statement was quickly and justly criticised by Boltzmann and Poincaré; and more recently Lord Rayleigh † has shown very clearly that my simple test-case was quite indecisive. This last article of Rayleigh's has led me to resume the consideration of several classes of dynamical problems, which had occupied me more or less at various times during the last twenty years, each presenting exceedingly interesting features in connection with the double question: Is this a case which admits of the application of the Boltzmann-Maxwell doctrine; and if so, is the doctrine true for it?

When, for example, is the motion of a single particle bouncing around in a fixed region ergodic? He considers first an ellipse, and proves that, no, there isn't in general ergodicity there:

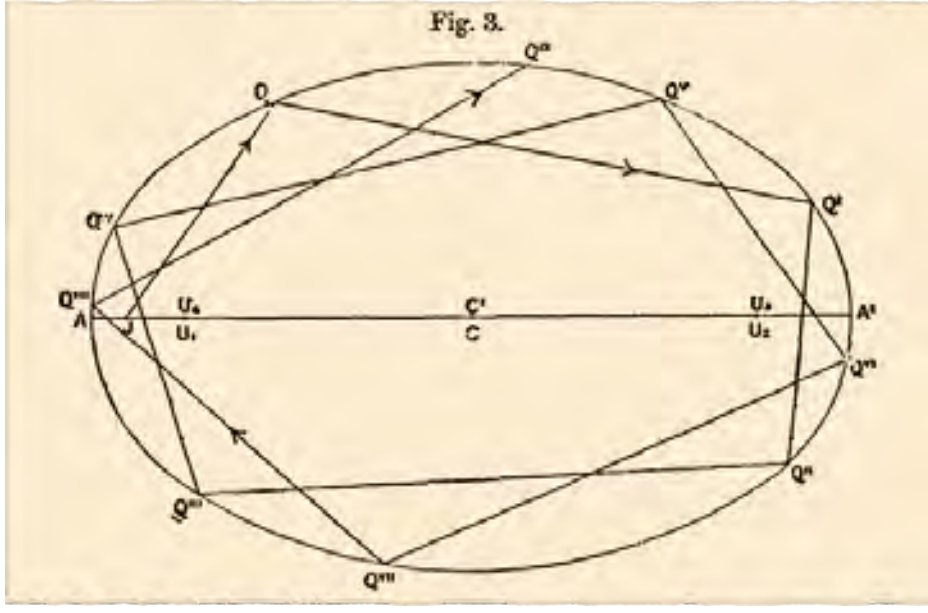

Then he goes on to the much more complicated case

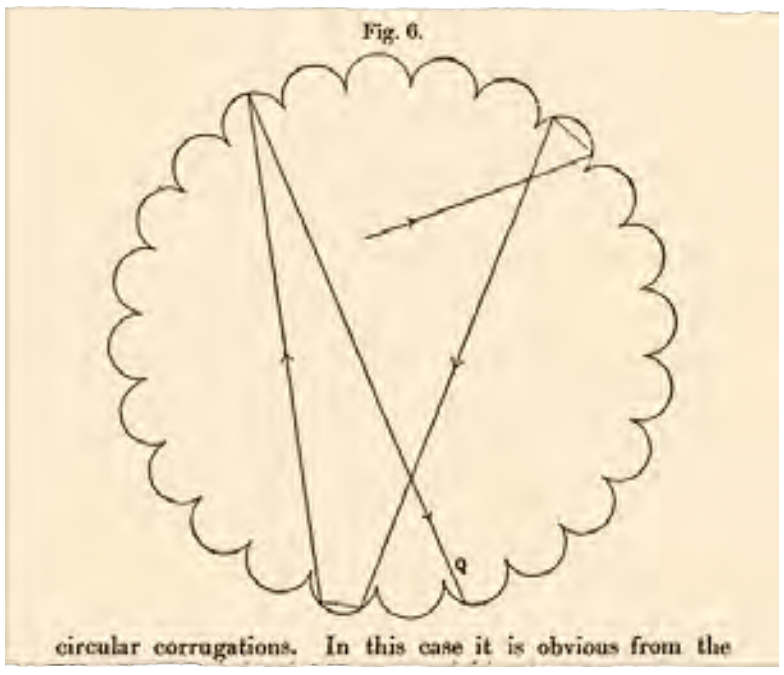

and now he does an "experiment" (with a rather Monte Carlo flavor):

circular corrugations. In this case it is obvious from the symmetry that the time-integral of kinetic energy of component motion parallel to any straight line must, in the long run, be equal to that parallel to any other. But the Boltzmann-Maxwell doctrine asserts, that the time-integrals of the kinetic energies of the two components, radial and transversal, according to polar coordinates, would be equal. To test this, I have taken the case of an infinite number of the semicircular corrugations, so that in the time-integral it is not necessary to include the times between successive impacts of the particle on any one of the semicircles. In this case the geometrical construction would, of course, fail to show the precise point Q at which the free path would cut the diameter AB of the semicircular hollow to which it is approaching; and I have evaded the difficulty in a manner thoroughly suitable for thermodynamic application such as the kinetic theory of gases. I arranged to draw lots for 1 out of the 199 points dividing AB into 200 equal parts. This was done by taking 100 cards*, 0, 1 . . . . . 98, 99, to represent distances from the middle point, and, by the toss of a coin, determining on which side of the middle point it was to be (plus or minus for head or tail, frequently changed to avoid possibility of error by bias). The draw for one of the hundred numbers (0 . . . . 99) was taken after very thorough shuffling of the cards in each case. The point of entry having been found, a large-scale geometrical construction was used to determine the successive points of impact and the inclination $\theta$ of the emergent path to the diameter AB. The inclination of the entering path to the diameter of the semicircular hollow struck at the end of the flight, has the same value $\theta$. If we call the diameter of the large circle unity, the length of each flight is $\sin\theta$. Hence, if the velocity is unity and the mass of the particle 2, the time-integral of the whole kinetic energy is $\sin\theta$; and it is easy to prove that the time-integrals of the components of the velocity, along and perpendicular to the line from each point of the path to the centre of the large circle, are respectively $\theta\cos\theta$, and $\sin\theta-\theta\cos\theta$. The excess of the latter above the former is $\sin\theta-2\theta\cos\theta$. By summation for 143 flights we have found,

$$\Sigma\sin\theta=121{\cdot}3\ ;\ 2\Sigma\theta\cos\theta=108{\cdot}3\ ;$$

whence,

$$\Sigma\sin\theta-2\Sigma\theta\cos\theta=13{\cdot}0.$$

This is a notable deviation from the Boltzmann-Maxwell doctrine, which makes $\Sigma(\sin\theta-\theta\cos\theta)$ equal to $\Sigma\theta\cos\theta$. We have found the former to exceed the latter by a difference which amounts to 10·7 of the whole $\Sigma\sin\theta$.

Out of fourteen sets of ten flights, I find that the time-integral of the transverse component is less than half the whole in twelve sets, and greater in only two. This seems to prove beyond doubt that the deviation from the Boltzmann-Maxwell doctrine is genuine; and that the time-integral of the transverse component is certainly smaller than the time-integral of the radial component.

Kelvin considers a few other examples

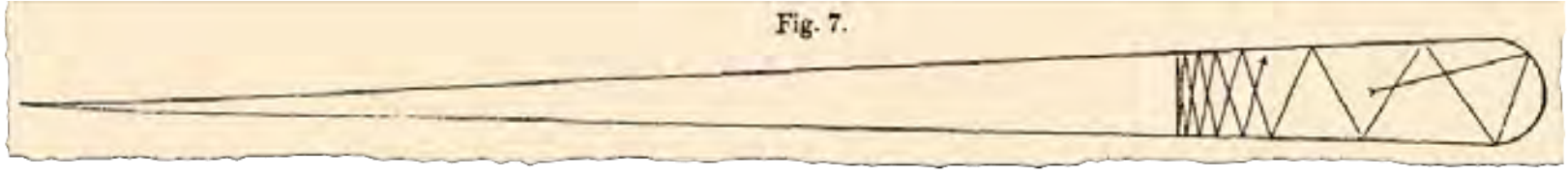

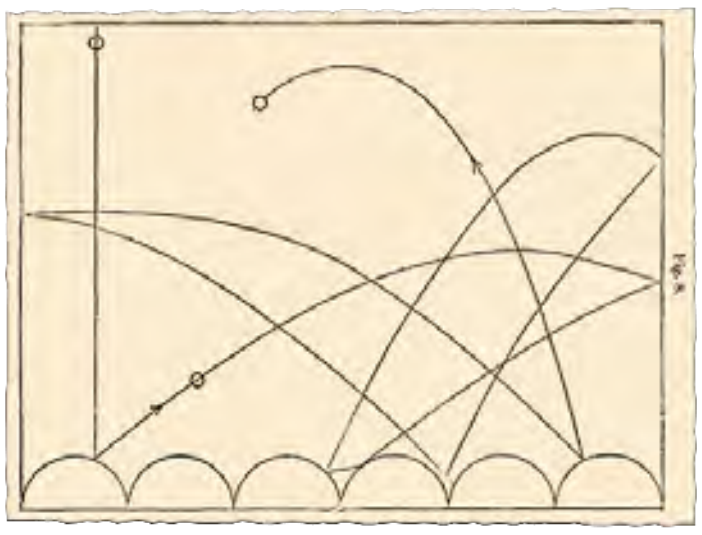

but mostly concludes that he can't tell in general about ergodicity—and that probably something else is needed, or as he puts it (somehow wrapping the theory of light into the story as well):

The simplest way of arriving at this desired result is to deny the conclusion; and so, in the beginning of the twentieth century, to lose sight of a cloud which has obscured the brilliance of the molecular theory of heat and light during the last quarter of the nineteenth century.

## But What about Reversibility?

Had Boltzmann's 1872 *H* theorem proved the Second Law? Was the Second Law—with its rather downbeat implication about the heat death of the universe—even true? One skeptic was Boltzmann's friend and former teacher, the chemist Josef Loschmidt (1821–1895), who in 1866 had used kinetic theory to (rather accurately) estimate the size of air molecules. And in 1876 Loschmidt wrote a paper entitled "On the State of Thermal Equilibrium in a System of Bodies with Consideration of Gravity" in which he claimed to show that when gravity was taken into account, there wouldn't be uniform thermal equilibrium, the Maxwell distribution, or the Second Law—and thus, as he poetically explained:

> The terroristic nimbus of the Second Law is destroyed, a nimbus which makes that Second Law appear as the annihilating principle of all life in the universe—and at the same time we are confronted with the comforting perspective that, as far as the conversion of heat into work is concerned, mankind will not solely be dependent on the intervention of coal or of the Sun, but will have available an inexhaustible resource of convertible heat at all times.

His main argument revolves around a thought experiment involving molecules in a gravitational field:

zontalschichten auch für die Gasarten als fraglich erscheinen lässt.

Fig. 9.

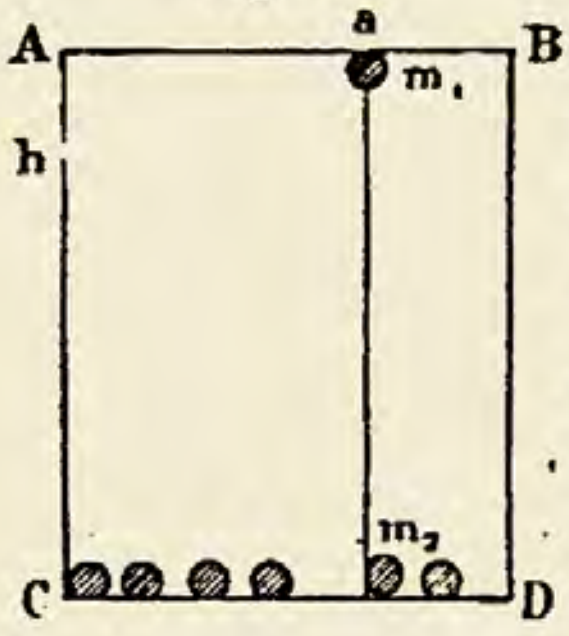

In dem prismatischen Gefässe *ABCD* Fig. 9 befinden sich *n* einfache Atome, jedes von der Masse *m*. Die Molecularkraft zwischen den Atomen und den Gefässwänden specialisiren wir als die der elastischen Kugeln, und als einzige äussere Kraft nehmen wir die Schwerkraft an. Die Anfangszustände des Systemes aber setzen wir auf folgende Weise fest.

Von den *n* Atomen heben wir ein einziges $m_1$ an die obere Gefässwand *AB*,

Over the next couple of years, despite Loschmidt's progressively more elaborate constructions

Ich nehme also zwei gleiche prismatische Gefässe $V_1$ und $V_2$ Fig. 1, vertical übereinandergestellt. In jedem befindet sich eine grosse Anzahl schwerer elastischer Kugeln, d. h. beide sind mit demselben Gase gefüllt. Diese beiden Gefässe denken wir uns durch die ver-

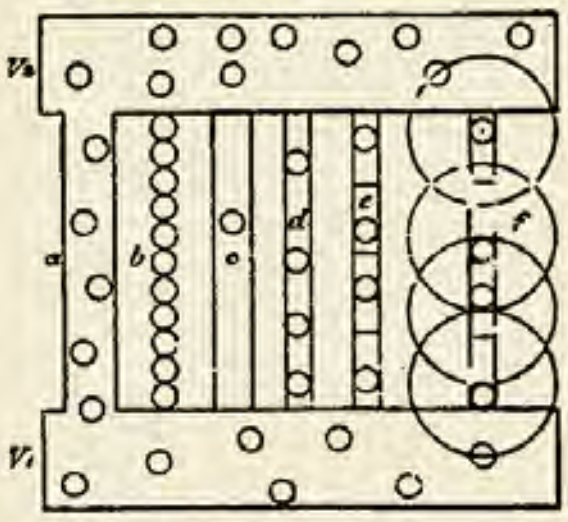

Boltzmann and Maxwell will debunk this particular argument—even though to this day the role of gravity in relation to the Second Law remains incompletely resolved.

But what's more important for our narrative about Loschmidt's original paper are a couple of paragraphs tucked away at the end of one section (that in fact Kelvin had basically anticipated in 1874):

> [Consider what would happen if] after a time *t* sufficiently long for the stationary state to obtain, we suddenly reversed the velocities of all atoms. Initially we would be in a state that would look like the stationary state. This would be true for some time, but in the long run the stationary state would deteriorate and after the time *t* we would inevitably return to the initial state...
>
> It is clear that in general in any system one can revert the entire course of events by suddenly inverting the velocities of all the elements of the system. This doesn't give a solution to the problem of undoing everything that happens [in the universe] but it does give a simple prescription: just suddenly revert the instantaneous velocities of all atoms of the universe.

How did this relate to the *H* theorem? The underlying molecular equations of motion that Boltzmann had assumed in his proof were reversible in time. Yet Boltzmann claimed that *H* was always going to a minimum. But why couldn't one use Loschmidt's argument to construct an equally possible "reverse evolution" in which *H* was instead going to a maximum?

It didn't take Boltzmann long to answer, in print, tucked away in a section of his paper "Remarks on Some Problems in the Mechanical Theory of Heat". He admits that Loschmidt's argument "has great seductiveness". But he claims it is merely "an interesting sophism"—and then says he will "locate the source of the fallacy". He begins with a classic setup: a collection of hard spheres in a box.

> Suppose that at time zero the distribution of spheres in the box is not uniform; for example, suppose that the density of spheres is greater on the right than on the left ... The sophism now consists in saying that, without reference to the initial conditions, it cannot be proved that the spheres will become uniformly mixed in the course of time.

But then he rather boldly claims that with the actual initial conditions described, the spheres will "almost always [become] uniform" at a future time *t*. Now he imagines (following Loschmidt) reversing all the velocities in this state at time *t*. Then, he says:

> ... the spheres would sort themselves out as time progresses, and at [the analog of] time 0, they would have a completely nonuniform distribution, even though the [new] initial distribution [one had used] was almost uniform.

But now he says that, yes—given this counterexample—it won't be possible to prove that the final distribution of spheres will always be uniform.

> This is in fact a consequence of probability theory, for any nonuniform distribution, no matter how improbable it may be, is still not absolutely impossible. Indeed it is clear that any individual uniform distribution, which might arise after a certain time from some particular initial state, is just as improbable as an individual nonuniform distribution; just as in the game of Lotto, any individual set of five numbers is as improbable as the set 1, 2, 3, 4, 5. It is only because there are many more uniform distributions than nonuniform ones that the distribution of states will become uniform in the course of time. One therefore cannot prove that, whatever may be the positions and velocities of the spheres at the beginning, the distribution must become uniform after a long time; rather one can only prove that infinitely many more initial states will lead to a uniform one after a definite length of time than to a nonuniform one.

He adds:

> One could even calculate, from the relative numbers of the different state distributions, their probabilities, which might lead to an interesting method for the calculation of thermal equilibrium.

And indeed within a few months Boltzmann has followed up on that "interesting method" to produce his classic paper on the probabilistic interpretation of entropy.

But in his earlier paper he goes on to argue:

> Since there are infinitely many more uniform than nonuniform distributions of states, the latter case is extraordinarily improbable [to arise] and can be considered impossible for practical purposes; just as it may be considered impossible that if one starts with oxygen and nitrogen mixed in a container, after a month one will find chemically pure oxygen in the lower half and nitrogen in the upper half, although according to probability theory this is merely very improbable but not impossible.

He talks about how interesting it is that the Second Law is intimately connected with probability while the First Law is not. But at the end he does admit:

> Perhaps this reduction of the Second Law to the realm of probability makes its application to the entire universe appear dubious, but the laws of probability theory are confirmed by all experiments carried out in the laboratory.

At this point it's all rather unconvincing. The *H* theorem had purported to prove the Second Law. But now he's just talking about probability theory. He seems to have given up on proving the Second Law. And he's basically just saying that the Second Law is true because it's observed to be true—like other laws of nature, but not like something that can be "proved", say from underlying molecular dynamics.

For many years not much attention was paid to these issues, but by the late 1880s there were attempts to clarify things, particularly among the rather active British circle of kinetic theorists. A published 1894 letter from the Irish mathematician Edward Culverwell (1855–1931) (who also wrote about ice ages and Montessori education) summed up some of the confusions that were circulating:

**Boltzmann's Minimum Theorem.**

THE remarkable differences of opinion as to what the H-theorem *is*, and how it can be proved, show how necessary is the discussion elicited by my letter on the oversight in Dr. Watson's proof. Each of the four authorities who have replied takes a different view.

Dr. Larmor enforces the view I put forward at the close of my letter, and says that the theorem *is* what I said appeared an *à priori* possibility; and I may here point out that his letter is a complete answer to the argument I used in the *Phil. Mag.* 1890, p. 95, urging that, as there were as many configurations which receded from the permanent state as approached it, there was an *à priori* improbability that a permanent state would ever be reached. This argument was criticised at some length, not really answered, in Messrs. Larmor and Bryan's Report on Thermodynamics (British Association Report, 1891), but the suggestive remarks there given helped me, I think, to arrive (independently) at the complete answer given in Dr. Larmor's recent letter. But my present use of the argument is not that which Dr. Larmor criticises; I now use it as a test of a particular proof of the H-theorem. I say that if that proof does not somewhere or other introduce some assumption about averages, probability, or irreversibility, it cannot be valid.

Mr. Burbury appears to consider that the theorem can only be proved if we assume that some element of the distribution does tend to an average (quite a different position from Dr. Larmor's), and he is as yet unable to state the appropriate assumption except for the case of hard elastic spherical particles colliding or "encountering" (for since $a$ is constant in his last letter, it seems as if the $q_1 \ldots q_{n-2}$ coordinates are really dummies). Yet Mr. Burbury has already given what purports to be a *general* proof of the theorem for any number of degrees of freedom.

Mr. Bryan thinks that a condition which excludes the reversed motion is implied in Dr. Watson's proof, for he says that in taking unaccented letters $F f$ as proportional to the number of molecules passing from one configuration to another in the reversed motion, I make a less "natural" supposition than Dr. Watson, who takes accented letters $F' f'$. I cannot see what virtue there is in putting accents on or leaving them off, and after a very careful study of Mr. Bryan's letter, I can only think that he has fallen into some confusion owing to the way in which he uses at one time *accented* and at another time *unaccented differentials*, although (as he himself remarks) there is no difference whatever between their accented and unaccented *products*. But even if Mr. Bryan be right, would he put us any "forrarder"? What we want is a *proof* that the collisions will make H decrease, and we can hardly be satisfied with a proof which depends on the previous assumption that the particles do "naturally" tend to move in the desired way.

Dr. Watson meets my reversibility argument by saying that H decreases even in the reversed motion, when the system is confessedly *receding* from its permanent state. No other correspondent agrees with him in this view, which would indeed *take away all physical meaning from the H theorem*, for the decrease of H would then be quite unconnected with the approach to a permanent state. As to the other point, Dr. Watson does not amend his proof himself, but says it is "easy" to do, and so does Mr. Bryan. Yet one has an instinctive distrust of things which are said to be "easily seen," and at all events Dr. Watson's reference to the case in which the theorem is *applied* does not help one in the *proof*, where it is necessary to express *separately* the products of the differentials expressed by the small and capital letters respectively in his "Kinetic Theory."

Mr. Burbury asks why I say the error law has been proved for the case of hard spheres without the use of Boltzmann's Minimum Theorem. I thought Tait had done so (*Trans.* R.S.E. 1886), and at all events I thought the ordinary investigation showed that there was but *one* solution, that of the error law, in that case; but perhaps I am mistaken.

Mr. Bryan says Lorenz gives the clearest account of the assumptions in Boltzmann's theorem. He would earn our gratitude if he would state them in his next letter.

EDW. P. CULVERWELL.

Trinity College, Dublin, December 29, 1894.

At a lecture in England the next year, Boltzmann countered (conveniently, in English):

§ 2. Mr. Culverwell's objections against my Minimum Theorem bear the closest connections to what I pointed out in the second part of my paper „Bemerkungen über einige Probleme der mechanischen Wärmetheorie", Wien. Ber. 75. 1877.[1]) There I pointed out that my Minimum Theorem, as well as the so-called Second Law of Thermodynamics, are only theorems of probability. The Second Law can never be proved mathematically by means of the equations of dynamics alone.

He goes on, but doesn't get much more specific:

Though interesting and striking at the first moment, Mr. Culverwell's arguments rest, as I think, only upon a mistake of my assumptions. It can never be proved from the equations of motion alone, that the minimum function *H* must always decrease. It can only be deduced from the laws of probability, that if the initial state is not specially arranged for a certain purpose, but haphazard governs freely, the probability that *H* decreases is always greater than that it increases. It is well known that the theory of probability is

He then makes an argument that will be repeated many times in different forms, saying that there will be fluctuations, where *H* deviates temporarily from its minimum value, but these will be rare:

shall beexcluded. During the greater part of this time *H* will be very nearly equal to its minimum value *H* (min.). Let us construct the *H*-curve, *i. e.* let us take the time as axis of abscissæ and draw the curve, whose ordinates are the corresponding values of *H*. The greater majority of the ordinates of this curve are very nearly equal to *H* (min.). But because greater values of *H* are not mathematically impossible, but only very improbable, the curve has certain, though very few, summits or maximum ordinates which rise to a greater height than *H* (min.).

Later he's talking about what he calls the "*H* curve" (a plot of *H* as a function of time), and he's trying to describe its limiting form:

remark. Not for every curve, but only for the particular form of the $H$-curve, disymmetrical in the upward and downward direction, can it be proved that $H$ has a tendency to decrease. This particular form is very well illustrated by Mr. Culverwell's suggestion of an inverted tree. The $H$-curve is composed of a succession of such trees. Almost all these trees are extremely low, and have branches very nearly horizontal. Here $H$ has nearly the minimum value. Only very few trees are higher, and have branches inclined to the axis of abscissæ, and the improbability of such a tree increases enormously with its height. The difficulty consists only in imagining all these branches infinitely short.

Finally there is the difference between the ordinary cases, where $H$ decreases or is near to its minimum value, and the very rare cases, where $H$ is far from the minimum value and still increasing. In the last cases, $H$ will reach, probably in a very short time, a maximum value. Then it will decrease from that value to the well-known minimum value.

And he even refers to Weierstrass's recent work on nondifferentiable functions:

increase of $x$ for all points, whose ordinates are $= 1$. The $P$-curve belongs to the large class of curves which have nowhere a uniquely defind tangent. Even at the top of each summit the tangent is not parallel to the $x$-axis, but is undefined. In other words, the chord joining two points of the curve does not tend toward a difinite limiting position when one of the two points approaches and ultimately coincides with the other.[1]) The same applies to the $H$-curve in the Theory of Gases. If I find a certain negative value for $dH/dt$, that does not define the tangent of the curve in the ordinary sense, but it is only an average value.

But he doesn't pursue this, and instead ends his "rebuttal" with a more philosophical—and in some sense anthropic—argument that he attributes to his former assistant Ignaz Schütz (1867–1927):

§ 3. Mr. Culverwell says that my theorem cannot be true because if it were true every atom of the universe would have the same average *vis viva*, and all energy would be dissipated. I find, on the contrary, that this argument only tends to confirm my theorem, which requires only that in the course of time the universe must tend to a state where the average *vis viva* of every atom is the same and all energy is dissipated, and that is indeed the case. But if we ask why this state is not yet reached, we again come to a „Salisburian mystery".

I will conclude this paper with an idea of my old assistant, Dr. Schuetz.

We assume that the whole universe is, and rests for ever, in thermal equilibrium. The probability that one (only one) part of the universe is in a certain state, is the smaller the further this state is from thermal equilibrium; but this probability is greater, the greater the universe itself is. If we assume the universe great enough we can make the probability of one relatively small part being in any given state (however far from the state of thermal equilibrium), as great as we please. We can also make the probability great that, though the whole universe is in thermal equilibrium, our world is in its present state. It may be sayd that the world is so far from thermal equilibrium that we cannot imagine the improbability of such a state. But can we imagine, on the other side, how small a part of the whole universe this world is? Assuming the universe great enough, the probability that such a small part of it as our world should be in its present state, is no longer small.

If this assumption were correct, our world would return more and more to thermal equilibrium; but because the whole universe is so great, it might be probable that at some future time some other world might deviate as far from thermal equilibrium as our world does at present. Then the aforementioned *H*-curve would form a representation of what takes place in the universe. The summits of the curve would represent the worlds where visible motion and life exist.

It's an argument that we'll see in various forms repeated over the century and a half that follows. In essence what it's saying is that, yes, the Second Law implies that the universe will end up in thermal equilibrium. But there'll always be fluctuations. And in a big enough universe there'll be fluctuations somewhere that are large enough to correspond to the world as we experience it, where "visible motion and life exist".

But regardless of such claims, there's a purely formal question about the *H* theorem. How exactly is it that from the Boltzmann transport equation—which is supposed to describe reversible mechanical processes—the *H* theorem manages to prove that the *H* function irreversibly decreases? It wasn't until 1895—fully 25 years after Boltzmann first claimed to prove the *H* theorem—that this issue was even addressed. And it first came up rather circuitously through Boltzmann's response to comments in a textbook by Gustav Kirchhoff (1824–1887) that had been completed by Max Planck.

The key point is that Boltzmann's equation makes an implicit assumption, that's essentially the same as Maxwell made back in 1860: that before each collision between molecules, the molecules are statistically uncorrelated, so that the probability for the collision has the factored form $f(v_1)\,f(v_2)$. But what about after the collision? Inevitably the collision itself will lead to correlations. So now there's an asymmetry: there are no correlations before each collision, but there are correlations after. And that's why the behavior of the system doesn't have to be symmetrical—and the *H* theorem can prove that *H* irreversibly decreases.

In 1895 Boltzmann wrote a 3–page paper (after half in footnotes) entitled "More about Maxwell's Distribution Law for Speeds" where he explained what he thought was going on:

> [The reversibility of the laws of mechanics] has been recently applied in judging the assumptions necessary for a proof of [the *H* theorem]. This proof requires the hypothesis that the state of the gas is and remains molecularly disordered, namely, that the molecules of a given class do not always or predominantly collide in a specific manner and that, on the contrary, the number of collisions of a given kind can be found by the laws of probability.
>
> Now, if we assume that in general a state distribution never remains molecularly ordered for an unlimited time and also that for a stationary state–distribution every velocity is as probable as the reversed velocity, then it follows that by inversion of all the velocities after an infinitely long time every stationary state–distribution remains unchanged. After the reversal, however, there are exactly as many collisions occurring in the reversed way as there were collisions occurring in the direct way. Since the two state distributions are identical, the probability of direct and indirect collisions must be equal for each of them, whence follows Maxwell's distribution of velocities.

Boltzmann is introducing what we'd now call the "molecular chaos" assumption (and what Ehrenfest would call the *Stosszahlansatz*)—giving a rather self–fulfilling argument for why the assumption should be true. In Boltzmann's time there wasn't really anything better to do. By the 1940s the BBGKY hierarchy at least let one organize the hierarchy of correlations between molecules—though it still didn't give one a tractable way to assess what correlations should exist in practice, and what not.

Boltzmann knew these were all complicated issues. But he wrote about them at a technical level only a few more times in his life. The last time was in 1898 when, responding to a request from the mathematician Felix Klein (1849–1925), he wrote a paper about the *H* curve for mathematicians. He begins by saying that although this curve comes from the theory of gases, the essence of it can be reproduced by a process based on accumulating balls randomly picked from an urn. He then goes on to outline what amounts to a story of random walks and fractals. In another paper, he actually sketches the curve

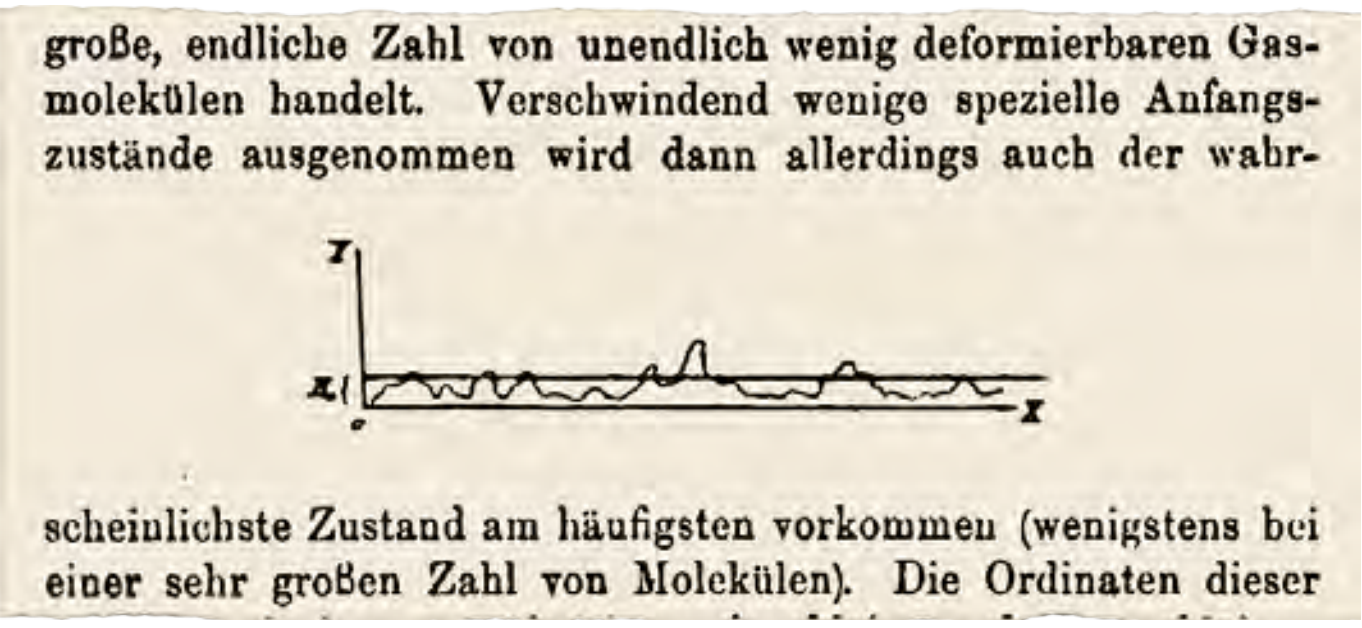
große, endliche Zahl von unendlich wenig deformierbaren Gasmolekülen handelt. Verschwindend wenige spezielle Anfangszustände ausgenommen wird dann allerdings auch der wahrscheinlichste Zustand am häufigsten vorkommen (wenigstens bei einer sehr großen Zahl von Molekülen). Die Ordinaten dieser saying that his drawing "should be taken with a large grain of salt", noting—in a remarkably fractal-reminiscent way—that "a zincographer [i.e. an engraver of printing plates] would not have been able to produce a real figure since the *H*-curve has a very large number of maxima and minima on each finite segment, and hence defies representation as a line of continuously changing direction."

Of course, in modern times it's easy to produce an approximation to the *H* curve according to his prescription:

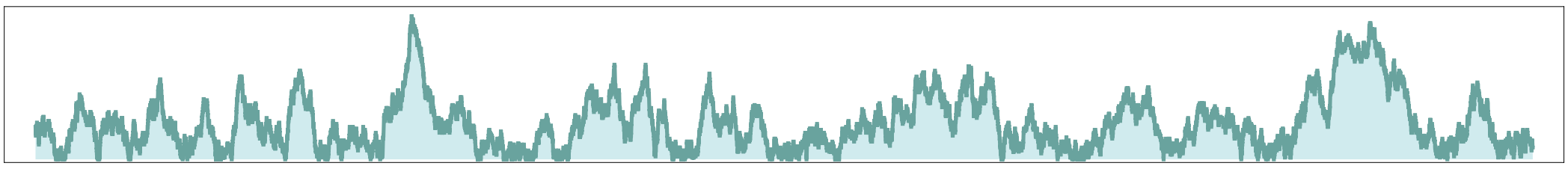

But at the end of his "mathematical" paper he comes back to talking about gases. And first he makes the claim that the effective reversibility seen in the *H* curve will never be seen in actual physical systems because, in essence, there are always perturbations from outside. But then he ends, in a statement of ultimate reversibility that casts our everyday observation of irreversibility as tautological:

> There is no doubt that it is just as conceivable to have a world in which all natural processes take place in the wrong chronological order. But a person living in this upside-down world would have feelings no different than we do: they would just describe what we call the future as the past and vice versa.

## The Recurrence Objection

Probably the single most prominent research topic in mathematical physics in the 1800s was the three-body problem—of solving for the motion under gravity of three bodies, such as the Earth, Moon and Sun. And in 1890 the French mathematician Henri Poincaré (1854–1912) (whose breakout work had been on the three-body problem) wrote a paper entitled "On the Three-Body Problem and the Equations of Dynamics" in which, as he said:

> It is proved that there are infinitely many ways of choosing the initial conditions such that the system will return infinitely many times as close as one wishes to its initial position. There are also an infinite number of solutions that do not have this property, but it is shown that these unstable solutions can be regarded as "exceptional" and may be said to have zero probability.

This was a mathematical result. But three years later Poincaré wrote what amounted to a philosophy paper entitled "Mechanism and Experience" which expounded on its significance for the Second Law:

> In the mechanistic hypothesis, all phenomena must be reversible; for example, the stars might traverse their orbits in the retrograde sense without violating Newton's law; this would be true for any law of attraction whatever. This is therefore not a fact peculiar to astronomy; reversibility is a necessary consequence of all mechanistic hypotheses.
>
> Experience provides on the contrary a number of irreversible phenomena. For example, if one puts together a warm and a cold body, the former will give up its heat to the latter; the opposite phenomenon never occurs. Not only will the cold body not return to the warm one the heat which it has taken away when it is in direct contact with it; no matter what artifice one may employ, using other intervening bodies, this restitution will

> be impossible, at least unless the gain thereby realized is compensated by an equivalent or large loss. In other words, if a system of bodies can pass from state A to state B by a certain path, it cannot return from B to A, either by the same path or by a different one. It is this circumstance that one describes by saying that not only is there not direct reversibility, but also there is not even indirect reversibility.

But then he continues:

> A theorem, easy to prove, tells us that a bounded world, governed only by the laws of mechanics, will always pass through a state very close to its initial state. On the other hand, according to accepted experimental laws (if one attributes absolute validity to them, and if one is willing to press their consequences to the extreme), the universe tends toward a certain final state, from which it will never depart. In this final state, which will be a kind of death, all bodies will be at rest at the same temperature.

But in fact, he says, the recurrence theorem shows that:

> This state will not be the final death of the universe, but a sort of slumber, from which it will awake after millions of millions of centuries. According to this theory, to see heat pass from a cold body to a warm one ... it will suffice to have a little patience. [And we may] hope that some day the telescope will show us a world in the process of waking up, where the laws of thermodynamics are reversed.

By 1903, Poincaré was more strident in his critique of the formalism around the Second Law, writing (in English) in a paper entitled "On Entropy":

8. *Conclusions.*

A. The entropy is a function of the co-ordinates.

B. Is not defined by the equation

$$d\varphi = \int \frac{dH}{\theta}.$$

This equation, arising from *another* definition of entropy, can be demonstrated for reversible changes.

C. It is wrong for *all* irreversible changes, and not only for those where there is exchange of heat in the narrow sense of the word.

D. In an irreversible change in which there is no exchange of heat the entropy increases.

E. It increases, for instance, in the case of a mixture of gases, and the increase can be calculated by the artifice I have discussed at the end of section 3.

F. If the universe is regarded as an isolated system, it can never come back to its original state; for its entropy is always growing, and this entropy being a function of the cŏ-ordinates, would come back to its original value if the universe came back to its original state.

But back in 1896, Boltzmann and the *H* theorem had another critic: Ernst Zermelo (1871–1953), a recent German math PhD who was then working with Max Planck on applied mathematics—though would soon turn to foundations of mathematics and become the "Z" in ZFC set theory. Zermelo's attack on the *H* theorem began with a paper entitled "On a Theorem of Dynamics and the Mechanical Theory of Heat". After explaining Poincaré's recurrence theorem, Zermelo gives some "mathematician-style" conditions (the gas must be in a finite region, must have no infinite energies, etc.), then says that even though there must exist states that would be non-recurrent and could show irreversible behavior, there would necessarily be infinitely more states that "would periodically repeat themselves ...

with arbitrarily small variations". And, he argues, such repetition would affect macroscopic quantities discernable by our senses. He continues:

> In order to retain the general validity of the Second Law, we therefore would have to assume that just those initial states leading to irreversible processes are realized in nature, their small number notwithstanding, while the other ones, whose probability of existence is higher, mathematically speaking, do not actually occur.

And he concludes that the Poincaré recurrence phenomenon means that:

> ... it is certainly impossible to carry out a mechanical derivation of the Second Law on the basis of the existing theory without specializing the initial states.

Boltzmann responded promptly but quite impatiently:

> I have pointed out particularly often, and as clearly as I possibly could ... that the Second Law is but a principle of probability theory as far as the molecular-theoretic point of view is concerned. ... While the theorem by Poincaré that Zermelo discusses in the beginning of his paper is of course correct, its application to heat theory is not.

Boltzmann talks about the *H* curve, and first makes rather a mathematician-style point about the order of limits:

> If we first take the number of gas molecules to be infinite, as was clearly done in [my 1896 proof], and only then let the time grow very large, then, in the vast majority of cases, we obtain a curve asymptotically [always close to zero]. Moreover, as can easily be seen, Poincaré's theorem is not applicable in this case.
> If, however, we take the time [span] to be infinitely great and, in contrast, the number of molecules to be very great but not absolutely infinite, then the *H*-curve has a different character. It almost always runs very close to [zero], but in rare cases it rises above that, in what we shall call a "hump" ... at which significant deviations from the Maxwell velocity distribution can occur ...

Boltzmann then argues that even if you start "at a hump", you won't stay there, and "over an enormously long period of time" you'll see something infinitely close to "equilibrium behavior". But, he says:

> ... it is [always] possible to reach again a greater hump of the *H*-curve by further extending the time ... In fact, it is even the case that the original state must return, provided only that we continue to sufficiently extend the time...

He continues:

> Mr. Zermelo is therefore right in claiming that, mathematically speaking, the motion is periodic. He has by no means succeeded, however, in refuting my theorems, which, in fact, are entirely consistent with this periodicity.

After giving arguments about the probabilistic character of his results, and (as we would now say it) the fact that a 1D random walk is certain to repeatedly return to the origin, Boltzmann says that:

> ... we must not conclude that the mechanical approach has to be modified in any way. This conclusion would be justified only if the approach had a consequence that runs contrary to experience. But this would be the case only if Mr. Zermelo were able to prove that the duration of the period within which the old state of the gas must recur in accordance with Poincaré's theorem has an observable length...

He goes on to imagine "a trillion tiny spheres, each with a [certain initial velocity] ... in the one corner of a box" (and by "trillion" he means million million million, or today's quintillion) and then says that "after a short time the spheres will be distributed fairly evenly in the box", but the period for a "Poincaré recurrence" in which they all will return to their original corner is "so great that nobody can live to see it happen". And to make this point more forcefully, Boltzmann has an appendix in which he tries to get an actual approximation to the recurrence time, concluding that its numerical value "has many trillions of digits".

He concludes:

> If we consider heat as a motion of molecules that occurs in accordance with the general equations of mechanics and assume that the arrangement of bodies that we perceive is currently in a highly improbable state, then a theorem follows that is in agreement with the Second Law for all phenomena so far observed.
>
> Of course, this theorem can no longer hold once we observe bodies of so small a scale that they only contain a few molecules. Since, however, we do not have at hand any experimental results on the behavior of bodies so small, this assumption does not run counter to previous experience. In fact, certain experiments conducted on very small bodies in gases seem rather to support the assumption, although we are still far from being able to assert its correctness on the basis of experimental proof.

But then he gives an important caveat—with a small philosophical flourish:

> Of course, we cannot expect natural science to answer the question as to why the bodies surrounding us currently exist in a highly improbable state, just as we cannot expect it to answer the question as to why there are any phenomena at all and why they adhere to certain given principles.

Unsurprisingly—particularly in view of his future efforts in the foundations of mathematics—Zermelo is unconvinced by all of this. And six months later he replies again in print. He admits that a full Poincaré recurrence might take astronomically long, but notes that (where, by "physical state", he means one that we perceive):

> ... we are after all always concerned only with the "physical state", which can be realized by many different combinations, and hence can recur much sooner.

Zermelo zeroes in on many of the weaknesses in Boltzmann's arguments, saying that the thing he particularly "contests ... is the analogy that is supposed to exist between the properties of the *H* curve and the Second Law". He claims that irreversibility cannot be explained from "mechanical suppositions" without "new physical assumptions"—and in particular criteria for choosing appropriate initial states. He ends by saying that:

> From the great successes of the kinetic theory of gases in explaining the relationships among states we must not deduce its ... applicability also to temporal processes. ... [For in this case I am] convinced that it necessarily fails in the absence of entirely new assumptions.

Boltzmann replies again—starting off with the strangely weak argument:

> The Second Law receives a mechanical explanation by virtue of the assumption, which is of course unprovable, that the universe, when considered as a mechanical system, or at least a very extensive part thereof surrounding us, started out in a highly improbable state and still is in such a state.

And, yes, there's clearly something missing in the understanding of the Second Law. And even as Zermelo pushes for formal mathematician-style clarity, Boltzmann responds with physicist-style "reasonable arguments". There's lots of rhetoric:

> The applicability of the calculus of probabilities to a particular case can of course never be proved with precision. If 100 out of 100,000 objects of a particular sort are consumed by fire per year, then we cannot infer with certainty that this will also be the case next year. On the contrary! If the same conditions continue to obtain for $10^{10}$ years, then it will often be the case during this period that the 100,000 objects are all consumed by fire at once on a single day, and even that not a single object suffers damage over the course of an entire year. Nevertheless, every insurance company places its faith in the calculus of probabilities.

Or, in justification of the idea that we live in a highly improbable "low-entropy" part of the universe:

> I refuse to grant the objection that a mental picture requiring so great a number of dead parts of the universe for the explanation of so small a number of animated parts is wasteful, and hence inexpedient. I still vividly remember someone who adamantly refused to believe that the Sun's distance from the Earth is 20 million miles on the ground that it would simply be foolish to assume so vast a space only containing light ether alongside so small a space filled with life.

Curiously—given his apparent reliance on "commonsense" arguments—Boltzmann also says:

> I myself have repeatedly cautioned against placing excessive trust in the extension of our mental pictures beyond experience and issued reminders that the pictures of contemporary mechanics, and in particular the conception of the smallest particles of bodies as material points, will turn out to be provisional.

In other words, we don't know that we can think of atoms (even if they exist at all) as points, and we can't really expect our everyday intuition to tell us about how they work. Which presumably means that we need some kind of solid, "formal" argument if we're going to explain the Second Law.

Zermelo didn't respond again, and moved on to other topics. But Boltzmann wrote one more paper in 1897 about "A Mechanical Theorem of Poincaré" ending with two more why-it-doesn't-apply-in-practice arguments:

> Poincaré's theorem is of course never applicable to terrestrial bodies which we can hold in our hands as none of them is entirely closed. Nor it is applicable to an entirely closed gas of the sort considered by the kinetic theory if first the number of molecules and only then the quotients of the intervals between two neighboring collisions in the observation time is allowed to become infinite.

## Ensembles, and an Effort to Make Things Rigorous

Boltzmann—and Maxwell before him—had introduced the idea of using probability theory to discuss the emergence of thermodynamics and potentially the Second Law. But it wasn't until around 1900—with the work of J. Willard Gibbs (1839–1903)—that a principled mathematical framework for thinking about this developed. And while we can now see that this framework distracts in some ways from several of the key issues in understanding the foundations of the Second Law, it's been important in framing the discussion of what the Second Law really says—as well as being central in defining the foundations for much of what's been done over the past century or so under the banner of "statistical mechanics".

Gibbs seems to have first gotten involved with thermodynamics around 1870. He'd finished his PhD at Yale on the geometry of gears in 1863—getting the first engineering PhD awarded in the US. After traveling in Europe and interacting with various leading mathematicians and physicists, he came back to Yale (where he stayed for the remaining 34 years of his life) and in 1871 became professor of mathematical physics there.

His first papers (published in 1873 when he was already 34 years old) were in a sense based on taking seriously the formalism of equilibrium thermodynamics defined by Clausius and Maxwell—treating entropy and internal energy, just like pressure, volume and temperature, as variables that defined properties of materials (and notably whether they were solids, liquids or gases). Gibbs's main idea was to "geometrize" this setup, and make it essentially a story of multivariate calculus:

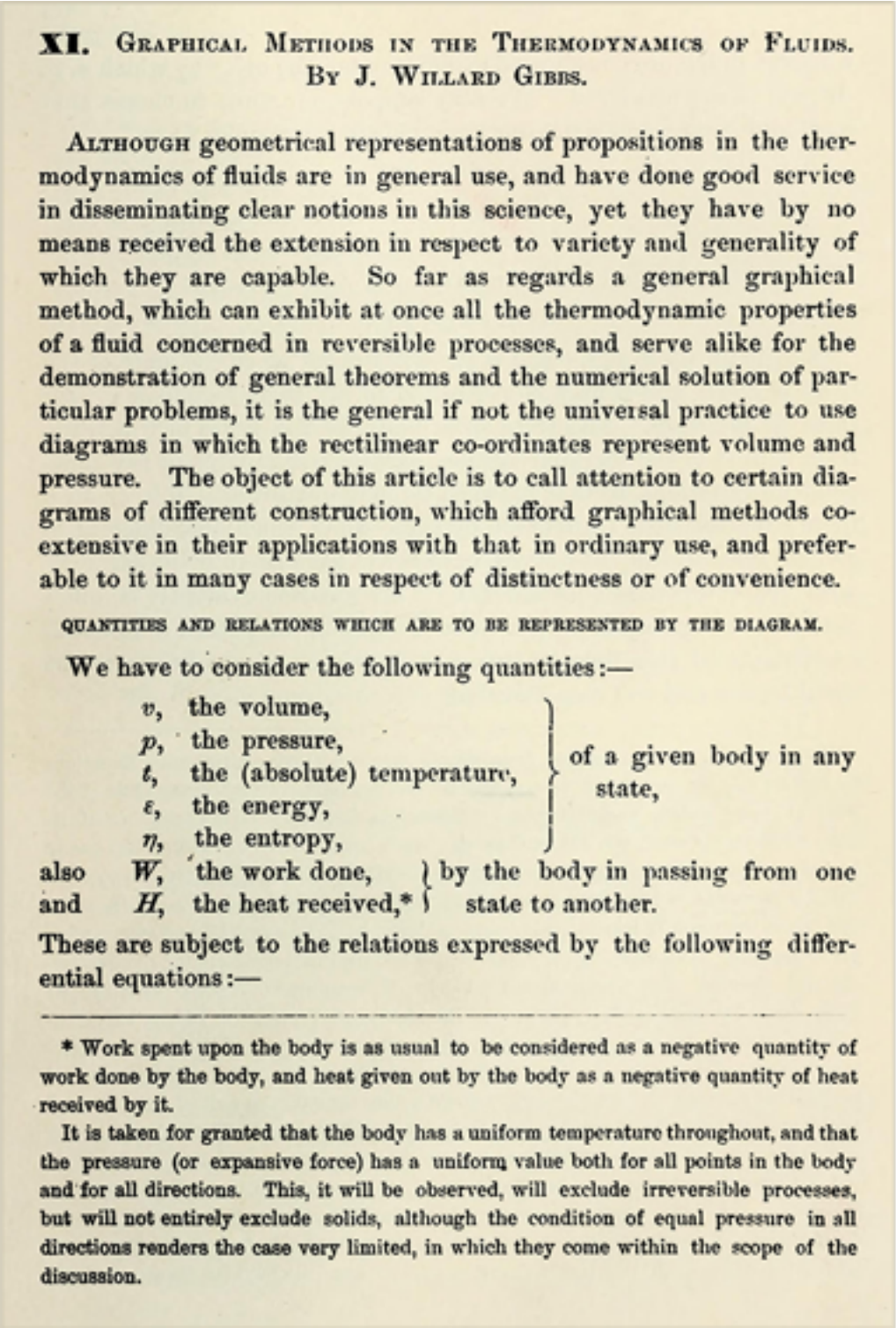

**XI.** GRAPHICAL METHODS IN THE THERMODYNAMICS OF FLUIDS.
BY J. WILLARD GIBBS.

ALTHOUGH geometrical representations of propositions in the thermodynamics of fluids are in general use, and have done good service in disseminating clear notions in this science, yet they have by no means received the extension in respect to variety and generality of which they are capable. So far as regards a general graphical method, which can exhibit at once all the thermodynamic properties of a fluid concerned in reversible processes, and serve alike for the demonstration of general theorems and the numerical solution of particular problems, it is the general if not the universal practice to use diagrams in which the rectilinear co-ordinates represent volume and pressure. The object of this article is to call attention to certain diagrams of different construction, which afford graphical methods coextensive in their applications with that in ordinary use, and preferable to it in many cases in respect of distinctness or of convenience.

QUANTITIES AND RELATIONS WHICH ARE TO BE REPRESENTED BY THE DIAGRAM.

We have to consider the following quantities:—

$v$, the volume,
$p$, the pressure,
$t$, the (absolute) temperature,
$\varepsilon$, the energy,
$\eta$, the entropy,
(of a given body in any state,)

also $W$, the work done,
and $H$, the heat received,*
(by the body in passing from one state to another.)

These are subject to the relations expressed by the following differential equations:—

* Work spent upon the body is as usual to be considered as a negative quantity of work done by the body, and heat given out by the body as a negative quantity of heat received by it.

It is taken for granted that the body has a uniform temperature throughout, and that the pressure (or expansive force) has a uniform value both for all points in the body and for all directions. This, it will be observed, will exclude irreversible processes, but will not entirely exclude solids, although the condition of equal pressure in all directions renders the case very limited, in which they come within the scope of the discussion.

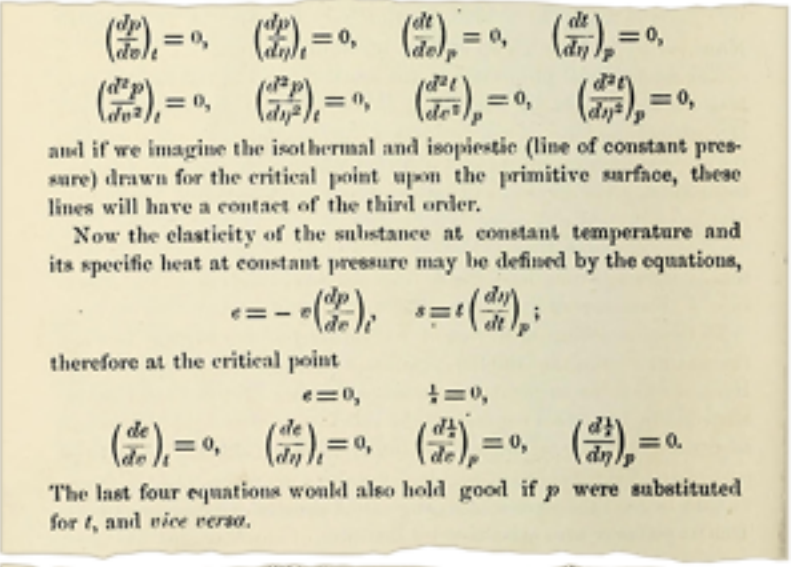

$$\left(\frac{dp}{dv}\right)_t = 0, \quad \left(\frac{dp}{d\eta}\right)_t = 0, \quad \left(\frac{dt}{dv}\right)_p = 0, \quad \left(\frac{dt}{d\eta}\right)_p = 0,$$

$$\left(\frac{d^2p}{dv^2}\right)_t = 0, \quad \left(\frac{d^2p}{d\eta^2}\right)_t = 0, \quad \left(\frac{d^2t}{dv^2}\right)_p = 0, \quad \left(\frac{d^2t}{d\eta^2}\right)_p = 0,$$

and if we imagine the isothermal and isopiestic (line of constant pressure) drawn for the critical point upon the primitive surface, these lines will have a contact of the third order.

Now the elasticity of the substance at constant temperature and its specific heat at constant pressure may be defined by the equations,

$$e = -v\left(\frac{dp}{dv}\right)_t, \quad s = t\left(\frac{d\eta}{dt}\right)_p;$$

therefore at the critical point

$$e = 0, \quad \frac{1}{s} = 0,$$

$$\left(\frac{de}{dv}\right)_t = 0, \quad \left(\frac{de}{d\eta}\right)_t = 0, \quad \left(\frac{d\frac{1}{s}}{dv}\right)_p = 0, \quad \left(\frac{d\frac{1}{s}}{d\eta}\right)_p = 0.$$

The last four equations would also hold good if $p$ were substituted for $t$, and *vice versa*.

Fig. 2.

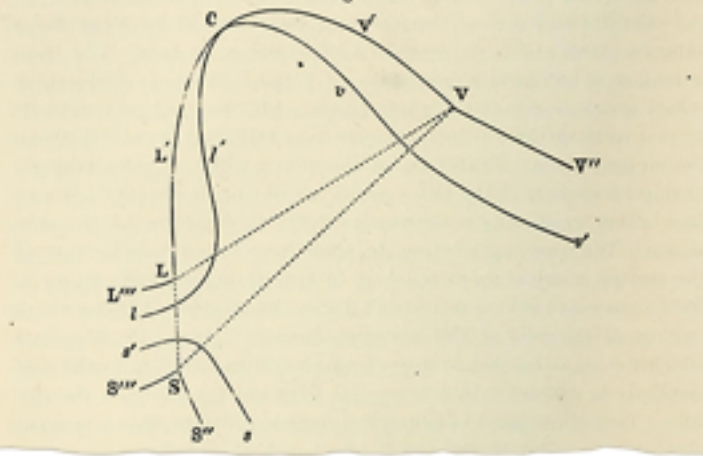

Unlike the European developers of thermodynamics, Gibbs didn't interact deeply with other scientists—with the possible exception of Maxwell, who (a few years before his death in 1879) made a 3D version of Gibbs's thermodynamic surface out of clay—and supplemented his 2D thermodynamic diagrams in the first edition of his textbook *Theory of Heat* with renderings of 3D versions:

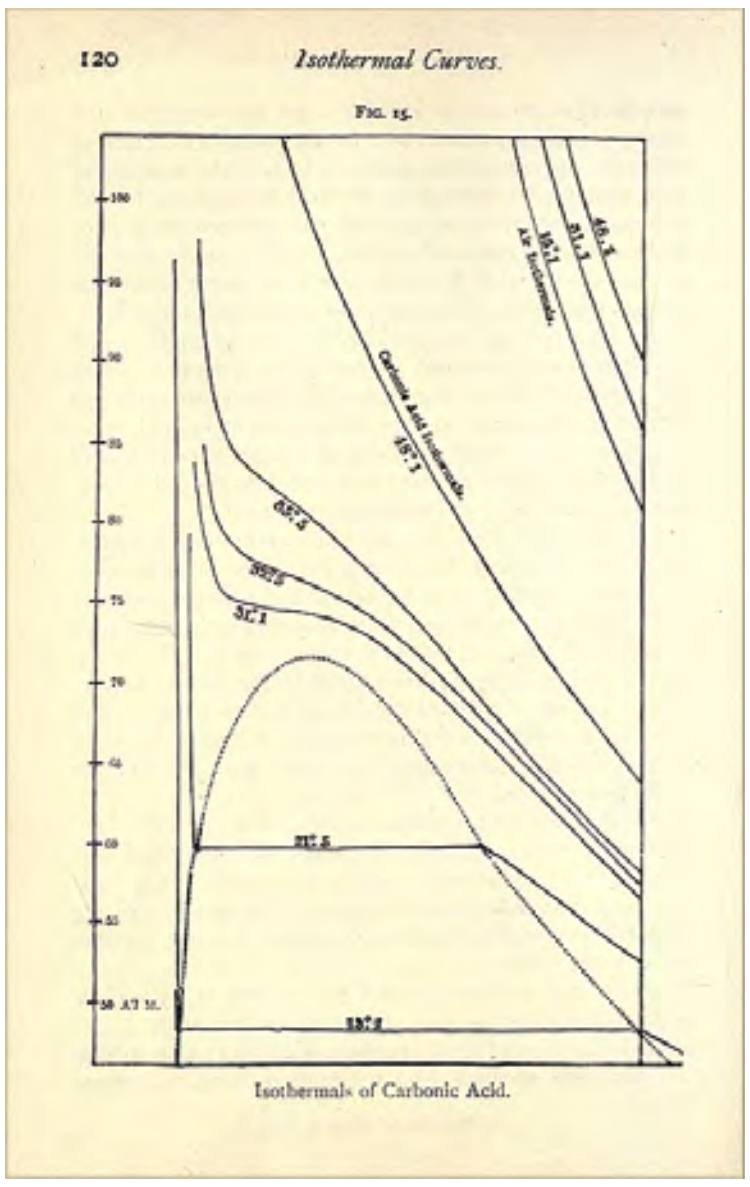

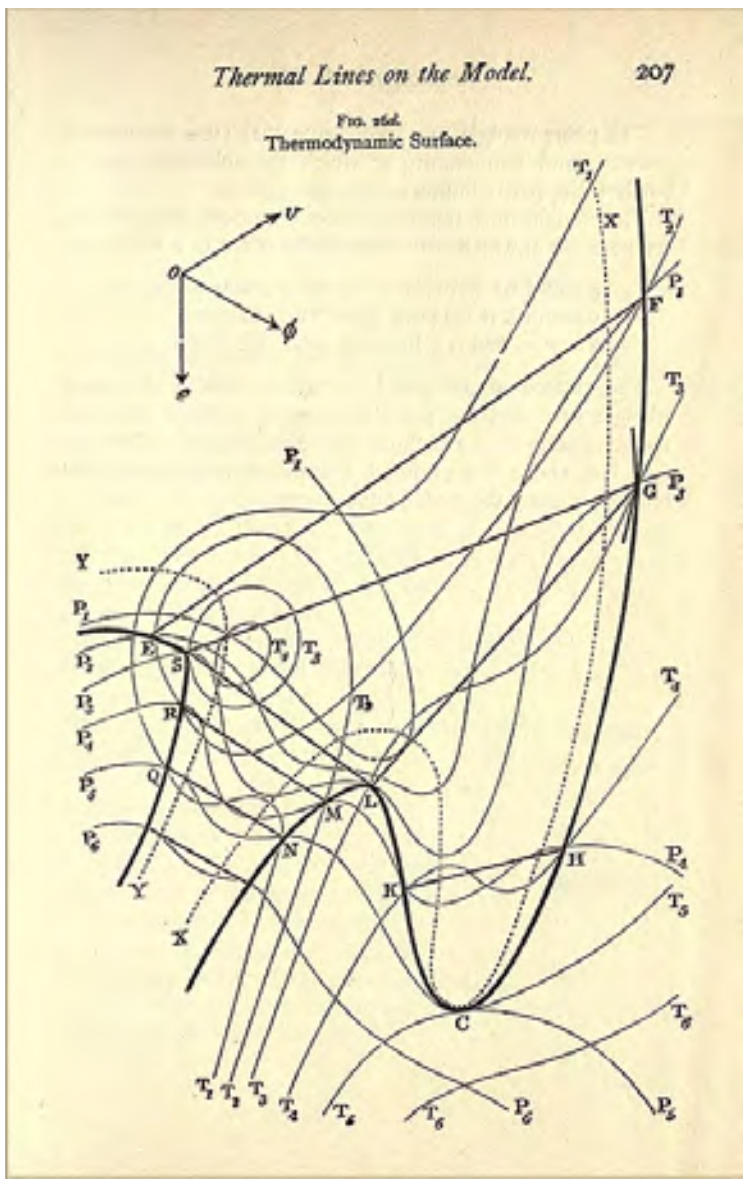

Three years later, Gibbs began publishing what would be a 300-page work defining what has become the standard formalism for equilibrium chemical thermodynamics. He began with a quote from Clausius:

V. On the Equilibrium of Heterogeneous Substances.
By J. Willard Gibbs.

"Die Energie der Welt ist constant.
Die Entropie der Welt strebt einem Maximum zu."
Clausius.*

The comprehension of the laws which govern any material system is greatly facilitated by considering the energy and entropy of the system in the various states of which it is capable. As the difference of the values of the energy for any two states represents the combined amount of work and heat received or yielded by the system when it is brought from one state to the other, and the difference of entropy is the limit of all the possible values of the integral $\int \frac{dQ}{t}$, ($dQ$ denoting the element of the heat received from external sources, and $t$ the temperature of the part of the system receiving it,) the

In the years that followed, Gibbs's work—stimulated by Maxwell—mostly concentrated on electrodynamics, and later quaternions and vector analysis. But Gibbs published a few more small papers on thermodynamics—always in effect taking equilibrium (and the Second Law) for granted.

In 1882—a certain Henry Eddy (1844–1921) (who in 1879 had written a book on thermodynamics, and in 1890 would become president of the University of Cincinnati), claimed that "radiant heat" could be used to violate the Second Law:

*Radiant Heat an Exception to the Second Law of Thermodynamics.*
*By H. T. Eddy, Ph.D., University of Cincinnati.*

(*Read before the American Philosophical Society, June 16, 1882.*)

Since the radiation of heat takes places by propagation through space at a certain finite velocity and not instantaneously, it is quite possible for occurrences to intervene during the exchange of radiations between two bodies such as to essentially change the distribution of heat which would otherwise have ultimately taken place.

To make this evident, let us employ first a mechanical analogy. In the accompanying figure, let there be three parallel screens, $a$, $b$ and $c$, the latter between the two former and all three perpendicular to the plane of the paper. Let them be pierced respectively by series of equidistant apertures $a_1\ a_2\ .\ .\ .\ a_n$, $b_1\ b_2\ .\ .\ .\ b_n$, $c_1\ c_2\ .\ .\ .\ c_n$, situated in the plane of the paper, and let these apertures be so placed that $a_1\ b_1\ c_1$ are upon one straight line, not quite at right angles to the screens; then are $a_2\ b_2\ c_2$, etc., and $a_n\ b_n\ c_n$ upon lines parallel to $a_1\ b_1\ c_1$. Now conceive the screens $a\ b\ c$ to have a common uniform velocity $u$ in the direction from the $c_2$ to $c_1$.

Also let a series of projectiles be discharged from any fixed position $A$ at the left of the screen $a$ at such instants as to pass the first one through the aperture $a_1$, the second through $a_2$, etc., and let the direction of discharge be perpendicular to the screens, and the velocity $v$ such that

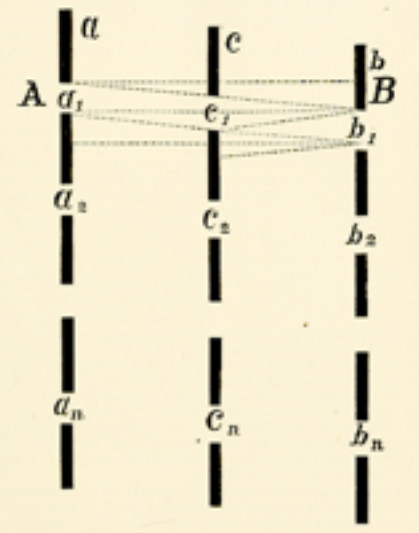

Gibbs soon published a 2-page rebuttal (in the 6th-ever issue of *Science* magazine):

***ON AN ALLEGED EXCEPTION TO THE SECOND LAW OF THERMODYNAMICS.***

ACCORDING to the received doctrine of radiation, heat is transmitted with the same intensity in all directions and at all points within any space which is void of ponderable matter and entirely surrounded by stationary bodies of the same temperature. We may apply this principle to the arrangement recently proposed by Prof. H. T. Eddy[1] for transferring heat from a colder body A to a warmer B without expenditure of work.

Then in 1889 Clausius died, and Gibbs wrote an obituary—praising Clausius but making it clear he didn't think the kinetic theory of gases was a solved problem:

The origin of the kinetic theory of gases is lost in remote antiquity, and its completion the most sanguine cannot hope to see. But a single generation has seen it advance from the stage of vague surmises to an extensive and well established body of doctrine. This is mainly the work of three men, Clausius, Maxwell, and Boltzmann, of whom Clausius was the earliest in the field, and has been called by Maxwell the principal founder of the science.* We may regard his That same year Gibbs announced a short course that he would teach at Yale on "The *a priori* Deduction of Thermodynamic Principles from the Theory of Probabilities". After a decade of work, this evolved into Gibbs's last publication—an original and elegant book that's largely defined how the Second Law has been thought about ever since:

> ELEMENTARY PRINCIPLES
> IN
> STATISTICAL MECHANICS
> DEVELOPED WITH ESPECIAL REFERENCE TO
> THE RATIONAL FOUNDATION OF THERMODYNAMICS
>
> BY
> J. WILLARD GIBBS
> *Professor of Mathematical Physics in Yale University*
>
> NEW YORK: CHARLES SCRIBNER'S SONS
> LONDON: EDWARD ARNOLD
> 1902

The book begins by explaining that mechanics is about studying the time evolution of single systems:

> THE usual point of view in the study of mechanics is that where the attention is mainly directed to the changes which take place in the course of time in a given system. The principal problem is the determination of the condition of the system with respect to configuration and velocities at any required time, when its condition in these respects has been given for some one time, and the fundamental equations are those which express the changes continually taking place in the system. Inquiries of this kind are often simplified by

But Gibbs says he is going to do something different: he is going to look at what he'll call an ensemble of systems, and see how the distribution of their characteristics changes over time:

> For some purposes, however, it is desirable to take a broader view of the subject. We may imagine a great number of systems of the same nature, but differing in the configurations and velocities which they have at a given instant, and differing not merely infinitesimally, but it may be so as to embrace every conceivable combination of configuration and velocities. And here we may set the problem, not to follow a particular system through its succession of configurations, but to determine how the whole number of systems will be distributed among the various conceivable configurations and velocities at any required time, when the distribution has been given for some one time. The fundamental equation

He explains that these "inquiries" originally arose in connection with deriving the laws of thermodynamics:

> Such inquiries have been called by Maxwell *statistical*. They belong to a branch of mechanics which owes its origin to the desire to explain the laws of thermodynamics on mechanical principles, and of which Clausius, Maxwell, and Boltzmann are to be regarded as the principal founders. The first inquiries in this field were indeed somewhat narrower in their scope than that which has been mentioned, being applied to the particles of a system, rather than to independent systems.

But he argues that this area—which he's calling statistical mechanics—is worth investigating even independent of its connection to thermodynamics:

> But although, as a matter of history, statistical mechanics owes its origin to investigations in thermodynamics, it seems eminently worthy of an independent development, both on account of the elegance and simplicity of its principles, and because it yields new results and places old truths in a new light in departments quite outside of thermodynamics. More-

Still, he expects this effort will be relevant to the foundations of thermodynamics:

> 
> over, the separate study of this branch of mechanics seems to afford the best foundation for the study of rational thermodynamics and molecular mechanics.

He immediately then goes on to what he'll claim is the way to think about the relation of "observed thermodynamics" to his exact statistical mechanics:

> The laws of thermodynamics, as empirically determined, express the approximate and probable behavior of systems of a great number of particles, or, more precisely, they express the laws of mechanics for such systems as they appear to beings who have not the fineness of perception to enable them to appreciate quantities of the order of magnitude of those which relate to single particles, and who cannot repeat their experiments often enough to obtain any but the most probable results. The laws of statistical mechanics apply to

Soon he makes the interesting—if, in the light of history, very overly optimistic—claim that "the laws of thermodynamics may be easily obtained from the principles of statistical mechanics":

> that the effect of the quantities and circumstances neglected will be negligible in the result. The laws of thermodynamics may be easily obtained from the principles of statistical mechanics, of which they are the incomplete expression, but they make a somewhat blind guide in our search for those laws. This is perhaps the principal cause of the slow progress of rational thermodynamics, as contrasted with the rapid deduction of the consequences of its laws as empirically established. To this must be added that the rational foundation

At first the text of the book reads very much like a typical mathematical work on mechanics:

> CHAPTER I.
>
> GENERAL NOTIONS. THE PRINCIPLE OF CONSERVATION OF EXTENSION-IN-PHASE.
>
> We shall use Hamilton's form of the equations of motion for a system of $n$ degrees of freedom, writing $q_1, \ldots q_n$ for the (generalized) coördinates, $\dot{q}_1, \ldots \dot{q}_n$ for the (generalized) velocities, and
>
> $$F_1\, dq_1 + F_2\, dq_2 \ldots + F_n\, dq_n \qquad (1)$$
>
> for the moment of the forces. We shall call the quantities $F_1, \ldots F_n$ the (generalized) forces, and the quantities $p_1 \ldots p_n$, defined by the equations
>
> $$p_1 = \frac{d\epsilon_p}{d\dot{q}_1}, \quad p_2 = \frac{d\epsilon_p}{d\dot{q}_2}, \quad \text{etc.,} \qquad (2)$$
>
> where $\epsilon_p$ denotes the kinetic energy of the system, the (generalized) momenta. The kinetic energy is here regarded as a function of the velocities and coördinates. We shall usually

But soon it's "going statistical", talking about the "density" of systems in "phase" (i.e. with respect to the variables defining the configuration of the system). And a few pages in, he's proving the fundamental result that the density of "phase fluid" satisfies a continuity equation (which we'd now call the Liouville equation):

> *In an ensemble of mechanical systems identical in nature and subject to forces determined by identical laws, but distributed in phase in any continuous manner, the density-in-phase is constant in time for the varying phases of a moving system; provided, that the forces of a system are functions of its co-ordinates, either alone or with the time.**
>
> This may be called the principle of *conservation of density-in-phase.* It may also be written
>
> $$\left(\frac{dD}{dt}\right)_{a,\ldots h} = 0, \qquad (22)$$

It's all quite elegant, and all very rooted in the calculus-based mathematics of its time. He's thinking about a collection of instances of a system. But while with our modern computational paradigm we'd readily be able to talk about a discrete list of instances, with his calculus-based approach he has to consider a continuous collection of instances—whose treatment inevitably seems more abstract and less explicit.

He soon makes contact with the "theory of errors", discussing in effect how probability distributions over the space of possible states evolve. But what probability distributions should one consider? By chapter 4, he's looking at what he calls (and is still called) the "canonical distribution":

> CHAPTER IV.
>
> ON THE DISTRIBUTION IN PHASE CALLED CANONICAL, IN WHICH THE INDEX OF PROBABILITY IS A LINEAR FUNCTION OF THE ENERGY.
>
> LET us now give our attention to the statistical equilibrium of ensembles of conservation systems, especially to those cases and properties which promise to throw light on the phenomena of thermodynamics.

He gives a now-classic definition for the probability as a function of energy $\epsilon$:

> or
>
> $$P = e^{\frac{\psi - \epsilon}{\Theta}}, \qquad (91)$$
>
> where $\Theta$ and $\psi$ are constants, and $\Theta$ positive, seems to represent the most simple case conceivable, since it has the property that when the system consists of parts with separate energies, the laws of the distribution in phase of the separate parts are of the same nature,— a property which enormously simplifies the discussion, and is the foundation of extremely important relations to thermodynamics. The case is not rendered less He observes that this distribution combines nicely when independent parts of a system are brought together, and soon he's noting that:

> The modulus Θ has properties analogous to those of temperature in thermodynamics. Let the system $A$ be defined as

But so far he's careful to just talk about how things are "analogous", without committing to a true connection:

> How far, or in what sense, the similarity of these equations constitutes any demonstration of the thermodynamic equations, or accounts for the behavior of material systems, as described in the theorems of thermodynamics, is a question of which we shall postpone the consideration until we have further investigated the properties of an ensemble of systems distributed in phase according to the law which we are considering. The analogies which have been pointed out will at least supply the motive for this investigation, which will naturally commence with the determination of the average values in the ensemble of the most important quantities relating to the systems, and to the distribution of the ensemble with respect to the different values of these quantities.

More than halfway through the book he's defined certain properties of his probability distributions that "may ... correspond to the thermodynamic notions of entropy and temperature":

> Now we have already noticed a certain correspondence between the quantities Θ and $\bar{\eta}$ and those which in thermodynamics are called temperature and entropy. The property just demonstrated, with those expressed by equation (336), therefore suggests that the quantities $\phi$ and $d\epsilon/d\phi$ may also correspond to the thermodynamic notions of entropy and temperature. We leave the discussion of this point to a subsequent chapter, and only mention it here to justify the somewhat detailed investigation of the relations of these quantities.

Next he's on to the concept of a "microcanonical ensemble" that includes only states of a given energy. For him—with his continuum-based setup—this is a slightly elaborate thing to define; in our modern computational framework it actually becomes more straightforward than his "canonical ensemble". Or, as he already says:

> From a certain point of view, the microcanonical distribution may seem more simple than the canonical, and it has perhaps been more studied, and been regarded as more closely related to the fundamental notions of thermodynamics. To this last point we shall return in a subsequent chapter. It is sufficient here to remark that analytically the canonical distribution is much more manageable than the microcanonical.

But what about the Second Law? Now he's getting a little closer:

> CHAPTER XI.
>
> MAXIMUM AND MINIMUM PROPERTIES OF VARIOUS DISTRIBUTIONS IN PHASE.

When he says "index of probability" he's talking about the log of a probability in his ensemble, so this result is about the fact that this quantity is extremized when all the elements of the ensemble have equal probability:

> original ensembles are all identical.
>
> *Theorem IX.* A uniform distribution of a given number of systems within given limits of phase gives a less average index of probability of phase than any other distribution.

Soon he's discussing whether he can use his index as a way—like Boltzmann tried to do with his version of entropy—to measure deviations from "statistical equilibrium":

> It would seem, therefore, that we might find a sort of measure of the deviation of an ensemble from statistical equilibrium in the excess of the average index above the minimum which is consistent with the condition of the invariability of the distribution with respect to the constant functions of phase. But we have seen that the index of probability is constant in time for each system of the ensemble. The average index is therefore constant, and we find by this method no approach toward statistical equilibrium in the course of time.

But now Gibbs has hit one of the classic gotchas of his approach: if you look in perfect detail at the evolution of an ensemble of systems, there'll never be a change in the value of his index—essentially because of the overall conservation of probability. Gibbs brings in what amounts to a commonsense physics argument to handle this. He says to consider putting "coloring matter" in a liquid that one stirs. And then he says that even though the liquid (like his phase fluid) is microscopically conserved, the coloring matter will still end up being "uniformly mixed" in the liquid:

> distributed with a variable density. If we give the liquid any motion whatever, subject only to the hydrodynamic law of incompressibility, — it may be a steady flux, or it may vary with the time, — the density of the coloring matter at any same point of the liquid will be unchanged, and the average square of this density will therefore be unchanged. Yet no fact is more familiar to us than that stirring tends to bring a liquid to a state of uniform mixture, or uniform densities of its components, which is characterized by minimum values of the average squares of these densities. It is quite true that

He talks about how the conclusion about whether mixing happens in effect depends on what order one takes limits in. And while he doesn't put it quite this way, he's essentially realized that there's a competition between the system "mixing things up more and more finely" and the observer being able to track finer and finer details. He realizes, though, that not all systems will show this kind of mixing behavior, noting for example that there are mechanical systems that'll just keep going in simple cycles forever.

He doesn't really resolve the question of why "practical systems" should show mixing, more or less ending with a statement that even though his underlying mechanical systems are reversible, it's somehow "in practice" difficult to go back:

> But while the distinction of prior and subsequent events may be immaterial with respect to mathematical fictions, it is quite otherwise with respect to the events of the real world. It should not be forgotten, when our ensembles are chosen to illustrate the probabilities of events in the real world, that
>
> while the probabilities of subsequent events may often be determined from the probabilities of prior events, it is rarely the case that probabilities of prior events can be determined from those of subsequent events, for we are rarely justified in excluding the consideration of the antecedent probability of the prior events.

Despite things like this, Gibbs appears to have been keen to keep the majority of his book "purely mathematical", in effect proving theorems that necessarily followed from the setup he had given. But in the penultimate chapter of the book he makes what he seems to have viewed as a less–than–satisfactory attempt to connect what he's done with "real thermodynamics". He doesn't really commit to the connection, though, characterizing it more as an "analogy":

> CHAPTER XIV.
>
> DISCUSSION OF THERMODYNAMIC ANALOGIES.
>
> If we wish to find in rational mechanics an *a priori* foundation for the principles of thermodynamics, we must seek mechanical definitions of temperature and entropy. The

But he soon starts to be pretty clear that he actually wants to prove the Second Law:

> We have also to enunciate in mechanical terms, and to prove, what we call the tendency of heat to pass from a system of higher temperature to one of lower, and to show that this tendency vanishes with respect to systems of the same temperature.

He quickly backs off a little, in effect bringing in the observer to soften the requirements:

> At least, we have to show by *a priori* reasoning that for such systems as the material bodies which nature presents to us, these relations hold with such approximation that they are sensibly true for human faculties of observation. This indeed is all that is really necessary to establish the science of thermodynamics on an *a priori* basis. Yet we will naturally desire to find the exact expression of those principles of which the laws of thermodynamics are the approximate expression.

But then he fires his best shot. He says that the quantities he's defined in connection with his canonical ensemble satisfy the same equations as Clausius originally set up for temperature and entropy:

> Now we have defined what we have called the *modulus* ($\Theta$) of an ensemble of systems canonically distributed in phase, and what we have called the index of probability ($\eta$) of any phase in such an ensemble. It has been shown that between
>
> the modulus ($\Theta$), the external coördinates ($a_1$, etc.), and the average values in the ensemble of the energy ($\epsilon$), the index of probability ($\eta$), and the external forces ($A_1$, etc.) exerted by the systems, the following differential equation will hold:
>
> $$d\bar{\epsilon} = -\Theta\, d\bar{\eta} - \bar{A}_1\, da_1 - \bar{A}_2\, da_2 - \text{etc.} \qquad (483)$$
>
> This equation, if we neglect the sign of averages, is identical in form with the thermodynamic equation (482), the modulus ($\Theta$) corresponding to temperature, and the index of probability of phase with its sign reversed corresponding to entropy.*

He adds that fluctuations (or "anomalies", as he calls them) become imperceptible in the limit of a large system:

> We have also shown that the average square of the anomalies of $\epsilon$, that is, of the deviations of the individual values from the average, is in general of the same order of magnitude as the reciprocal of the number of degrees of freedom, and therefore to human observation the individual values are indistinguishable from the average values when the number of degrees of freedom is very great.† In this case also the anomalies of $\eta$

But in physical reality, why should one have a whole collection of systems as in the canonical ensemble? Gibbs suggests it would be more natural to look at the microcanonical ensemble—and in fact to look at a "time ensemble", i.e. an averaging over time rather than an averaging over different possible states of the system:

> The definitions and propositions which we have been considering relate essentially to what we have called a canonical ensemble of systems. This may appear a less natural and simple conception than what we have called a microcanonical ensemble of systems, in which all have the same energy, and which in many cases represents simply the *time-ensemble*, or ensemble of phases through which a single system passes in the course of time.

Gibbs has proved some results (e.g. related to the virial theorem) about the relation between time and ensemble averages. But as the future of the subject amply demonstrates, they're not nearly strong enough to establish any general equivalence. Still, Gibbs presses on.

In the end, though, as he himself recognized, things weren't solved—and certainly the canonical ensemble wasn't the whole story:

> It is certainly in the quantities relating to a canonical ensemble, $\bar{\epsilon}$, $\Theta$, $\bar{\eta}$, $\bar{A}_1$, etc., $a_1$, etc., that we find the most complete correspondence with the quantities of the thermodynamic equation (482). Yet the conception itself of the canonical ensemble may seem to some artificial, and hardly germane to a natural exposition of the subject; and the quantities $\epsilon$, $\frac{d\epsilon}{d \log V}$, $\log V$, $\overline{A_1}|_\epsilon$, etc., $a_1$, etc., or $\epsilon$, $\frac{d\epsilon}{d\phi}$, $\phi$, $\left(\frac{d\epsilon}{da_1}\right)_{\phi, a}$, etc., $a_1$, etc., which are closely related to ensembles of constant energy, and to average and most probable values in such ensembles, and most of which are defined without reference to any ensemble, may appear the most natural analogues of the thermodynamic quantities.

He discusses the tradeoff between having a canonical ensemble "heat bath" of a known temperature, and having a microcanonical ensemble with known energy. At one point he admits that it might be better to consider the time evolution of a single state, but basically decides that—at least in his continuous-probability-distribution-based formalism—he can't really set this up:

> arbitrary, will represent better than any one time-ensemble the effect of the bath. Indeed a single time-ensemble, when it is not also a microcanonical ensemble, is too ill-defined a notion to serve the purposes of a general discussion. We will therefore direct our attention, when we suppose the body placed in a bath, to the microcanonical ensemble of phases thus obtained.

Gibbs definitely encourages the idea that his "statistical mechanics" has successfully "derived" thermodynamics. But he's ultimately quite careful and circumspect in what he actually says. He mentions the Second Law only once in his whole book—and then only to note that he can get the same "mathematical expression" from his canonical ensemble as Clausius's form of the Second Law. He doesn't mention Boltzmann's *H* theorem anywhere in the book, and—apart from one footnote concerning "difficulties long recognized by

physicists"—he mentions only Boltzmann's work on theoretical mechanics.

One can view the main achievement of Gibbs's book as having been to define a framework in which precise results about the statistical properties of collections of systems could be defined and in some cases derived. Within the mathematics and other formalism of the time, such ensemble results represented in a sense a distinctly "higher-order" description of things. Within our current computational paradigm, though, there's much less of a distinction to be made: whether one's looking at a single path of evolution, or a whole collection, one's ultimately still just dealing with a computation. And that makes it clearer that—ensembles or not—one's thrown back into the same kinds of issues about the origin of the Second Law. But even so, Gibbs provided a language in which to talk with some clarity about many of the things that come up.

## Maxwell's Demon

In late 1867 Peter Tait (1831–1901)—a childhood friend of Maxwell's who was by then a professor of "natural philosophy" in Edinburgh—was finishing his sixth book. It was entitled *Sketch of Thermodynamics* and gave a brief, historically oriented and not particularly conceptual outline of what was then known about thermodynamics. He sent a draft to Maxwell, who responded with a fairly long letter:

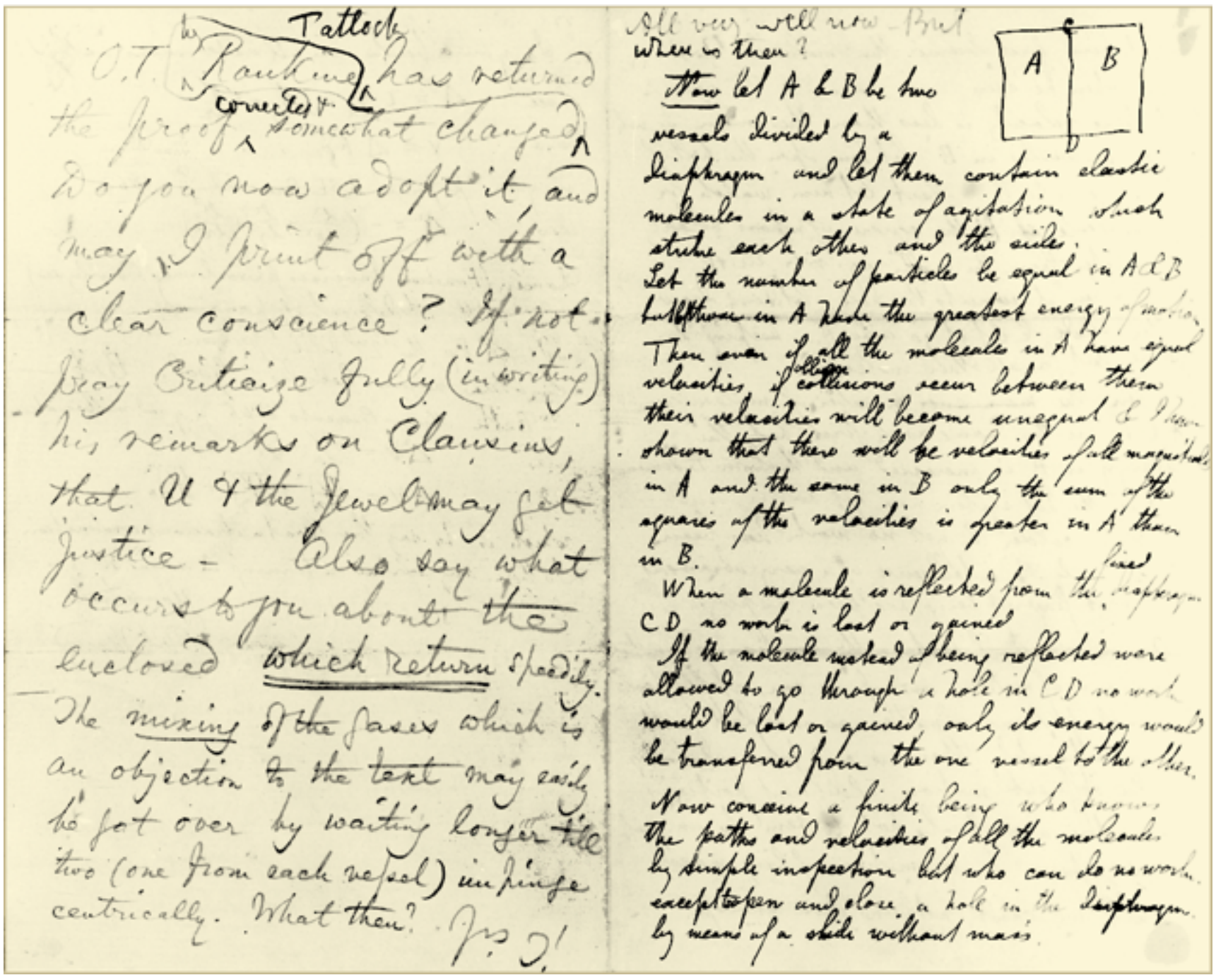
Tatlock

O.T. Rankine has returned the proof, corrected & somewhat changed. Do you now adopt it, and may I print off with a clear conscience? If not, pray Criticise fully (in writing) his remarks on Clausius, that U & the Jewel may get justice. Also say what occurs to you about the enclosed which return speedily.

The mixing of the gases which is an objection to the text may easily be got over by waiting longer till the two (one from each vessel) impinge centrically. What then? Yrs. T!

All very well now. But where is then?

Now let A & B be two vessels divided by a diaphragm and let them contain elastic molecules in a state of agitation which strike each other and the sides. Let the number of particles be equal in A & B but let those in A have the greatest energy of motion. Then even if all the molecules in A have equal velocities, if oblique collisions occur between them their velocities will become unequal & I have shown that there will be velocities of all magnitudes in A and the same in B only the sum of the squares of the velocities is greater in A than in B.

When a molecule is reflected from the fixed diaphragm CD no work is lost or gained.

If the molecule instead of being reflected were allowed to go through a hole in CD no work would be lost or gained, only its energy would be transferred from the one vessel to the other.

Now conceive a finite being who knows the paths and velocities of all the molecules by simple inspection but who can do no work except open and close a hole in the diaphragm by means of a slide without mass.

The letter begins:

> I do not know in a controversial manner the history of thermodynamics ... [and] I could make no assertions about the priority of authors ...
>
> Any contributions I could make ... [involve] picking holes here and there to ensure strength and stability.

Then he continues (with "ΘΔcs" being his whimsical Greekified rendering of the word "thermodynamics"):

> To pick a hole—say in the 2nd law of ΘΔcs, that if two things are in contact the hotter cannot take heat from the colder without external agency.
>
> Now let A and B be two vessels divided by a diaphragm ... Now conceive a finite being who knows the paths and velocities of all the molecules by simple inspection but who can do no work except open and close a hole in the diaphragm by means of a slide without mass.  Let him ... observe the molecules in A and when he sees one coming ... whose velocity is less than the mean [velocity] of the molecules in B let him open the hole and let it go into B [and vice versa].
>
> Then the number of molecules in A and B are the same as at first, but the energy in A is increased and that in B diminished, that is, the hot system has got hotter and the cold colder and yet no work has been done, only the intelligence of a very observant and neat-fingered being has been employed.
>
> Or in short [we can] ... restore a uniformly hot system to unequal temperatures...   Only we can't, not being clever enough.

And so it was that the idea of "Maxwell's demon" was launched.  Tait must at some point have shown Maxwell's letter to Kelvin, who wrote on it:

> Very good. Another way is to reverse the motion of every particle of the Universe and to preside over the unstable motion thus produced.

But the first place Maxwell's demon idea appeared in print was in Maxwell's 1871 textbook *Theory of Heat*:

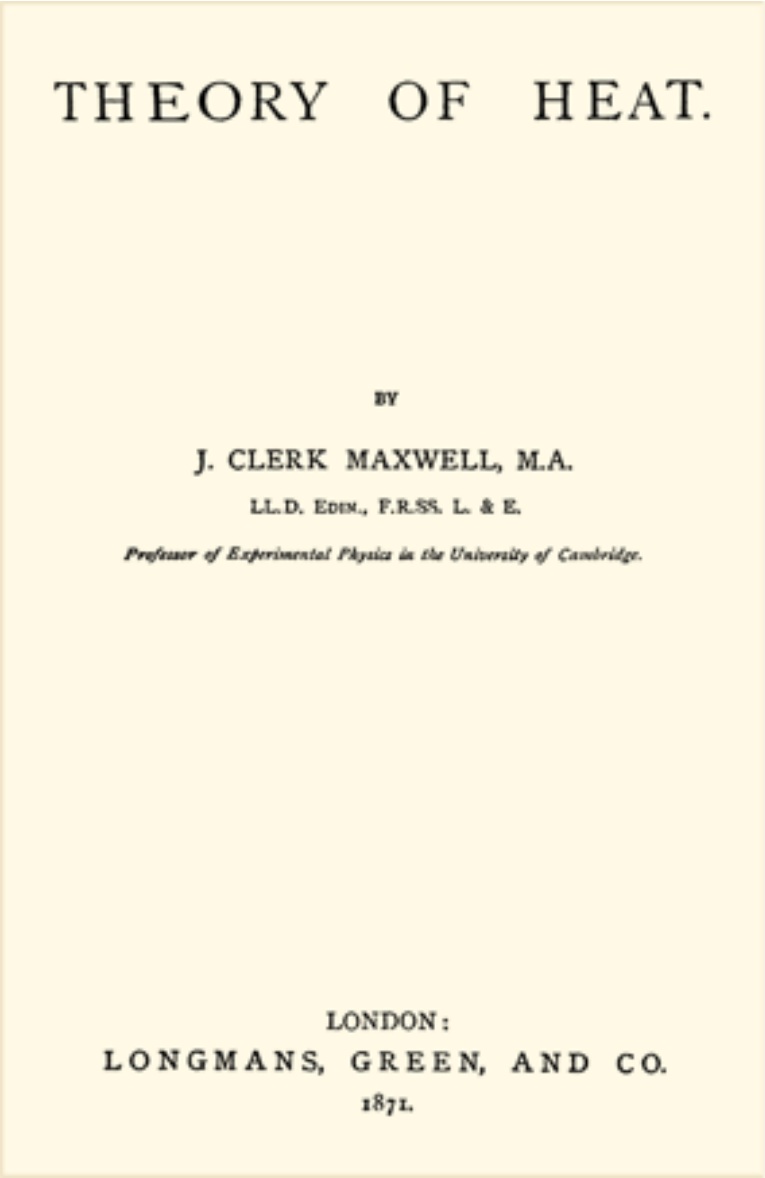

THEORY OF HEAT.

BY

J. CLERK MAXWELL, M.A.

LL.D. EDIN., F.R.SS. L. & E.

*Professor of Experimental Physics in the University of Cambridge.*

LONDON:
LONGMANS, GREEN, AND CO.
1871.

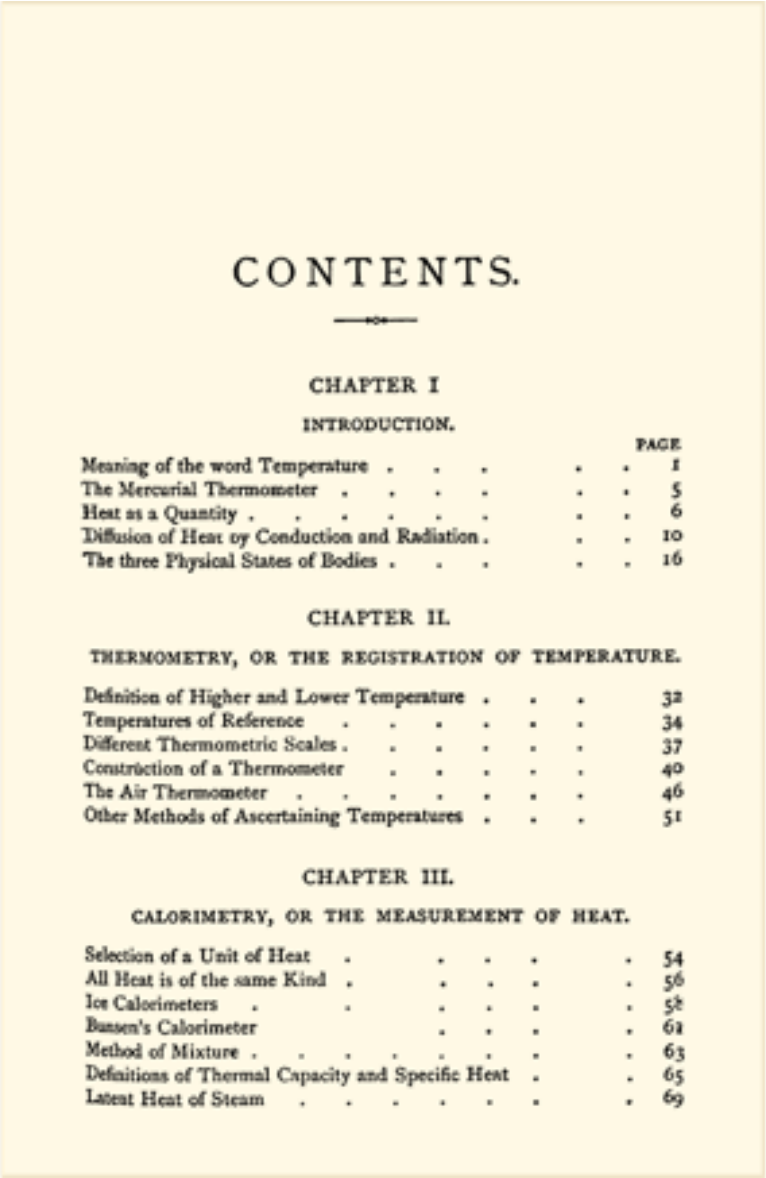

CONTENTS.

CHAPTER I

INTRODUCTION.

CHAPTER II.

THERMOMETRY, OR THE REGISTRATION OF TEMPERATURE.

CHAPTER III.

CALORIMETRY, OR THE MEASUREMENT OF HEAT.

Much of the book is devoted to what was by then quite traditional, experimentally oriented thermodynamics. But Maxwell included one final chapter:

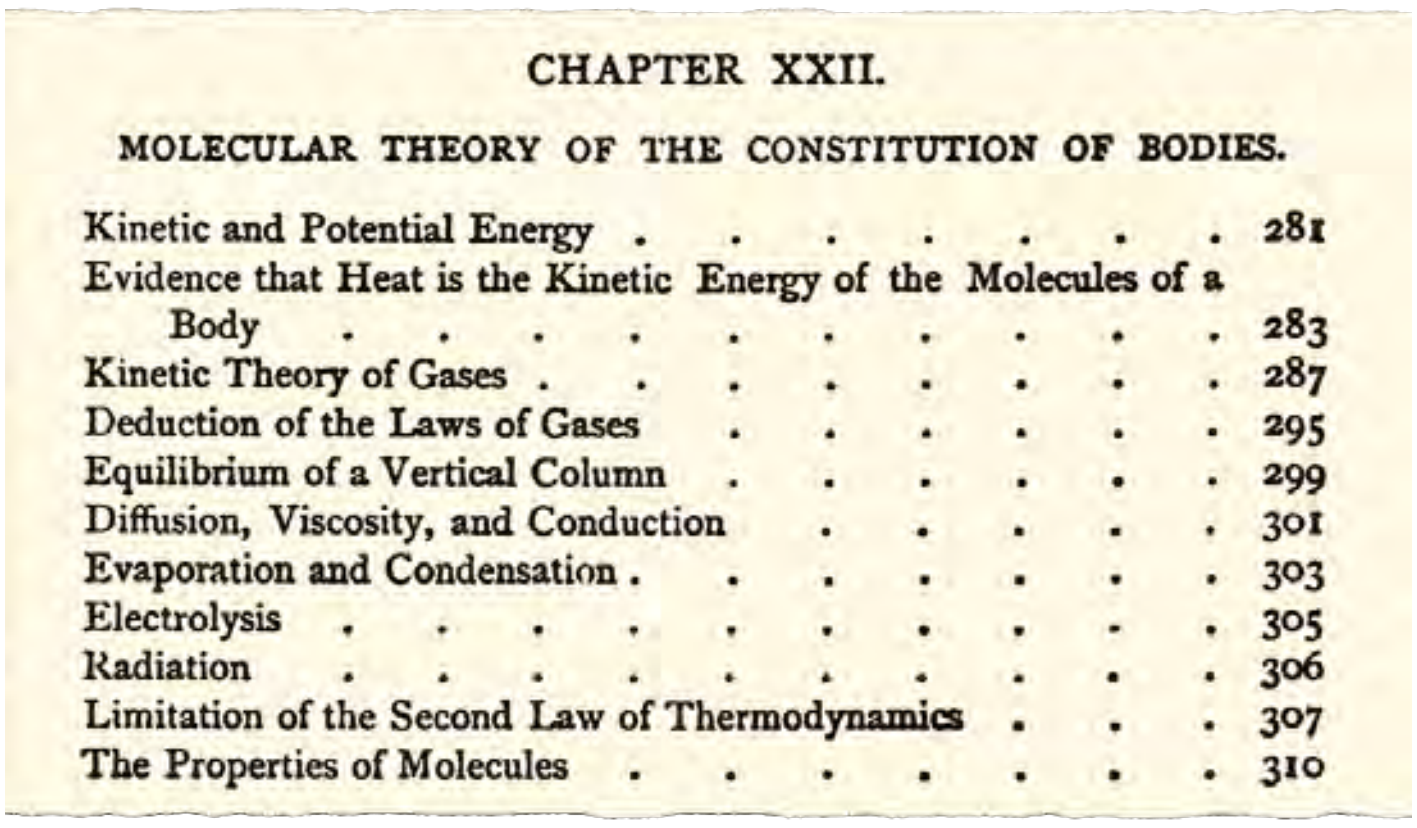

CHAPTER XXII.

MOLECULAR THEORY OF THE CONSTITUTION OF BODIES.

Even in 1871, after all his work on kinetic theory, Maxwell is quite circumspect in his discussion of molecules:

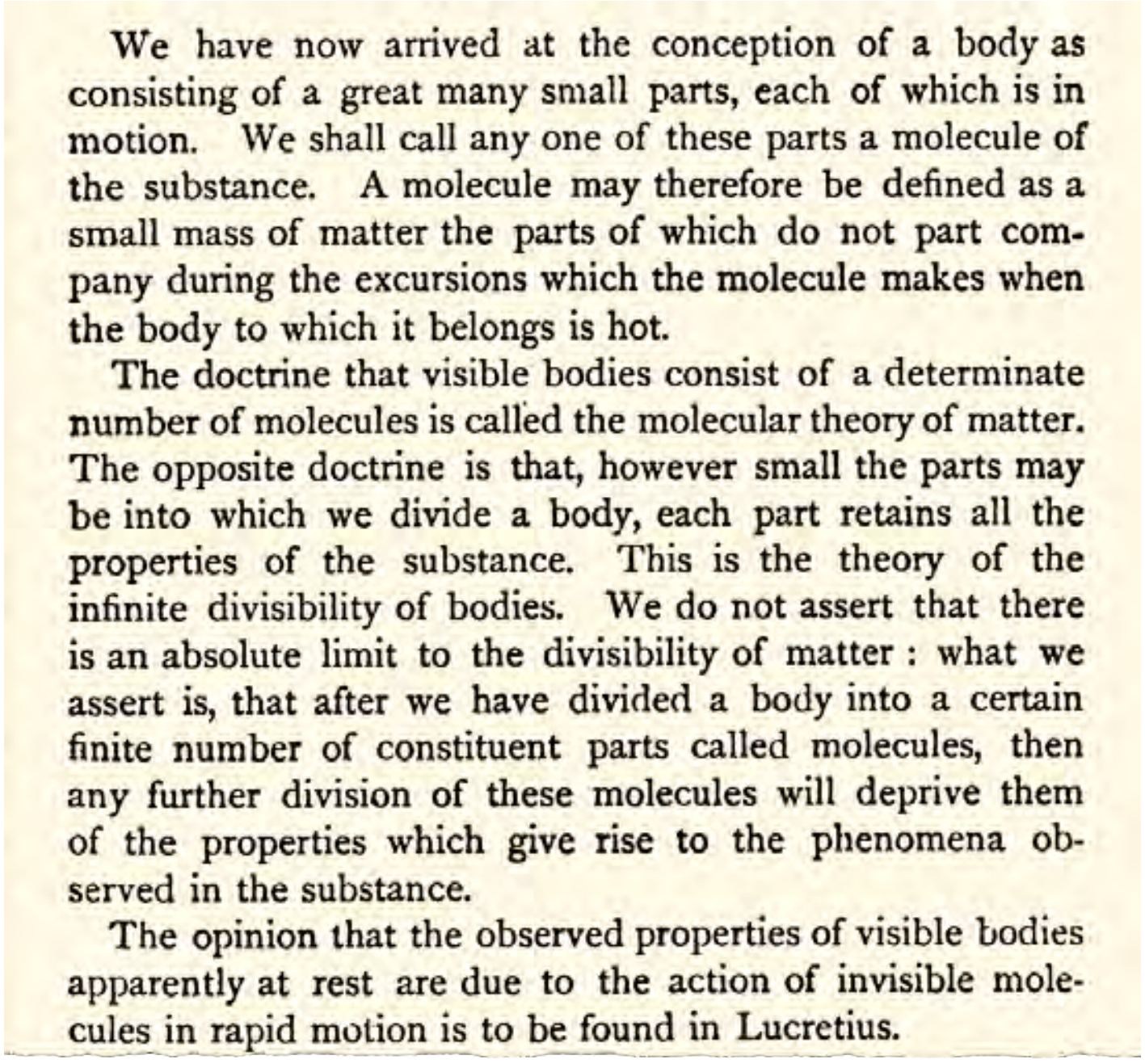

We have now arrived at the conception of a body as consisting of a great many small parts, each of which is in motion. We shall call any one of these parts a molecule of the substance. A molecule may therefore be defined as a small mass of matter the parts of which do not part company during the excursions which the molecule makes when the body to which it belongs is hot.

The doctrine that visible bodies consist of a determinate number of molecules is called the molecular theory of matter. The opposite doctrine is that, however small the parts may be into which we divide a body, each part retains all the properties of the substance. This is the theory of the infinite divisibility of bodies. We do not assert that there is an absolute limit to the divisibility of matter : what we assert is, that after we have divided a body into a certain finite number of constituent parts called molecules, then any further division of these molecules will deprive them of the properties which give rise to the phenomena observed in the substance.

The opinion that the observed properties of visible bodies apparently at rest are due to the action of invisible molecules in rapid motion is to be found in Lucretius.

But Maxwell's textbook goes through a series of standard kinetic theory results, much as a modern textbook would. The second-to-last section in the whole book sounds a warning, however:

LIMITATION OF THE SECOND LAW OF THERMODYNAMICS.

Before I conclude, I wish to direct attention to an aspect of the molecular theory which deserves consideration.

One of the best established facts in thermodynamics is that it is impossible in a system enclosed in an envelope which permits neither change of volume nor passage of heat, and in which both the temperature and the pressure are everywhere the same, to produce any inequality of temperature or of pressure without the expenditure of work. This is the second law of thermodynamics, and it is undoubtedly true as long as we can deal with bodies only in mass, and have no power of perceiving or handling the separate molecules of which they are made up. But if we conceive a being whose faculties are so sharpened that he can follow every molecule in its course, such a being, whose attributes are still as essentially finite as our own, would be able to do what is at present impossible to us. For we have seen that the molecules in a vessel full of air at uniform temperature are moving with velocities by no means uniform, though the mean velocity of any great number of them, arbitrarily selected, is almost exactly uniform. Now let us suppose that such a vessel is divided into two portions, A and B, by a division in which there is a small hole, and that a being, who can see the individual molecules, opens and closes this hole, so as to allow only the swifter molecules to pass from A to B, and only the slower ones to pass from B to A. He will thus, without expenditure of work, raise the temperature of B and lower that of A, in contradiction to the second law of thermodynamics.

Interestingly, Maxwell continues, somewhat in anticipation of what Gibbs will say 30 years later:

This is only one of the instances in which conclusions which we have drawn from our experience of bodies consisting of an immense number of molecules may be found not to be applicable to the more delicate observations and experiments which we may suppose made by one who can perceive and handle the individual molecules which we deal with only in large masses.

In dealing with masses of matter, while we do not perceive the individual molecules, we are compelled to adopt what I have described as the statistical method of calculation, and to abandon the strict dynamical method, in which we follow every motion by the calculus.

But then there's a reminder that this is being written in 1871, several decades before any clear observation of molecules was made. Maxwell says:

> I do not think, however, that the perfect identity which we observe between different portions of the same kind of matter can be explained on the statistical principle of the stability of the averages of large numbers of quantities each of which may differ from the mean. For if of the

In other words, if there are water molecules, there must be something other than a law of averages that makes them all appear the same. And, yes, it's now treated as a fundamental fact of physics that, for example, all electrons have exactly—not just statistically—the same properties such as mass and charge. But back in 1871 it was much less clear what characteristics molecules—if they existed as real entities at all—might have.

Maxwell included one last section in his book that to us today might seem quite wild:

> NATURE AND ORIGIN OF MOLECULES.
>
> We have thus been led by our study of visible things to a theory that they are made up of a finite number of parts or molecules, each of which has a definite mass, and possesses other properties. The molecules of the same substance are all exactly alike, but different from those of other substances. There is not a regular gradation in the mass of molecules from that of hydrogen, which is the least of those known to us, to that of bismuth ; but they all fall into a limited number of classes or species, the individuals of each species being exactly similar to each other, and no intermediate links are found to connect one species with another by a uniform gradation.
>
> We are here reminded of certain speculations concerning the relations between the species of living things. We find that in these also the individuals are naturally grouped into species, and that intermediate links between the species are wanting. But in each species variations occur, and there is a perpetual generation and destruction of the individuals of which the species consist.
>
> Hence it is possible to frame a theory to account for the present state of things by means of generation, variation, and discriminative destruction.
>
> In the case of the molecules, however, each individual is permanent ; there is no generation or destruction, and no variation, or rather no difference, between the individuals of each species.
>
> Hence the kind of speculation with which we have become so familiar under the name of theories of evolution is quite inapplicable to the case of molecules.

In other words, aware of Darwin's (1809–1882) 1859 *Origin of Species*, he's considering a kind of "speciation" of molecules, along the lines of the discrete species observed in biology. But then he notes that unlike biological organisms, molecules are "permanent", so their "selection" must come from some kind of pure separation process:

> In speculating on the cause of this equality we are debarred from imagining any cause of equalization, on account of the immutability of each individual molecule. It is difficult, on the other hand, to conceive of selection and elimination of intermediate varieties, for where can these eliminated molecules have gone to if, as we have reason to believe, the hydrogen, &c., of the fixed stars is composed of molecules identical in all respects with our own? The time required to eliminate from the whole of the visible universe every molecule whose mass differs from that of some one of our so-called elements, by processes similar to Graham's method of dialysis, which is the only method we can conceive of at present, would exceed the utmost limits ever demanded by evolutionists as many times as these exceed the period of vibration of a molecule.

And at the very end he suggests that if molecules really are all identical, that suggests a level of fundamental order in the world that we might even be able to flow through to "exact principles of distributive justice" (presumably for people rather than molecules):

> But if we suppose the molecules to be made at all, or if we suppose them to consist of something previously made, why should we expect any irregularity to exist among them? If they are, as we believe, the only material things which still remain in the precise condition in which they first began to exist, why should we not rather look for some indication of that spirit of order, our scientific confidence in which is never shaken by the difficulty which we experience in tracing it in the complex arrangements of visible things, and of which our moral estimation is shown in all our attempts to think and speak the truth, and to ascertain the exact principles of distributive justice?

Maxwell has described rather clearly his idea of demons. But the actual name "demon" first appears in print in a paper by Kelvin in 1874:

> *KINETIC THEORY OF THE DISSIPATION OF ENERGY*
>
> IN abstract dynamics an instantaneous reversal of the motion of every moving particle of a system causes the system to move backwards, each particle of it along its old path, and at the same speed as before when again in the same position—that is to say, in mathematical language, any solution remains a solution when *t* is changed into $-t$. In physical dynamics, this simple and

> This process of diffusion could be perfectly prevented by an army of Maxwell's "intelligent demons"* stationed at the surface, or interface as we may call it with Prof. James Thomson, separating the hot from the cold part of the bar.

> * The definition of a "demon," according to the use of this word by Maxwell, is an intelligent being endowed with free will, and fine enough tactile and perceptive organisation to give him the faculty of observing and influencing individual molecules of matter

It's a British paper, so—in a nod to future nanomachinery—it's talking about (molecular) cricket bats:

> Now, suppose the weapon of the ideal army to be a club,
>
> or, as it were, a molecular cricket-bat; and suppose for convenience the mass of each demon with his weapon to be several times greater than that of a molecule. Every time he strikes a molecule he is to send it away with the same energy as it had immediately before. Each demon is to keep as nearly as possible to a certain station, making only such excursions from it as the execution of his orders requires. He is to experience no forces except such as result from collisions with molecules, and mutual forces between parts of his own mass, including his weapon: thus his voluntary movements cannot influence the position of his centre of gravity, otherwise than by producing collision with molecules.
>
> The whole interface between hot and cold is to be divided into small areas, each allotted to a single demon. The duty of each demon is to guard his allotment, turning molecules back or allowing them to pass through from either side, according to certain definite orders. First,

Kelvin's paper—like his note written on Maxwell's letter—imagines that the demons don't just "sort" molecules; they actually reverse their velocities, thus in effect anticipating Loschmidt's 1876 "reversibility objection" to Boltzmann's *H* theorem.

In an undated note, Maxwell discusses demons, attributing the name to Kelvin—and then starts considering the "physicalization" of demons, simplifying what they need to do:

> Concerning Demons.
> I. Who gave them this name? Thomson.
> 2. What were they by nature? Very small BUT lively beings incapable of doing work but able to open and shut valves which move without friction or inertia.
> 3. What was their chief end? To show that the 2nd Law of Thermodynamics has only a statistical certainty.
> 4. Is the production of an inequality of temperature their only occupation? No, for less intelligent demons can produce a difference in pressure as well as temperature by merely allowing all particles going in one direction while stopping all those going the other way. This reduces the demon to a valve. As such value him. Call him no more a demon but a valve like that of the hydraulic ram, suppose.

It didn't take long for Maxwell's demon to become something of a fixture in expositions of thermodynamics, even if it wasn't clear how it connected to other things people were saying about thermodynamics. And in 1879, for example, Kelvin gave a talk all about Maxwell's "sorting demon" (like other British people of the time he referred to Maxwell as "Clerk Maxwell"):

**87. The Sorting Demon of Maxwell.**

[Abstract of Royal Institution Lecture, Feb. 28, 1879, *Roy. Institution Proc.* Vol. IX. p. 113. Reprinted in *Popular Lectures and Addresses*, Vol. I. pp. 137—141.]

The word "demon," which originally in Greek meant a supernatural being, has never been properly used to signify a real or ideal personification of malignity.

Clerk Maxwell's "demon" is a creature of imagination having certain perfectly well defined powers of action, purely mechanical in their character, invented to help us to understand the "Dissipation of Energy" in nature.

He is a being with no preternatural qualities, and differs from real living animals only in extreme smallness and agility. He can at pleasure stop, or strike, or push, or pull any single atom of matter, and so moderate its natural course of motion. Endowed ideally with arms and hands and fingers—two hands and ten fingers suffice—he can do as much for atoms as a pianoforte player can do for the keys of the piano—just a little more, he can push or pull each atom *in any direction.*

Kelvin describes—without much commentary, and without mentioning the Second Law—some of the feats of which the demon would be capable. But he adds:

> The conception of the "sorting demon" is purely mechanical, and is of great value in purely physical science. It was not invented to help us to deal with questions regarding the influence of life and of mind on the motions of matter, questions essentially beyond the range of mere dynamics.

The description of the lecture ends:

> The discourse was illustrated by a series of experiments.

Presumably no actual Maxwell's demon was shown—or Kelvin wouldn't have continued for the rest of his life to treat the Second Law as an established principle.

But in any case, Maxwell's demon has always remained something of a fixture in discussions of the foundations of the Second Law. One might think that the observability of Brownian motion would make something like a Maxwell's demon possible. And indeed in 1912 Marian Smoluchowski (1872–1917) suggested experiments that one could imagine would "systematically harvest" Brownian motion—but showed that in fact they couldn't. In later years, a sequence of arguments were advanced that the mechanism of a Maxwell's demon just couldn't work in practice—though even today microscopic versions of what amount to Maxwell's demons are routinely being investigated.

## What Happened to Those People?

We've finally now come to the end of the story of how the original framework for the Second Law came to be set up. And, as we've seen, only a fairly small number of key players were involved:

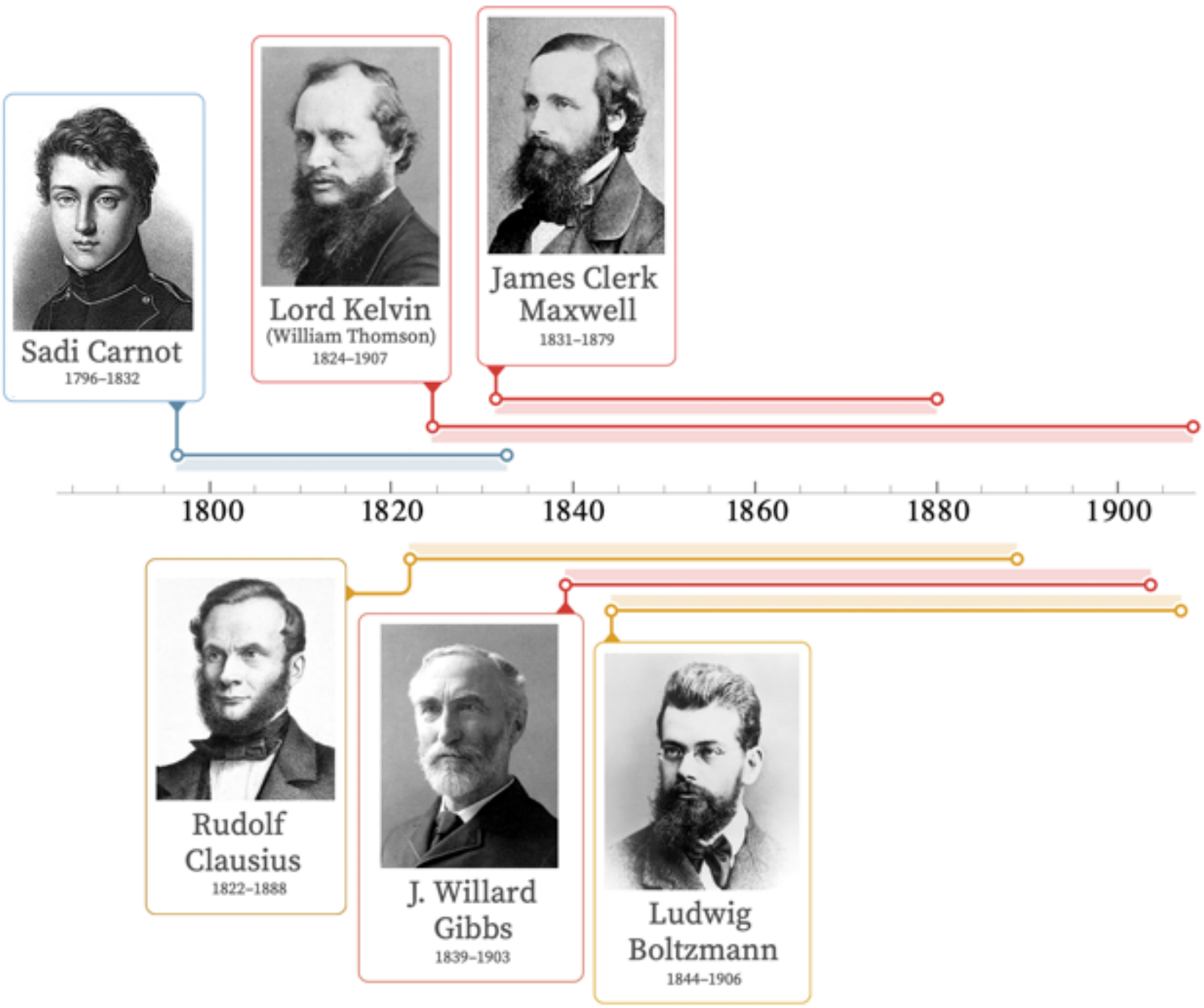

So what became of these people? Carnot lived a generation earlier than the others, never made a living as a scientist, and was all but unknown in his time. But all the others had distinguished careers as academic scientists, and were widely known in their time. Clausius, Boltzmann and Gibbs are today celebrated mainly for their contributions to thermodynamics; Kelvin and Maxwell also for other things. Clausius and Gibbs were in a sense "pure professors"; Boltzmann, Maxwell and especially Kelvin also had engagement engaged with the more general public.

All of them spent the majority of their lives in the countries of their birth—and all (with the exception of Carnot) were able to live out the entirety of their lives without time-consuming disruptions from war or other upheavals:

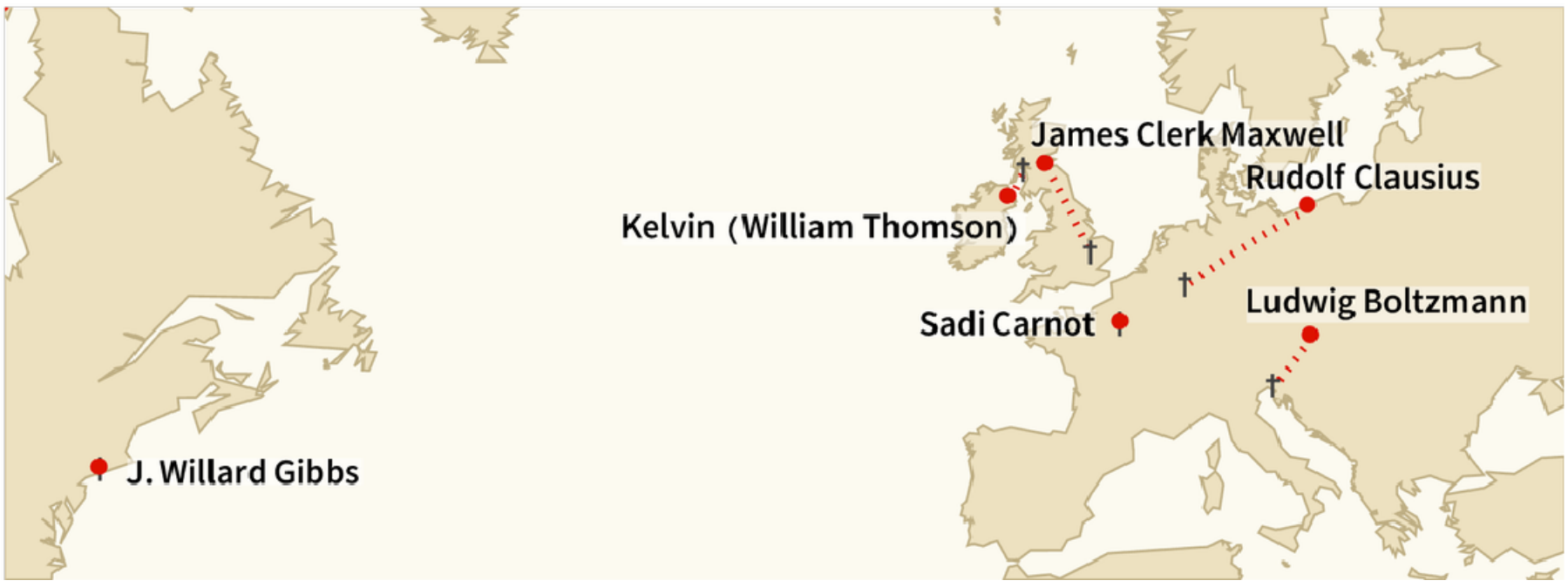

## *Sadi Carnot (1796–1832)*

Almost all of what is known about Sadi Carnot as a person comes from a single biographical note written nearly half a century after his death by his younger brother Hippolyte Carnot (who was a distinguished French politician—and sometime education minister—and father of the Sadi Carnot who would become president of France). Hippolyte Carnot began by saying that:

> As the life of Sadi Carnot was not marked by any notable event, his biography would have occupied only a few lines; but a scientific work by him, after remaining long in obscurity, brought again to light many years after his death, has caused his name to be placed among those of great inventors.

The Carnots' father was close to Napoleon, and Hippolyte explains that when Sadi was a young child he ended up being babysat by "Madame Bonaparte"—but one day wandered off, and was found inspecting the operation of a nearby mill, and quizzing the miller about it. For the most part, however, throughout his life, Sadi Carnot apparently kept very much to himself—while with quiet intensity showing a great appetite for intellectual pursuits from mathematics and science to art, music and literature, as well as practical engineering and the science of various sports.

Even his brother Hippolyte can't explain quite how Sadi Carnot—at the age of 28—suddenly "came out" and in 1824 published his book on thermodynamics. (As we discussed above, it no doubt had something to do with the work of his father, who died two years earlier.) Sadi Carnot funded the publication of the book himself—having 600 copies printed (at least some of which remained unsold a decade later). But after the book was published, Carnot appears to have returned to just privately doing research, living alone, and never publishing again in his lifetime. And indeed he lived only another eight years, dying (apparently after some months of ill health) in the same Paris cholera outbreak that claimed General Lamarque of *Les Misérables* fame.

Twenty-three pages of unpublished personal notes survive from the period after the publication of Carnot's book. Some are general aphorisms and life principles:

> Speak little of what you know, and not at all of what you do not know.
>
> Why try to be witty? I would rather be thought stupid and modest than witty and pretentious.
>
> God cannot punish man for not believing when he could so easily have enlightened and convinced him.
>
> The belief in an all-powerful Being, who loves us and watches over us, gives to the mind great strength to endure misfortune.
>
> When walking, carry a book, a notebook to preserve ideas, and a piece of bread in order to prolong the walk if need be.

But others are more technical—and in fact reveal that Carnot, despite having based his book on caloric theory, had realized that it probably wasn't correct:

> When a hypothesis no longer suffices to explain phenomena, it should be abandoned. This is the case with the hypothesis which regards caloric as matter, as a subtile fluid.
>
> The experimental facts tending to destroy this theory are as follows: The development of heat by percussion or friction of bodies ... The elevation of temperature which takes place [when] air [expands into a] vacuum ...

He continues:

> At present, light is generally regarded as the result of a vibratory movement of the ethereal fluid. Light produces heat, or at least accompanies radiating heat, and moves with the same velocity as heat. Radiating heat is then a vibratory movement. It would be ridiculous to suppose that it is an emission of matter while the light which accompanies it could be only a movement.
>
> Could a motion (that of radiating heat) produce matter (caloric)? No, undoubtedly; it can only produce a motion. Heat is then the result of a motion.

And then—in a rather clear enunciation of what would become the First Law of thermodynamics:

> Heat is simply motive power, or rather motion which has changed form. It is a movement among the particles of bodies. Wherever there is destruction of motive power there is, at the same time, production of heat in quantity exactly proportional to the quantity of motive power destroyed. Reciprocally, wherever there is destruction of heat, there is production of motive power.

Carnot also wonders:

> Liquefaction of bodies, solidification of liquids, crystallization—are they not forms of combinations of integrant molecules? Supposing heat due to a vibratory movement, how can the passage from the solid or the liquid to the gaseous state be explained ?

There is no indication of how Carnot felt about this emerging rethinking of thermodynamics, or of how it might affect the results in his book. But Carnot clearly hoped to do experiments (as outlined in his notes) to test what was really going on. But as it was, he presumably didn't get around to any of them—and his notes, ahead of their time as they were, did not resurface for many decades, by which time the ideas they contained had already been discovered by others.

## *Rudolf Clausius (1822–1888)*

Rudolf Clausius was born in what's now Poland (and was then Prussia), one of more than 14 children of an education administrator and pastor. He went to university in Berlin, and, after considering doing history, eventually specialized in math and physics. After graduating in 1844 he started teaching at a top high school in Berlin (which he did for 6 years), and meanwhile earned his PhD in physics. His career took off after his breakout paper on thermodynamics appeared in 1850. For a while he was a professor in Berlin, then for 12 years in Zürich, then briefly in Würzburg, then—for the remaining 19 years of his life—in Bonn.

He was a diligent—if, one suspects, somewhat stiff—professor, notable for the clarity of his lectures, and his organizational care with students. He seems to have been a competent administrator, and late in his career he spent a couple of years as the president ("rector") of his university. But first and foremost, he was a researcher, writing about a hundred papers over the course of his career. Most physicists of the time devoted at least some of their efforts to doing actual physics experiments. But Clausius was a pioneer in the idea of being a "pure theoretical physicist", inspired by experiments and quoting their results, but not doing them himself.

The majority of Clausius's papers were about thermodynamics, though late in his career his emphasis shifted more to electrodynamics. Clausius's papers were original, clear, incisive and often fairly mathematically sophisticated. But from his very first paper on thermodynamics in 1850, he very much adopted a macroscopic approach, talking about what he considered to be "bulk" quantities like energy, and later entropy. He did explore some of the potential mechanics of molecules, but he never really made the connection between molecular phenomena and entropy—or the Second Law. He had a number of run-ins about academic credit with Kelvin, Tait, Maxwell and Boltzmann, but he didn't seem to ever pay much attention to, for example, Boltzmann's efforts to find molecular-based probabilistic derivations of Clausius's results.

It probably didn't help that after two decades of highly productive work, two misfortunes befell Clausius. First, in 1870, he had volunteered to lead an ambulance corps in the Franco-Prussian war, and was wounded in the knee, leading to chronic pain (as well as to his habit of riding to class on horseback). And then, in 1875, Clausius's wife died in the birth of their sixth child—leaving him to care for six young children (which apparently he did with great conscientiousness). Clausius nevertheless continued to pursue his research—even to the end of his life—receiving many honors along the way (like election to no less than 40 professional societies), but it never again rose to the level of significance of his early work on thermodynamics and the Second Law.

## Kelvin (William Thomson) (1824–1907)

Of the people we're discussing here, by far the most famous during their lifetime was Kelvin. In his long career he wrote more than 600 scientific papers, received dozens of patents, started several companies and served in many administrative and governmental roles. His father was a math professor, ultimately in Glasgow, who took a great interest in the education of his children. Kelvin himself got an early start, effectively going to college at the age of 10, and becoming a professor in Glasgow at the age of 22—a position in which he continued for 53 years.

Kelvin's breakout work, done in his twenties, was on thermodynamics. But over the years he also worked on many other areas of physics, and beyond, mixing theory, experiment and engineering. Beginning in 1854 he became involved in a technical megaproject of the time: the attempt to lay a transatlantic telegraph cable. He wound up very much on the front lines, helping out as a just-in-time physicist + engineer on the cable-laying ship. The first few attempts didn't work out, but finally in 1866—in no small part through Kelvin's contributions—a cable was successfully laid, and Kelvin (or William Thomson, as he then was) became something of a celebrity. He was made "Sir William Thomson" and—along with two other techies—formed his first company, which had considerable success in exploiting telegraph-cable-related engineering innovations.

Kelvin's first wife died after a long illness in 1870, and Kelvin, with no children and already enthusiastic about the sea, bought a fairly large yacht, and pursued a number of nautical-related projects. One of these—begun in 1872—was the construction of an analog computer for calculating tides (basically with 10 gears for adding up 10 harmonic tide components), a device that, with progressive refinements, continued to be used for close to a century.

Being rather charmed by Kelvin's physicist-with-a-big-yacht persona, I once purchased a letter that Kelvin wrote in 1877 on the letterhead of "Yacht Lalla Rookh":

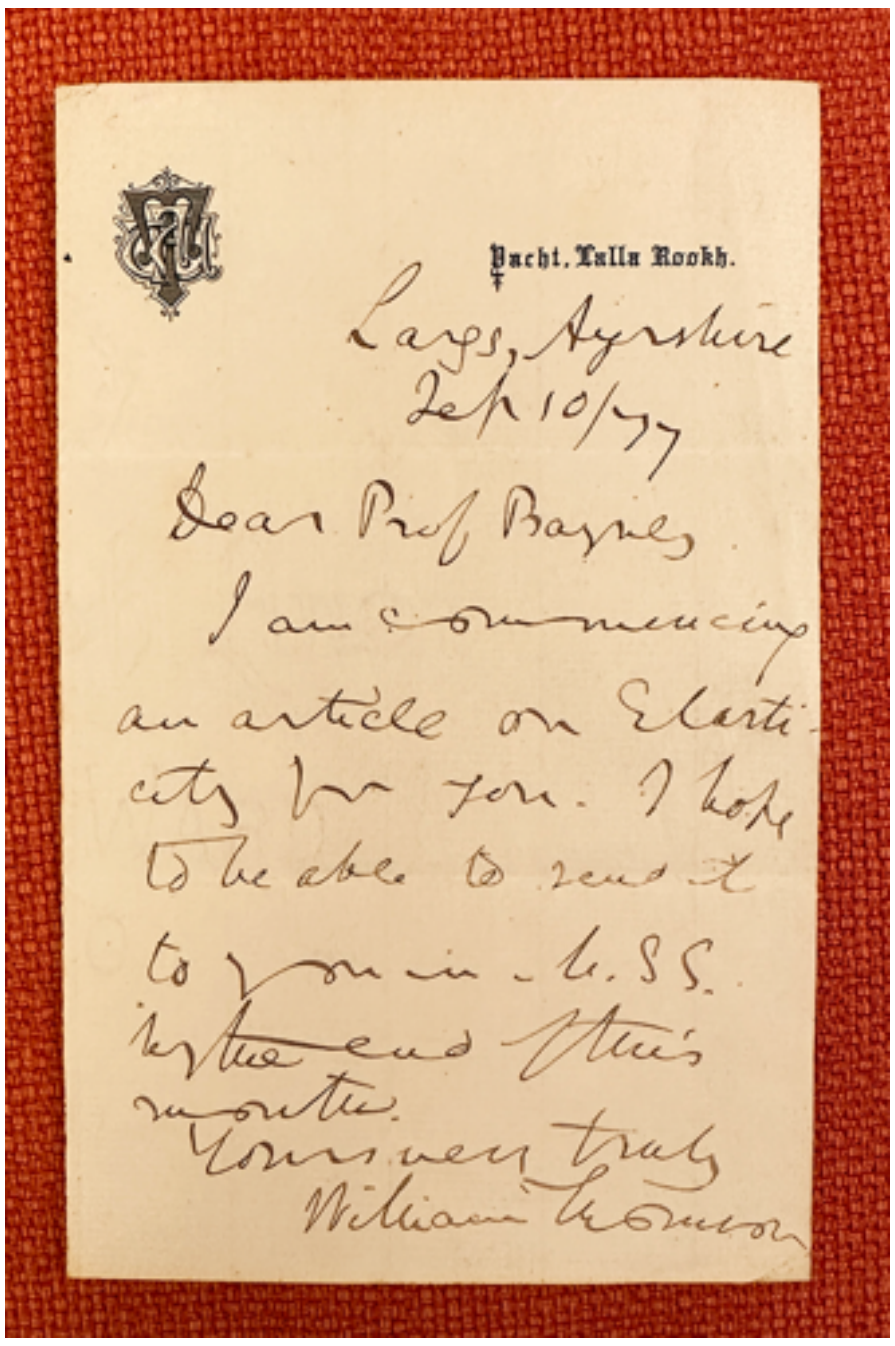
Yacht. Lalla Rookh.

Largs, Ayrshire
Sep 10/77

Dear Prof Baynes,
I am commencing an article on Elasticity for you. I hope to be able to send it to you in MSS. by the end of this month.
Yours very truly
William Thomson

The letter—in true academic style—promises that Kelvin will soon send an article he's been asked to write on elasticity theory. And in fact he did write the article, and it was an expository one that appeared in the 9th edition of the *Encyclopedia Britannica*.

Kelvin was a prolific (if, to modern ears, sometimes rather pompous) writer, who took exposition seriously. And indeed—finding the textbooks available to him as a professor inadequate—he worked over the course of a dozen years (1855—1867) with his (and Maxwell's) friend Peter Guthrie Tait to produce the influential *Treatise on Natural Philosophy*. Kelvin explored many topics and theories, some more immediately successful than others. In the 1870s he suggested that perhaps atoms might be knotted vortices in the (luminiferous) aether (causing Tait to begin developing knot theory)—a hypothesis that's in some sense a Victorian prelude to modern ideas about particles in our Physics Project.

Throughout his life, Kelvin was a devout Christian, writing that "The more thoroughly I conduct scientific research, the more I believe science excludes atheism." And indeed this belief seems to make an appearance in his implication that humans—presumably as a result of their special relationship with God—might avoid the Second Law. But more significant at the time was Kelvin's skepticism about Charles Darwin's 1859 theory of natural selection, believing that there must in the end be a "continually guiding and controlling intelligence". Despite being somewhat ridiculed for it, Kelvin talked about the possibility that life might have come to Earth from elsewhere via meteorites, believing that his estimates of the age of the Earth (which didn't take into account radioactivity) made it too young for the things Darwin described to have occurred.

By the 1870s, Kelvin had become a distinguished man of science, receiving all sorts of honors, assignments and invitations. And in 1876, for example, he was invited to Philadelphia to chair the committee judging electrical inventions at the US Centennial International Exhibition, notably reporting, in the terms of the time:

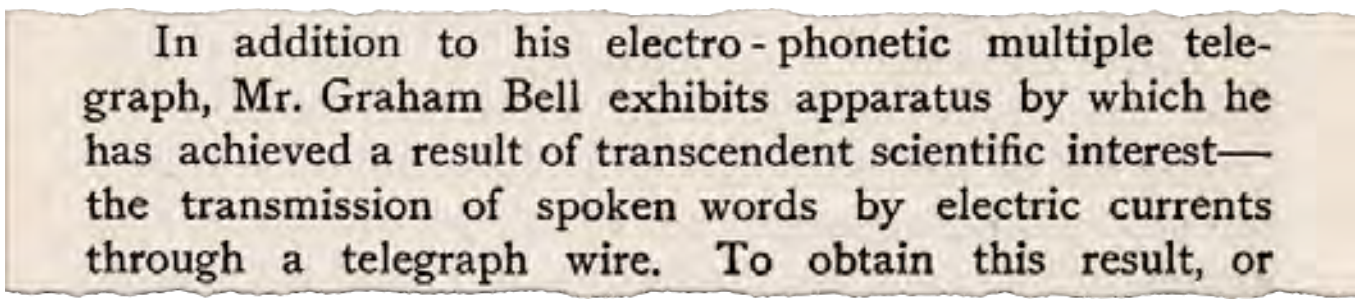
In addition to his electro-phonetic multiple telegraph, Mr. Graham Bell exhibits apparatus by which he has achieved a result of transcendent scientific interest—the transmission of spoken words by electric currents through a telegraph wire. To obtain this result, or

Then in 1892 a "peerage of the realm" was conferred on him by Queen Victoria. His wife (he had remarried) and various friends (including Charles Darwin's son George) suggested he pick the title "Kelvin", after the River Kelvin that flowed by the university in Glasgow. And by the end of his life "Lord Kelvin" had accumulated enough honorifics that they were just summarized with "…" (the MD was an honorary degree conferred by the University of Heidelberg because "it was the only one at their disposal which he did not already possess"):

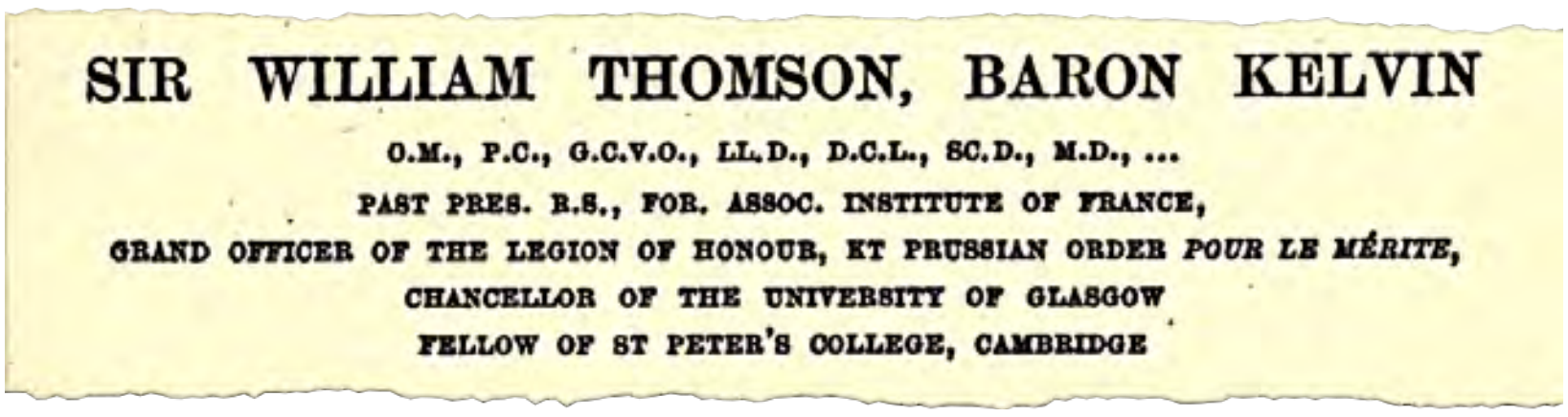
SIR WILLIAM THOMSON, BARON KELVIN

O.M., P.C., G.C.V.O., LL.D., D.C.L., SC.D., M.D., ...
PAST PRES. R.S., FOR. ASSOC. INSTITUTE OF FRANCE,
GRAND OFFICER OF THE LEGION OF HONOUR, KT PRUSSIAN ORDER *POUR LE MÉRITE*,
CHANCELLOR OF THE UNIVERSITY OF GLASGOW
FELLOW OF ST PETER'S COLLEGE, CAMBRIDGE

And when Kelvin died in 1907 he was given a state funeral and buried in Westminster Abbey near Newton and Darwin.

## James Clerk Maxwell (1831–1879)

James Clerk Maxwell lived only 48 years but in that time managed to do a remarkable amount of important science. His early years were spent on a 1500-acre family estate (inherited by his father) in a fairly remote part of Scotland—to which he would return later. He was an only child and was homeschooled—initially by his mother, until she died, when he was 8. At 10 he went to an upscale school in Edinburgh, and by the age of 14 had written his first scientific paper. At 16 he went as an undergraduate to the University of Edinburgh, then, effectively as a graduate student, to Cambridge—coming second in the final exams ("Second Wrangler") to a certain Edward Routh, who would spend most of his life coaching other students on those very same exams.

Within a couple of years, Maxwell was a professor, first in Aberdeen, then in London. In Aberdeen he married the daughter of the university president, who would soon be his "Observer K' (for "Katherine") in his classic work on color vision. But after nine fairly strenuous years as a professor, Maxwell in 1865 "retired" to his family estate, supervising a house renovation, and in "rural solitude" (recreationally riding around his estate on horse-back with his wife) having the most scientifically productive time of his life. In addition to his work on things like the kinetic theory of gases, he also wrote his 2–volume *Treatise on Electricity and Magnetism,* which ultimately took 7 years to finish, and which, with considerable clarity, described his approach to electromagnetism and what are now called "Maxwell's Equations". Occasionally, there were hints of his "country life"—like his 1870 "On Hills and Dales" that in his characteristic mathematicize–everything way gave a kind of "pre–topological" analysis of contour maps (perhaps conceived as he walked half a mile every day down to the mailbox at which journals and correspondence would arrive):

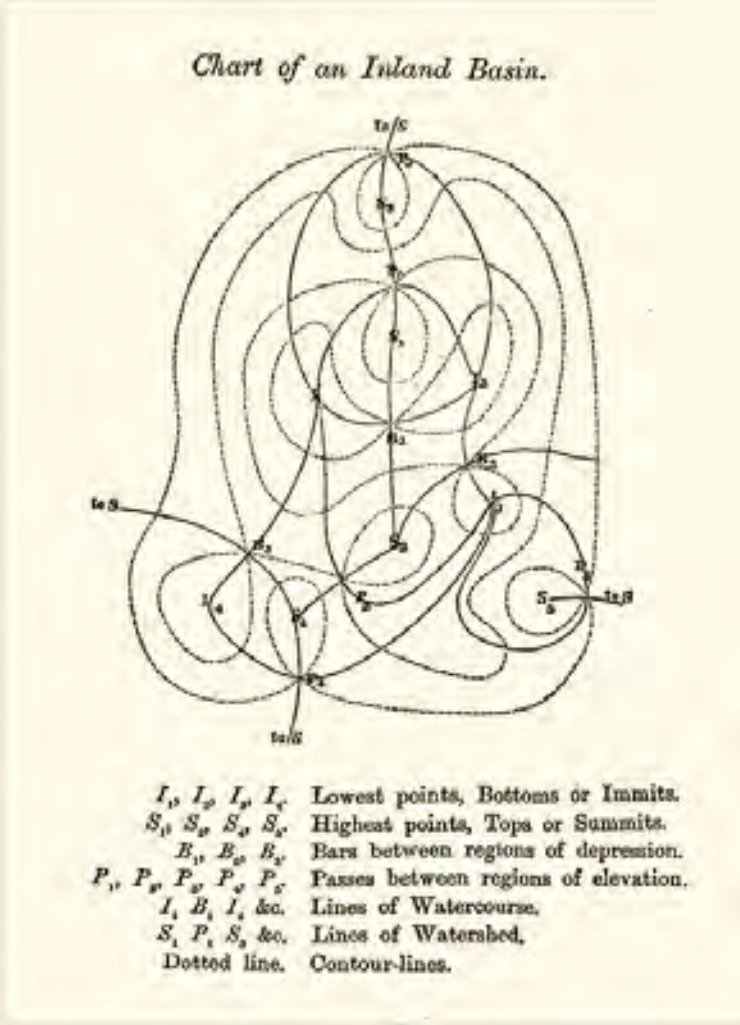

As a person, Maxwell was calm, reserved and unassuming, yet cheerful and charming—and given to writing (arguably sometimes sophomoric) poetry:

THE SONG OF THE ATLANTIC TELEGRAPH COMPANY.

I.

2(ʊ)
Mark how the telegraph motions to me,
2(ʊ)
Signals are coming along,
With a wag, wag, wag;
The telegraph needle is vibrating free,
And every vibration is telling to me
How they drag, drag, drag,
The telegraph cable along,

II.

2(ʊ)
No little signals are coming to me,
2(ʊ)
Something has surely gone wrong,
And it's broke, broke, broke;
What is the cause of it does not transpire,
But something has broken the telegraph wire
With a stroke, stroke, stroke,
Or else they've been pulling too strong.

With a certain sense of the absurd, he would occasionally publish satirical pieces in *Nature*, signing them *dp*/*dt*, which in the thermodynamic notation created by his friend Tait was equal to JCM, which were his initials. Maxwell liked games and tricks, and spinning tops featured prominently in some of his work. He enjoyed children, though never had any of his own. As a lecturer, he prepared diligently, but often got too sophisticated for his audience. In writing, though, he showed both great clarity and great erudition, for example freely quoting Latin and Greek in articles he wrote for the 9th edition of the *Encyclopedia Britannica* (of which he was scientific co–editor) on topics such as "Atom" and "Ether".

As we mentioned above, Maxwell was quite an enthusiast of diagrams and visual presentation (even writing an article on "Diagrams" for the *Encyclopedia Britannica*). He was also a capable experimentalist, making many measurements (sometimes along with his wife), and in 1861 creating the first color photograph.

In 1871 William Cavendish, 7th Duke of Devonshire, who had studied math in Cambridge, and was now chancellor of the university, agreed to put up the money to build what became the Cavendish Laboratory and to endow a new chair of experimental physics. Kelvin having turned down the job, it was offered to the still–rather–obscure Maxwell, who somewhat reluctantly accepted—with the result that for several years he spent much of his time supervising the design and building of the lab.

The lab was finished in 1874, but then William Cavendish dropped on Maxwell a large collection of papers from his great uncle Henry Cavendish, who had been a wealthy "gentleman scientist" of the late 1700s and (among other things) the discoverer of hydrogen. Maxwell liked history (as some of us do!), noticed that Cavendish had discovered Ohm's law 50 years before Ohm, and in the end spent several years painstakingly editing and annotating the papers into a 500–page book. By 1879 Maxwell was finally ready to energetically concentrate on physics research again, but, sadly, in the fall of that year his health failed, and he died at the age of 48—having succumbed to stomach cancer, as his mother also had at almost the same age.

## J. Willard Gibbs (1839–1903)

Gibbs was born near the Yale campus, and died there 64 years later, in the same house where he had lived since he was 7 years old (save for three years spent visiting European universities as a young man, and regular summer "out–in–nature" vacations). His father (who, like, "our Gibbs" was named "Josiah Willard"—making "our Gibbs" be called "Willard") came from an old and distinguished intellectual and religious New England family, and was a professor of sacred languages at Yale. Willard Gibbs went to college and graduate school at Yale, and then spent his whole career as a professor at Yale.

He was, it seems, a quiet, modest and rather distant person, who radiated a certain serenity, regularly attended church, had a small circle of friends and lived with his two sisters (and the husband and children of one of them). He diligently discharged his teaching responsibilities, though his lectures were very sparsely attended, and he seems not to have been thought forceful enough in dealing with people to have been called on for many administrative

tasks—though he became the treasurer of his former high school, and himself was careful enough with money that by the end of his life he had accumulated what would now be several million dollars.

He had begun his academic career in practical engineering, for example patenting an "improved [railway] car-brake", but was soon drawn in more mathematical directions, favoring a certain clarity and minimalism of formulation, and a cleanliness, if not brevity, of exposition. His work on thermodynamics (initially published in the rather obscure *Transactions of the Connecticut Academy*) was divided into two parts: the first, in the 1870s, concentrating on macroscopic equilibrium properties, and second, in the 1890s, concentrating on microscopic "statistical mechanics" (as Gibbs called it). Even before he started on thermodynamics, he'd been interested in electromagnetism, and between his two "thermodynamic periods", he again worked on electromagnetism. He studied Maxwell's work, and was at first drawn to the then-popular formalism of quaternions—but soon decided to invent his own approach and notation for vector analysis, which at first he presented only in notes for his students, though it later became widely adopted.

And while Gibbs did increasingly mathematical work, he never seems to have identified as a mathematician, modestly stating that "If I have had any success in mathematical physics, it is, I think, because I have been able to dodge mathematical difficulties." His last work was his book on statistical mechanics, which—with considerable effort and perhaps damage to his health—he finished in time for publication in connection with the Yale bicentennial in 1901 (an event which notably also brought a visit from Kelvin), only to die soon thereafter.

Gibbs had a few graduate students at Yale, a notable one being Lee de Forest, inventor of the vacuum tube (triode) electronic amplifier, and radio entrepreneur. (de Forest's 1899 PhD thesis was entitled "Reflection of Hertzian Waves from the Ends of Parallel Wires".) Another student of Gibbs was Lynde Wheeler, who became a government radio scientist, and who wrote a biography of Gibbs, of which I have a copy bought years ago at a used bookstore—that I was now just about to put back on a shelf when I opened its front cover and found an inscription:

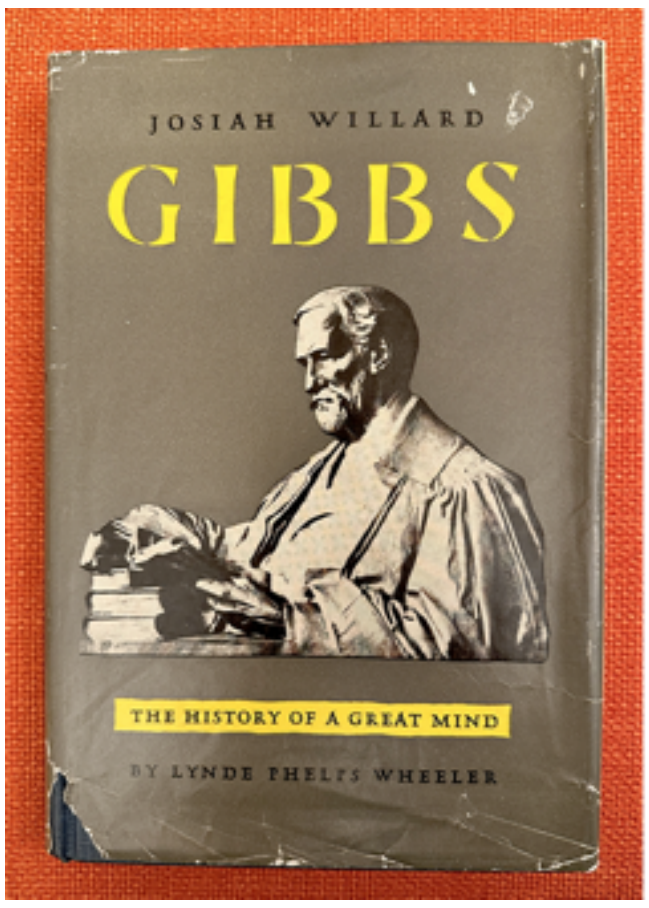

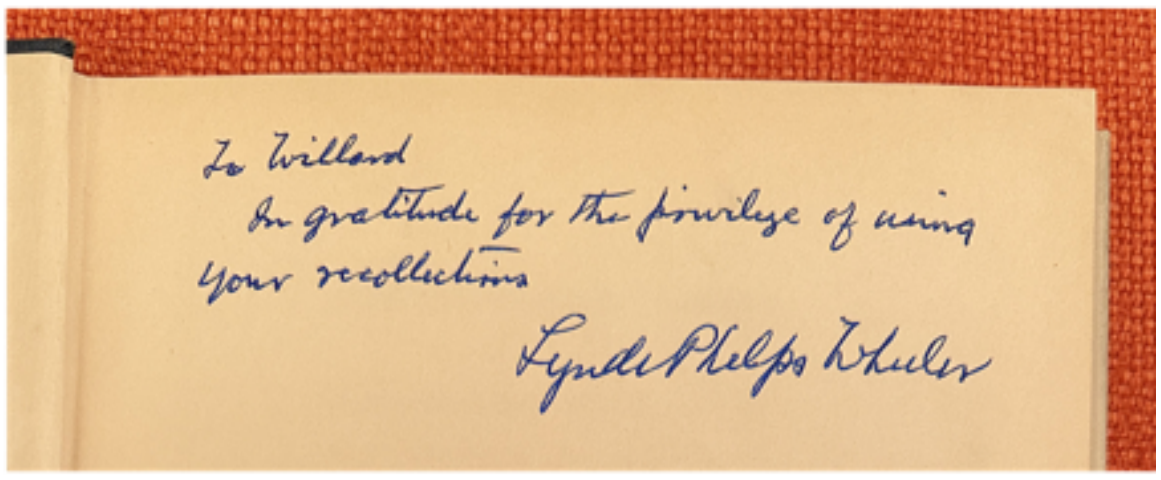

And, yes, it's a small world, and "To Willard" refers to Gibbs's sister's son (Willard Gibbs Van Name, who became a naturalist and wrote a 1929 book about national park deforestation).

## Ludwig Boltzmann (1844–1906)

Of the people we're discussing, Boltzmann is the one whose career was most focused on the Second Law. Boltzmann grew up in Austria, where his father was a civil servant (who died when Boltzmann was 15) and his mother was something of an heiress. Boltzmann did his PhD at the University of Vienna, where his professor notably gave him a copy of some of Maxwell's papers, together with an English grammar book. Boltzmann started publishing his own papers near the end of his PhD, and soon landed a position as a professor of mathematical physics in Graz. Four years later he moved to Vienna as a professor of mathematics, soon moving back to Graz as a professor of "general and experimental physics"—a position he would keep for 14 years.

He'd married in 1876, and had 5 children, though a son died in 1889, leaving 3 daughters and another son. Boltzmann was apparently a clear and lively lecturer, as well as a spirited and eager debater. He seems, at least in his younger years, to have been a happy and gregarious person, with a strong taste for music—and some charming do-it-your-own-way tendencies. For example, wanting to provide fresh milk for his children, he decided to just buy a cow, which he then led from the market through the streets—though had to consult his colleague, the professor of zoology, to find out how to milk it. Boltzmann was a capable experimental physicist, as well as a creator of gadgets, and a technology enthusiast—promoting the idea of airplanes (an application for gas theory!) and noting their potential power as a means of transportation.

Boltzmann had always had mood swings, but by the early 1890s he claimed they were getting worse. It didn't help that he was worn down by administrative work, and had worsening asthma and increasing nearsightedness (that he'd thought might be a sign of going blind). He moved positions, but then came back to Vienna, where he embarked on writing what would become a 2-volume book on *Gas Theory*—in effect contextualizing his life's work. The introduction to the first volume laments that "gas theory has gone out of fashion in Germany". The introduction to the second volume, written in 1898 when Boltzmann was 54, then says that "attacks on the theory of gases have begun to increase", and continues:

> ... it would be a great tragedy for science if the theory of gases were temporarily thrown into oblivion because of a momentary hostile attitude toward it, as, for example, was the wave theory [of light] because of Newton's authority.
>
> I am conscious of being only an individual struggling weakly against the stream of time. But it still remains in my power to contribute in such a way that, when the theory of gases is again revived, not too much will have to be rediscovered.

But even as he was writing this, Boltzmann had pretty much already wound down his physics research, and had basically switched to exposition, and to philosophy. He moved jobs again, but in 1902 again came back to Vienna, but now also as a professor of philosophy. He gave an inaugural lecture, first quoting his predecessor Ernst Mach (1838–1916) as saying "I do not believe that atoms exist", then discussing the philosophical relations between reality, perception and models. Elsewhere he discussed things like his view of the different philosophical character of models associated with differential equations and with atomism—and he even wrote an article on the general topic of “Models” for *Encyclopedia Britannica* (which curiously also talks about “in pure mathematics, especially geometry, models constructed of papier-mâché and plaster”). Sometimes Boltzmann’s philosophy could be quite polemical, like his attack on Schopenhauer, that ends by saying that “men [should] be freed from the spiritual migraine that is called metaphysics”.

Then, in 1904, Boltzmann addressed the Vienna Philosophical Society (a kind of predecessor of the Vienna Circle) on the subject of a “Reply to a Lecture on Happiness by Professor Ostwald”. Wilhelm Ostwald (1853–1932) (a chemist and social reformer, who was a personal friend of Boltzmann’s, but intellectual adversary) had proposed the concept of “energy of will” to apply mathematical physics ideas to psychology. Boltzmann mocked this, describing its faux formalism as “dangerous for science”. Meanwhile, Boltzmann gives his own Darwinian theory for the origin of happiness, based essentially on the idea that unhappiness is needed as a way to make organisms improve their circumstances in the struggle for survival.

Boltzmann himself was continuing to have problems that he attributed to then-popular but very vague “diagnosis” of “neurasthenia”, and had even briefly been in a psychiatric hospital. But he continued to do things like travel. He visited the US three times, in 1905 going to California (mainly Berkeley)—which led him to write a witty piece entitled “A German Professor’s Trip to El Dorado” that concluded:

> Yes, America will achieve great things. I believe in these people, even after seeing them at work in a setting where they’re not at their best: integrating and differentiating at a theoretical physics seminar...

In 1905 Einstein published his Boltzmann-and-atomism-based results on Brownian motion and on photons. But it’s not clear Boltzmann ever knew about them. For Boltzmann was sinking further. Perhaps he’d overexerted himself in California, but by the spring of 1906 he said he was no longer able to teach. In the summer he went with his family to an Italian seaside resort in an attempt to rejuvenate. But a day before they were to return to Vienna he failed to join his family for a swim, and his youngest daughter found him hanged in his hotel room, dead at the age of 62.

## Coarse-Graining and the “Modern Formulation”

After Gibbs’s 1902 book introducing the idea of ensembles, most of the language used (at least until now!) to discuss the Second Law was basically in place. But in 1912 one additional term—representing a concept already implicit in Gibbs’s work—was added: coarse-graining. Gibbs had discussed how the phase fluid representing possible states of a system could be elaborately mixed by the mechanical time evolution of the system. But realistic practical measurements could not be expected to probe all the details of the distribution of phase fluid; instead one could say that they would only sample “coarse-grained” aspects of it.

The term “coarse-graining” first appeared in a survey article entitled “The Conceptual

Foundations of the Statistical Approach in Mechanics", written for the German–language *Encyclopaedia of the Mathematical Sciences* by Boltzmann's former student Paul Ehrenfest, and his wife Tatiana Ehrenfest–Afanassjewa:

> Let us divide the Γ-space somehow into very small, but finite cells $\Omega$: $\Omega_1, \Omega_2, \cdots, \Omega_\lambda, \cdots$, which might be, for instance, cubes of equal size. The average value which the "fine-grained" density $\rho(q, p, t)$ has at time $t$ over the cell $\Omega_\lambda$ we will call "coarse-grained" density $P_\lambda(t)$ (read: capital $\rho$) of this cell. Because of Eq. (54) we have

The article also introduced all sorts of now–standard notation, and in many ways can be read as a final summary of what was achieved in the original development around the foundations of thermodynamics and the Second Law. (And indeed the article was sufficiently "final" that when it was republished as a book in 1959 it could still be presented as usefully summarizing the state of things.)

Looking at the article now, though, it's notable how much it recognized was not at all settled about the Second Law and its foundations. It places Boltzmann squarely at the center, stating in its preface:

> Since 1876 numerous papers have called attention to these foundations. In these papers the Boltzmann $H$-theorem, a central theorem of the kinetic theory of gases, was attacked. Without exception all studies so far published dealing with the connection of mechanics with probability theory grew out of the synthesis of these polemics and of Boltzmann's replies. These discussions will therefore be referred to frequently in our report.

The section titles are already revealing:

And soon they're starting to talk about "loose ends", and lots of them. Ergodicity is something one can talk about, but there's no known example (and with this definition it was later proved that there couldn't be):

> However, the existence of ergodic systems (i.e., the consistency of their definition) is doubtful. So far, not even one example is known of a mechanical system for which the single $G$-path approaches arbitrarily closely each point of the corresponding energy surface.[97] More-

But, they point out, it's something Boltzmann needed in order to justify his results:

> In order to get from this ensemble average to the time average, which is what Boltzmann wants, one needs the following chain of equalities:
>
> (33) Ensemble average = the time average of the ensemble average
> = the ensemble average of the time average
> = time average.
>
> The first equality follows from the stationary character of the phase distribution,[110] the second because the forming of averages is commutative. The third equality, however, is based on the statement that all motions of the set in question give the same time average for $\psi(q, p)$.
>
> It is at this point that the hypothesis about the gas model being ergodic enters. Because of the doubts about the internal consistency of the ergodic hypothesis, this investigation cannot be considered free of objection

Soon they're talking about Boltzmann's sloppiness in his discussion of the *H* curve:

> 3. In particular, he promoted this misunderstanding by always calling these maxima of the $H$-curves "humps," which makes one think almost necessarily of a maximum with a horizontal tangent.

And then they're on to talking about Gibbs, and the gaps in his reasoning:

> realize the following: While in all these cases Chapters XI and XIII prove at most a certain change in the direction of the canonical distribution, in Chapter XIV the analogies with thermodynamics are discussed as if it had been proved that the canonical distribution will be reached, at least approximately, in time. Such a jump

In the end they conclude:

> The foregoing account dealt chiefly with the conceptual foundations of statistico-mechanical investigations. Accordingly we had to emphasize that in these investigations a large number of loosely formulated and perhaps even inconsistent statements occupy a central position. In fact, we encounter here an incompleteness which from the logical point of view is serious and which appears in other branches of mechanics to a much smaller extent. This incompleteness, however, does not seem to have influenced the physicists in their evaluation of the statistico-mechanical investigations. In particular, the last few years have seen a sudden and wide dissemination of Boltzmann's ideas (the *H*-theorem, the Maxwell-Boltzmann distribution, the equipartition of energy, the relationship between entropy and probability, etc.). However, one cannot point at a corresponding progress in the conceptual clarification of Boltzmann's system to which one can ascribe this turn of affairs.

In other words, even though people now seem to be buying all these results, there are still plenty of issues with their foundations. And despite people's implicit assumptions, we can in no way say that the Second Law has been "proved".

## Radiant Heat, the Second Law and Quantum Mechanics

It was already realized in the 1600s that when objects get hot they emit "heat radiation"—which can be transferred to other bodies as "radiant heat". And particularly following Maxwell's work in the 1860s on electrodynamics it came to be accepted that radiant heat was associated with electromagnetic waves propagating in the "luminiferous aether". But unlike the molecules from which it was increasingly assumed that one could think of matter as being made, these electromagnetic waves were always treated—particularly on the basis of their mathematical foundations in calculus—as fundamentally continuous.

But how might this relate to the Second Law? Could it be, perhaps, that the Second Law should ultimately be attributed not to some property of the large-scale mechanics of discrete molecules, but rather to a feature of continuous radiant heat?

The basic equations assumed for mechanics—originally due to Newton—are reversible. But what about the equations for electrodynamics? Maxwell's equations are in and of themselves also reversible. But when one thinks about their solutions for actual electromagnetic radiation, there can be fundamental irreversibility. And the reason is that it's natural to describe the emission of radiation (say from a hot body), but then to assume that, once emitted, the radiation just "escapes to infinity"—rather than ever reversing the process of emission by being absorbed by some other body.

All the various people we've discussed above, from Clausius to Gibbs, made occasional remarks about the possibility that the Second Law—whether or not it could be "derived mechanically"—would still ultimately work, if nothing else, because of the irreversible emission of radiant heat.

But the person who would ultimately be most intimately connected to these issues was Max Planck—though in the end the somewhat-confused connection to the Second Law would recede in importance relative to what emerged from it, which was basically the raw material that led to quantum theory.

As a student of Helmholtz's in Berlin, Max Planck got interested in thermodynamics, and in 1879 wrote a 61-page PhD thesis entitled "On the Second Law of Mechanical Heat Theory". It was a traditional (if slightly streamlined) discussion of the Second Law, very much based on Clausius's approach (and even with the same title as Clausius's 1867 paper)—and without any mention whatsoever of Boltzmann:

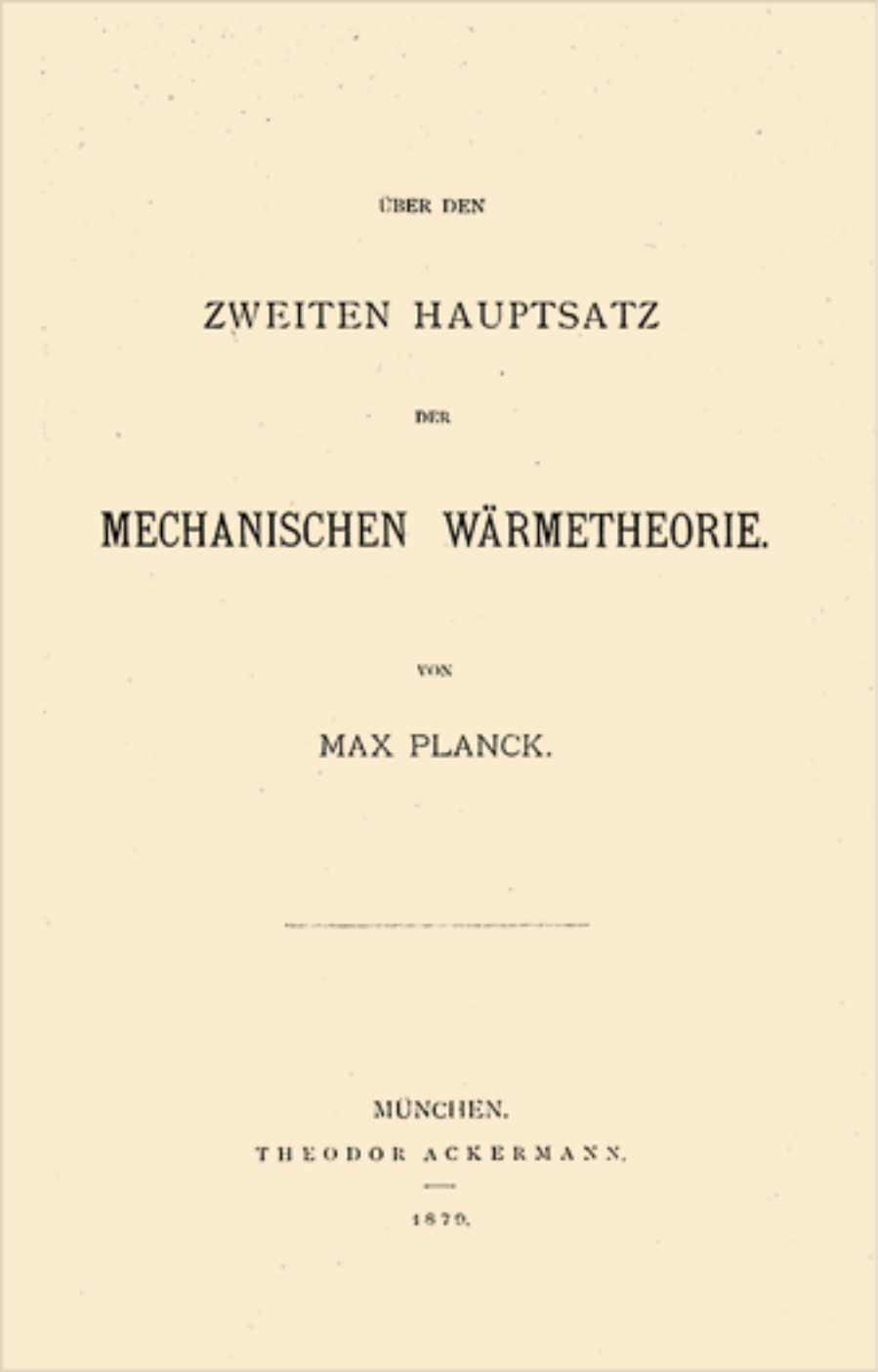

ÜBER DEN

ZWEITEN HAUPTSATZ

DER

MECHANISCHEN WÄRMETHEORIE.

VON

MAX PLANCK.

MÜNCHEN.

THEODOR ACKERMANN.

1879.

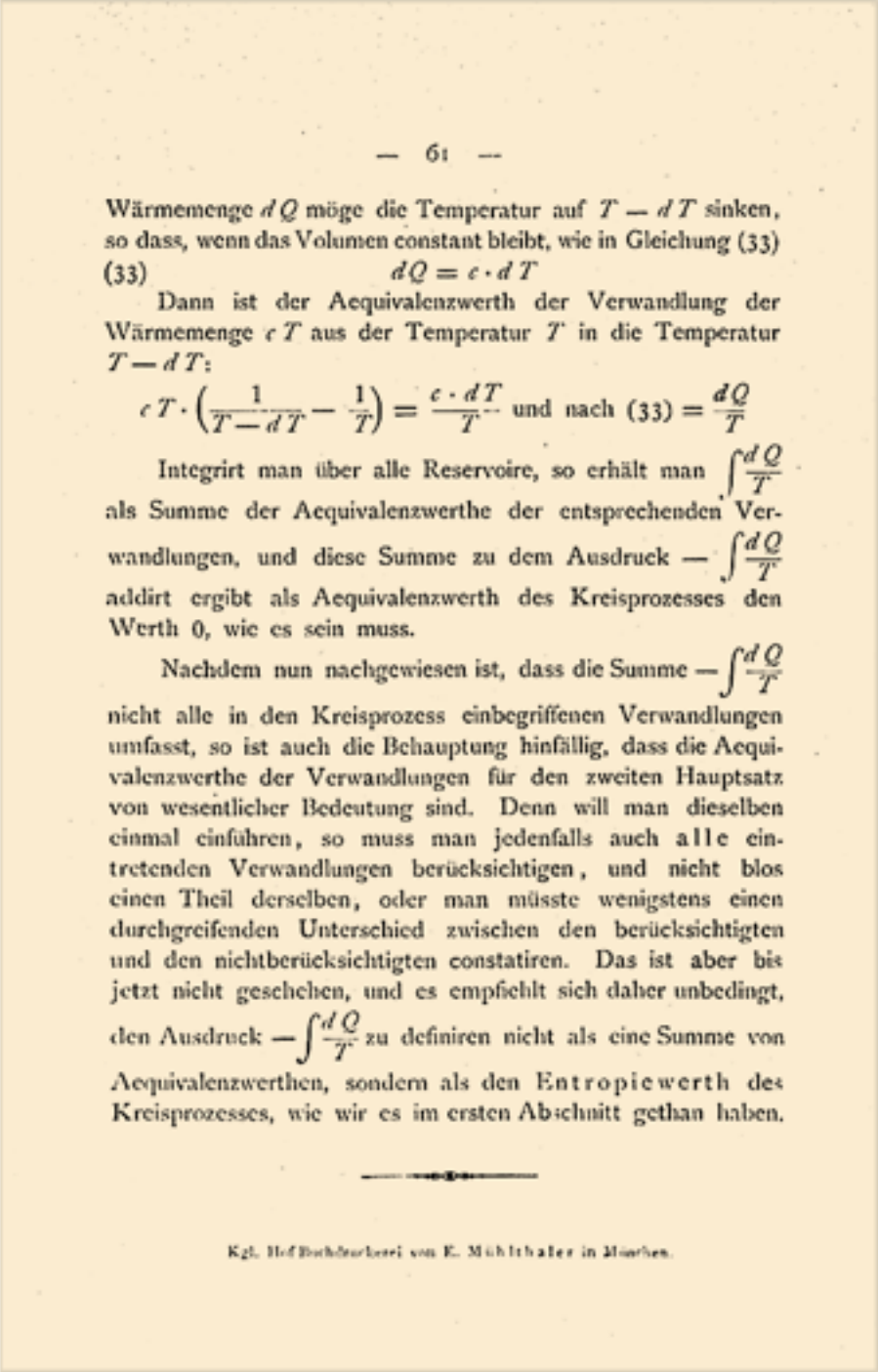

— 61 —

Wärmemenge $dQ$ möge die Temperatur auf $T - dT$ sinken, so dass, wenn das Volumen constant bleibt, wie in Gleichung (33)

(33) $$dQ = c \cdot dT$$

Dann ist der Aequivalenzwerth der Verwandlung der Wärmemenge $cT$ aus der Temperatur $T$ in die Temperatur $T - dT$:

$$cT \cdot \left(\frac{1}{T - dT} - \frac{1}{T}\right) = \frac{c \cdot dT}{T} \text{ und nach (33) } = \frac{dQ}{T}$$

Integrirt man über alle Reservoire, so erhält man $\int \frac{dQ}{T}$ als Summe der Aequivalenzwerthe der entsprechenden Verwandlungen, und diese Summe zu dem Ausdruck $-\int \frac{dQ}{T}$ addirt ergibt als Aequivalenzwerth des Kreisprozesses den Werth 0, wie es sein muss.

Nachdem nun nachgewiesen ist, dass die Summe $-\int \frac{dQ}{T}$ nicht alle in den Kreisprozess einbegriffenen Verwandlungen umfasst, so ist auch die Behauptung hinfällig, dass die Aequivalenzwerthe der Verwandlungen für den zweiten Hauptsatz von wesentlicher Bedeutung sind. Denn will man dieselben einmal einführen, so muss man jedenfalls auch alle eintretenden Verwandlungen berücksichtigen, und nicht blos einen Theil derselben, oder man müsste wenigstens einen durchgreifenden Unterschied zwischen den berücksichtigten und den nichtberücksichtigten constatiren. Das ist aber bis jetzt nicht geschehen, und es empfiehlt sich daher unbedingt, den Ausdruck $-\int \frac{dQ}{T}$ zu definiren nicht als eine Summe von Aequivalenzwerthen, sondern als den Entropiewerth des Kreisprozesses, wie wir es im ersten Abschnitt gethan haben.

For most of the two decades that followed, Planck continued to use similar methods to study the Second Law in various settings (e.g. elastic materials, chemical mixtures, etc.)—and meanwhile ascended the German academic physics hierarchy, ending up as a professor of theoretical physics in Berlin. Planck was in many ways a physics traditionalist, not wanting to commit to things like "newfangled" molecular ideas—and as late as 1897 (with his assistant Zermelo having made his "recurrence objection" to Boltzmann's work) still saying that he would "abstain completely from any definite assumption about the nature of heat". But regardless of its foundations, Planck was a true believer in the Second Law, for example in 1891 asserting that it "must extend to all forces of nature … not only thermal and chemical, but also electrical and other".

And in 1895 he began to investigate how the Second Law applied to electrodynamics—and in particular to the "heat radiation" that it had become clear (particularly through Heinrich Hertz's (1857–1894) experiments) was of electromagnetic origin. In 1896 Wilhelm Wien (1864–1928) suggested that the heat radiation (or what we now call blackbody radiation) was in effect produced by tiny Hertzian oscillators with velocities following a Maxwell distribution.

Planck, however, had a different viewpoint, instead introducing the concept of "natural radiation"—a kind of intrinsic thermal equilibrium state for radiation, with an associated intrinsic entropy. He imagined "resonators" interacting through Maxwell's equations with this radiation, and in 1899 invented a (rather arbitrary) formula for the entropy of these resonators, that implied (through the laws of electrodynamics) that overall entropy would increase—just like the Second Law said—and when the entropy was maximized it gave the same result as Wien for the spectrum of blackbody radiation. In early 1900 he sharpened his treatment and began to suggest that with his approach Wien's form of the blackbody spectrum would emerge as a provable consequence of the universal validity of the Second Law.

But right around that time experimental results arrived that disagreed with Wien's law. And by the end of 1900 Planck had a new hypothesis, for which he finally began to rely on ideas from Boltzmann. Planck started from the idea that he should treat the behavior of his resonators statistically. But how then could he compute their entropy? He quotes (for the first time ever) his simplification of Boltzmann's formula for entropy:

dem Logarithmus der Wahrscheinlichkeit $W$ dafür, dass die $N$ Resonatoren insgesamt die Energie $U_N$ besitzen, also:

$$S_N = k \log W + \text{const.} \tag{3}$$

Diese Festsetzung kommt nach meiner Meinung im Grunde auf eine Definition der genannten Wahrscheinlichkeit $W$ hinaus;

As he explains it—claiming now, after years of criticizing Boltzmann, that this is a "theorem":

> We now set the entropy *S* of the system proportional to the logarithm of its probability *W* ... In my opinion this actually serves as a definition of the probability *W*, since in the basic assumptions of electromagnetic theory there is no definite evidence for such a probability. The suitability of this expression is evident from the outset, in view of its simplicity and close connection with a theorem from kinetic gas theory.

But how could he figure out the probability for a resonator to have a certain energy, and thus a certain entropy? For this he turns directly to Boltzmann—who, as a matter of convenience in his 1877 paper had introduced discrete values of energy for molecules. Planck simply states that it's "necessary" (i.e. to get the experimentally right answer) to treat the resonator energy "not as a continuous, infinitely divisible quantity, but as a discrete quantity composed of an integral number of finite equal parts". As an example of how this works he gives a table just like the one in Boltzmann's paper from nearly a quarter of a century earlier:

elemente, so erhält man für jede Complexion ein Symbol von folgender Form:

| 1 | 2 | 3 | 4 | 5 | 6 | 7 | 8 | 9 | 10 |
|---|---|---|---|---|---|---|---|---|---|
| 7 | 38 | 11 | 0 | 9 | 2 | 20 | 4 | 4 | 5 |

Hier ist $N = 10$, $P = 100$ angenommen. Die Anzahl $\mathfrak{R}$ aller

Pretty soon he's deriving the entropy of a resonator as a function of its energy, and its discrete energy unit $\epsilon$:

Also nach (2) die Entropie $S$ eines Resonators als Function seiner Energie $U$:

$$S = k\left\{\left(1+\frac{U}{\varepsilon}\right)\log\left(1+\frac{U}{\varepsilon}\right) - \frac{U}{\varepsilon}\log\frac{U}{\varepsilon}\right\}. \tag{6}$$

Connecting this to blackbody radiation he claims that each resonator's energy unit is connected to its frequency according to

der Schwingungszahl $\nu$ sein muss, also:

$$\varepsilon = h \cdot \nu$$

so that its entropy is

und somit:

$$S = k\left\{\left(1+\frac{U}{h\nu}\right)\log\left(1+\frac{U}{h\nu}\right) - \frac{U}{h\nu}\log\frac{U}{h\nu}\right\}.$$

Hierbei sind $h$ und $k$ universelle Constante.

"[where] $h$ and $k$ are universal constants".

In a similar situation Boltzmann had effectively taken the limit $\epsilon \to 0$, because that's what he believed corresponded to ("calculus-based") physical reality. But Planck—in what he later described as an "act of desperation" to fit the experimental data—didn't do that. So in computing things like average energies he's evaluating Sum[x Exp[-a x], {x, 0, ∞}] rather than Integrate[x Exp[-a x], {x, 0, Infinity}]. And in doing this it takes him only a few lines to derive what's now called the Planck spectrum for blackbody radiation (i.e. for "radiation in equilibrium"):

und aus (8) folgt dann das gesuchte Energieverteilungsgesetz:

$$u = \frac{8\pi h \nu^3}{c^3} \cdot \frac{1}{e^{\frac{h\nu}{k\vartheta}} - 1} \tag{12}$$

And then by fitting this result to the data of the time he gets "Planck's constant" (the correct result is 6.62):

$$(15) \qquad h = 6{,}55 \,.\, 10^{-27}\,\text{erg}\,.\,\text{sec}\,,$$

And, yes, this was essentially the birth of quantum mechanics—essentially as a spinoff from an attempt to extend the domain of the Second Law. Planck himself didn't seem to internalize what he'd done for at least another decade. And it was really Albert Einstein's 1905 analysis of the photoelectric effect that made the concept of the quantization of energy that Planck had assumed (more as a calculational hypothesis than anything else) seem to be something of real physical significance—that would lead to the whole development of quantum mechanics, notably in the 1920s.

## Are Molecules Real? Continuous Versus Discrete

As we discussed at the very beginning above, already in antiquity there was a notion that at least things like solids and liquids might not ultimately be continuous (as they seemed), but might instead be made of large numbers of discrete "atomic" elements. By the 1600s there was also the idea that light might be "corpuscular"—and, as we discussed above, gases too. But meanwhile, there were opposing theories that espoused continuity—like the caloric theory of heat. And particularly with the success of calculus, there was a strong tendency to develop theories that showed continuity—and to which calculus could be applied.

But in the early 1800s—notably with the work of John Dalton (1766–1844)—there began to be evidence that there were discrete entities participating in chemical reactions. Meanwhile, as we discussed above, the success of the kinetic theory of gases gave increasing evidence for some kind of—at least effectively—discrete elements in gases. But even people like Boltzmann and Maxwell were reluctant to assert that gases really were made of molecules. And there were plenty of well-known scientists (like Ernst Mach (1838–1916)) who "opposed atomism", often effectively on the grounds that in science one should only talk about things one can actually see or experience—not things like atoms that were too small for that.

But there was something else too: with Newton's theory of gravitation as a precursor, and then with the investigation of electromagnetic phenomena, there emerged in the 1800s the idea of a "continuous field". The interpretation of this was fairly clear for something like an elastic solid or a fluid that exhibited continuous deformations.

Mathematically, things like gravity, magnetism—and heat—seemed to work in similar ways. And it was assumed that this meant that in all cases there had to be some fluid-like "carrier" for the field. And this is what led to ideas like the luminiferous aether as the "carrier" of electromagnetic waves. And, by the way, the idea of an aether wasn't even obviously incompatible with the idea of atoms; Kelvin, for example, had a theory that atoms were vortices (perhaps knotted) in the aether.

But how does this all relate to the Second Law? Well, particularly through the work of Boltzmann there came to be the impression that given atomism, probability theory could essentially "prove" the Second Law. A few people tried to clarify the formal details (as we discussed above), but it seemed like any final conclusion would have to await the validation (or not) of atomism, which in the late 1800s was still a thoroughly controversial theory.

By the first decade of the 1900s, however, the fortunes of atomism began to change. In 1897 J. J. Thomson (1856–1940) discovered the electron, showing that electricity was fundamentally "corpuscular". And in 1900 Planck had (at least calculationally) introduced discrete quanta of energy. But it was the three classic papers of Albert Einstein in 1905 that—in their different ways—began to secure the ultimate success of atomism.

First there was his paper "On a Heuristic View about the Production and Transformation of Light", which began:

> Maxwell's theory of electromagnetic [radiation] differs in a profound, essential way from the current theoretical models of gases and other matter. We consider the state of a material body to be completely determined by the positions and velocities of a finite number of atoms and electrons, albeit a very large number. But the electromagnetic state of a region of space is described by continuous functions ...

He then points out that optical experiments look only at time-averaged electromagnetic fields, and continues:

> In particular, blackbody radiation, photoluminescence, [the photoelectric effect] and other phenomena associated with the generation and transformation of light seem better modeled by assuming that the energy of light is distributed discontinuously in space. According to this picture, the energy of a light wave emitted from a point source is not spread continuously over ever larger volumes, but consists of a finite number of energy quanta that are spatially localized at points of space, move without dividing and are absorbed or generated only as a whole.

In other words, he's suggesting that light is "corpuscular", and that energy is quantized. When he begins to get into details, he's soon talking about the "entropy of radiation"—and, then, in three core sections of his paper, he's basing what he's doing on "Boltzmann's principle":

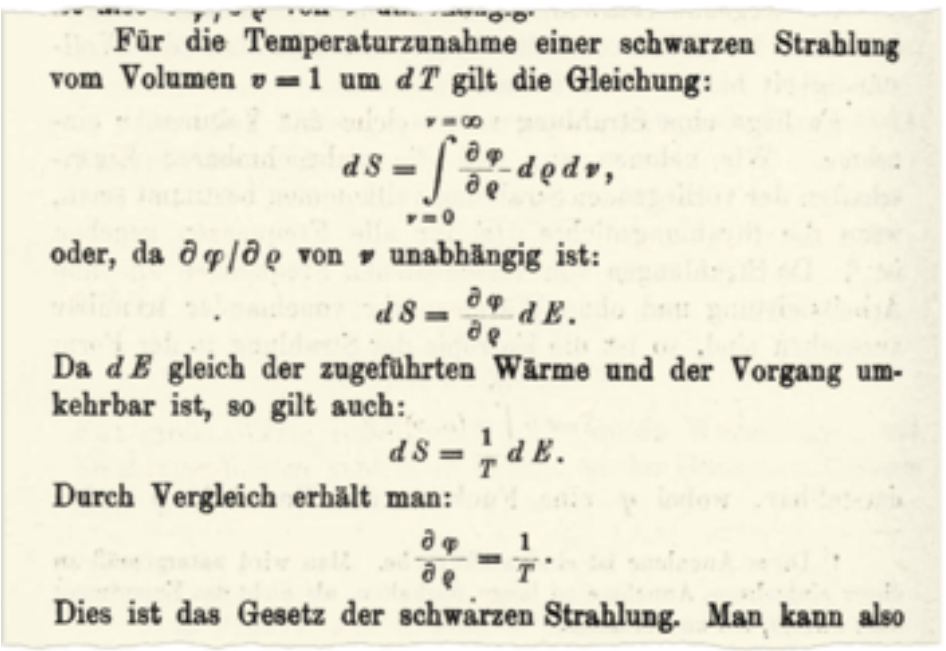

Für die Temperaturzunahme einer schwarzen Strahlung vom Volumen $v = 1$ um $dT$ gilt die Gleichung:

$$dS = \int_{\nu=0}^{\nu=\infty} \frac{\partial\varphi}{\partial\varrho} d\varrho \, d\nu,$$

oder, da $\partial\varphi/\partial\varrho$ von $\nu$ unabhängig ist:

$$dS = \frac{\partial\varphi}{\partial\varrho} dE.$$

Da $dE$ gleich der zugeführten Wärme und der Vorgang umkehrbar ist, so gilt auch:

$$dS = \frac{1}{T} dE.$$

Durch Vergleich erhält man:

$$\frac{\partial\varphi}{\partial\varrho} = \frac{1}{T}.$$

Dies ist das Gesetz der schwarzen Strahlung. Man kann also

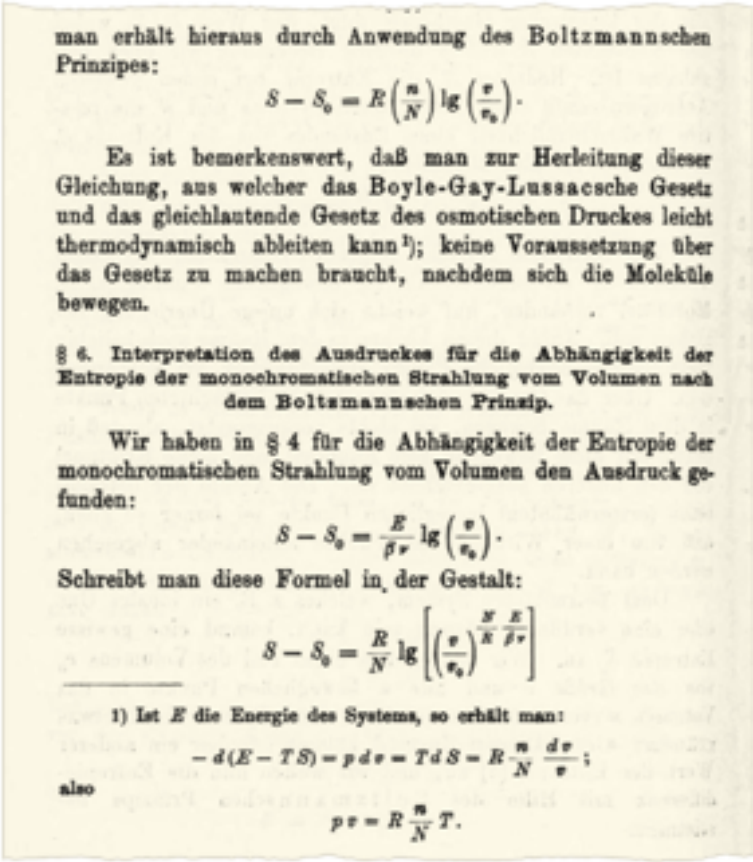

man erhält hieraus durch Anwendung des Boltzmannschen Prinzipes:

$$S - S_0 = R\left(\frac{n}{N}\right) \lg\left(\frac{v}{v_0}\right).$$

Es ist bemerkenswert, daß man zur Herleitung dieser Gleichung, aus welcher das Boyle-Gay-Lussacsche Gesetz und das gleichlautende Gesetz des osmotischen Druckes leicht thermodynamisch ableiten kann[1]; keine Voraussetzung über das Gesetz zu machen braucht, nachdem sich die Moleküle bewegen.

§ 6. Interpretation des Ausdruckes für die Abhängigkeit der Entropie der monochromatischen Strahlung vom Volumen nach dem Boltzmannschen Prinzip.

Wir haben in § 4 für die Abhängigkeit der Entropie der monochromatischen Strahlung vom Volumen den Ausdruck gefunden:

$$S - S_0 = \frac{E}{\beta\nu} \lg\left(\frac{v}{v_0}\right).$$

Schreibt man diese Formel in der Gestalt:

$$S - S_0 = \frac{R}{N} \lg\left[\left(\frac{v}{v_0}\right)^{\frac{N}{R}\frac{E}{\beta\nu}}\right]$$

1) Ist $E$ die Energie des Systems, so erhält man:

$$-d(E - TS) = p\,dv = T\,dS = R\frac{n}{N}\frac{dv}{v};$$

also

$$pv = R\frac{n}{N}T.$$

Two months later, Einstein produced another paper: "Investigations on the Theory of Brownian Motion". Back in 1827 the British botanist Robert Brown (1773–1858) had seen under a microscope tiny grains (ejected by pollen) randomly jiggling around in water. Einstein began his paper:

> In this paper it will be shown that according to the molecular–kinetic theory of heat, bodies of microscopically visible size suspended in a liquid will perform movements of such magnitude that they can be easily observed in a microscope, on account of the molecular motions of heat.

He doesn't explicitly mention Boltzmann in this paper, but there's Boltzmann's formula again:

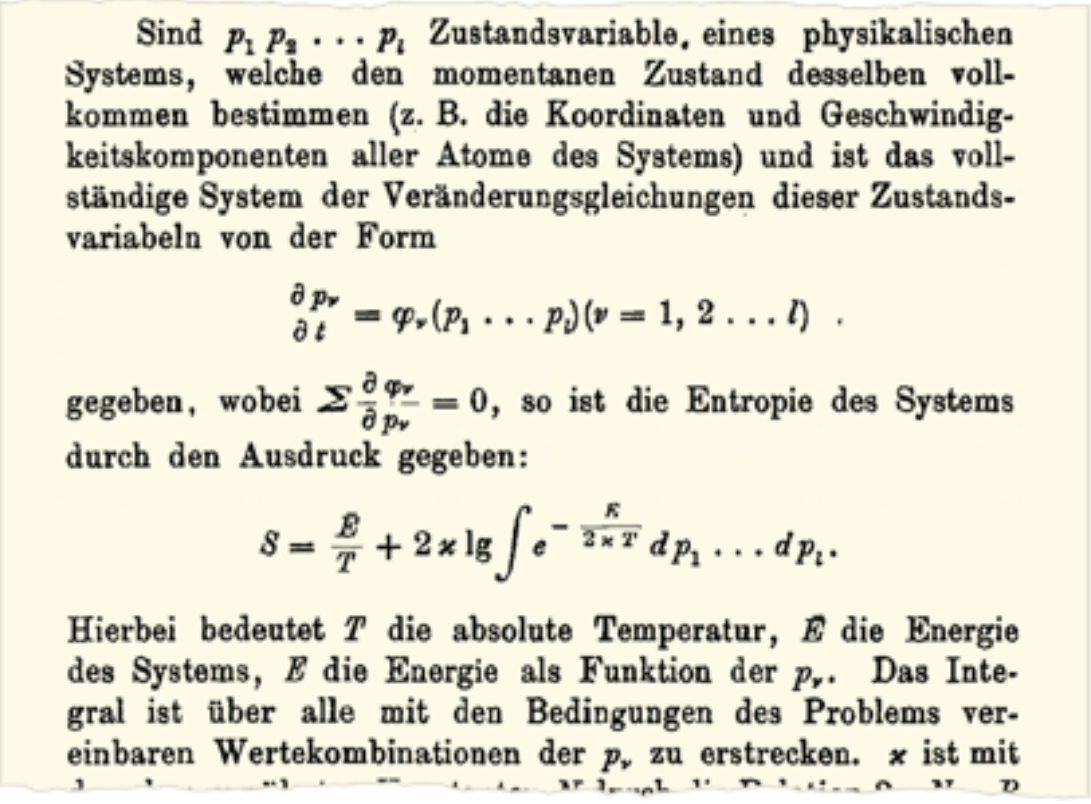

Sind $p_1\, p_2 \ldots p_l$ Zustandsvariable, eines physikalischen Systems, welche den momentanen Zustand desselben vollkommen bestimmen (z. B. die Koordinaten und Geschwindigkeitskomponenten aller Atome des Systems) und ist das vollständige System der Veränderungsgleichungen dieser Zustandsvariabeln von der Form

$$\frac{\partial p_\nu}{\partial t} = \varphi_\nu(p_1 \ldots p_l)(\nu = 1, 2 \ldots l)\ .$$

gegeben, wobei $\Sigma \frac{\partial \varphi_\nu}{\partial p_\nu} = 0$, so ist die Entropie des Systems durch den Ausdruck gegeben:

$$S = \frac{\bar{E}}{T} + 2\varkappa \lg \int e^{-\frac{E}{2\varkappa T}} dp_1 \ldots dp_l .$$

Hierbei bedeutet $T$ die absolute Temperatur, $\bar{E}$ die Energie des Systems, $E$ die Energie als Funktion der $p_\nu$. Das Integral ist über alle mit den Bedingungen des Problems vereinbaren Wertekombinationen der $p_\nu$ zu erstrecken. $\varkappa$ ist mit

And by the next year it's become clear experimentally that, yes, the jiggling Robert Brown had seen was in fact the result of impacts from discrete, real water molecules.

Einstein's third 1905 paper, "On the Electrodynamics of Moving Bodies"—in which he introduced relativity theory—wasn't so obviously related to atomism. But in showing that the luminiferous aether will (as Einstein put it) "prove superfluous" he was removing what was (almost!) the last remaining example of something continuous in physics.

In the years after 1905, the evidence for atomism mounted rapidly, segueing in the 1920s into the development of quantum mechanics. But what happened with the Second Law? By the time atomism was generally accepted, the generation of physicists that had included Boltzmann and Gibbs was gone. And while the Second Law was routinely invoked in expositions of thermodynamics, questions about its foundations were largely forgotten. Except perhaps for one thing: people remembered that "proofs" of the Second Law had been controversial, and had depended on the controversial hypothesis of atomism. But—they appear to have reasoned—now that atomism isn't controversial anymore, it follows that the Second Law is indeed "satisfactorily proved". And, after all, there were all sorts of other things to investigate in physics.

There are a couple of "footnotes" to this story. The first has to do with Einstein. Right before Einstein's remarkable series of papers in 1905, what was he working on? The answer is: the Second Law! In 1902 he wrote a paper entitled "Kinetic Theory of Thermal Equilibrium and of the Second Law of Thermodynamics". Then in 1903: "A Theory of the Foundations of Thermodynamics". And in 1904: "On the General Molecular Theory of Heat". The latter paper claims:

> I derive an expression for the entropy of a system, which is completely analogous to the one found by Boltzmann for ideal gases and assumed by Planck in his theory of radiation. Then I give a simple derivation of the Second Law.

But what's actually there is not quite what's advertised:

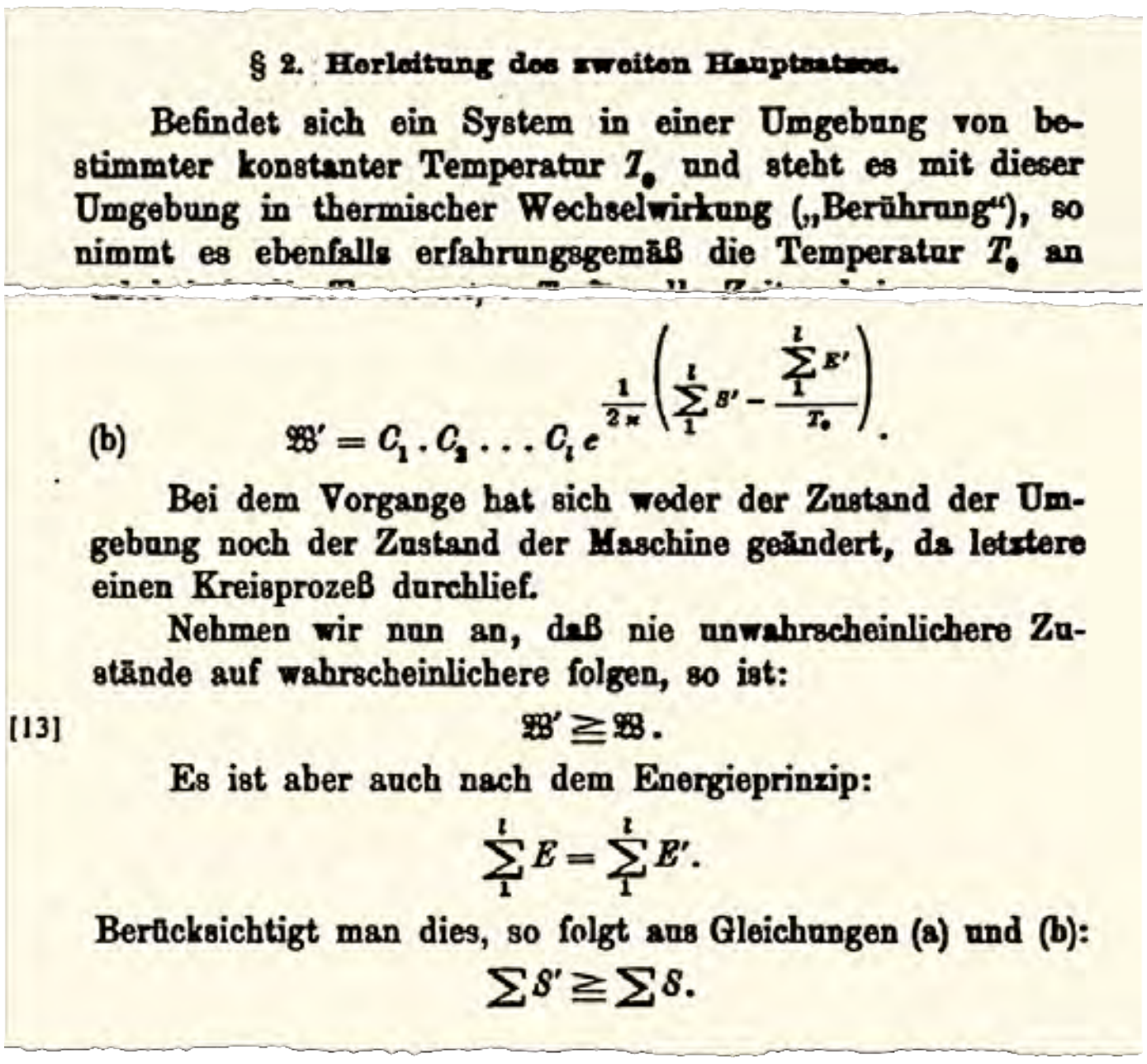

§ 2. Herleitung des zweiten Hauptsatzes.

Befindet sich ein System in einer Umgebung von bestimmter konstanter Temperatur $T_0$ und steht es mit dieser Umgebung in thermischer Wechselwirkung („Berührung"), so nimmt es ebenfalls erfahrungsgemäß die Temperatur $T_0$ an

(b) $$\mathfrak{W}' = C_1 . C_2 \dots C_l \, e^{\frac{1}{2\varkappa}\left(\sum_1^l S' - \frac{\sum_1^l E'}{T_0}\right)}.$$

Bei dem Vorgange hat sich weder der Zustand der Umgebung noch der Zustand der Maschine geändert, da letztere einen Kreisprozeß durchlief.

Nehmen wir nun an, daß nie unwahrscheinlichere Zustände auf wahrscheinlichere folgen, so ist:

[13] $$\mathfrak{W}' \geqq \mathfrak{W}.$$

Es ist aber auch nach dem Energieprinzip:

$$\sum_1^l E = \sum_1^l E'.$$

Berücksichtigt man dies, so folgt aus Gleichungen (a) und (b):

$$\sum S' \geqq \sum S.$$

It's a short argument—about interactions between a collection of heat reservoirs. But in a sense it already assumes its answer, and certainly doesn't provide any kind of fundamental "derivation of the Second Law". And this was the last time Einstein ever explicitly wrote about deriving the Second Law. Yes, in those days it was just too hard, even for Einstein.

There's another footnote to this story too. As we said, at the beginning of the twentieth century it had become clear that lots of things that had been thought to be continuous were in fact discrete. But there was an important exception: space. Ever since Euclid (~300 BC), space had almost universally been implicitly assumed to be continuous. And, yes, when quantum mechanics was being built, people did wonder about whether space might be discrete too (and even in 1916 Einstein expressed the opinion that eventually it would turn out to be). But over time the idea of continuous space (and time) got so entrenched in the fabric of physics that when I started seriously developing the ideas that became our Physics Project based on space as a discrete network (or what—in homage to the dynamical theory of heat one might call the "dynamical theory of space") it seemed to many people quite shocking. And looking back at the controversies of the late 1800s around atomism and its application to the Second Law it's charming how familiar many of the arguments against atomism seem. Of course it turns out they were wrong—as they seem again to be in the case of space.

## The Twentieth Century

The foundations of thermodynamics were a hot topic in physics in the latter half of the nineteenth century—worked on by many of the most prominent physicists of the time. But by the early twentieth century it'd been firmly eclipsed by other areas of physics. And going forward it'd receive precious little attention—with most physicists just assuming it'd "somehow been solved", or at least "didn't need to be worried about".

As a practical matter, thermodynamics in its basic equilibrium form nevertheless became very widely used in engineering and in chemistry. And in physics, there was steadily increasing interest in doing statistical mechanics—typically enumerating states of systems (quantum or otherwise), weighted as they would be in idealized thermal equilibrium. In mathematics, the field of ergodic theory developed, though for the most part it concerned itself with systems (such as ordinary differential equations) involving few variables—making it relevant to the Second Law essentially only by analogy.

There were a few attempts to "axiomatize" the Second Law, but mostly only at a macroscopic level, not asking about its microscopic origins. And there were also attempts to generalize the Second Law to make robust statements not just about equilibrium and the fact that it would be reached, but also about what would happen in systems driven to be in some manner away from equilibrium. The fluctuation-dissipation theorem about small perturbations from equilibrium—established in the mid-1900s, though anticipated in Einstein's work on Brownian motion—was one example of a widely applicable result. And there were also related ideas of "minimum entropy production"—as well as "maximum entropy production". But for large deviations from equilibrium there really weren't convincing general results, and in practice most investigations basically used phenomenological models that didn't have obvious connections to the foundations of thermodynamics, or derivations of the Second Law.

Meanwhile, through most of the twentieth century there were progressively more elaborate mathematical analyses of Boltzmann's equation (and the $H$ theorem) and their relation to rigorously derivable but hard-to-manage concepts like the BBGKY hierarchy. But despite occasional claims to the contrary, such approaches ultimately never seem to have been able to make much progress on the core problem of deriving the Second Law.

And then there's the story of entropy. And in a sense this had three separate threads. The first was the notion of entropy—essentially in the original form defined by Clausius—being used to talk quantitatively about heat in equilibrium situations, usually for either engineering or chemistry. The second—that we'll discuss a little more below—was entropy as a qualitative characterization of randomness and degradation. And the third was entropy as a general and formal way to measure the "effective number of degrees of freedom" in a system, computed from the log of the number of its achievable states.

There are definitely correspondences between these different threads. But they're in no sense "obviously equivalent". And much of the mystery—and confusion—that developed around entropy in the twentieth century came from conflating them.

Another piece of the story was information theory, which arose in the 1940s. And a core question in information theory is how long an "optimally compressed" message will be. And (with various assumptions) the average such length is given by a $\sum p \log p$ form that has essentially the same structure as Boltzmann's expression for entropy. But even though it's

"mathematically like entropy" this has nothing immediately to do with heat—or even physics; it's just an abstract consequence of needing log Ω bits (i.e. log Ω degrees of freedom) to specify one of Ω possibilities. (Still, the coincidence of definitions led to an "entropy branding" for various essentially information-theoretic methods, with claims sometimes being made that, for example, the thing called entropy must always be maximized "because we know that from physics".)

There'd been an initial thought in the 1940s that there'd be an "inevitable Second Law" for systems that "did computation". The argument was that logical gates (like And and Or) take 2 bits of input (with 4 overall states 11, 10, 01, 00) but give only 1 bit of output (1 or 0), and are therefore fundamentally irreversible. But in the 1970s it became clear that it's perfectly possible to do computation reversibly (say with 2-input, 2-output gates)—and indeed this is what's used in the typical formalism for quantum circuits.

As I've mentioned elsewhere, there were some computer experiments in the 1950s and beyond on model systems—like hard sphere gases and nonlinear springs—that showed some sign of Second Law behavior (though less than might have been expected). But the analysis of these systems very much concentrated on various regularities, and not on the effective randomness associated with Second Law behavior.

In another direction, the 1970s saw the application of thermodynamic ideas to black holes. At first, it was basically a pure analogy. But then quantum field theory calculations suggested that black holes should produce thermal radiation as if they had a certain effective temperature. By the late 1990s there were more direct ways to "compute entropy" for black holes, by enumerating possible (quantum) configurations consistent with the overall characteristics of the black hole. But such computations in effect assume (time-invariant) equilibrium, and so can't be expected to shed light directly on the Second Law.

Talking about black holes brings up gravity. And in the course of the twentieth century there were scattered efforts to understand the effect of gravity on the Second Law. Would a self-gravitating gas achieve "equilibrium" in the usual sense? Does gravity violate the Second Law? It's been difficult to get definitive answers. Many specific simulations of $n$-body gravitational systems were done, but without global conclusions for the Second Law. And there were cosmological arguments, particularly about the role of gravity in accounting for entropy in the early universe—but not so much about the actual evolution of the universe and the effect of the Second Law on it.

Yet another direction has involved quantum mechanics. The standard formalism of quantum mechanics—like classical mechanics—is fundamentally reversible. But the formalism for measurement introduced in the 1930s—arguably as something of a hack—is fundamentally irreversible, and there've been continuing arguments about whether this could perhaps "explain the Second Law". (I think our Physics Project finally provides more clarity about what's going on here—but also tells us this isn't what's "needed" for the Second Law.)

From the earliest days of the Second Law, there had always been scattered but ultimately unconvincing assertions of exceptions to the Second Law—usually based on elaborately constructed machines that were claimed to be able to achieve perpetual motion "just powered by heat". Of course, the Second Law is a claim about large numbers of molecules, etc.—and shouldn't be expected to apply to very small systems. But by the end of the twentieth century it was starting to be possible to make micromachines that could operate on small numbers of molecules (or electrons). And with the right control systems in place, it was argued that such machines could—at least in principle—effectively be used to set up Maxwell's demons that would systematically violate the Second Law, albeit on a very small scale.

And then there was the question of life. Early formulations of the Second Law had tended to talk about applying only to "inanimate matter"—because somehow living systems didn't seem to follow the same process of inexorable "dissipation to heat" as inanimate, mechanical systems. And indeed, quite to the contrary, they seemed able to take disordered input (like food) and generate ordered biological structures from it. And indeed, Erwin Schrödinger (1887–1961), in his 1944 book *What Is Life?* talked about "negative entropy" associated with life. But he—and many others since—argue that life doesn't really violate the Second Law because it's not operating in a closed environment where one should expect evolution to equilibrium. Instead, it's constantly being driven away from equilibrium, for example by "organized energy" ultimately coming from the Sun.

Still, the concept of at least locally "antithermodynamic" behavior is often considered to be a potential general signature of life. But already by the early part of the 1900s, with the rise of things like biochemistry, and the decline of concepts like "life force" (which seemed a little like "caloric"), there developed a strong belief that the Second Law must at some level always apply, even to living systems. But, yes, even though the Second Law seemed to say that one can't "unscramble an egg", there was still the witty rejoinder: "unless you feed it to a chicken".

What about biological evolution? Well, Boltzmann had been an enthusiast of Darwin's idea of natural selection. And—although it's not clear he made this connection—it was pointed out many times in the twentieth century that just as in the Second Law reversible underlying dynamics generate an irreversible overall effect, so also in Darwinian evolution effectively reversible individual changes aggregate to what at least Darwin thought was an "irreversible" progression to things like the formation of higher organisms.

The Second Law also found its way into the social sciences—sometimes under names like "entropy pessimism"—most often being used to justify the necessity of "Maxwell's–demon–like" active intervention or control to prevent the collapse of economic or social systems into random or incoherent states.

But despite all these applications of the Second Law, the twentieth century largely passed without significant advances in understanding the origin and foundations of the Second Law. Though even by the early 1980s I was beginning to find results—based on computational ideas—that seemed as if they might finally give a foundational understanding of what's really happening in the Second Law, and the extent to which the Second Law can in the end be "derived" from underlying "mechanical" rules.

# What the Textbooks Said: The Evolution of Certainty

Ask a typical physicist today about the Second Law and they're likely to be very sure that it's "just true". Maybe they'll consider it "another law of nature" like the conservation of energy, or maybe they'll think it is something that was "proved long ago" from basic principles of mathematics and mechanics. But as we've discussed here, there's really nowhere in the history of the Second Law that should give us this degree of certainty. So where did all the certainty come from? I think in the end it's a mixture of a kind of don't–question–this–it–comes–from–sophisticated–science mystique about the Second Law, together with a century and a half of "increasingly certain" textbooks. So let's talk about the textbooks.

While early contributions to what we now call thermodynamics (and particularly those from continental Europe) often got published as monographs, the first "actual textbooks" of thermodynamics already started to appear in the 1860s, with three examples (curiously, all in French) being:

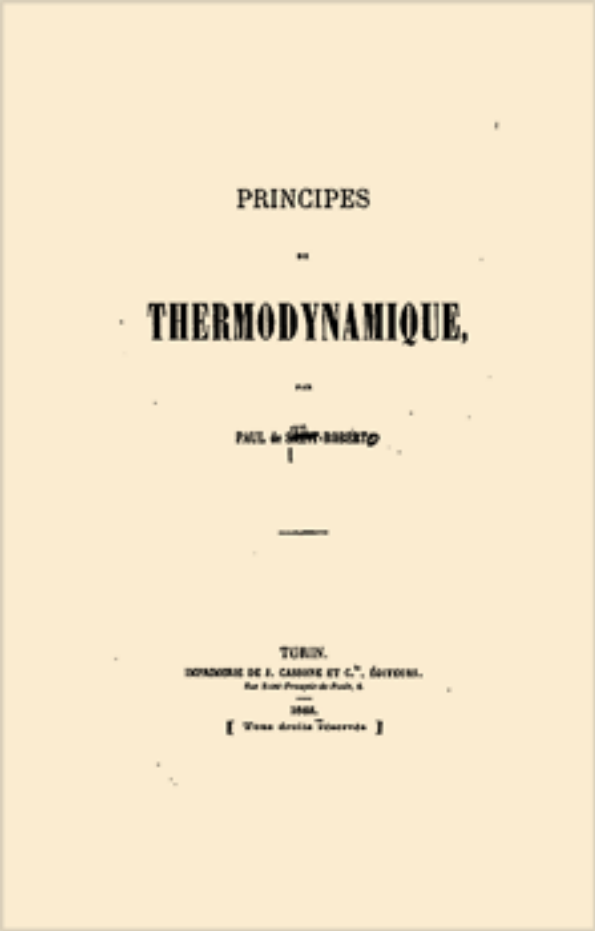
PRINCIPES

THERMODYNAMIQUE,

TURIN.

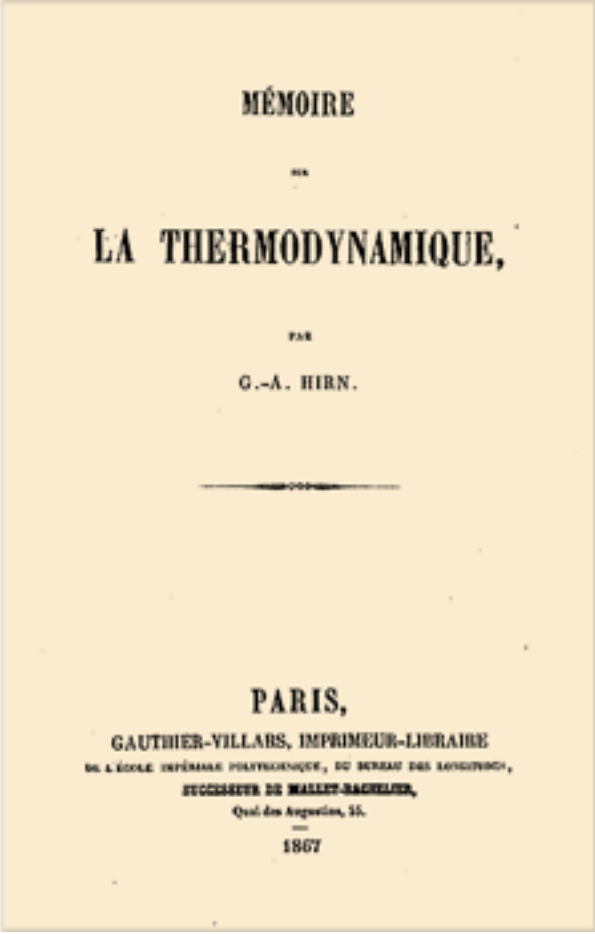
MÉMOIRE

LA THERMODYNAMIQUE,

G.-A. HIRN.

PARIS,

GAUTHIER-VILLARS, IMPRIMEUR-LIBRAIRE

1867

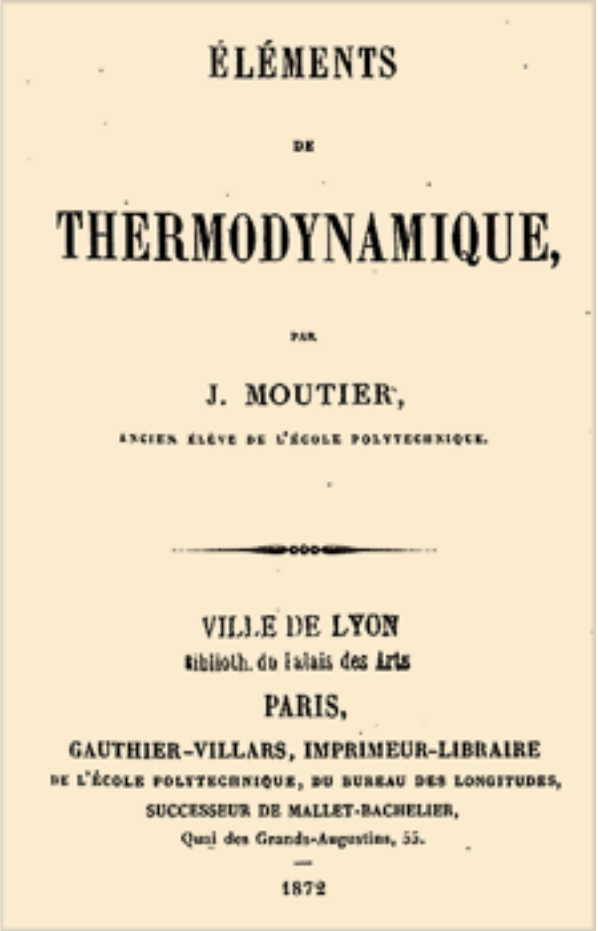
ÉLÉMENTS

DE

THERMODYNAMIQUE,

PAR

J. MOUTIER,

ANCIEN ÉLÈVE DE L'ÉCOLE POLYTECHNIQUE.

VILLE DE LYON

Biblioth. du Palais des Arts

PARIS,

GAUTHIER-VILLARS, IMPRIMEUR-LIBRAIRE

DE L'ÉCOLE POLYTECHNIQUE, DU BUREAU DES LONGITUDES,

SUCCESSEUR DE MALLET-BACHELIER,

Quai des Grands-Augustins, 55.

1872

And in these early textbooks what one repeatedly sees is that the Second Law is simply cited—without much comment—as a "principle" or "axiom" (variously attributed to Carnot, Kelvin or Clausius, and sometimes called "the Principle of Carnot"), from which theory will be developed. By the 1870s there's a bit of confusion starting to creep in, because people are talking about the "Theorem of Carnot". But, at least at first, by this they mean not the Second Law, but the result on the efficiency of heat engines that Carnot derived from this.

Occasionally, there are questions in textbooks about the validity of the Second Law. A notable one, that we discussed above when we talked about Maxwell's demon, shows up under the title "Limitation of the Second Law of Thermodynamics" at the end of Maxwell's 1871 *Theory of Heat*.

Tait's largely historical 1877 *Sketch of Thermodynamics* notes that, yes, the Second Law hasn't successfully been proved from the laws of mechanics:

> modynamics. A great many attempts have of late been made to show that this second law is merely a form of Hamilton's Principle of *Varying Action*—a pure principle of dynamics. It is obvious, from the fact that if we could at any moment exactly reverse the motion of every particle, we should make a dynamical system (however complex) go back through all its previous states of motion, that such a deduction from Hamilton's principle can only be made by a method of averages which virtually *assumes* the degradation of energy, a consequence of the law to be proved.

In 1879, Eddy's *Thermodynamics* at first shows even more skepticism

> Various attempts* have been made to derive the fundamental equation of thermodynamics resulting from the second law from the first law, by known mechanical principles.
>
> But Tait says,† respecting these attempts, that they virtually assume, in course of the demonstration, a consequence of the law to be proved, and hence are inconclusive.
>
> Clausius incidentally referred to this matter in his address at the 41st meeting of German Naturalists and Physicists at Frankfort in words to this effect:
>
> "Besides, there is a second principle, which is not yet contained in the first, but requires a special demonstration."

but soon he's talking about how "Rankine's theory of molecular vortices" has actually "proved the Second Law":

> Rankine however states‡ that "Carnot's principle (*i.e.*, the second principle) is not an independent principle in the theory of heat, but is deducible as a consequence from the equations of the mutual conversion of heat and expansive power" (*i.e.*, from the first principle). The demonstration of this which he has given rests upon his hypothesis of molecular vortices.

He goes on to give some standard "phenomenological" statements of the Second Law, but then talks about "molecular hypotheses from which Carnot's principle has been derived":

Besides these axioms, various molecular hypotheses have been advanced, from which Carnot's principle has been derived.

1°. Hypothesis of Molecular Vortices. (Rankine, 1849).

2°. Hypothesis of Circulating Streams of any figure whatever. (Rankine, 1851 and 1855) †.

3°. Hypothesis of Periodic Motions. (Boltzmann, 1866.‡)

4°. Hypothesis of Quasi-Periodic Mo-

Pretty soon there's confusion like the section in Alexandre Gouilly's (1842–1906) 1877 *Mechanical Theory of Heat* that's entitled "Second Fundamental Theorem of Thermodynamics or the Theorem of Carnot":

**Deuxième théorème fondamental de la thermodynamique ou théorème de Carnot.**

More textbooks on thermodynamics follow, but the majority tend to be practical expositions (that are often incredibly similar to each other) with no particular theoretical discussion of the Second Law, its origins or validity.

In 1891 there's an "official report about the Second Law" commissioned by the British Association for the Advancement of Science (and written by a certain George Bryan (1864–1928) who would later produce a thermodynamics textbook):

*Report of a Committee, consisting of* Messrs. J. LARMOR *and* G. H. BRYAN, *on the present state of our knowledge of Thermodynamics, specially with regard to the Second Law.*

[Ordered by the General Committee to be printed among the Reports.]

PART I.—RESEARCHES RELATING TO THE CONNECTION OF THE SECOND LAW WITH DYNAMICAL PRINCIPLES. DRAWN UP BY G. H. BRYAN.

*Introduction.*

1. The present report treats exclusively of the attempts that have been made to deduce the Second Law of Thermodynamics from purely mechanical principles.

There's an enumeration of approaches so far:

I. The Hypothesis of 'Stationary' or 'Quasi-Periodic' Motions as adopted by Clausius and Szily.

II. The Hypothesis of 'Monocyclic Systems' of von Helmholtz, and similar hypotheses.

III. The Statistical Hypothesis of Boltzmann, Clerk Maxwell, and other writers on the Kinetic Theory of Gases.

Somewhat confusingly it talks about a "proof of the Second Law"—actually referring to an already–in–equilibrium result:

46. *Applications to the Second Law.*—The simplest proof of the Second Law of Thermodynamics based on the hypothesis of the Boltzmann-Maxwell law of distribution of speed is that due to Mr. S. H. Burbury.[3] The proof is too well known to need description here. It leads to the same form for the entropy as Boltzmann's original investigation for the case of a system of point-atoms.[4] Although Watson and Burbury

There's talk of mechanical instability leading to irreversibility:

50. If we regard the whole matter as one of probabilities, the argument derived from reversing the system may be met without an appeal to the luminiferous æther. Although a conservative dynamical system is always reversible, the reversed motion may not unfrequently be dynamically unstable in the highest degree. One of the best illustrations in point is afforded by the impossibility of riding a bicycle backwards (*i.e.* with the steering wheel behind); here the forward motion is stable, but the reversed motion is highly unstable.

The conclusions say that, yes, the Second Law isn't proved "yet"

51. Although many of the researches mentioned in this report are not unfrequently called dynamical proofs of the Second Law, yet to prove the Second Law, about which we know something, by means of molecules, about which we know much less, would not be in consonance with the sentiments expressed at the end of the last paragraph. The most conclusive evidence for regarding Carnot's principle as a theorem in molecular dynamics lies in the remarkable agreement between the results obtained by the methods described in the three different sections of this report, all of which are based on different fundamental hypotheses. It

but imply that if only we knew more about molecules that might be enough to nail it:

In conclusion we may reasonably hope that future researches in the domain of molecular science will still further strengthen the bond of connection which we suppose to exist between the Second Law of Thermodynamics and Newton's Laws of Motion.

But back to textbooks. In 1895 Boltzmann published his *Lectures on Gas Theory*, which includes a final chapter about the *H* theorem and its relation to the Second Law. Boltzmann goes through his mathematical derivations for gases, then (rather over-optimistically) asserts that they'll also work for solids and liquids:

> We have looked mainly at processes in gases and have calculated the function H for this case. Yet the laws of probability that govern atomic motion in the solid and liquid states are clearly not qualitatively different ... from those for gases, so that the calculation of the function *H* corresponding to the entropy would not be more difficult in principle, although to be sure it would involve greater mathematical difficulties.

But soon he's discussing the more philosophical aspects of things (and by the time Boltzmann wrote this book, he was a professor of philosophy as well as physics). He says that the usual statement of the Second Law is "asserted phenomenologically as an axiom" (just as he says the infinite divisibility of matter also is at that time):

> ... the Second Law is formulated in such a way that the unconditional irreversibility of all natural processes is asserted as an axiom, just as general physics based on a purely phenomenological standpoint asserts the unconditional divisibility of matter without limit as an axiom.

One might then expect him to say that actually the Second Law is somehow provable from basic physical facts, such as the First Law. But actually his claims about any kind of "general derivation" of the Second Law are rather subdued:

> Since however the probability calculus has been verified in so many special cases, I see no reason why it should not also be applied to natural processes of a more general kind. The applicability of the probability calculus to the molecular motion in gases cannot of course be rigorously deduced from the differential equations for the motion of the molecules. It follows rather from the great number of the gas molecules and the length of their paths, by virtue of which the properties of the position in the gas where a molecule undergoes a collision are completely independent of the place where it collided the previous time.

But he still believes in the ultimate applicability of the Second Law, and feels he needs to explain why—in the face of the Second Law—the universe as we perceive it "still has interesting things going on":

> ... small isolated regions of the universe will always find themselves "initially" in an improbable state. This method seems to me to be the only way in which one can understand the Second Law—the heat death of each single world—without a unidirectional change of the entire universe from a definite initial state to a final state.

Meanwhile, he talks about the idea that elsewhere in the universe things might be different—and that, for example, entropy might be systematically decreasing, making (he suggests) perceived time run backwards:

> In the entire universe, the aggregate of all individual worlds, there will however in fact occur processes going in the opposite direction. But the beings who observe such processes will simply reckon time as proceeding from the less probable to the more probable states, and it will never be discovered whether they reckon time differently from us, since they are separated from us by eons of time and spatial distances $10^{10^{10}}$ times the distance of Sirius—and moreover their language has no relation to ours.

Most other textbook discussions of thermodynamics are tamer than this, but the rather anthropic-style argument that "we live in a fluctuation" comes up over and over again as an ultimate way to explain the fact that the universe as we perceive it isn't just a featureless maximum-entropy place.

It's worth noting that there are roughly three general streams of textbooks that end up discussing the Second Law. There are books about rather practical thermodynamics (of the type pioneered by Clausius), that typically spend most of their time on the equilibrium case. There are books about kinetic theory (effectively pioneered by Maxwell), that typically spend most of their time talking about the dynamics of gas molecules. And then there are books about statistical mechanics (as pioneered by Gibbs) that discuss with various degrees of mathematical sophistication the statistical characteristics of ensembles.

In each of these streams, many textbooks just treat the Second Law as a starting point that can be taken for granted, then go from there. But particularly when they are written by physicists with broader experience, or when they are intended for a not–totally–specialized audience, textbooks will quite often attempt at least a little justification or explanation for the Second Law—though rather often with a distinct sleight of hand involved.

For example, when Planck in 1903 wrote his *Treatise on Thermodynamics* he had a chapter in his discussion of the Second Law, misleadingly entitled "Proof". Still, he explains that:

> The second fundamental principle of thermodynamics [Second Law] being, like the first, an empirical law, we can speak of its proof only in so far as its total purport may be deduced from a single self–evident proposition. We, therefore, put forward the following proposition as being given directly by experience *It is impossible to construct an engine which will work in a complete cycle, and produce no effect except the raising of a weight and the cooling of a heat–reservoir*.

In other words, his "proof" of the Second Law is that nobody has ever managed to build a perpetual motion machine that violates it. (And, yes, this is more than a little reminiscent of P ≠ NP, which, through computational irreducibility, is related to the Second Law.) But after many pages, he says:

> In conclusion, we shall briefly discuss the question of the possible limitations to the Second Law. If there exist any such limitations—a view still held by many scientists and philosophers—then this [implies an error] in our starting point: the impossibility of perpetual motion ...

(In the 1905 edition of the book he adds a footnote that frankly seems bizarre in view of his—albeit perhaps initially unwilling—role in the initiation of quantum theory five years earlier: "The following discussion, of course, deals with the meaning of the Second Law only insofar as it can be surveyed from the points of view contained in this work avoiding all atomic hypotheses.")

He ends by basically saying "maybe one day the Second Law will be considered necessarily true; in the meantime let's assume it and see if anything goes wrong":

> Presumably the time will come when the principle of the increase of the entropy will be presented without any connection with experiment. Some metaphysicians may even put it forward as being *a priori* valid. In the meantime, no more effective weapon can be used by both champions and opponents of the Second Law than the indefatigable endeavour to follow the real purport of this law to the utmost consequences, taking the latter one by one to the highest court of appeal experience. Whatever the decision may be, lasting gain will accrue to us from such a proceeding, since thereby we serve the chief end of natural science—the enlargement of our stock of knowledge.

Planck's book came in a sense from the Clausius tradition. James Jeans's (1877–1946) 1904 book *The Dynamical Theory of Gases* came instead from the Maxwell + Boltzmann tradition.

He says at the beginning—reflecting the fact the existence of molecules had not yet been firmly established in 1904—that the whole notion of the molecular basis of heat "is only a hypothesis":

> The essential feature of the Kinetic Theory is that it interprets heat in matter as a manifestation of a motion of the molecules which compose the matter. It need hardly be said that this identification of heat and motion is only a hypothesis: it never has been, and from the nature of things never can be, proved. At the same time this hypothesis shews an ability to explain and even to predict natural phenomena, such that there can be little doubt that it rests upon a foundation of truth.

Later he argues that molecular-scale processes are just too "fine-grained" to ever be directly detected:

> The principal lesson to be learned from the foregoing figures is that the mechanism of the Kinetic Theory is extremely "fine-grained" when measured by ordinary standards. Molecules are, in fact, not infinitely small,

> human observation go, from those of a continuous medium. It is for this reason that the hypothesis upon which the Kinetic Theory rests is, and probably will always remain, an unproved hypothesis.

But soon Jeans is giving a derivation of Boltzmann's *H* theorem, though noting some subtleties:

> **67.** The assumption of molecular chaos (corrected, if necessary, in accordance with § 66) will therefore give correct results, provided it is interpreted with reference to the basis of probability supplied by our generalised space, and provided it is understood that it gives probable, and not certain, results. If we wish to obtain strictly accurate results the

His take on the "reversibility objection" is that, yes, the *H* function will be symmetric at every maximum, but, he argues, it'll also be discontinuous there:

> **70.** It may, perhaps, still be thought paradoxical that $dH/dt$ is not zero at each of these maxima. The explanation is that the variation of $H$ is not governed by the laws of the differential calculus, since this variation is not, strictly speaking, continuous. The value of $H$ is constant between collisions of the molecules, and changes abruptly at every collision. When the number

And in the time–honored tradition of saying "it is clear" right when an argument is questionable, he then claims that an "obvious averaging" will give irreversibility and the Second Law:

> It is therefore clear that, averaged over all systems which have a given $f$, $dH/dt$ will be negative except when $f$ is the law of distribution for the normal state, in which case it is zero. This result is now in agreement with that of Chapter II.

Later in his book Jeans simply quotes Maxwell and mentions his demon:

> of work. This is the second law of thermodynamics, and it is undoubtedly true so long as we can deal with bodies only in mass and have no power of perceiving or handling the separate molecules of which they are made up. But if we conceive a being whose faculties are so sharpened that he can

Then effectively just tells readers to go elsewhere:

> The reader who wishes to study the question of irreversibility further is referred to the following works:
>
> (i) "Report on the Present State of our knowledge of Thermodynamics, specially with regard to the Second Law," by J. Larmor and G. H. Bryan, *British Association Report*, 1891 (Cardiff), p. 85.
>
> (ii) *Elementary Principles of Statistical Mechanics*, J. Willard Gibbs (Scribners, New York), 1902.
>
> (iii) *Vorlesungen über Gastheorie*, Boltzmann.

In 1907 George Bryan (whose 1891 report we mentioned earlier) published *Thermodynamics, an Introductory Treatise Dealing Mainly with First Principles and Their Direct Applications*. But despite its title, Bryan has now "walked back" the hopes of his earlier report and is just treating the Second Law as an "axiom":

> We are thus led to assume the following axiom which may be regarded as the simplest form of the Second Law of Thermodynamics:
>
> *Energy in the form of mechanical work is always wholly convertible into any other forms of energy to which the present theory is applicable, but the converse processes are not in general possible.*

And—presumably from his interactions with Boltzmann—is saying that the Second Law is basically an empirical fact of our particular experience of the universe, and thus not something fundamentally derivable:

> The necessity of the appeal to experience is manifest from the following considerations: If in our Universe events occur in a certain definite sequence, it is possible to conceive a universe in which events occur in the opposite sequence, by merely reversing the scale of *time*. In such a universe the transformations of energy would be exactly the opposite to those of which we have experience, and the forms of energy which are least capable of being converted into other forms in our Universe would become the most convertible. In stating this it is assumed that the individuals living in either universe possess the power of influencing the progress only of future events and possess a knowledge only of past events. This assumption is implicitly involved in all our ideas relating to irreversibility.

As the years went by, many thermodynamics textbooks appeared, increasingly with an emphasis on applications, and decreasingly with a mention of foundational issues—typically treating the Second Law essentially just as an absolute empirical "law of nature" analogous to the First Law.

But in other books—including some that were widely read—there were occasional mentions of the foundations of the Second Law. A notable example was in Arthur Eddington's (1882–1944) 1929 *The Nature of the Physical World*—where now the Second Law is exalted as having the "supreme position among the laws of Nature":

> dence. The law that entropy always increases—the second law of thermodynamics—holds, I think, the supreme position among the laws of Nature. If someone points out to you that your pet theory of the universe is in disagreement with Maxwell's equations—then so much the worse for Maxwell's equations. If it is found to be contradicted by observation—well, these experimentalists do bungle things sometimes. But if your theory is found to be against the second law of thermodynamics I can give you no hope; there is nothing for it but to collapse in deepest humiliation. This exaltation

Although Eddington does admit that the Second Law is probably not "mathematically derivable":

> The question whether the second law of thermodynamics and other statistical laws are mathematical deductions from the primary laws, presenting their results in a conveniently usable form, is difficult to answer; but I think it is generally considered that there is an unbridgeable hiatus. At the bottom of all the questions settled by secondary law there is an elusive conception of "*a priori* probability of states of the world" which involves an essentially different attitude to knowledge from that presupposed in the construction of the scheme of primary law.

And even though in the twentieth century questions about thermodynamics and the Second Law weren't considered "top physics topics", some top physicists did end up talking about them, if nothing else in general textbooks they wrote. Thus, for example, in the 1930s and 1940s people like Enrico Fermi (1901–1954) and Wolfgang Pauli (1900–1958) wrote in some detail about the Second Law—though rather strenuously avoided discussing foundational issues about it.

Lev Landau (1908–1968), however, was a different story. In 1933 he wrote a paper "On the Second Law of Thermodynamics and the Universe" which basically argues that our everyday experience is only possible because "the world as a whole does not obey the laws of thermodynamics"—and suggests that perhaps relativistic quantum mechanics (which he says, quoting Niels Bohr (1885–1962), could be crucial in the center of stars) might fundamentally violate the Second Law. (And yes, even today it's not clear how "relativistic temperature" works.)

But this kind of outright denial of the Second Law had disappeared by the time Lev Landau and Evgeny Lifshitz (1915–1985) wrote the 1951 version of their book *Statistical Mechanics*—though they still showed skepticism about its origins:

> There is no doubt that the foregoing simple formulations [of the Second Law] accord with reality; they are confirmed by all our everyday observations. But when we consider more closely the problem of the physical nature and origin of these laws of behaviour, substantial difficulties arise, which to some extent have not yet been overcome.

Their book continues, discussing Boltzmann's fluctuation argument:

> Firstly, if we attempt to apply statistical physics to the entire universe ... we immediately encounter a glaring contradiction between theory and experiment. According to the results of statistics, the universe ought to be in a state of complete statistical equilibrium. ... Everyday experience shows us, however, that the properties of Nature bear no resemblance to those of an equilibrium system; and astronomical results show that the same is true throughout the vast region of the Universe accessible to our observation.

> We might try to overcome this contradiction by supposing that the part of the Universe which we observe is just some huge fluctuation in a system which is in equilibrium as a whole. The fact that we have been able to observe this huge fluctuation might be explained by supposing that the existence of such a fluctuation is a necessary condition for the existence of an observer (a condition for the occurrence of biological evolution). This argument, however, is easily disproved, since a fluctuation within, say, the volume of the solar system only would be very much more probable, and would be sufficient to allow the existence of an observer.

What do they think is the way out? The effect of gravity:

> ... in the general theory of relativity, the Universe as a whole must be regarded not as a closed system but as a system in a variable gravitational field. Consequently the application of the law of increase of entropy does not prove that statistical equilibrium must necessarily exist.

But they say this isn't the end of the problem, essentially noting the reversibility objection. How should this be overcome? First, they suggest the solution might be that the observer somehow "artificially closes off the history of a system", but then they add:

> Such a dependence of the laws of physics on the nature of an observer is quite inadmissible, of course.

They continue:

> At the present time it is not certain whether the law of increase of entropy thus formulated can be derived on the basis of classical mechanics. ... It is more reasonable to suppose that the law of increase of entropy in the above general formulation arises from quantum effects.

They talk about the interaction of classical and quantum systems, and what amounts to the explicit irreversibility of the traditional formalism of quantum measurement, then say that if quantum mechanics is in fact the ultimate source of irreversibility:

> ... there must exist an inequality involving the quantum constant $\hbar$ which ensures the validity of the law and is satisfied in the real world...

What about other textbooks? Joseph Mayer (1904–1983) and Maria Goeppert Mayer's (1906–1972) 1940 *Statistical Mechanics* has the rather charming

**4e. The Limits of Validity of the Second Law of Thermodynamics**

"What, never?" "No, never!"
"What, never?" "Well, hardly ever!"
*H. M. S. Pinafore.*

The second law of thermodynamics can be stated in the form: the entropy of an isolated system never decreases. We have now claimed that this fundamental law is a consequence of the theorems of mechanics. It is appropriate at this time to investigate the extent to which exceptions to this law might conceivably be observed.

though in the end they sidestep difficult questions about the Second Law by basically making convenient definitions of what $S$ and $\Omega$ mean in $S = k \log \Omega$.

For a long time one of the most cited textbooks in the area was Richard Tolman's (1881–1948) 1938 *Principles of Statistical Mechanics*. Tolman (basically following Gibbs) begins by explaining that statistical mechanics is about making predictions when all you know are probabilistic statements about initial conditions:

> theoretically possible. The principles of ordinary mechanics may be regarded as allowing us to make precise predictions as to the future state of a mechanical system from a precise knowledge of its initial state.† On the other hand, the principles of statistical mechanics are to be regarded as permitting us to make reasonable predictions as to the future condition of a system, which may be expected to hold on the average, starting from an incomplete knowledge of its initial state.

Tolman continues:

> Since our actual contacts with the physical world are such that we never do have the maximal knowledge of systems regarded as theoretically allowable, the idea of the precise state of a system is in any case an abstract limiting concept. Hence the methods of ordinary mechanics really apply to somewhat highly idealized situations, and the methods of statistical mechanics provide a significant supplement in the direction of decreased abstraction and closer correspondence between theoretical methods and actual experience. Even in the case of simple systems of only a few degrees of freedom, where our lack of maximal knowledge is not due to difficulties arising from the complexity of the system, the methods of statistical mechanics may be applied to a system whose initial state is not completely specified.‡

He notes that, historically, statistical mechanics was developed for studying systems like gases, where (in a vague foreshadowing of the concept of computational irreducibility) "it is evident that we should be quickly lost in the complexities of our computations" if we try to trace every molecule, but where, he claims, statistical mechanics can still accurately tell us "statistically" what will happen:

> mechanics. For example, in the case of a gas, consisting, say, of a large number of simple classical particles, even if we were given at some initial time the positions and velocities of all the particles so that we could foresee the collisions that were about to take place, it is evident that we should be quickly lost in the complexities of our computations if we tried to follow the results of such collisions through any extended length of time. Nevertheless, a system such as a gas composed of many molecules is actually found to exhibit perfectly definite regularities in its behaviour, which we feel must be ultimately traceable to the laws of mechanics even though the detailed application of these laws defies our powers. For the treatment of such regularities in the behaviour of complicated systems of many degrees of freedom, the methods of statistical mechanics are adequate and especially appropriate.

But where exactly should we get the probability distributions for initial states from? Tolman says he's going to consider the kinds of mathematically defined ensembles that Gibbs discusses. And tucked away at the end of a chapter he admits that, well, yes, this setup is really all just a postulate—set up so as to make the results of statistical mechanics "merely a matter for computation":

> From the point of view adopted in the present book, the hypothesis, of equal *a priori* probabilities for equal regions in the phase space, must in any case be regarded as an essential element of statistical mechanics, which has to be introduced by postulation and which is then sufficient to provide the statistical methods used. *Without* this postulate there would be nothing to correspond to the circumstance that nature does not have any tendency to present us with systems in conditions which we regard as mechanically entirely possible but statistically improbable; and *with* the postulate the use of statistical mechanics for the determination of averages and fluctuations then becomes merely a matter for computation.

On this basis Tolman then derives Boltzmann's $H$ theorem, and his $\bar{\bar{H}}$ "coarse-grained" generalization (where, yes, the coarse-graining ultimately operates according to his postulate). For 530 pages, there's not a single mention of the Second Law. But finally, on page 558 Tolman is at least prepared to talk about an "analog of the Second Law":

> **130. The analogue of the second law of thermodynamics**
>
> In preceding sections we have discussed the statistical mechanical analogues for a considerable number of thermodynamic processes that involve the second law of thermodynamics. These processes included

And basically what Tolman argues is that his $\bar{\bar{H}}$ can reasonably be identified with thermodynamic entropy $S$. In the end, the argument is very similar to Boltzmann's, though Tolman seems to feel that it has achieved more:

> In concluding this chapter, it is hoped that due appreciation will be felt for the importance and significance of the great achievement of statistical mechanics in providing a fundamental, mechanical interpretation and explanation of the principles of thermodynamics.

Very different in character from Tolman's book, another widely cited book is Percy Bridgman's (1882–1961) largely philosophical 1943 *The Nature of Thermodynamics*. His chapter on the Second Law begins:

### THE SECOND LAW OF THERMODYNAMICS

THERE have been nearly as many formulations of the second law as there have been discussions of it. Although many of these formulations are doubtless roughly equivalent, and the proof that they are equivalent has been considered to be one of the tasks of a thermodynamic analysis, I question whether any really rigorous examination has been attempted from the postulational point of view and I question whether such an examination would be of great physical interest. It does seem obvious, however, that not all these formulations can be exactly equivalent, but it is possible to distinguish stronger and weaker forms.

A decade earlier Bridgman had discussed outright violations of the Second Law, saying that he'd found that the younger generation of physicists at the time seemed to often think that "it may be possible some day to construct a machine which shall violate the Second Law on a scale large enough to be commercially profitable"—perhaps, he said, by harnessing Brownian motion:

### STATISTICAL MECHANICS AND THE SECOND LAW OF THERMODYNAMICS[1]

By Dr. P. W. BRIDGMAN

HARVARD UNIVERSITY

ONE thing that has much impressed me in recent conversations with physicists, particularly those of the younger generation, is the frequency of the conviction that it may be possible some day to construct a machine which shall violate the second law of thermodynamics on a scale large enough to be commercially profitable. This constitutes a striking reversal of the attitude of the founders of thermodynamics, Kelvin and Clausius, who postulated the impossibility of perpetual motion of the second kind as a generalization from the uniformly unsuccessful attempts of the entire human race to realize it. Paradoxically, one very important factor in bringing about this change in attitude is the feeling of better understanding of the second law which the present generation enjoys, and which is largely due to the universal acceptance of the explanation of the second law in statistical terms, for which Gibbs was in so large a degree responsible. Statistical mechanics reduces the second law from a law of ostensibly absolute validity to a statement about high probabilities, leaving open the possibility that once in a great while there may be important violations. Doubtless another most important factor in present scepticism as to the ultimate commercial validity of the second law is the discovery of the importance in many physical phenomena of those fluctuation effects which are demanded by sta-

[1] The Ninth Josiah Willard Gibbs Lecture, delivered at New Orleans, December 29, 1931, under the auspices of the American Mathematical Society, at a joint meeting of the society with the American Physical Society, and Section A of the American Association for the Advancement of Science.

At a philosophical level, a notable treatment of the Second Law appeared in Hans Reichenbach's (1891–1953) (unfinished–at–his–death) 1953 work *The Direction of Time*. Wanting to make use of the Second Law, but concerned about the reversibility objections, Reichenbach introduces the notion of "branch systems"—essentially parts of the universe that can eventually be considered isolated, but which were once connected to other parts that were responsible for determining their ("nonrandom") effective initial conditions:

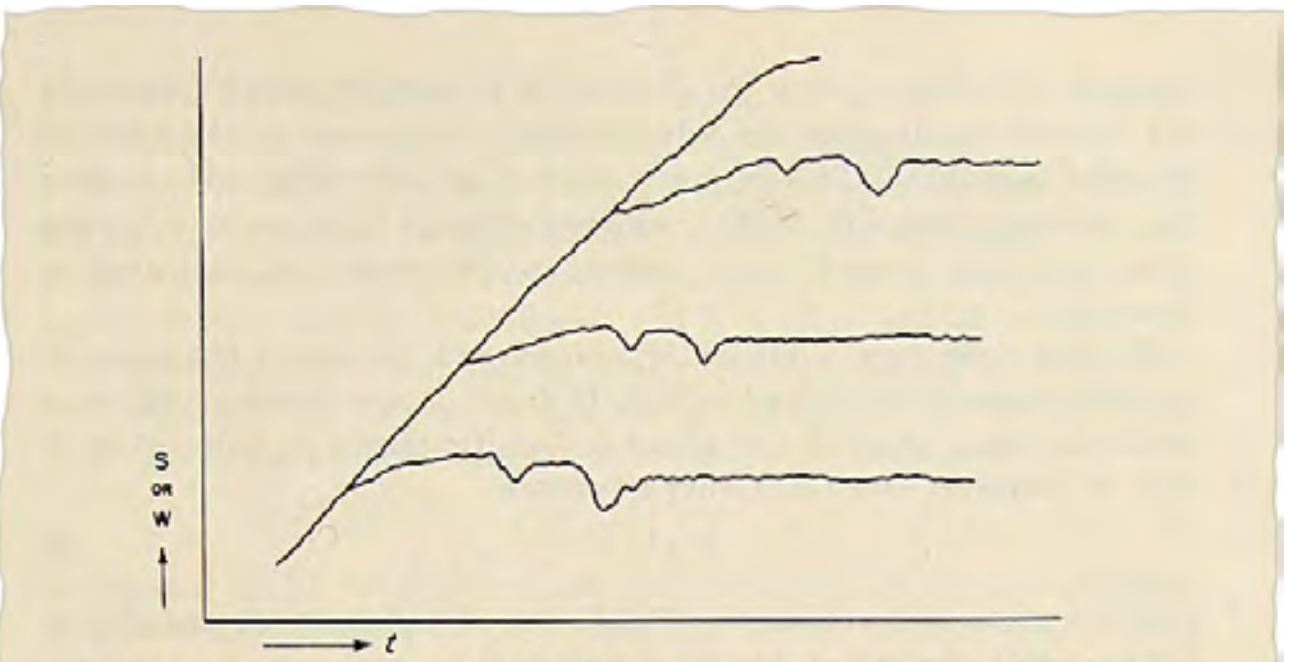

Fig. 20. The initial upgrade of the entropy curve of the universe. Some isolated systems branch off, remaining isolated for an infinite time. (Simplified schematization.)

Most textbooks that cover the Second Law use one of the formulations that we've already discussed. But there is one more formulation that also sometimes appears, usually associated with the name "Carathéodory" or the term "axiomatic thermodynamics".

Back in the first decade of the twentieth century—particularly in the circle around David Hilbert (1862–1943)—there was a lot of enthusiasm for axiomatizing things, including physics. And in 1908 the mathematician Constantin Carathéodory (1873–1950) suggested an axiomatization of thermodynamics. His essential idea—that he developed further in the 1920s—was to consider something like Gibbs's phase fluid and then roughly to assert that it gets (in some measure-theoretic sense) "so mixed up" that there aren't "experimentally doable" transformations that can unmix it. Or, in his original formulation:

> In any arbitrary neighborhood of an arbitrarily given initial point there is a state that cannot be arbitrarily approximated by adiabatic changes of state.

There wasn't much pickup of this approach—though Max Born (1882–1970) supported it, Max Planck dismissed it, and in 1939 S. Chandrasekhar (1910–1995) based his exposition of stellar structure on it. But in various forms, the approach did make it into a few textbooks. An example is Brian Pippard's (1920–2008) otherwise rather practical 1957 *The Elements of Classical Thermodynamics*:

> A third formulation, due to Carathéodory, is not so clearly related:
>
> *In the neighbourhood of any equilibrium state of a system there are states which are inaccessible by an adiathermal process.*
>
> Each formulation has its enthusiastic supporters. For the engineer or the practically minded physicist Clausius's or Kelvin's formulations are more directly meaningful, and, moreover, the derivation from them of the important consequences of the law may be made without a great deal of mathematics. On the other hand, Carathéodory's formulation is undoubtedly more economical, in that it demands the impossibility of a rather simpler type of process than that considered
>
> accomplished adiathermally. Carathéodory's law appears to differ in outlook from the others. The average physicist is prepared to take Clausius's and Kelvin's laws as reasonable generalizations of common experience, but Carathéodory's law (at any rate in the author's opinion) is not immediately acceptable except in the trivial cases, of which Joule's experiment is one; it is neither intuitively obvious nor supported by a mass of experimental evidence. It may be argued therefore that the further development of thermodynamics should not be made to rest on this basis, but that Carathéodory's law should be regarded, in view of the fact that it leads to the same conclusions as the others, as a statement of the minimal postulate which is needed in order to achieve the desired end. It bears somewhat the same relation to the other statements as Hamilton's principle bears to Newton's laws of motion.

Yet another (loosely related) approach is the "postulatory formulation" on which Herbert Callen's (1919–1993) 1959 textbook *Thermodynamics* is based:

> The postulatory formulation of thermodynamics features states, rather than processes, as fundamental constructs. Statements about Carnot cycles and about the impossibility of perpetual motion of various kinds do not appear in the postulates, but state functions, energy, and entropy become the fundamental concepts. An enormous simplification in the mathematics is obtained, for processes then enter simply as differentials of the state functions. The conventional method proceeds inversely from processes to state functions by the relatively difficult procedure of integration of partial differential equations.

In effect this is now "assuming the result" of the Second Law:

> **Postulate II.** *There exists a function (called the entropy S) of the extensive parameters of any composite system, defined for all equilibrium states and having the following property. The values assumed by the extensive parameters in the absence of an internal constraint are those that maximize the entropy over the manifold of constrained equilibrium states.*

Though in an appendix he rather tautologically states:

> *The entropy of the system is defined as*
>
> $$S(X) = k \ln N(X) \qquad \text{(B.1)}$$
>
> *in which k is Boltzman's constant. Consequently, the macrostate (X)*
>
> *observed is that corresponding to the maximum value of the entropy, consistent with the imposed constraints. The entropy of any macrostate is proportional to the logarithm of the number of microstates associated with that macrostate.*
>
> The foregoing statements, although made in reference to a highly artificial model of a system, are actually general. They are valid for any closed thermodynamic system.

So what about other textbooks? A famous set are Richard Feynman's (1918–1988) 1963 *Lectures on Physics*. Feynman starts his discussion of the Second Law quite carefully, describing it as a "hypothesis":

> hot body. Now, the hypothesis of Carnot, the second law of thermodynamics, is sometimes stated as follows: heat cannot, of itself, flow from a cold to a hot object.

Feynman says he's not going to go very far into thermodynamics, though quotes (and criticizes) Clausius's statements:

> The two laws of thermodynamics are often stated this way:
>
> *First law:* the energy of the universe is always constant.
>
> *Second law:* the entropy of the universe is always increasing.
>
> That is not a very good statement of the second law; it does not say, for example, that in a reversible cycle the entropy stays the same, and it does not say exactly what the entropy is. It is just a clever way of remembering the two laws, but it

But then he launches into a whole chapter on "Ratchet and pawl":

**46**

***Ratchet and pawl***

**46–1 How a ratchet works**

In this chapter we discuss the ratchet and pawl, a very simple device which allows a shaft to turn only one way. The possibility of having something turn only one way requires some detailed and careful analysis, and there are some very interesting consequences.

The plan of the discussion came about in attempting to devise an elementary explanation, from the molecular or kinetic point of view, for the fact that there is a maximum amount of work which can be extracted from a heat engine. Of course we have seen the essence of Carnot's argument, but it would be nice to find an explanation which is elementary in the sense that we can see what is happening

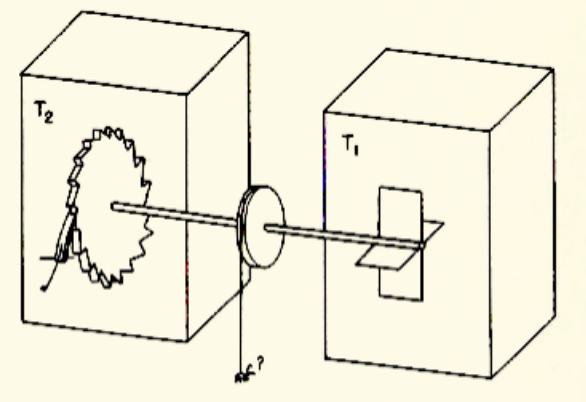

Fig. 46–1. The ratchet and pawl machine.

His goal, he explains, is to analyze a device (similar to what Marian Smoluchowski had considered in 1912) that one might think by its one–way ratchet action would be able to "harvest random heat" and violate the Second Law. But after a few pages of analysis he claims that, no, if the system is in equilibrium, thermal fluctuations will prevent systematic "one–way" mechanical work from being achieved, so that the Second Law is saved.

But now he applies this to Maxwell's demon, claiming that the same basic argument shows that the demon can't work:

It turns out, if we build a finite-sized demon, that the demon himself gets so warm that he cannot see very well after a while. The simplest possible demon, as an example, would be a trap door held over the hole by a spring. A fast molecule comes through, because it is able to lift the trap door. The slow molecule cannot get through, and bounces back. But this thing is nothing but our ratchet and pawl in another form, and ultimately the mechanism will heat up. If we assume that the specific heat of the demon is not infinite, it must heat up. It has but a finite number of internal gears and wheels, so it cannot get rid of the extra heat that it gets from observing the molecules. Soon it is shaking from Brownian motion so much that it cannot tell whether it is coming or going, much less whether the molecules are coming or going, so it does not work.

But what about reversibility? Feynman first discusses what amounts to Boltzmann's fluctuation idea:

Thus one possible explanation of the high degree of order in the present-day world is that it is just a question of luck. Perhaps our universe happened to have had a fluctuation of some kind in the past, in which things got somewhat separated, and now they are running back together again. This kind of theory is not un-

But then he opts instead for the argument that for some reason—then unknown—the universe started in a "low–entropy" state, and has been "running down" ever since:

> This is not to say that we understand the logic of it. For some reason, the universe at one time had a very low entropy for its energy content, and since then the entropy has increased. So that is the way toward the future. That is the origin of all irreversibility, that is what makes the processes of growth and decay, that makes us remember the past and not the future, remember the things which are closer to that moment in the history of the universe when the order was higher than now, and why we are not able to remember things where the disorder is higher than now, which we call the future. So, as we commented in an earlier chapter, the

By the beginning of the 1960s an immense number of books had appeared that discussed the Second Law. Some were based on macroscopic thermodynamics, some on kinetic theory and some on statistical mechanics. In all three of these cases there was elegant mathematical theory to be described, even if it never really addressed the ultimate origin of the Second Law.

But by the early 1960s there was something new on the scene: computer simulation. And in 1965 that formed the core of Fred Reif's (1927–2019) textbook *Statistical Physics*:

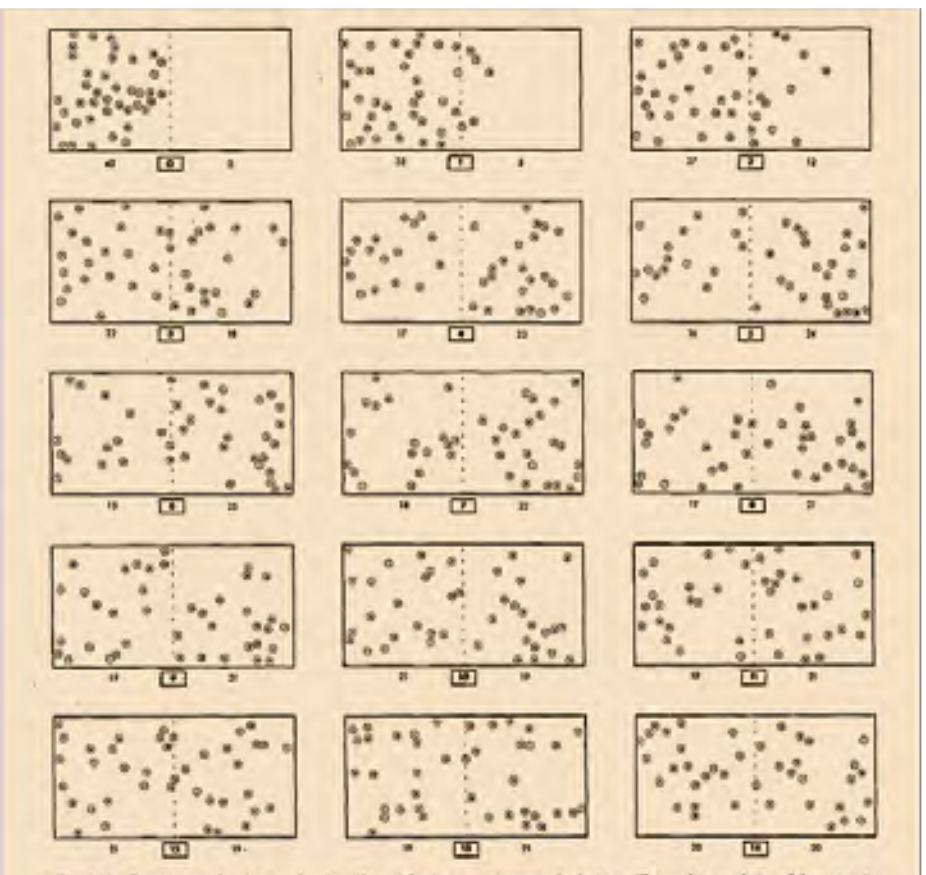

*Fig. 1.15* Computer-made pictures showing 40 particles in a box. The pictures were constructed by starting with the special situation where all the particles are in the left half of the box in the positions shown in the frame $j = 0$ and are given some arbitrary assumed velocities. The resulting evolution of the system in time is then shown by the sequence of frames $j = 0, 1, 2, \ldots, 14$. The number of particles located in each half of the box is printed directly beneath that half. No velocities are indicated.

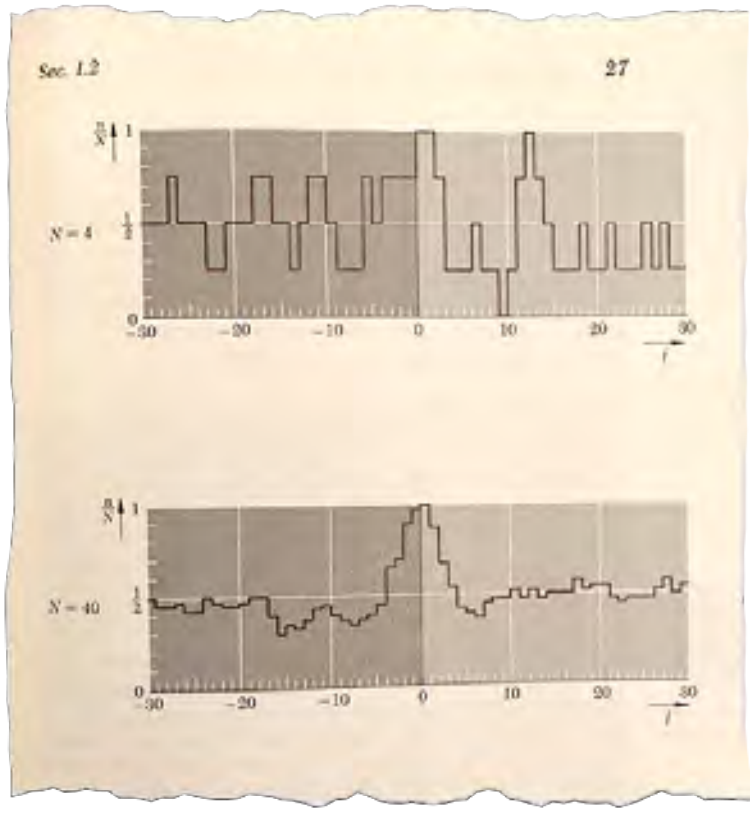

In a sense the book is an exploration of what simulated hard sphere gases do—as analyzed using ideas from statistical mechanics. (The simulations had computational limitations, but they could go far enough to meaningfully see most of the basic phenomena of statistical mechanics.)

Even the front and back covers of the book provide a bold statement of both reversibility and the kind of randomization that's at the heart of the Second Law:

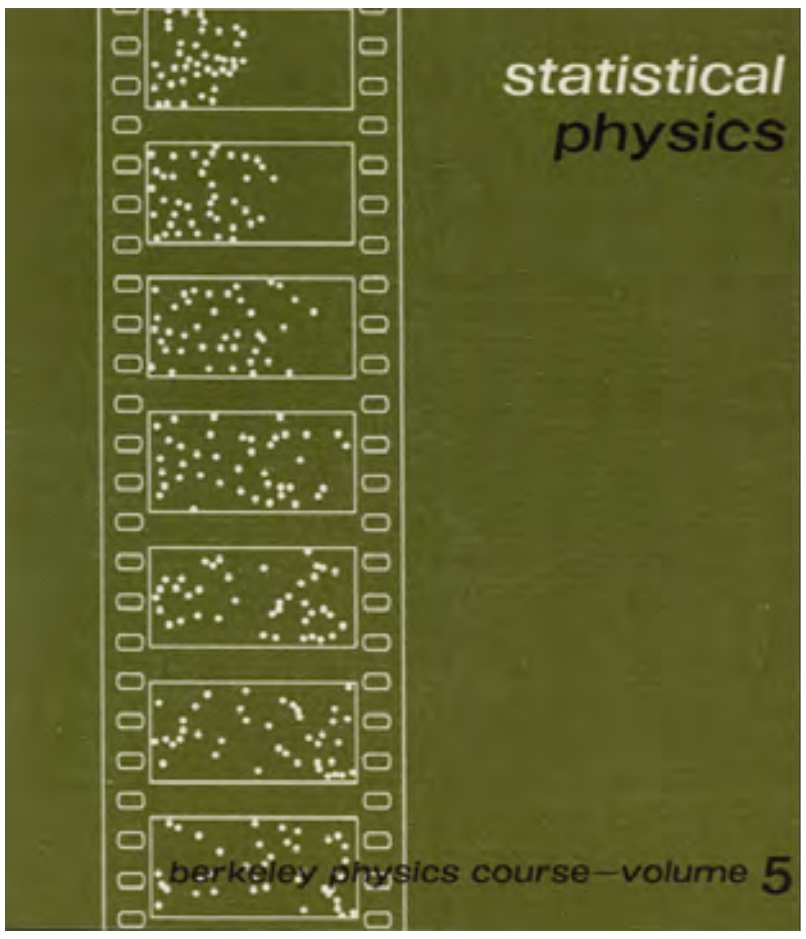

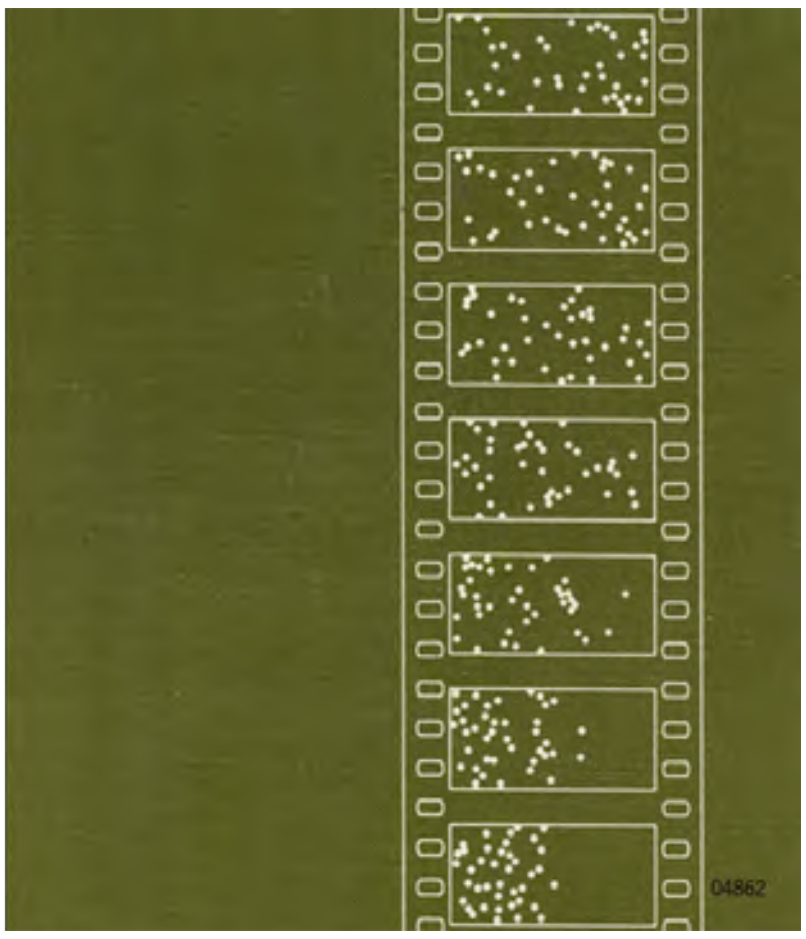

But inside the book the formal concept of entropy doesn't appear until page 147—where it's defined very concretely in terms of states one can explicitly enumerate:

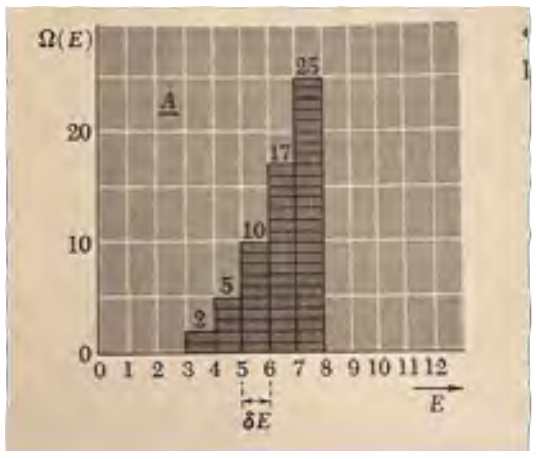

where we have introduced the quantity $S$ defined by

$$\boxed{S \equiv k \ln \Omega.} \qquad (14)$$

This quantity $S$ is called the *entropy* of the system under consideration. It has the dimensions of energy because its definition involves the constant $k$. According to its definition (14), the entropy of a system is merely a logarithmic measure of the number of states accessible to the system. In accordance with the comments at the end of Sec. 3.6, the entropy thus provides a quantitative measure of the degree of randomness of the system.†

And finally, on page 283—after all necessary definitions have been built up—there's a rather prosaic statement of the Second Law, almost as a technical footnote:

***Statement 2***

We have seen that the number of states accessible to a system (or, equivalently, its entropy) is a quantity of fundamental importance in describing the macrostate of a system. In Sec. 7.2 we showed that changes in the entropy of a system can be related by (32) to the heat absorbed by this system. In Sec. 3.6 we also showed that an *isolated* system tends to approach a situation of greater probability where the number of states accessible to it (or equivalently, its entropy) is larger than initially. (As a special case, when the system is initially already in its most probable situation, it remains in equilibrium and its entropy remains unchanged.) Thus we arrive at the following statement:

*Second law of thermodynamics*

An equilibrium macrostate of a system can be characterized by a quantity $S$ (called *entropy*) which has the following properties:

(i) In any infinitesimal quasi-static process in which the system absorbs heat $đQ$, its entropy changes by an amount

$$dS = \frac{đQ}{T} \qquad (61)$$

where $T$ is a parameter characteristic of the macrostate of the system and is called its *absolute temperature*.

(ii) In any process in which a thermally isolated system changes from one macrostate to another, its entropy tends to increase, i.e.,

$$\Delta S \geq 0. \qquad (62)$$

Looking though many textbooks of thermodynamics and statistical mechanics it's striking how singular Reif's "show–don't–tell" computer–simulation approach is. And, as I've described in detail elsewhere, for me personally it has a particular significance, because this is the book that in 1972, at the age of 12, launched me on what has now been a 50–year journey to understand the Second Law and its origins.

When the first textbooks that described the Second Law were published nearly a century and a half ago they often (though even then not always) expressed uncertainty about the Second Law and just how it was supposed to work. But it wasn't long before the vast majority of books either just "assumed the Second Law" and got on with whatever they wanted to apply it to, or tried to suggest that the Second Law had been established from underlying principles, but that it was a sophisticated story that was "out of the scope of this book" but to be found elsewhere. And so it was that a strong sense emerged that the Second Law was something whose ultimate character and origins the typical working scientist didn't need to question—and should just believe (and protect) as part of the standard canon of science.

## So Where Does This Leave the Second Law?

The Second Law is now more than 150 years old. But—at least until now—I think it's fair to say that the fundamental ideas used to discuss it haven't materially changed in more than a century. There's a lot that's been written about the Second Law. But it's always tended to follow lines of development already defined over a century ago—and mostly those from Clausius, or Boltzmann, or Gibbs.

Looking at word clouds of titles of the thousands of publications about the Second Law over the decades we see just a few trends, like the appearance of the "generalized Second Law" in the 1990s relating to black holes:

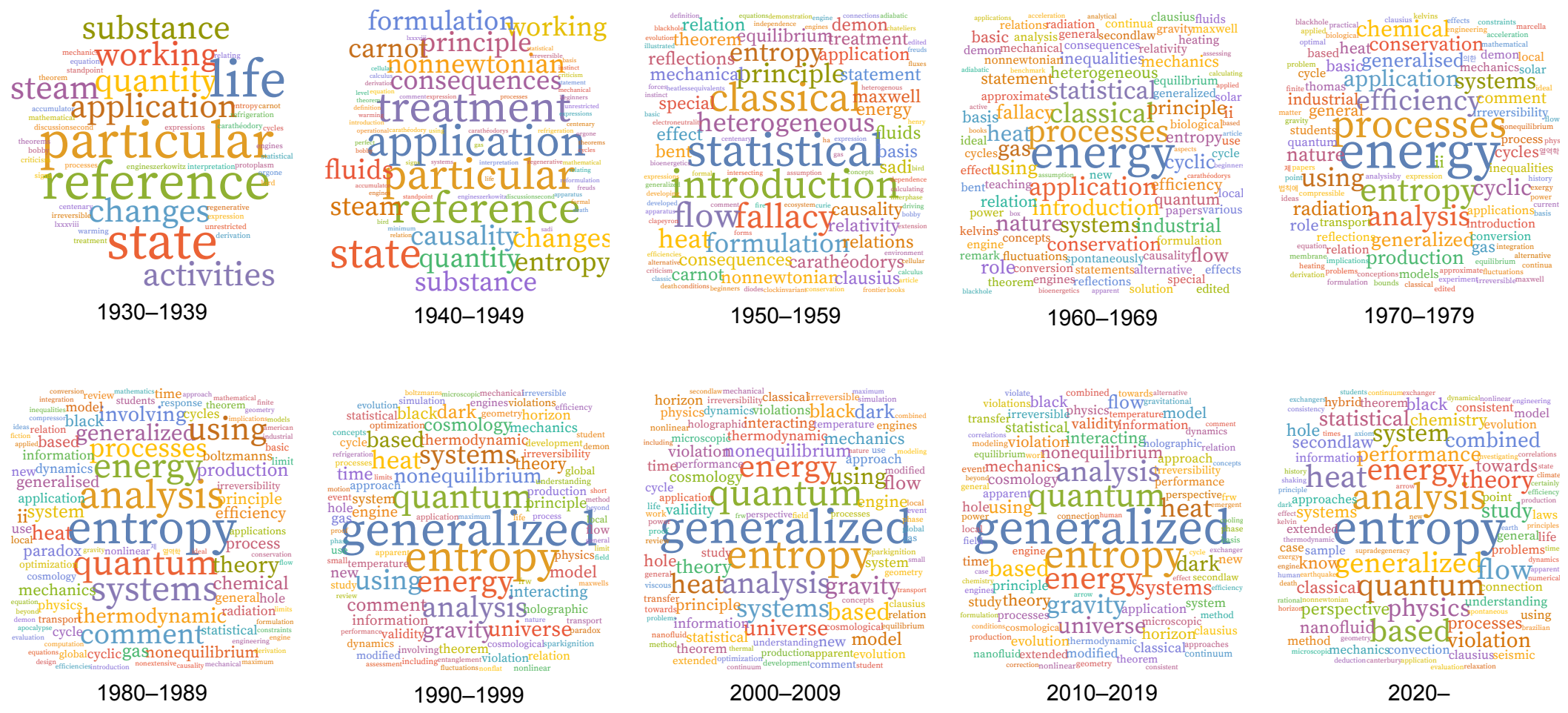

But with all this activity why hasn't more been worked out about the Second Law? How come after all this time we still don't really even understand with clarity the correspondence between the Clausius, Boltzmann and Gibbs approaches—or how their respective definitions of "entropy" are ultimately related?

In the end, I think the answer is that it needs a new paradigm—that, yes, is fundamentally based on computation and on ideas like computational irreducibility. A little more than a century ago—with people still actively arguing about what Boltzmann was saying—I don't think anyone would have been too surprised to find out that to make progress would need a new way of looking at things. (After all, just a few years earlier Boltzmann and Gibbs had needed to bring in the new idea of using probability theory.)

But as we discussed, by the beginning of the twentieth century—with other areas of physics heating up—interest in the Second Law was waning. And even with many questions unresolved people moved on. And soon several academic generations had passed. And as is typical in the history of science, by that point nobody was questioning the foundations anymore. In the particular case of the Second Law there was some sense that the uncertainties had to do with the assumption of the existence of molecules, which had by then been established. But more important, I think, was just the passage of "academic time" and the fact that what might once have been a matter of discussion had now just become a statement in the textbooks—that future academic generations should learn and didn't need to question.

One of the unusual features of the Second Law is that at the time it passed into the "standard canon of science" it was still rife with controversy. How did those different approaches relate? What about those "mathematical objections"? What about the thought experiments that seemed to suggest exceptions? It wasn't that these issues were resolved. It was just that after enough time had passed people came to assume that "somehow that must have all been worked out ages ago".

And it wasn't that there was really any pressure to investigate foundational issues. The Second Law—particularly in its implications for thermal equilibrium—seemed to work just fine in all its standard applications. And it even seemed to work in new domains like black holes. Yes, there was always a desire to extend it. But the difficulties encountered in trying to do so didn't seem in any obvious way related to issues about its foundations.

Of course, there were always a few people who kept wondering about the Second Law. And indeed I've been surprised at how much of a *Who's Who* of twentieth-century physics this seems to have included. But while many well-known physicists seem to have privately thought about the foundations of the Second Law they managed to make remarkably little progress—and as a result left very few visible records of their efforts.

But—as is so often the case—the issue, I believe, is that a fundamentally new paradigm was needed in order to make real progress. When the "standard canon" of the Second Law was formed in the latter part of the nineteenth century, calculus was the primary tool for physics—with probability theory a newfangled addition introduced specifically for studying the Second Law. And from that time it would be many decades before even the beginnings of the computational paradigm began to emerge, and nearly a century before phenomena like computational irreducibility were finally discovered. Had the sequence been different I have no doubt that what I have now been able to understand about the Second Law would have been worked out by the likes of Boltzmann, Maxwell and Kelvin.

But as it is, we've had to wait more than a century to get to this point. And having now studied the history of the Second Law—and seen the tangled manner in which it developed—I believe that we can now be confident that we have indeed successfully been able to resolve many of the core issues and mysteries that have plagued the Second Law and its foundations over the course of nearly 150 years.

## Note

Almost all of what I say here is based on my reading of primary literature, assisted by modern tools and by my latest understanding of the Second Law. About some of what I discuss, there is—sometimes quite extensive—existing scholarship; some references are given in the bibliography.

## Annotated Bibliography

*For a discussion of the personal and other background history of this work, see:*

S. Wolfram (2023), "A 50-Year Quest: My Personal Journey with the Second Law of Thermodynamics". writings.stephenwolfram.com/2023/02/a-50-year-quest-my-personal-journey-with-the-second-law-of-thermodynamics.

*Pointers to specific references are included as hyperlinks in the online version of this piece.*

### Development of the Approach Described Here

*The concept of computational irreducibility was described in:*

S. Wolfram (1985), "Undecidability and Intractability in Theoretical Physics", *Physical Review Letters* 54, 735–738. content.wolfram.com/undecidability-intractability-theoretical-physics.pdf.

*An early description of the computational character of the Second Law was given in:*

S. Wolfram (1985), "Origins of Randomness in Physical Systems", *Physical Review Letters* 55, 449–452. content.wolfram.com/origins-randomness-physical-systems.pdf.

*Further development was done in:*

S. Wolfram (2002), "Irreversibility and the Second Law of Thermodynamics", in *A New Kind of Science*, Wolfram Media, 441–457. wolframscience.com/nks/chap-9--fundamental-physics/#sect-9-3--irreversibility-and-the-second-law-of-thermodynamics.

*The Wolfram Physics Project is described in:*

S. Wolfram (2020), "A Class of Models with the Potential to Represent Fundamental Physics". arXiv:2004.08210.

*The "particle cellular automaton" used here was introduced in:*

S. Wolfram (1986), "Minimal Cellular Automaton Approximations to Continuum Systems", presented at *Cellular Automata '86;* reprinted in *Cellular Automata and Complexity: Collected Papers* (2019)*,* Addison-Wesley. content.wolfram.com/cellular-automaton-continuum-systems.pdf.

# Other Works on the Second Law

## *Classic Original Sources*

S. Carnot (1824), *Réflexions sur la puissance motrice du feu et sur les machines propres à développer cette puissance* (in French), Bachelier. ark:/13960/t7rn68p52. (Translated by R. Thurston (1890), as *Reflections on the Motive Power of Heat*, reprinted in *Reflections on the Motive Power of Fire* (1988), Dover. ark:/13960/t0jv2661s.)

R. Clausius (1857), "Über die Art der Bewegung die wir Wärme nennen" (in German), *Annalen der Physik* 100, 353–380. ark:/13960/t9679978p. (Translated as "On The Nature of the Motion which we call Heat", *Philosophical Magazine* 14, 108–127 (1857). ark:/13960/t3jw8w07q.

J. Maxwell (1858), "Illustrations of the Dynamical Theory of Gases", *Philosophical Magazine* 19, 19–32. ark:/13960/t4sj20460.

J. Maxwell (1866), "On the Dynamical Theory of Gases", *Philosophical Transactions of the Royal Society of London* 157, 49–88. ark:/13960/t06x5c78d.

L. Boltzmann (1872), "Weitere Studien über das Wärmegleichgewicht unter Gasmolekülen" (in German), *Sitzungsberichte Akademie der Wissenschaften* 66, 275–370. ark:/13960/t4pk0sf66. (Translated as "Further Studies on the Thermal Equilibrium of Gas Molecules", in *The Kinetic Theory of Gases: An Anthology of Classic Papers with Historical Commentary* (2003), S. Brush (ed.)*,* Imperial College Press, 262–349. doi: 10.1142/p281.)

W. Thomson (1874), "The Kinetic Theory of the Dissipation of Energy", *Proceedings of the Royal Society of Edinburgh* 8, 325–334. ark:/13960/t2b85fc5t.

L. Boltzmann (1877), "Über die Beziehung zwischen dem zweiten Hauptsatze der mechanischen Wärmetheorie und der Wahrscheinlichkeitsrechnung respective den Sätzen über das Wärmegleichgewicht" (in German), *Sitzungberichte Akademie der Wissenschaften* 76, 373–435. doi: 10.1017/CBO9781139381437.011. (Translated as "On the Relationship between the Second Fundamental Theorem of the Mechanical Theory of Heat and Probability Calculations Regarding the Conditions for Thermal Equilibrium", *Entropy* 17, 1971–2009 (2015). doi: 10.3390/e17041971.)

J. Gibbs (1902), *Elementary Principles in Statistical Mechanics*, Charles Scribner's Sons. ark:/13960/t6rz5sz8r.

## Notable Collections, etc.

R. Clausius (1864), *Abhandlungen über die mechanische Wärmetheorie* (in German), Friedrich Viewweg und Sohn. ark:/13960/t53f4qn28. (Translated by W. Browne (1879), as *The mechanical theory of heat*, Macmillan and Co. ark:/13960/t1wd4fr9c.)

W. Thomson (1882–1911), *Mathematical and physical papers* vols. 1–6, Cambridge University Press. ark:/13960/t20d2r137.

A. Tuckerman (1890), “Index to the literature of thermodynamics”, *Smithsonian Miscellaneous Collections* 34, iii–239. repository.si.edu/handle/10088/23167.

W. Niven (ed.) (1890), *The Scientific Papers of James Clerk Maxwell* vols. 1 & 2, Dover. ark:/13960/t3pv6kt6d.

F. Hasenöhrl (ed.) (1909), *Wissenschaftliche Abhandlungen von Ludwig Boltzmann* (in German) [*Scientific Works of Ludwig Boltzmann*] vols. 1–3, Johann Ambrosius Barth. (Available on HathiTrust.)

F. Donnan and A. Haas (eds.) (1936), *A Commentary on the Scientific Writings of J. Willard Gibbs* vols. 1 & 2, Yale University Press. ark:/13960/t6sx6c24j.

J. Stachel, D. Cassidy, J. Renn and R. Schulmann (eds.) (1990), *The Collected Papers of Albert Einstein* vol. 2, Princeton University Press. einsteinpapers.press.princeton.edu/vol2-doc.

H. Ebbinghaus, C. Fraser and A. Kanamori (eds.) (2010), *Ernst Zermelo - Collected Works/Gesammelte Werke* vols. 1 & 2, Springer. doi: 10.1007/978-3-540-79384-7.

## Notable Textbooks

J. Maxwell (1871), *Theory of Heat,* Longmans, Green, and Co. (Available on Wikimedia.org.) (Reprinted as *Theory of Heat* (1872), D. Aplleton and Co. ark:/13960/t26976m7w.)

L. Boltzmann (1896), *Vorlesungen über Gastheorie* (in German), Johann Ambrosius Barth. ark:/13960/t18k7bb6k. (Translated by S. Brush (1995), as *Lectures on Gas Theory*, Dover. ark:/13960/t40s8nn1q.)

M. Planck (1897), *Vorlesungen über Thermodynamik* (in German), Veit & Comp. ark:/13960/t1hh6k62v. (Translated by A. Ogg (1905), as *Treatise On Thermodynamics* 3rd edition, Dover. ark:/13960/t06x4mn0f.)

J. Jeans (1904), *The Dynamical Theory of Gases*, Cambridge University Press. ark:/13960/t8pc2w43r.

P. Ehrenfest and T. Ehrenfest (1912), “Begriffliche Grundlagen der Statistischen Auffasung in der Mechanik” (in German), *Encyklopädie der mathematischen Wissenschaften* 4, part 4, 1–90. lorentz.leidenuniv.nl/IL-publications/sources/Ehrenfest_1911b.pdf. (Translated by M. Moravcsik (1959), as *The Conceptual Foundations of the Statistical Approach in Mechanics*, Cornell University Press. ark:/13960/t8jf2nx05.)

L. Landau and E. Lifshitz (1938), *Statisticheskaia fizika* (in Russian). (Translated by E. Peierls and R. Peierls (1958), as *Statistical Physics*, Pergamon Press. ark:/13960/t07x01p71.)

R. Tolman (1938), *The Principles of Statistical Mechanics*, Oxford University Press. ark:/13960/t9w11hx7r.

A. Sommerfeld (1956), *Thermodynamics and Statistical Mechanics*, Academic Press. ark:/13960/t53g3bb8j.

H. Callen (1960), *Thermodynamics: An Introduction to the Physical Theories of Equilibrium Thermostatics and Irreversible Thermodynamics*, Wiley & Sons. ark:/13960/t77t65w24.

K. Huang (1963), *Statistical Mechanics*, Wiley & Sons. ark:/13960/t14n0bb0g.

F. Reif (1965), *Statistical Physics*, McGraw Hill Company. ark:/13960/t44q8ww6q.

D. Ruelle (1969), *Statistical Mechanics: Rigorous Results*, Imperial College Press. ark:/13960/t6rz72b6s.

L. Landau and E. Lifshitz (1981), *Physical Kinetics*, Pergamon Press. ark:/13960/t20d2fj56.

G. Gallavotti (1999), *Statistical Mechanics: A Short Treatise*, Springer. doi: 10.1007/978-3-662-03952-6.

## Surveys about Foundations

P. Bridgman (1941), *The Nature of Thermodynamics*, Harvard University Press. ark:/13960/t7vm43w9f.

H. Reichenbach (1956), *The Direction of Time*, University of California Press. ark:/13960/t8sc1wj45.

P. Landsberg (1956), “Foundations of Thermodynamics”, *Reviews of Modern Physics* 28, 363–392. doi: 10.1103/RevModPhys.28.363.

E. Cohen (1962), *Fundamental Problems in Statistical Mechanics*, North-Holland.

E. Jaynes (1965), “Gibbs vs Boltzmann Entropies”, *American Journal of Physics* 33, 391–398. doi: 10.1119/1.1971557.

O. Penrose (1970), *Foundations of Statistical Mechanics: A Deductive Treatment*, Dover.

P. Davies (1977), *The Physics of Time Asymmetry*, University of California Press. ark:/13960/t5n99zw8p.

C. Bennett (1987), “Demons, Engines and the Second Law”, *Scientific American* 257, 108–117. doi: 10.1038/SCIENTIFICAMERICAN1187-108.

M. Mackey (1992), *Time’s Arrow: The Origins of Thermodynamic Behavior*, Springer. doi: 10.1007/978-1-4613-9524-9.

J. Lebowitz (1993), “Macroscopic laws, microscopic dynamics, time’s arrow and Boltzmann’s entropy”, *Physica A* 194, 1–27. doi: 10.1016/0378-4371(93)90336-3.

J. Halliwell (1996), *Physical Origins of Time Asymmetry*, Cambridge University Press.

L. Schulman (1997), *Time’s Arrows and Quantum Measurement*, Cambridge University Press. doi: 10.1017/CBO9780511622878.

R. Peierls (1997), “Time reversal and the second law of thermodynamics”, in *Selected Scientific Papers of Sir Rudolf Peierls*, R. Dalitz and R. Peierls (eds.), World Scientific, 563–570. doi: 10.1142/9789812795779_ 0055.

J. Uffink (2001), “Bluff Your Way in the Second Law of Thermodynamics”, *Studies in History and Philosophy of Science Part B: Studies in History and Philosophy of Modern Physics* 32, 305–394. arXiv:cond-mat/0005327.

J. Uffink (2007), “Compendium of the foundations of classical statistical physics”, in *Philosophy of Physics, Handbook of the Philosophy of Science*, J. Butterfield pdf and J. Earman (eds.), Elsevier, 923–1074.

W. Grandy Jr. (2008), *Entropy and the Time Evolution of Macroscopic Systems*, Oxford University Press.

J. Parrondo, J. Horowitz and T. Sagawa (2015), “Thermodynamics of information”, *Nature Physics* 11, 131–139. doi: 10.1038/nphys3230.

D. Lairez (2022), “What Entropy Really Is: the Contribution of Information Theory”. arXiv:2204.05747.

## Historical Surveys & Collections

J. Kestin (ed.) (1976), *The Second Law of Thermodynamics*, Cambridge University Press.

S. Brush (1976), *The Kind of Motion We Call Heat: A History of the Kinetic Theory of Gases in the Nineteenth Century* vols. 1 & 2, North-Holland.

J. Kestin (ed.) (1977), *The Second Law of Thermodynamics*, Dowden, Hutchinson and Ross.

C. Truesdell (1980), *The Tragicomical History of Thermodynamics, 1822–1854*, Springer. archive.org/details/tragicomicalhist0000unse.

E. Jaynes (1984), *The Evolution of Carnot’s Principle*, Springer. doi: 10.1007/978-94-009-3049-0_ 15.

J. Plato (1991), “Boltzmann’s Ergodic Hypothesis”, *Archive for History of Exact Sciences* 42, 71–89. doi: 10.1007/BF00384333.

S. Brush (ed.) (2003), *The Kinetic Theory of Gases: An Anthology of Classic Papers with Historical Commentary*, Imperial College Press. doi: 10.1142/p281.

I. Müller (2006), *A History of Thermodynamics: The Doctrine of Energy and Entropy*, Springer.

SklogWiki (2007–). sklogwiki.org (website).

H. Leff and Andrew Rex (eds.) (2016), *Maxwell’s Demon: Entropy, Information, Computing*. Princeton University Press.

O. Darrigol (2018), *Atoms, Mechanics, and Probability: Ludwig Boltzmann’s Statistico-Mechanical Writings – An Exegesis*, Oxford University Press.

## *Less Technical Presentations*

J. Maxwell (1878), "Atom", "Attraction", "Constitution of bodies", "Diagrams", "Ether", "Molecule", *Encyclopedia Britannica* 9th edition. ark:/13960/s22210225hh. (Reprinted in *The Scientific Papers of James Clerk Maxwell* vol. 2 (1890). ark:/13960/t56f4vm7t.)

H. Helmholtz (1885), *Popular Lectures on Scientific Subjects*, Appleton and Company. urn:oclc:record:669327403.

L. Boltzmann (1895), *Theoretical Physics and Philosophical Problems: Selected Writings*, D. Reidel Publishing Company.

E. Schrödinger (1945), *What is Life? The Physical Aspect of the Living Cell*, The Macmillan Company. ark:/13960/t4qk7tk6c.

H. Bent (1965), *The Second Law: An Introduction to Classical and Statistical Thermodynamics*, Oxford University Press. ark:/13960/t1gj7793z.

P. Atkins (1984), *The Second Law*, Freeman and Company. ark:/13960/t2h78530v.

S. Berry (2019), *Three Laws of Nature: A Little Book on Thermodynamics*, Yale University Press.